\documentclass[prb,noeprint,aps,nofootinbib,amssymb,twocolumn,superscriptaddress,10pt]{revtex4-2}
\usepackage[T1]{fontenc}
\usepackage{graphicx}
\usepackage[dvipsnames,table]{xcolor}
\usepackage{dcolumn}
\usepackage{comment}
\usepackage{amsmath,amssymb,bbold,bm}
\usepackage{hyperref}
\usepackage{amsfonts}
\usepackage{float,latexsym}
\usepackage{multirow}
\usepackage{color}
\usepackage[normalem]{ulem}
\usepackage{cleveref}
\usepackage{physics}
\usepackage{tabularx}
\usepackage{array}
\usepackage[caption=false]{subfig}
\usepackage{siunitx}
\usepackage{gensymb}
\usepackage{pifont}
\usepackage{placeins}
\usepackage{mathtools}
\usepackage{longtable}
\usepackage{multirow}
\usepackage{nicematrix}
\usepackage{tikz}
\usepackage{xstring}
\usepackage{makecell}
\usepackage{ifthen}
\usepackage{combelow}
\usetikzlibrary{decorations.markings}
\usetikzlibrary{positioning}
\usetikzlibrary{arrows.meta}
\usepackage{tikz-feynman}
\usetikzlibrary{calc}
\tikzfeynmanset{warn luatex=false}
\let\vec\mathbf

\usepackage{chemformula}
\usepackage{centernot}

\makeatletter
\def\maketag@@@#1{\hbox{\m@th\normalfont\normalsize#1}}
\makeatother

\DeclarePairedDelimiterX\set[1]\lbrace\rbrace{#1}

\crefname{appendix}{Appendix}{Appendices}
\crefname{equation}{Eq.}{Eqs.}
\crefname{figure}{Fig.}{Figs.}
\crefname{table}{Table}{Tables}
\crefname{section}{Section}{Sections}
\crefname{enumi}{Point}{Points}
\creflabelformat{appendix}{[#2#1#3]}
\crefmultiformat{figure}{Figs.~#2#1\xdef\mycreffirstarg{#1}#3}%
{ and~#2#1#3}{, #2#1#3}{ and~#2#1#3}

\makeatletter     
\renewcommand\onecolumngrid{
	\do@columngrid{one}{\@ne}%
	\def\set@footnotewidth{\onecolumngrid}
	\def\footnoterule{\kern-6pt\hrule width 1.5in\kern6pt}%
}

\makeatletter
\AtBeginDocument{\let\LS@rot\@undefined}
\makeatother

\usepackage{xr}
\makeatletter
\newcommand*{\addFileDependency}[1]{
	\typeout{(#1)}
	\@addtofilelist{#1}
	\IfFileExists{#1}{}{\typeout{No file #1.}}
}
\makeatother

\crefname{appendix}{Appendix}{Appendices}
\crefname{equation}{Eq.}{Eqs.}
\crefname{figure}{Fig.}{Figs.}
\crefname{table}{Table}{Tables}
\crefname{section}{Section}{Sections}
\creflabelformat{appendix}{[#2#1#3]}

\makeatletter
\AtBeginDocument{\let\LS@rot\@undefined}
\makeatother

\makeatletter
\renewcommand\onecolumngrid{\do@columngrid{one}{\@ne}\def\set@footnotewidth{\onecolumngrid}\def\footnoterule{\kern-6pt\hrule width 1.5in\kern6pt}}

\newcommand{\siSection}{appendix}

\allowdisplaybreaks

\begin{document}
\title{Mixed-dimensional quantum Monte Carlo studies of M-point moiré materials}
\author{Dumitru~C\u{a}lug\u{a}ru}
\affiliation{Rudolf Peierls Centre for Theoretical Physics, University of Oxford, Oxford OX1 3PU, United Kingdom}
	\author{Konstantinos~Vasiliou}
	\affiliation{Rudolf Peierls Centre for Theoretical Physics, University of Oxford, Oxford OX1 3PU, United Kingdom}
	\author{Haoyu~Hu}
	\affiliation{Department of Physics, Princeton University, Princeton, New Jersey 08544, USA}
    \affiliation{Department of Physics, University of Science and Technology of China, Hefei, Anhui 230026, China}
	\author{B.~Andrei~Bernevig}
	\affiliation{Department of Physics, Princeton University, Princeton, New Jersey 08544, USA}
	\affiliation{Donostia International Physics Center, P. Manuel de Lardizabal 4, 20018 Donostia-San Sebastián, Spain}
	\affiliation{IKERBASQUE, Basque Foundation for Science, Bilbao, Spain}
	\author{Werner~Krauth}
	\affiliation{Laboratoire de Physique de l’Ecole normale	sup\'erieure, ENS, Universit\'e PSL, CNRS, Sorbonne Universit\'e, Universit\'e de Paris Cit\'e, Paris, France}
	\affiliation{Rudolf Peierls Centre for Theoretical Physics, University of Oxford, Oxford OX1 3PU, United Kingdom}
	\author{S.~A.~Parameswaran}
	\email{sid.parameswaran@physics.ox.ac.uk}
	\affiliation{Rudolf Peierls Centre for Theoretical Physics, University of Oxford, Oxford OX1 3PU, United Kingdom}

\let\oldaddcontentsline\addcontentsline

\begin{abstract}
A new moiré-material platform has recently been proposed based on twisting two-dimensional triangular-lattice monolayers whose low-energy states lie at the three M points of the Brillouin zone. Continuum models derived from extensive \textit{ab initio}{} simulations suggest that electrons in the conduction bands of one such M-point moir\'e material, twisted AA-stacked \ch{SnSe2}, realize a three-orbital Hubbard model with orbitally-selective, quasi-one-dimensional (quasi-1D) hopping, protected by a projective mirror symmetry. Here, we show that the resulting ``mixed-dimensional'' limit -- in which the hopping is exactly quasi-1D in each valley, while the valleys are coupled by interactions into a fully two-dimensional network -- can be sampled with Stochastic Series Expansion (SSE) quantum Monte Carlo (QMC) without a sign problem \emph{at any filling}. We develop an efficient new SSE QMC algorithm that combines custom global updates with parallel tempering to overcome the equilibration challenges posed by the mixed-dimensional setting. We then use this algorithm to explore the phase diagram of M-point twisted AA-stacked \ch{SnSe2}. Over extended and realistic ranges of twist angles and interaction strengths, we find that at integer fillings the system supports correlated insulators whose nature and strength depend strongly on angle. At certain commensurate fractional fillings, we further find evidence for Wigner-Mott insulators. We analytically account for the main features observed numerically using a strong-coupling description. Finally, we discuss perturbations away from the mixed-dimensional limit and the possibility of applying our method to other realizations of mixed-dimensional Hubbard models.
\end{abstract}
\maketitle

\section{Introduction}\label{sec:intro}

Experimental and theoretical developments in quantum condensed matter physics frequently go hand-in-hand: emerging material platforms stimulate new theoretical ideas, which in turn guide deeper experimental investigations. Historical examples include the formulation of the renormalization group to tackle thorny questions raised by the Kondo effect and critical phenomena~\cite{KON64,WIL72,WIL74,WIL75a}, and the introduction of topological ideas to understand the remarkable robustness at the heart of the integer and fractional quantum Hall effect~\cite{THO82,TSU82a,LAU83a}. Often, theoretical advances are rooted in identifying aspects or limits of experimentally-relevant problems that are amenable to systematic and controlled numerical investigation: for example, the local approximation inherent in dynamical mean-field theory~\cite{MET89a,GEO92a,GEO96}, or the one-dimensionality that facilitates efficient density-matrix renormalization group calculations~\cite{WHI92,SCH11a}. An ideal scenario involves three ingredients whose confluence is rare: an experimental system, a tractable limit thereof, and an algorithm designed to exploit the latter. 

This work identifies and explores a new instance of such a confluence: we show that the newly proposed class of M-point moir\'e materials~\cite{CAL25b,MAH24,JIA24b} admits a realistic limit that can be \emph{exactly} studied using sign-problem-free quantum Monte Carlo (QMC) simulations. The materials in question are built by stacking two-dimensional (2D) atomically thin layers with a small relative twist. Although in isolation each layer hosts three species of low-energy electrons each dispersing in 2D, the electronic structure is strongly reconstructed in the moiré setting. In many cases, the resulting low-energy physics is closely approximated by a limit wherein each species of electron is forced to disperse only along a single direction along one-dimensional ``chains''. The different species propagate in distinct directions and are coupled by interactions into a fully 2D network. Remarkably, this seemingly artificial limit is nevertheless \emph{natural} in that it is controlled by an emergent approximate symmetry so that deviations from it remain small even in experimentally-relevant parameter regimes. We show that this special structure can be exploited so as to eliminate the sign problem from Markov-chain Monte Carlo sampling via the stochastic series expansion (SSE) method~\cite{SAN99,SEN02,SAN19}. In doing so, we also introduce a significant methodological innovation: we devise a new global update scheme for SSE that facilitates inter-chain charge transfer processes, which we show are essential in order to achieve equilibration and hence accurate sampling in the mixed-dimensional setting.

The goals of this paper are therefore twofold. On the one hand, we numerically elucidate and analytically characterize the weak-, intermediate-, and strong-coupling physics of a specific member of the M-point moir\'e material family in the mixed-dimensional limit, at any filling, while also validating earlier strong-coupling ideas~\cite{LI25} relevant to integer filling of each moir\'e unit cell. On the other hand, we provide a new, experimentally realistic, example of a fermionic system that is free of a sign problem at any electron density. The accompanying SSE framework, including the inter-chain update we devise, turns this sign-problem-free formulation into a practical simulation method and may be useful more broadly for mixed-dimensional Hubbard-like systems.

Before proceeding to specifics, we orient our work within the active and evolving moir\'e setting. Moir\'e materials are engineered by stacking either 2D materials with a small relative twist, or by simply layering materials with a slight lattice mismatch~\cite{AND21,KEN21,NUC24}. As a result, the translation symmetry of the individual monolayers is reduced to a longer-range emergent \emph{moir\'e} translation symmetry. The low-energy states of the constituent monolayers are strongly reorganized by interlayer tunneling, endowing the resulting heterostructure with properties that can differ dramatically from those of the individual layers. Compared to conventional crystals, the main advantage of moir\'e materials is their tunability: the twist angle, as well as applied electric and magnetic fields, can strongly reshape the single-particle band structure and therefore serve as experimental tuning knobs. This makes it possible to access regimes in which interaction effects dominate and strongly correlated physics can be explored. Perhaps even more importantly, the enlarged crystalline unit cell allows one to dope the system up to densities of a few electrons per moir\'e unit cell by electrostatic gating, unlike conventional materials, which typically require chemical doping. Finally, the two-dimensional nature of these materials enables them to be studied by a range of means beyond transport~\cite{CAO18,CAO18a}: surface scanning probes that interrogate samples through tunneling~\cite{KER19,XIE19,CHO19} or capacitive coupling~\cite{TOM19,ZON20}, photoemission~\cite{LIS21,CHE24b}, optical spectroscopy~\cite{SEY19,TRA19,JIN19}, and newer momentum-resolved techniques that implement \textit{in situ} twisting~\cite{INB23,XIA25}. This has enabled experiments to map out comprehensive phase diagrams and explore the rich physics of these systems at a remarkable pace.

Despite being assembled from a relatively limited set of building blocks -- mostly triangular monolayers whose low-energy states lie at the K point of the Brillouin zone, as realized experimentally with graphene and certain transition metal dichalcogenides (TMDs), or at the $\Gamma$ point, as proposed theoretically~\cite{ANG21,CLA22a,PI26} -- moir\'e materials have proven to be remarkably versatile. Combined with their tunability, this has established moir\'e systems as veritable analog quantum simulators of prototypical condensed-matter models~\cite{KEN21}, some of which have so far found material realization only in the moir\'e setting. Twisted bilayer graphene, predicted to host flat bands~\cite{BIS11,SUA10,LOP07} at the so-called magic angle, has realized a plethora of correlated phases, most notably correlated insulators~\cite{CAO18} at integer fillings with superconductivity emerging in between~\cite{CAO18a}. From a theoretical perspective, moir\'e graphene has also been mapped to a topological heavy-fermion model~\cite{SON22}. In multilayer settings, heterostructures based on rhombohedral graphene multilayers and hexagonal boron nitride have gone on to realize fractional Chern insulators and potentially chiral superconductivity~\cite{LU24,DON23,DON24,HER24b,HAN25}. TMD heterobilayers can emulate the Hubbard model on a triangular lattice~\cite{WU18c,TAN20,XU20a,LI21e}. Twisted \ch{WTe2} exhibits signatures of a one-dimensional Luttinger liquid~\cite{WAN22}, although its theoretical description remains challenging because of the complex monolayer band structure. Beyond these examples, a growing body of theoretical and experimental work has explored other exotic phases in moir\'e TMDs, including Kondo lattice physics~\cite{ZHA23c}, superconductivity~\cite{XIA24a,GUO24,XIA26}, and both integer and fractional Chern insulator states~\cite{WU19b,DEV21,ZEN23,WAN24a,YU24a,JIA24,ZHA24a,SHE24}.

Recently, the class of moir\'e materials was expanded by the introduction of a new platform based on triangular monolayers whose low-energy states lie at the M point of the Brillouin zone~\cite{CAL25b,MAH24,JIA24b}, followed by additional material proposals~\cite{ING25,BAO25}. M-point moir\'e materials differ from their K- or $\Gamma$-point counterparts in several ways already apparent in preliminary studies and toy models~\cite{KAR19,FUJ22,KAR23}. One such difference is the presence of three valleys with intra-valley time-reversal symmetry, which prevents the existence of single-particle Chern bands in each valley. A more subtle distinction is the emergence of momentum-space nonsymmorphic, or projective, symmetries~\cite{CAL25b}. Beyond their conceptual significance as the first realization of projective symmetries in a crystalline system without a magnetic field, these symmetries have direct consequences for the interacting physics of certain M-point moir\'e materials. In twisted AA-stacked \ch{SnSe2} (AA t-\ch{SnSe2}{}), the emergent projective in-plane mirror symmetry $\tilde{M}_z$, which survives lattice relaxation effects, renders the corresponding single-particle Hamiltonian quasi-one-dimensional (quasi-1D) within each valley, while interactions couple the system into a fully 2D network~\cite{CAL25b,LI25}. Upon projection into the first moir\'e conduction band, the leading kinetic term yields one-dimensional electron motion along electrostatically coupled ``chains'' within each valley, in a picture reminiscent of a sliding Luttinger liquid~\cite{EME00,VIS01,MUK01a}. Taken together across all valleys, the three copies resemble a \SI{120}{\degree} version of a crossed sliding Luttinger liquid~\cite{MUK01}. The corresponding interacting Hamiltonian was shown to possess exact analytical ground states at integer electron fillings per moir\'e unit cell in the strong-coupling limit~\cite{LI25}, supplemented by Hartree-Fock calculations~\cite{LI25,BEU25}. As in other strongly interacting condensed-matter systems, understanding the physics beyond these idealized limits -- {\it i.e.}{}, in the intermediate-coupling regime and away from integer fillings -- remains an essential open question. Beyond its direct relevance to forthcoming experiments on this new platform, addressing this challenge may offer insight into a broader class of mixed-dimensional quantum many-body systems, where predominantly one-dimensional electronic motion competes with genuinely two-dimensional electron correlations.
\begin{figure}[!t]
	\begin{tikzpicture}[x=1.15cm,y=1.0cm,>=stealth]
\draw[->,line width=0.9pt] (0,0) -- (6.5,0);
		\draw[->,line width=0.9pt] (0,0) -- (0,3.25);
		
\node[left] at (0,2.05) {$t_{\perp}$};
		\node[below] at (3,-0.6) {$\nu$};
		
\foreach \x in {0,1,2,3,4,5,6}{
			\draw[line width=0.8pt] (\x,0.1) -- (\x,-0.1);
			\node[below] at (\x,-0.14) {\x};
  		}
		
\node[below] at (0,-0.55) {BI};
		\node[below] at (6,-0.55) {BI};
		
		\def\x{0}
		\def\y{0}
		\def\xwidth{6}
		\def\ywidth{0.15}
		\draw[line width=1.0pt, rounded corners=2pt, red, dash pattern=on 4pt off 2pt]
		(\x-\ywidth,\y-\ywidth) -- (\x+\xwidth+\ywidth,\y-\ywidth) -- (\x+\xwidth+\ywidth,\y+\ywidth) -- (\x-\ywidth,\y+\ywidth) --	 cycle;
		
\draw[blue,line width=1.0pt] (3,0) -- (3,2.75);
		\def\x{3}
		\def\y{0}
		\def\xwidth{0.15}
		\def\ywidth{2.75}
		\draw[line width=1.0pt, rounded corners=2pt, blue, dash pattern=on 4pt off 2pt]
		(\x-\xwidth,\y-\xwidth) -- (\x-\xwidth,\y+\ywidth+\xwidth) -- (\x+\xwidth,\y+\ywidth+\xwidth) -- (\x+\xwidth,\y-\xwidth) --	 cycle;
		
\node[blue] at (3,3.1) {half-filling};
		
\node[blue,align=center] (A) at (1.4,1.75)
		{\textbf{DQMC}\\(companion work~\cite{VAS26})};
		\draw[blue,line width=0.9pt,->]
		(A.south) .. controls (1.5,1) and (2,1) .. (2.85,1);

\node[red,align=center,inner sep=1pt,outer sep=0pt] (B) at (5,1.5)
		{\textbf{SSE QMC}\\(this work)};
		
\draw[red,line width=0.9pt,->]
		(B.south) .. controls (5,0.75) and (3.5,0.75) .. (3.5,0.15);

\coordinate (C) at (3,0);
		\draw[blue,line width=0.9pt,<->]
		($(C)+(220:1)$) arc[start angle=220,end angle=320,radius=1];
		\node[blue,align=center] at (3,-1.5) {particle-hole\\symmetry};	
	\end{tikzpicture}
	\caption{Parameter regions without a sign problem for the AA t-\ch{SnSe2}{} Hamiltonian. We highlight the regions in the space of next-nearest-neighbor inter-chain hopping $t_{\perp}$ and electron filling per moir\'e unit cell $\nu$ in which the AA t-\ch{SnSe2}{} Hamiltonian can be sampled within QMC without a sign problem. Here, $\nu$ denotes the filling of the six conduction-band flavors of AA t-\ch{SnSe2}{}, arising from three valleys and two spin species, so that $0 \leq \nu \leq 6$. The limits $\nu=0$ and $\nu=6$ correspond to empty and fully filled conduction bands, respectively, {\it i.e.}{} band insulators (BIs).}
	\label{fig:sign_free}
\end{figure}
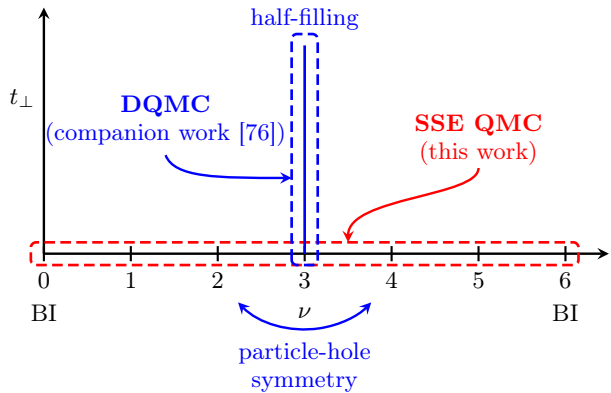

In this work and the companion paper~\cite{VAS26}, we take fundamental steps toward solving this problem by showing that, remarkably, the projective mirror symmetry $\tilde{M}_z$ renders the interacting Hamiltonian of AA t-\ch{SnSe2}{} amenable to sign-problem-free QMC sampling in two ``orthogonal'' parameter regimes, as illustrated schematically in \cref{fig:sign_free}. In Ref.~\cite{VAS26}, we show that, unusually for a triangular-lattice model, at half-filling of the first moir\'e conduction band -- equivalently, at a filling of $\nu = 3$ electrons per moir\'e unit cell -- the $\tilde{M}_z$ symmetry enforces a hidden bipartite structure of the system, making it amenable to determinantal QMC (DQMC)~\cite{BLA81,SCA81,WHI89,ASS08} even in the presence of subleading next-nearest-neighbor inter-chain hopping terms. Complementarily, in the present work, we show that in the ``mixed-dimensional'' limit -- valley-selective one-dimensional hopping combined with 2D Coulomb repulsion -- the AA t-\ch{SnSe2}{} Hamiltonian can be sampled without a sign problem within SSE QMC~\cite{SAN92,SEN02,SAN19} at \emph{any} filling $\nu$, whether fractional or integer.

It is worth pausing here to remind the reader that QMC is an ideal unbiased method for strongly correlated systems. For fermionic systems, however, QMC suffers from the infamous sign problem~\cite{TRO05,LI19b}. Multiple attempts have been made to circumvent this problem in certain contrived models, within DQMC~\cite{WU05,BER12,LI19b,CHR20a,WAN21h}, within SSE~\cite{XU15}, and with Majorana QMC~\cite{LI15,LI16,HAN24}. Our work establishes AA t-\ch{SnSe2}{} as a rare example, on the crystalline-material side of condensed matter, of a strongly interacting fermionic system that both admits a direct experimental realization and has a limit amenable to unbiased numerical investigation using sign-problem-free QMC. 

However, we anticipate that the mixed-dimensional Monte Carlo method we develop will have wider applicability to a range of other mixed-dimensional Hubbard-like systems (in appropriate limits), including spin-orbital models of correlated oxides~\cite{KUG82,KHA05,STR17}, compass models~\cite{NUS15}, cold-atom realizations~\cite{GRU18,GRU20,SCH23,SCH24}, and related quantum simulator architectures.

The rest of the paper is organized as follows. In \cref{sec:model}, we review the model of AA t-\ch{SnSe2}{} in the mixed-dimensional limit, including a discussion of the subleading terms that we omit to reach the sign-free limit. We then review the SSE method, placing particular emphasis on the features specific to its implementation in the present system and the inter-chain updates that are required for equilibration. The results are presented and interpreted in \cref{sec:results}. Finally, \cref{sec:conclusions} summarizes our conclusions and discusses our results, the leading perturbing effects, and possible future directions. The main text is supplemented by comprehensive appendices, which make the paper both pedagogical and self-contained.

\section{Model}\label{sec:model}

In this work, we focus on the mixed-dimensional limit of twisted AA-stacked \ch{SnSe2} (AA t-\ch{SnSe2}{})~\cite{CAL25b,LI25}. This system features a triangular lattice generated by the moir\'e lattice vectors $\vec{a}_{M_1}$ and $\vec{a}_{M_2}$, as shown in \cref{fig:model}~\cite{CAL25b}. The low-energy Wannier model of its first moir\'e conduction band contains three spinful orbitals per unit cell, each associated with one of the three M-point valleys of the system, indexed by $\eta \in \left\{ 0, 1, 2\right\}$~\cite{LI25,VAS26}. These three orbitals are located \emph{almost} on top of one another within each unit cell, but are slightly displaced from the unit cell origin, since within a given valley no symmetry pins the orbital center to the origin.

In addition to emergent spin $\mathrm{SU} \left( {2} \right)$, valley-$\mathrm{U} \left( {1} \right)$, spinless time-reversal $\mathcal{T}$, and moir\'e translation symmetries, the system also possesses $C_{3z}$ and $C_{2x}$ symmetries, with $C_{3z}$ permuting the three orbitals within each unit cell. We denote by $\hat{d}^\dagger_{\vec{R},\eta,s}$ the creation operator for an electron of spin $s$ in unit cell $\vec{R}$ and valley $\eta$. The action of the $C_{3z}$ and $C_{2x}$ symmetries is given by $C_{3z} \hat{d}^\dagger_{\vec{R},\eta,s} C^{-1}_{3z} = \hat{d}^\dagger_{C_{3z} \vec{R},\eta+1,s}$ and $C_{2x} \hat{d}^\dagger_{\vec{R},\eta,s} C^{-1}_{2x} = \hat{d}^\dagger_{C_{2x} \vec{R},-\eta,s}$, respectively. A detailed review of the Wannier-model construction is provided in \cref{app:sec:model}.

\subsection{Single-particle Hamiltonian}\label{sec:model:sp}

\begin{figure}[!t]
	\centering
	\includegraphics[width=\columnwidth]{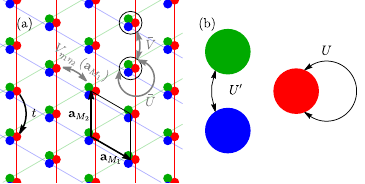}\subfloat{\label{fig:model:a}}\subfloat{\label{fig:model:b}}\caption{Tight-binding model of the first conduction band of AA t-\ch{SnSe2}{}. The lattice and the Wannier orbitals are shown in (a). The black parallelogram marks the conventional unit cell and the black arrows the moir\'e lattice vectors. Orbitals are shown as disks, with valley index $\eta$ encoded by color: red, green, and blue correspond to $\eta=0,1,2$, respectively. The leading quasi-1D hopping amplitude $t$ is shown for valley $\eta=0$, and the colored lines indicate the chains along which electrons tunnel within each valley. The interaction is fully 2D and of density-density form, with onsite three-orbital Hubbard interactions and a nearest-neighbor repulsion shown in gray. As a coarse characterization of its overall strength, we use the average onsite and nearest-neighbor Coulomb repulsions $\bar{U}$ and $\bar{V}$, respectively. (b) shows the orbitals within a single unit cell, illustrating the onsite intra-valley ($U$) and inter-valley ($U' \lesssim U$) Hubbard interactions.}
	\label{fig:model}
\end{figure}

\newcommand{\vE}{\vec{e}}

In the mixed-dimensional limit, the first conduction band of AA t-\ch{SnSe2}{} is described by the following Wannier many-body grand-canonical Hamiltonian
\begin{equation}
	\label{eqn:m_pt_ham}
	H = H_0 + H_I.
\end{equation}
The single-particle term $H_0$ is quasi-1D within each valley and is given by
\begin{equation}
	\label{eqn:sp_ham}
	H_{0} = -t\sum_{\vec{R},\eta,s} \left( \hat{d}^\dagger_{\vec{R} + \vE_\eta,\eta,s} \hat{d}_{\vec{R},\eta,s} + \text{h.c.} - \mu \hat{n}_{\vec{R},\eta,s} \right),
\end{equation}
where $\vE_\eta \equiv C^{\eta}_{3z} \vec{a}_{M_2}$ is the hopping direction within valley $\eta$, $t > 0$ is the leading nearest-neighbor hopping amplitude, ``$\text{h.c.}$'' denotes the Hermitian conjugate, $\mu$ is the chemical potential, and $\hat{n}_{\vec{R},\eta,s} \equiv \hat{d}^\dagger_{\vec{R},\eta,s}\hat{d}_{\vec{R},\eta,s}$ is the number operator corresponding to the $\hat{d}^\dagger_{\vec{R},\eta,s}$ fermion. The quasi-1D nature of $H_{0}$ originates from a projective in-plane mirror symmetry $\tilde{M}_z$~\cite{CAL25b}. As a result, within each valley electrons hop along one-dimensional ``chains'', as illustrated in \cref{fig:model:a}. In valley $\eta$, these chains run parallel to $\vE_\eta$.

\Cref{eqn:sp_ham} receives small corrections from next-nearest-neighbor $\tilde{M}_z$-preserving inter-chain hopping $t_{\perp}$ and from nearest-neighbor $\tilde{M}_z$-breaking inter-chain tunneling, the latter arising from in-plane lattice relaxation effects. Both corrections are at most $15-20\%$ of $t$ and are therefore dropped in our numerical simulations. Their effects are assessed at half-filling in our companion work~\cite{VAS26}. A discussion of these terms, together with their values and the full set of single-particle Hamiltonian parameters, is provided in \cref{app:sec:model}. In what follows, we choose energy units in which $t=1$.

\subsection{Interaction Hamiltonian}\label{sec:model:int}

While the single-particle term is quasi-1D, the interaction Hamiltonian $H_I$ is fully 2D and takes the form of an extended multi-orbital Hubbard repulsion featuring only density-density interactions~\cite{LI25}
\begin{equation}
	\label{eqn:int_ham}
	H_{I} = \frac12 \!\!\! \sum_{\substack{\vec{R}, \Delta \vec{R} \\ \eta_1,s_1,\eta_2,s_2}} \!\!\! V_{\eta_1 \eta_2} \left( \Delta \vec{R} \right) :\mathrel{\hat{n}_{\vec{R}, \eta_1, s_1}}: :\mathrel{\hat{n}_{\vec{R} + \Delta \vec{R}, \eta_2, s_2}}:.
\end{equation}
In \cref{eqn:int_ham}, $:\mathrel{...}:$ denotes normal ordering with respect to half filling of each orbital, {\it i.e.}{} $:\mathrel{\hat{n}_{\vec{R},\eta,s}}: \equiv \hat{n}_{\vec{R},\eta,s} - \frac12$, while $V_{\eta_1 \eta_2} \left( \Delta \vec{R} \right)$ is the interaction tensor parametrizing the interaction Hamiltonian. We adopt this convention because it makes the particle-hole symmetry of the model more transparent. Equivalently, one could normal order with respect to the empty lattice; the two conventions differ only by a chemical-potential term. The onsite part of the interaction tensor ({\it i.e.}{}, the terms with $\Delta \vec{R}=\vec{0}$ in \cref{eqn:int_ham}) is parametrized by the intra-(inter-)valley Hubbard repulsion $U \equiv V_{\eta \eta} \left( \vec{0} \right)$ ($U' \equiv V_{\eta \eta'} \left( \vec{0} \right) \lesssim U$ for $\eta \neq \eta'$), as shown in \cref{fig:model:b}. We truncate the interaction tensor by retaining only nearest-neighbor offsite terms, such that $V_{\eta \eta'} \left( \Delta \vec{R} \right) = 0$ for $\abs{\Delta \vec{R}} > \abs{\vec{a}_{M_1}}$. The nearest-neighbor interaction is then parametrized by six independent components, which are detailed in \cref{app:sec:model}.

The interaction Hamiltonian in \cref{eqn:int_ham} is obtained by projecting the dual-gated screened Coulomb interaction onto the first conduction band of AA t-\ch{SnSe2}{}~\cite{LI25}. With the sample located halfway between two metallic gates separated by the screening length $\xi$, the Coulomb interaction between two electrons in the sample separated by $\vec{r}$ is given by
\begin{equation}
	\label{eqn:dual_gate_interaction}
	V \left( \vec{r} \right) = \frac{e^2}{4 \pi \epsilon \epsilon_0} \sum_{n=-\infty}^{\infty} \frac{\left( -1 \right)^n}{\sqrt{r^2 + (n \xi)^2}},
\end{equation}
where $\epsilon$ is the (effective) dielectric constant of the setup, $e$ is the electronic charge, and $\epsilon_0$ is the vacuum permitivity. Upon projection into the Wannier orbital basis, the leading contribution to the interaction Hamiltonian is given by \cref{eqn:int_ham}, as we review in \cref{app:sec:model}. Beyond \cref{eqn:int_ham}, additional valley- and spin-conserving density-pair and pair-pair interaction terms arise from the finite extent of the Wannier orbitals of AA t-\ch{SnSe2}{}, but these terms are much smaller (around $1\%$ of $\bar{U}$)~\cite{LI25} and are therefore neglected in this work~\cite{LI25}. 

In addition to the valley- and spin-conserving contributions discussed above, a realistic Hamiltonian also contains interaction terms that break valley-spin conservation. These terms are significantly smaller than the former, and we derive and discuss them in detail in \cref{app:sec:model}. For instance, a ferromagnetic inter-valley Hund's coupling (a few percent of $\bar{U}$) originates from the Coulomb interaction and involves particles scattering between valleys rather than within a valley. Such processes are suppressed by a factor $\frac{V \left( \vec{b}_{M_1} \right)}{V \left( \vec{b}_1 \right)}$, where $\vec{b}_{M_1}$ ($\vec{b}_1$) is the moir\'e (monolayer) reciprocal lattice vector and $V\left( \vec{q} \right)$ is the Fourier transform of the dual-gate Coulomb repulsion from \cref{eqn:dual_gate_interaction}. Conversely, this Hund's coupling can be (partially) offset by an anti-ferromagnetic inter-valley anti-Hund's coupling arising, in principle, from phonon-mediated electron-electron interactions. Although electron-phonon effects are generically expected to be small, they may be relevant in \ch{SnSe2}, as suggested by \textit{ab initio}{} studies~\cite{KUD20,KAF20,FAN26} and by the superconducting~\cite{ZEN18} and pressure-induced charge-density-wave~\cite{YIN18} phases of the monolayer. In the present work, we therefore neglect both the Hund's and anti-Hund's contributions in $H_I$: they are individually small, difficult to estimate reliably, and act antagonistically. We note that a similar situation has been discussed in twisted bilayer graphene, where analogous Hund's and anti-Hund's terms are also believed to be of comparable magnitude and opposite sign, making their net effect difficult to assess quantitatively~\cite{KWA23,WAN24,LIU24e,YOU24,WAN25,WAN25a}. A careful investigation of their effects is left for future work.

\subsection{Model parameters}\label{sec:model:param}

Because the three orbitals in each unit cell are located almost at the same Wyckoff position within the unit cell, the intra-valley Hubbard repulsion $U$ is only slightly larger than the inter-valley repulsion $U'$. The interaction Hamiltonian $H_I$ is therefore close to a $\mathrm{U} \left( {6} \right)$-symmetric limit. For this reason, we introduce the average onsite $\bar{U}=\frac13 \left( U + 2 U' \right)$ and offsite 
\begin{equation}
	\bar{V} = \frac{1}{54} \sum_{\eta_1,\eta_2} \sum_{\substack{\Delta \vec{R} \\ \abs{\Delta \vec{R}} = \abs{\vec{a}_{M_1}}}} V_{\eta \eta'} \left( \Delta \vec{R} \right)
\end{equation} 
Coulomb repulsion parameters to characterize $H_I$ in realistic AA t-\ch{SnSe2}{} devices, where ``onsite'' refers to the full three-orbital moir\'e unit cell. The small relative displacement of the three orbitals reduces this symmetry to its valley-spin $\mathrm{U} \left( {2} \right) \otimes \mathrm{U} \left( {2} \right) \otimes \mathrm{U} \left( {2} \right)$ subgroup. The resulting breaking of the $\mathrm{U} \left( {6} \right)$ symmetry in the offsite interaction favors valley polarization at integer fillings~\cite{LI25}. Accordingly, rather than truncating the interaction directly, we choose the parameters of the nearest-neighbor truncated model so that the mean-field critical temperatures for valley polarization at integer fillings match those of the full interaction, as detailed in \cref{app:sec:model}.

\begin{figure}[!t]
	\centering
	\includegraphics[width=\columnwidth]{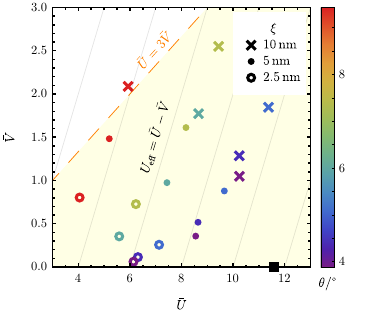}
	\caption{Parameters of the AA t-\ch{SnSe2}{} Hamiltonian in \cref{eqn:m_pt_ham}. The plot shows the average onsite ($\bar{U}$) and nearest-neighbor ($\bar{V}$) Coulomb repulsion parameters for the different twist angles $\theta$ and screening lengths $\xi$ studied in this work (see also \cref{app:sec:model}). The twist angle is encoded by color, while the marker type distinguishes the different screening lengths $\xi$. The oblique gray lines are contours of constant effective onsite Hubbard repulsion $U_{\text{eff}}$~\cite{SCH13}. The dashed orange line bounds the yellow region, which is free of a sign problem at half filling within DQMC~\cite{VAS26}. The idealized strong-coupling model with only onsite Hubbard repulsion is shown by the square black marker.}
	\label{fig:parameters}
\end{figure}

In \cref{fig:parameters}, we plot the average onsite and nearest-neighbor Coulomb repulsion as a function of the commensurate twist angles investigated in Ref.~\cite{CAL25b}, for different screening lengths $\xi$. The corresponding 18 AA t-\ch{SnSe2}{} Hamiltonians are studied numerically in this work. As the twist angle $\theta$ is decreased, the system becomes more strongly interacting ({\it i.e.}{}, $\bar{U}$ and $\bar{V}$ approximately increase with decreasing twist angle), while the ratio $\bar{V}/\bar{U}$ roughly increases with $\xi$. In addition to these 18 \emph{physical} parameter sets, we also investigate the strong-coupling regime~\cite{LI25}, which retains only the onsite Hubbard interaction with $U=12$ and either $U'=11.4$ or $U'=12$.

In obtaining the numerical values of the interaction parameters, we treat the dielectric constant of the device $\epsilon$ as a parameter, rather than fixing it to a calculated value, for three main reasons: (1) dielectric screening receives contributions beyond those of the encapsulating hexagonal boron nitride, including from remote bands, making a fully quantitative determination of $\epsilon$ difficult; (2) treating $\epsilon$ as a tunable parameter allows us to systematically scan the phase diagram of the AA t-\ch{SnSe2}{} Hamiltonian in \cref{eqn:m_pt_ham}; (3) it gives us finer control over the interaction strength in the intermediate-coupling regime, which is the main focus of this work and which is less accessible to either weak-coupling perturbative treatments or strong-coupling approaches~\cite{LI25}. In \cref{app:sec:model}, we list the assumed values of the dielectric constant $\epsilon$ together with the corresponding interaction parameters for the twist angles studied in this work. 

In addition to screening effects, which drive the system from the strong- to the intermediate-coupling regime, the nearest-neighbor Coulomb repulsion further reduces the effective onsite repulsion $U_{\text{eff}}$ of the model~\cite{SCH13,HUA14}. In this context, we define the strong-coupling regime as the range of parameters for which the system forms correlated insulators at integer fillings of the moir\'e unit cell, corresponding to sufficiently large values of $\bar{U}$. Using variational arguments, Ref.~\cite{SCH13} showed that an extended Hubbard model with onsite repulsion $\bar{U}$ and nearest-neighbor repulsion $\bar{V}$ can be approximated by a Hubbard model with purely onsite interactions and an effective repulsion $U_{\text{eff}} \approx \bar{U} - \bar{V}$. Since AA t-\ch{SnSe2}{} features a sizeable nearest-neighbor repulsion, this observation further motivates our focus on the intermediate-coupling regime, where offsite interactions remain significant and realistic AA t-\ch{SnSe2}{} devices are naturally expected to lie. Finally, the short-range nature of the gated screened Coulomb interaction ensures that, for all but one of the 18 parameter sets studied in this work, $\bar{V} \leq \frac{\bar{U}}{3}$, placing them in a regime where DQMC is also free of a sign problem at half filling~\cite{VAS26}. This is not an additional restriction imposed in our study, but simply reflects the fact that most realistic AA t-\ch{SnSe2}{} parameter sets fall into this range. By contrast, the SSE method employed here remains sign-problem-free without this constraint and at arbitrary filling, which in particular allows us to study the remaining parameter set that violates this inequality, as we explain in detail in \cref{sec:SSE}.

\subsection{Symmetries of the model}\label{sec:model:symmetries}

In addition to the crystalline symmetries of AA t-\ch{SnSe2}{}, the model from \cref{eqn:m_pt_ham} possesses a spin-charge $\mathrm{U} \left( {2} \right)$ symmetry along each one-dimensional chain. More precisely, for finite systems of size $\mathcal{N} \times \mathcal{N}$ considered here, the model exhibits a large $\mathrm{U} \left( {2} \right)^{\otimes 3 \mathcal{N}}$ symmetry. In addition, the Hamiltonian in \cref{eqn:m_pt_ham} is particle-hole symmetric (under a simultaneous flip of the chemical potential $\mu$), with the particle-hole symmetry operator defined by
\begin{equation}
	\mathcal{P} \hat{d}^\dagger_{\vec{R},\eta,s} \mathcal{P}^{-1} = (-1)^{C^{\eta}_{3z} \vec{b}_{M_2} \cdot \vec{R}} \hat{d}_{\vec{R},\eta,s}.
\end{equation}
As a result, it is sufficient to restrict ourselves to negative values of the chemical potential, $\mu \leq 0$, corresponding to fillings $0 \leq \nu \leq 3$. The filling $\nu$ ranges from 0 to 6 electrons per moir\'e unit cell.

\section{Mixed-Dimensional Monte Carlo via the Stochastic Series Expansion}\label{sec:SSE}

To study the AA t-\ch{SnSe2}{} Hamiltonian in \cref{eqn:m_pt_ham} in a numerically unbiased manner, we employ QMC using the SSE method~\cite{SAN19}, which has previously been applied to fermionic~\cite{SAN92,SEN02,XU15,MOR16a}, bosonic~\cite{DOR01,DOR02,MAJ16}, and spin systems~\cite{SAN91,SAN97a,SAN97,SAN99,DOR01,SYL02,SAN03,MEL07,GRO09}.

In this section, we briefly outline the aspects of SSE that are specific to the mixed-dimensional setting of the AA t-\ch{SnSe2}{} Hamiltonian from \cref{eqn:m_pt_ham}. Our algorithm builds on standard SSE implementations~\cite{SAN19}, but the quasi-1D structure of the model -- in particular, charge conservation along each chain -- requires additional ``inter-chain'' updates to ensure ergodicity at low temperatures. We briefly describe these updates together with the observables measured in our calculations. Technical details of the SSE formulation for the mixed-dimensional limit of AA t-\ch{SnSe2}{} are collected in \cref{app:sec:sse}, while \cref{app:sec:updates} further discusses both the conventional and new QMC updates and proves that they satisfy detailed balance. At very low temperatures ($\beta \gtrsim 12$), we also accelerate convergence by employing replica-exchange sampling within SSE~\cite{SWE86,HUK96}; its implementation is described in \cref{app:sec:sse}.

\subsection{The SSE method}\label{sec:SSE:review}

For the SSE formulation, we begin by decomposing the grand-canonical Hamiltonian in \cref{eqn:m_pt_ham} into local terms. As is standard in this method, we express it as a sum over nearest-neighbor ``bond'' operators. In this work, we focus exclusively on systems of size $\mathcal{N} \times \mathcal{N}$ with periodic boundary conditions, such that there are $g \mathcal{N}^2$ distinct nearest-neighbor bonds, where $g$ denotes half the coordination number of a triangular lattice site ($g=3$). Each bond is specified by a triplet $(\vec{R}^{b}_{1},\vec{R}^{b}_{2},\eta^{b})$ defined through $ \vec{R}^b_1 = x \vec{a}_{M_1} + y \vec{a}_{M_2}$, $\vec{R}^b_2 = \vec{R}^b_1 + \vE_m$, and $\eta^b = m$, which corresponds to the bond index $b=m\mathcal{N}^{2} + x \mathcal{N} + y + 1$, with $0\le m\le2$ and $0\le x,y \le \mathcal{N}-1$. With this convention, the Hamiltonian in \cref{eqn:m_pt_ham} can be decomposed, up to constant terms, as
\begin{equation}
	\label{eqn:SSE_hamiltonian_decomposition}
	H = - \sum_{b=1}^{g \mathcal{N}^2} \left( H_{1,b} + H_{2,b} + H_{3,b} + H_{4,b} \right),
\end{equation}
where the corresponding bond-Hamiltonian terms are
{\small \begin{align}
	&H_{1,b} = \mathcal{C} - \sum_{\eta_1, \eta_2} V_{\eta_1 \eta_2} \left( \vec{R}^b_1 - \vec{R}^b_2 \right) :\mathrel{\hat{N}_{\vec{R}^b_1, \eta_1}}: :\mathrel{\hat{N}_{\vec{R}^b_2, \eta_2}}: \nonumber \\
	&- \frac{1}{g} \sum_{i=1}^2 \left( \sum_{\eta_1,\eta_2} \frac{V_{\eta_1 \eta_2} \left( \vec{0} \right)}{2} :\mathrel{\hat{N}_{\vec{R}^b_i, \eta_1}}: :\mathrel{\hat{N}_{\vec{R}^b_i, \eta_2}}: - \mu \hat{N}_{\vec{R}^b_i} \right), \label{eqn:sse_hams_1b} \\
	&H_{2,b} = t \left( \hat{c}^\dagger_{\vec{R}^b_1,\eta^b,\uparrow}\hat{c}_{\vec{R}^b_2,\eta^b,\uparrow} + \text{h.c.} \right), \label{eqn:sse_hams_2b} \\
	&H_{3,b} = t \left( \hat{c}^\dagger_{\vec{R}^b_1,\eta^b,\downarrow}\hat{c}_{\vec{R}^b_2,\eta^b,\downarrow} + \text{h.c.} \right), \label{eqn:sse_hams_3b} \\
	&H_{4,b} = t. \label{eqn:sse_hams_4b}
\end{align}}In \cref{eqn:sse_hams_1b}, we have defined the number operator for an orbital, or equivalently a valley-site, as $\hat{N}_{\vec{R},\eta} = \sum_{s} \hat{n}_{\vec{R}, \eta, s}$, and the total number operator on a site as $\hat{N}_{\vec{R}} = \sum_{\eta} \hat{N}_{\vec{R},\eta}$. The overall negative sign in \cref{eqn:SSE_hamiltonian_decomposition} follows standard SSE conventions~\cite{SEN02}. The real constant $\mathcal{C}$ is chosen such that the operator $H_{1,b}$ has only non-negative matrix elements in the fermionic number basis ({\it i.e.}{}, the basis of electronic Fock states). Here, and in what follows, ``diagonal'' and ``off-diagonal'' always refer to this fermionic number basis. In particular, $H_{1,b}$ and $H_{4,b}$ are diagonal in that basis, whereas $H_{2,b}$ and $H_{3,b}$ are off-diagonal. For open boundary conditions, the operators $H_{2,b}$ and $H_{3,b}$ likewise have non-negative matrix elements.

As we explain in \cref{sec:SSE:review:partition}, it is precisely the fact that all operators $H_{a,b}$ have non-negative matrix elements in the fermionic number basis that renders the model sign-problem-free within SSE despite its fermionic nature. For the periodic boundary conditions employed throughout this work, one must additionally introduce a parity-dependent twist along each chain in order to preserve this property: antiperiodic (periodic) boundary conditions are imposed on chains containing an even (odd) number of particles. This is possible because, in the mixed-dimensional limit considered here, particle number is conserved separately on each chain, so particles do not hop between different chains. Intuitively, this means that we can use independent Jordan-Wigner transformations on each chain to map the fermionic problem to one of hard-core bosons, and thus despite the fermionic nature of the problem we can evade the sign problem for arbitrary $V_{\eta\eta'} \left( \Delta \vec{R} \right)$ and chemical potentials. (This intuitive argument can be adapted into a rigorous proof, but this is technically involved and is hence provided in \cref{app:sec:sse}.) By contrast, DQMC is sign-problem-free only at the particle-hole symmetric point ($\mu=0$) and only for a restricted range of density-density interactions, corresponding to the region below the dashed line in \cref{fig:parameters}. Finally, the diagonal operators $H_{4,b}$ are introduced so that they can be exchanged with the off-diagonal operators $H_{2,b}$ and $H_{3,b}$ during the SSE off-diagonal updates, as discussed below.

\subsubsection{From partition function to operator strings}\label{sec:SSE:review:partition}

The grand-canonical partition function can be expanded as a sum of strictly non-negative contributions; our goal is to sample these using SSE QMC. To that end, the expansion of $\Tr \left( e^{-\beta H} \right)$ is most conveniently constructed in the basis of electronic Fock states $\ket{\alpha} = \ket{\left \lbrace \xi_{\vec{R},\eta} \right\rbrace}$, in which $\xi_{\vec{R},\eta} \in\{ 0, \uparrow, \downarrow, \uparrow\downarrow\}$ denotes the local occupation state for each orbital $\left( \vec{R}, \eta \right)$ and $\{ \dots \}$ signifies a full set of site and valley labels. Working in this representation, we perform a high-temperature expansion of the partition function
\begin{align}
	Z=&\Tr \left( e^{-\beta H} \right)= \sum_{\alpha}\sum_{n=0}^{\infty} \frac{\beta^n}{n!}
	\mel**{\alpha}{\left( -H \right)^n}{\alpha} \nonumber \\ 
	=& \sum_{\alpha} \sum_{\tilde{S}_L} \frac{\beta^{n} \left( L-n \right)!}{L!}
	\mel**{\alpha}{\prod_{p=1}^{L} H_{\tilde{a}_{p},\tilde{b}_{p}}}{\alpha}, \label{eqn:sse_z}
\end{align}
where the chemical potential is included in $H$, and therefore also in the diagonal bond operators $H_{1,b}$. 
In the second line of \cref{eqn:sse_z}, each of the $n$ Hamiltonian operators appearing in the first line is further decomposed into bond-local terms according to \cref{eqn:SSE_hamiltonian_decomposition}. The result is a sum of infinitely many terms, each specified by a so-called ``initial state'' $\ket{\alpha}$ and a sequence $\tilde{S}_L$ of bond operators, or ``operator string'', whose form we discuss below and whose expectation value in that initial state is evaluated.

Clearly, the above expansion can be accomplished for {\it any} Hamiltonian once written in a local basis. A special feature of our problem, and others that can be efficiently studied using SSE~\cite{SAN19}, is that each term in the expansion in \cref{eqn:sse_z} is \emph{non-negative}. This allows us to importance-sample the series expansion: whence the technique draws its name.

To efficiently implement this sampling, it is convenient to manipulate the first line of \cref{eqn:sse_z} into the more convenient form of the second row. First, we impose a cutoff $L$ on the maximal number of operators in an operator sequence ({\it i.e.}{}, equivalently, a cutoff in the expansion order $n$). The error introduced by this truncation is exponentially small, provided that $L$ is chosen sufficiently large. The relevant scale is given by the typical order of the high-temperature expansion in \cref{eqn:sse_z}: from the first line, we can see that $n$ is distributed \emph{approximately} as a Poisson variable with mean $\beta \left\langle -H \right\rangle$, so that the typical term that contributes to the sum has $n\sim\beta \left\langle -H \right\rangle$. Thus, as the temperature is lowered, the typical expansion order increases linearly in $\beta$, and the required cutoff $L$ must be increased accordingly. In practice, $L$ is chosen adaptively during equilibration so that the probability of sampling terms with $n>L$ is negligible, and the omitted contributions are exponentially small in $L-n$.

After imposing the cutoff $L$ in the expansion order $n$, we make a standard book-keeping simplification: rather than working with operator strings of varying length $n$, each sequence is padded by adding $L-n$ identity operators into an operator string of fixed length $L$,
\begin{equation}
	\label{eqn:op_string}
	\tilde{S}_L=[\tilde{a},\tilde{b}]_1 [\tilde{a},\tilde{b}]_2\dots [\tilde{a},\tilde{b}]_L,
\end{equation}
with the shorthand $[\tilde{a},\tilde{b}]_p = \left[\tilde{a}_p, \tilde{b}_p \right]$. In addition to the bond operators from \crefrange{eqn:sse_hams_1b}{eqn:sse_hams_4b}, we have also included an identity operator $H_{0,0}$, corresponding to $[\tilde{a},\tilde{b}]_p=[0,0]$, such that $0 \leq \tilde{a}_p \leq 4$ in \cref{eqn:op_string}. The number $n$ is the number of non-identity operators appearing in the string, so that the expansion order is encoded directly in $\tilde{S}_L$, while the remaining $L-n$ positions are occupied by identity operators. The combinatorial factor $\frac{(L-n)!}{L!} = {L \choose n}^{-1}\frac{1}{n!}$ accounts for the different ways of placing the $L-n$ identity operators in a string of length $L$.

\begin{figure}[!t]
	\centering
	\includegraphics[width=\columnwidth]{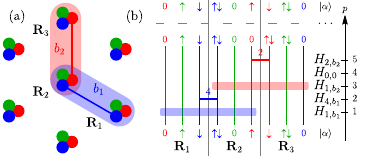}\subfloat{\label{fig:sse_config:a}}\subfloat{\label{fig:sse_config:b}}\caption{Graphical representation of an SSE configuration. (a) shows a cluster of seven sites from the lattice in \cref{fig:model:a}, with three sites labeled by $\vec{R}_1$, $\vec{R}_2$, and $\vec{R}_3$. Two bonds are highlighted, $b_1$ and $b_2$, with $\left( \vec{R}^{b_1}_1, \vec{R}^{b_1}_2, \eta^{b_1} \right) = \left( \vec{R}_2, \vec{R}_1, 2 \right)$ and $\left( \vec{R}^{b_2}_1, \vec{R}^{b_2}_2, \eta^{b_2} \right) = \left( \vec{R}_2, \vec{R}_3, 0 \right)$. The associated bond operators are shown in blue and red: $H_{1,b}$ operators as shaded ovals, and $H_{a,b}$ with $2 \leq a \leq 4$ as colored line segments. (b) shows part of a possible SSE configuration for the nine orbitals on $\vec{R}_i$ with $1 \leq i \leq 3$, with valley encoded by color. The vertical direction is the discrete propagation index $p$ encoding the order of bond operators in $\tilde{S}_L$, and the initial state $\ket{\alpha}$ is shown at the bottom. These operators are listed on the right at their corresponding values of $p$: $H_{1,b}$ appears as a 12-legged rectangle, while $H_{a,b}$ with $2 \leq a \leq 4$ appear as four-legged lines. The first five operators are shown explicitly and act on the world lines as indicated; the ellipsis denotes the remaining operators, which eventually return the propagated state to the initial one $\ket{\alpha}$.}	
	\label{fig:sse_config}\end{figure}

With these preliminaries, we can now formulate the problem of sampling the partition function by means of SSE. In SSE, one wishes to perform Monte Carlo sampling of ``configurations'', each specified by a tuple $\left( \alpha, \tilde{S}_L \right)$ consisting of an initial state $\ket{\alpha}$ and an operator sequence $\tilde{S}_L$ of fixed length $L$. These tuples are sampled with probabilities proportional to their weights in \cref{eqn:sse_z}. The fixed-length representation allows one to work with a common operator-string structure while updating diagonal, off-diagonal, and identity operators on equal footing. A useful graphical representation of such a configuration is shown in \cref{fig:sse_config}. Each orbital is ascribed a ``world line'' that tracks its occupation state as a function of the discrete SSE propagation index $p$. In this sense, $p$ plays a role analogous to a discretized imaginary-time coordinate, although it really just corresponds to the ordering index along the operator string~\cite{SAN19}. The operators in $\tilde{S}_L$ appear as vertices or ``gates'' inserted at definite values of $p$: the operators $H_{1,b}$ act on six orbitals and are therefore represented by 12-legged vertices, while the operators $H_{a,b}$ with $2 \leq a \leq 4$ act on two orbitals and are represented by four-legged vertices. Each vertex has two legs for each orbital on which it acts (one incoming and one outgoing). Diagonal operators, such as $H_{1,b}$, $H_{4,b}$, and $H_{0,0}$, do not modify the orbital occupations along the world lines, whereas the off-diagonal operators $H_{2,b}$ and $H_{3,b}$ do. The world lines satisfy periodic boundary conditions along the $p$ direction, since after the full application of the operator sequence $\tilde{S}_L$ the state must return to the initial state $\ket{\alpha}$. Each vertex carries a weight given by the corresponding non-negative matrix elements. We emphasize here that the world lines track the propagation history of orbitals along the operator string, rather than trajectories of particles in imaginary time.

In order to perform the sampling in a manner that satisfies detailed balance and is efficiently equilibrated at the lowest temperatures, we find that it is important to introduce two distinct classes of updates, distinguished by whether they transfer electrons between different chains that meet at a site\footnote{Note that the question of whether the sampling algorithm transfers electrons between chains is distinct from that of whether the Hamiltonian does so. $H$ does not involve any terms that transfer electrons between chains, but to equilibrate the system efficiently at low temperatures it is essential that we introduce updates that allow us to move between configurations that differ by such processes.} or not. The latter class of intra-chain updates is familiar from conventional SSE~\cite{SEN02,SAN19} and is summarized in the next subsection. The former class of inter-chain updates require a significant extension to SSE, and is central to the present mixed-dimensional generalization.

\subsubsection{Intra-chain updates}\label{sec:SSE:review:updates}

\begin{figure*}[!t]
	\centering
	\includegraphics[width=\textwidth]{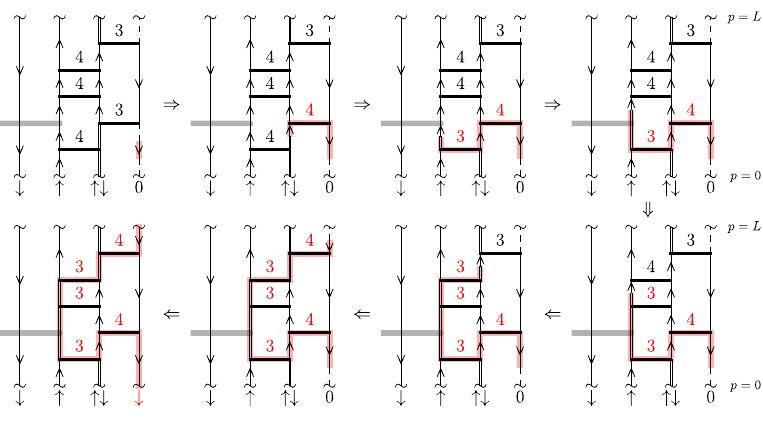}
	\caption{Off-diagonal loop update on a single chain. The top-left panel shows a fragment of an initial SSE configuration containing the full world lines of four orbitals. The orbital states in $\ket{\alpha}$ are indicated at the bottom; along the world lines they are represented schematically by dashed ($0$), single-arrow ($\uparrow$/$\downarrow$), and doubled ($\uparrow\downarrow$) lines. The subsequent panels, read clockwise, illustrate an off-diagonal update initiated by inserting a worm, highlighted in red, on one world line. The Fock state of the world line changes at the two worm ends (in the first panel, the state changes from $0$ to $\downarrow$ to $0$ along the world line containing the worm). One end of the worm, the worm head, is then propagated until the defect closes into a loop, thereby healing the discontinuity and updating the world line states, the state $\ket{\alpha}$, and the operator string along the highlighted path. At the four-legged vertices, corresponding to operators $H_{a,b}$ with $2 \leq a \leq 4$ and labeled by $a$ above each vertex, the worm head may turn, continue straight, or move sideways, as illustrated by the first two, third, and last two four-legged vertices it encounters here. At the 12-legged vertex (thin gray rectangle), the worm head can only continue straight or bounce back; in this example it continues straight.}
	\label{fig:worm_schematic}
\end{figure*}

We can view the intra-chain updates as those that modify states along the world lines of an individual chain while leaving the ones of all other chains unchanged. (Note that although the intra-chain updates only modify configurations on a single chain, the corresponding sampling probabilities are influenced by the states of other chains, which distinguishes it from SSE in the setting of an isolated chain~\cite{SEN02}.) As in the conventional SSE algorithm~\cite{SAN19}, we employ two main types of QMC updates to achieve this: diagonal updates and off-diagonal loop updates. A single QMC sweep includes one diagonal update, in which the operator sequence $\tilde{S}_L$ is traversed once. Whenever an identity operator $H_{0,0}$ is encountered, it is proposed to be exchanged with a non-identity diagonal operator, either $H_{1,b}$ or $H_{4,b}$ for a randomly chosen bond $b$, and vice versa. The details of this procedure, together with the corresponding acceptance probabilities, are given in \cref{app:sec:updates}. These acceptance probabilities depend only on the relevant operator matrix elements and combinatorial factors, and take the same form as in standard SSE implementations~\cite{SAN19}.

Because of the periodic boundary conditions along the world lines, {\it i.e.}{} along the index $p$, and the nontrivial connectivity imposed by the operator string, off-diagonal operators $H_{2,b}$ and $H_{3,b}$ cannot be inserted or removed independently, nor can the initial state $\ket{\alpha}$ be updated directly to obtain a new SSE configuration. Instead, both the off-diagonal operators and the initial state are updated through so-called ``off-diagonal loop updates''~\cite{SAN19}. Such updates are akin to worm algorithms~\cite{PRO01}. As exemplified in \cref{fig:worm_schematic}, a double discontinuity is first created on a world line by randomly changing its state to one of the other three allowed local states ({\it e.g.}{}, from $0$ to $\downarrow$). This creates a ``worm'' defect (shown in red in \cref{fig:worm_schematic}) with two ends at which the world line state is discontinuous. One of these ends is then propagated by modifying nearby vertices into other allowed operators, together with the corresponding world line states, in a manner that satisfies detailed balance. The worm is grown until its two ends meet again, thereby healing the double discontinuity and producing a new allowed configuration, {\it i.e.}{} a new Monte Carlo sample. In practice, we perform several such loop updates within a single QMC step (see \cref{app:sec:updates} for details). Because the world line states are modified during these off-diagonal updates, the initial state of the configuration, $\ket{\alpha}$, can also change in the process, provided the worm crosses the $p=0$ boundary, as in the example of \cref{fig:worm_schematic}.

During a single off-diagonal loop update, the propagation of the worm head is governed by the so-called ``vertex scattering'' rules: given a discontinuity on one leg of a vertex (termed the entrance leg), the worm head must either pass through the vertex, changing the corresponding operator to another allowed one and moving the discontinuity to another leg ({\it i.e.}{}, the exit leg), or ``bounce back'' and retrace its path. Given an entrance leg and a proposed change, the exit leg choice is not deterministic, but is chosen randomly in a detail-balance-obeying fashion. In general, the vertex scattering rules are not unique and depend on the expectation value of the corresponding operators within the SSE configuration, but can be chosen so as to minimize bounces (thereby decreasing autocorrelation times)~\cite{SAN99}. 

For the four-legged vertices, corresponding to the operators $H_{a,b}$ with $2 \leq a \leq 4$, the allowed exit legs depend on the occupations of the connected world lines and on the local change proposed by the worm head. However, all four-legged vertices have identical weights (expectation values) leading to particularly simple scattering rules: depending on the configuration, either a single new vertex is allowed following the change, in which case it is chosen with unit probability, or two new vertices are allowed, in which case each is chosen with probability one half. The detailed-balance condition can then be satisfied without introducing a bounce process (in which the vertex remains unchanged and the worm head backtracks). By contrast, for the 12-legged vertices, spin-valley-charge conservation, or equivalently the diagonal nature of the corresponding operators, leaves only two possibilities: the worm head either passes through the vertex changing the world line (and the expectation value of the corresponding $H_{1,b}$ operator) or bounces back. In this case, one of the two possibilities is sampled by a straightforward Metropolis acceptance filter based on the initial and final weights of the operator. A rigorous proof that these moves satisfy detailed balance is relegated to \cref{app:sec:updates}.

\subsection{Extending SSE to mixed dimensions: inter-chain updates}\label{sec:SSE:inter-chain}

\begin{figure*}[!t]
	\centering
	\subfloat{\label{fig:valley_flip_update:a}}\subfloat{\label{fig:valley_flip_update:b}}\subfloat{\label{fig:valley_flip_update:c}}\subfloat{\label{fig:valley_flip_update:d}}\includegraphics[width=\textwidth]{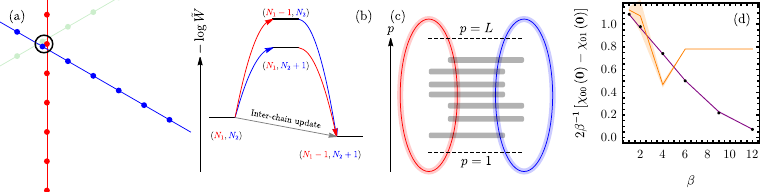}\caption{Inter-chain update. (a) shows two intersecting chains (red and blue) of the lattice from \cref{fig:model:a} between which a particle transfer is attempted at the site marked by the black circle. The third chain (green) passing through the same site is shown for completeness but is unaffected by the update. (b) presents a schematic weight ($\tilde{W}$) landscape of SSE configurations. Initially, the red (blue) chain contains $N_1$ ($N_2$) particles. Transferring a particle from the red to the blue chain changes the occupations to $\left( N_1-1 , N_2+1 \right)$. Conventional SSE updates accomplish this through two independent off-diagonal loop updates, which necessarily pass through one of the two intermediate configurations with strongly suppressed statistical weight. The inter-chain update directly connects the two high-weight configurations, avoiding these unfavorable intermediate states. (c) illustrates the two correlated off-diagonal loops that implement this transfer. The loops, shown in the colors of the chains they modify, correspond to the trajectories of the associated worms in the SSE configuration. If performed as two independent off-diagonal updates, the worm heads encounter a sequence of 12-legged diagonal vertices generated by the inter-chain repulsion terms at the intersection site. In the regime of suppressed charge fluctuations, these vertices typically cause the worms to bounce, making successful particle transfers exponentially unlikely. In the inter-chain update, the diagonal vertices are temporarily ignored during loop construction, allowing both worms to propagate freely; the resulting configuration is subsequently accepted or rejected using a final Metropolis test. (d) shows valley-imbalance fluctuations obtained from SSE with (purple) and without (orange) inter-chain updates, compared to DQMC results (black circles) for an idealized system with only onsite interaction $U = U' = 12$ and $\mathcal{N} = 8$. Including the inter-chain update restores agreement with DQMC and dramatically improves sampling efficiency. Error bars are shown in orange where visible. The valley-resolved susceptibility $\chi_{\eta\eta'}\left( \vec{k} \right)$ is defined in \cref{eqn:def_of_charge_susceptibility}.}
	\label{fig:valley_flip_update}
\end{figure*}

In the standard SSE algorithm applied to typical models~\cite{SAN19}, off-diagonal loop updates are, in principle, global, and a single accepted worm may modify many vertices of the SSE configuration. We might therefore expect the conventional SSE updates of \cref{sec:SSE:review:updates} to \emph{eventually} ensure ergodicity in the present problem. In practice, however, we find that, at sufficiently low temperatures, inter-chain equilibration times grow rapidly because the particle number on each chain becomes frozen in. The origin of this behavior is that, due to charge conservation along each chain, any single off-diagonal update can modify the particle occupations only on a single chain at a time. Since chains are updated independently, it also follows that a process in which one particle is ``transferred'' between two chains can only occur through two successive steps: the removal of a particle from one chain and the creation of a particle on another. At high temperatures this is straightforward, but at low temperatures, when correlations become significant, inter-chain interactions can strongly inhibit such processes through the ``environment'' created by the other chains, in a phenomenon akin to a ``Coulomb blockade''.

To restore inter-chain equilibration in this regime, we therefore introduce an ``inter-chain'' update, shown schematically in \cref{fig:valley_flip_update}. The idea behind this update is simple, although its implementation is technically complex: while particles cannot be transferred directly between chains, one can instead correlate a sequence of intra-chain updates on a pair of chains meeting at a site, such as the ones shown in \cref{fig:valley_flip_update:a}, so as to enhance the eventual acceptance probability of a move in which one chain gains a particle while the other loses one.

We now provide a graphical outline of the inter-chain update; additional details and proofs are relegated to \cref{app:sec:updates}, but the intuition and meaning of the updates should be clear from the following discussion. In \cref{fig:valley_flip_update:b}, we schematically show the statistical weight ($\tilde{W}$) landscape of SSE configurations relevant for two consecutive off-diagonal loops that, if accepted, would remove a particle from the blue chain and add a particle to the red chain (the green chain is simply shown for completeness). The paths of the corresponding worms are shown in \cref{fig:valley_flip_update:c}. At sufficiently low temperatures, however, there are many 12-legged vertices corresponding to classical repulsion terms $H_{1,b}$ between world lines on the two chains at a given site, and these inhibit the propagation of the corresponding worm heads. This is precisely because of spin-valley-charge conservation along each chain: when a worm head passes through such a vertex, it can only do so straight through by changing the weight of that vertex, and this change must be accepted by a Metropolis filter. Since many such 12-legged vertices must typically be crossed before a worm can wrap around the world line and change the particle occupation along a chain, as in the off-diagonal update of \cref{fig:worm_schematic}, the probability that the full process succeeds becomes very small. At the same time, after an inter-chain particle transfer, a particle should, in principle, only change its orbital, and hence the chain to which it belongs, at a given site. The weights of the corresponding 12-legged vertices should therefore remain nearly unchanged. The idea of the inter-chain update is to delay these intermediate Metropolis filters and replace them by a single final one: instead of many Metropolis steps required to climb an energy barrier, one performs one Metropolis step between the initial and final configurations, which are close in energy. 

To prevent the worm heads from repeatedly bouncing, we therefore temporarily ``remove'' the relevant 12-legged vertices from the SSE string during an inter-chain update. The two worm heads can then propagate through the SSE configuration unhindered, thereby transferring a particle between the two chains. A final Metropolis acceptance test is then performed based on the weight of the initially discarded 12-legged vertices, whose world lines were modified by the two consecutive off-diagonal loop updates. If this step is rejected, the entire move is undone. We expect, however, and indeed find numerically, that because the interaction is approximately $\mathrm{U} \left( {6} \right)$ symmetric, the weights of these 12-legged vertices with support on the two chains remain approximately unchanged after the particle transfer, so that the move is typically accepted. Put simply, rather than checking acceptance with respect to the 12-legged vertices ``on the fly'' as the update is constructed, we postpone this evaluation to a final step, thereby preserving detailed balance without inhibiting the intermediate steps needed to generate the correlated update.

In practice, we choose a site at random and two of the three valleys at that ``focus'' site, shown in black in \cref{fig:valley_flip_update:a}. A possible particle transfer between the corresponding chains is then proposed by performing two off-diagonal loop updates on the two intersecting chains. These two off-diagonal loop updates are carried out in the conventional ({\it i.e.}{} non-deterministic) manner discussed in \cref{sec:SSE:review:updates}, except that any diagonal operator with support on the focus site is ignored, {\it i.e.}{} the worm head is forced to pass straight through the corresponding 12-legged vertex instead of being subjected there to an intermediate Metropolis filter. After the two off-diagonal loop updates have been completed, the entire move is accepted or rejected with a final Metropolis based on the initially discarded 12-legged vertices. We find that the inter-chain updates substantially improve equilibration of the particle number between chains. Because the move is built entirely out of ordinary off-diagonal loop updates, the periodicity of the world lines along $p$ and the boundary conditions in the spatial directions are preserved automatically. 

As already evident from the benchmark shown in \cref{fig:valley_flip_update:d}, these inter-chain updates are essential for achieving equilibration in the regime of strongly suppressed charge fluctuations. Additional benchmarks and validation of the SSE implementation are provided in \cref{app:sec:benchmarking}.

\subsection{Observables}\label{sec:SSE:observables}

A number of observables can be estimated by sampling the grand-canonical partition function in \cref{eqn:sse_z} with the SSE method. Their evaluation has been discussed extensively in previous works~\cite{SAN97a,DOR01,SEN02,DOR02,SAN19}, while a comprehensive pedagogical review, together with its application to the Hamiltonian in \cref{eqn:m_pt_ham}, is provided in \cref{app:sec:sse}. The simplest observable is the filling of the system,
\begin{equation}
	\nu = \frac{1}{\mathcal{N}^2} \left\langle \sum_{\vec{R}} \hat{N}_{\vec{R}} \right\rangle,
\end{equation}
where $\left\langle \dots \right\rangle$ denotes grand-canonical expectation values at chemical potential $\mu$ and inverse temperature $\beta$. From the dependence of the filling on the chemical potential, we also estimate the inverse compressibility $\pdv{\mu}{\nu}$.

In addition, we study the spatial dependence of charge and spin correlation functions and extract the momentum- and valley-resolved charge-charge susceptibility
{\small \begin{equation}
	\label{eqn:def_of_charge_susceptibility}
	\chi_{\eta \eta'} \left( \vec{k} \right) = \frac{1}{\mathcal{N}^2} \sum_{\vec{R}} \int_{0}^{\beta} \dd{\tau} \left\langle \hat{N}_{\vec{R}',\eta} (\tau) \hat{N}_{\vec{R}' + \vec{R},\eta'} (0) \right\rangle_c e^{i \vec{k} \cdot \vec{R}},
\end{equation}}where the $(\tau)$ argument denotes imaginary-time evolution, $\hat{N}_{\vec{R},\eta} (\tau) \equiv e^{\tau H} \hat{N}_{\vec{R},\eta} e^{-\tau H}$, and $\left\langle \dots \right\rangle_c$ denotes connected correlators. We note that imaginary-time correlators can be estimated straightforwardly from correlations in the propagation index $p$~\cite{DOR02,SAN19}. For spin-spin correlations, the individual spin-rotation symmetry on each chain implies that spins of fermions belonging to different chains are uncorrelated. As a result, the spin-spin correlation function is fully specified by the one-dimensional function
\begin{equation}
	\label{eqn:def_spin_spin}
	\mathcal{S}(i) = \left\langle \hat{S}^z_{\vec{R},\eta} \hat{S}^z_{\vec{R} + i \vE_\eta,\eta} \right\rangle,
\end{equation} 
for any $\vec{R}$, $\eta$, and $0 \leq i \leq \mathcal{N}-1$, where the spin operators are given by $\hat{S}^z_{\vec{R},\eta} \equiv \frac{1}{2} \left( \hat{n}_{\vec{R},\eta,\uparrow} - \hat{n}_{\vec{R},\eta,\downarrow} \right)$.

In addition to these correlation functions, we also compute the charge stiffness and spin stiffnesses of the model~\cite{SCA92,SCA93,SAN97a,SEN02}. These are defined as the response of the system to a twist in the boundary conditions affecting either the charge or spin degrees of freedom. To define the two stiffnesses, we modify the kinetic Hamiltonian in \cref{eqn:sp_ham} to
\begin{align}
		\label{eqn:sp_ham_flux}
		H_{0} \left( \phi_{\text{C},\text{S}} \right)=& -t\sum_{\vec{R},\eta,s} \big( e^{i\phi_{\text{C},\text{S}} f^{\text{C},\text{S}}_{s} \vE_\eta \cdot \vec{\hat{x}}} \hat{d}^\dagger_{\vec{R},\eta,s} \hat{d}_{\vec{R} + \vE_\eta,\eta,s} \nonumber \\
		&\quad + \text{h.c.} - \mu \hat{n}_{\vec{R},\eta,s} \big),
\end{align}
where $\phi_{\text{C}}$ denotes a flux coupled to the charge sector, while $\phi_{\text{S}}$ denotes a flux coupled to the spin sector. The factor $f^{\text{C},\text{S}}_{s}$ specifies whether the flux couples to the charge or spin degrees of freedom,
\begin{equation}
	f^{\text{C}}_{s} = \delta_{s,\uparrow} + \delta_{s,\downarrow}, \quad
	f^{\text{S}}_{s} = \delta_{s,\uparrow} - \delta_{s,\downarrow}.
\end{equation}
Defining the grand potential in the presence of a flux as $\Phi \left( \phi \right)$, the charge and spin stiffnesses are given by
\begin{equation}
	\rho_{\text{C},\text{S}} \equiv \frac{1}{\mathcal{N}^2} \eval{\pdv[2]{\Phi \left( \phi_{\text{C},\text{S}} \right)}{\phi_{\text{C},\text{S}}}}_{\phi_{\text{C},\text{S}} = 0}.
\end{equation}
A non-vanishing charge (spin) stiffness provides a numerical diagnostic of whether the system transports charge (spin).

\section{Results}\label{sec:results}

Using SSE, we have numerically studied the AA t-\ch{SnSe2}{} Hamiltonian in \cref{eqn:m_pt_ham} across the 18 parameter sets shown in \cref{fig:parameters}, over a range of chemical potentials and inverse temperatures. In this section, we summarize the main findings of this work and illustrate them with representative examples drawn from this extensive parameter set. The complete set of results is presented in \cref{app:sec:numerical_results}, where we also discuss details of the numerical simulations. We begin by considering systems with relatively small nearest-neighbor repulsion and investigate the emergence of correlated insulators at integer fillings $\nu$ of the moir\'e unit cell (in between which the system remains metallic). We then turn to systems in which the nearest-neighbor repulsion is significant and show that the system forms Wigner-Mott states at certain commensurate fillings. Finally, using replica-exchange sampling~\cite{SWE86,HUK96} within SSE, we investigate the spin structure of the correlated insulators at integer filling in the most strongly correlated parameter set considered here, and show that upon cooling the system develops signatures of the spin-dimer phases found analytically in Ref.~\cite{LI25} in the strong coupling approximation.

\subsection{Correlated insulators}\label{sec:results:ci}

\begin{figure*}[!t]
	\centering
	\includegraphics[width=\textwidth]{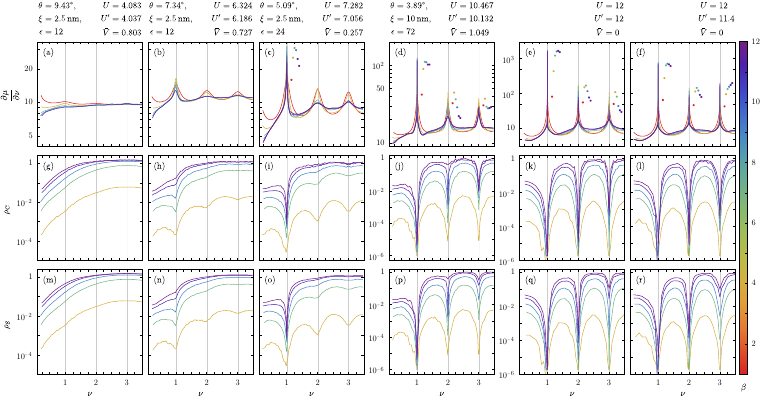}
	\caption{Formation of correlated insulators in AA t-\ch{SnSe2}{}. Rows (a) -- (f), (g) -- (l), and (m) -- (r) show, respectively, the inverse compressibility $\pdv{\mu}{\nu}$, the charge stiffness $\rho_{\text{C}}$, and the spin stiffness $\rho_{\text{S}}$ of the Hamiltonian in \cref{eqn:m_pt_ham} for a range of parameters. The color of each curve encodes the inverse temperature $\beta$ of the SSE simulations according to the color map on the right (with hot in red and cold in blue). Panels within the same column correspond to the same parameter set, indicated above each column. All results are for finite systems with $\mathcal{N} = 12$. In (c) -- (f), the peaks in the inverse compressibility at integer $\nu$ overlap for different inverse temperatures. To facilitate comparison of their heights, we additionally show scatter points next to each peak, using the same color coding as for the curves. These points are shifted only horizontally for visibility; their vertical position is exactly the height of the corresponding peak. The error bars are smaller than the marker sizes.}
	\label{fig:compress_simple}
\end{figure*}

\Cref{fig:compress_simple} shows the inverse compressibility, as well as the charge and spin stiffnesses, obtained for the AA t-\ch{SnSe2}{} Hamiltonian in \cref{eqn:m_pt_ham} over a range of parameters. We focus on fillings $\nu \leq 3$, since the results for $\nu \geq 3$ follow from particle-hole symmetry. At the largest twist angle we investigate, $\theta = \SI{9.43}{\degree}$, the system remains compressible at all fillings and is metallic, with finite charge and spin stiffnesses, as shown in the first column of \cref{fig:compress_simple}. As the angle is reduced to $\theta = \SI{7.34}{\degree}$, the system develops signatures of correlated behavior at integer fillings, namely small peaks in the inverse compressibility and dips in the charge and spin stiffnesses. Upon cooling, however, at this twist angle it does not form Mott insulators and instead remains metallic. The largest peak in the inverse compressibility, together with the most pronounced dip in the charge and spin stiffnesses, occurs at $\nu = 1$. Throughout this work, ``Mott insulator'' refers to a finite-temperature correlated insulating regime with strongly suppressed charge fluctuations and a pronounced reduction of the charge stiffness, even though the conductivity remains thermally activated rather than strictly zero.

Compared to the single-orbital model, in which a Mott insulator already forms at $U \sim 5-6$~\cite{ROZ92,GEO92a,GEO96,FLO02}, our system remains metallic for $\bar{U} \sim 6$ at all fillings, despite having a smaller bandwidth ($2t$ versus $4t$). This reflects the presence of valley fluctuations, which are absent in the single-orbital Hubbard model and were identified explicitly in Ref.~\cite{VAS26} at half-filling using DQMC. A similar enhancement of the critical interaction required for Mott-insulator formation has also been found in slave-rotor calculations~\cite{FLO02,FLO04} and Gutzwiller-projector treatments~\cite{LU94} of multiorbital Hubbard models with $\mathrm{U} \left( {N} \right)$ symmetry ($N > 2$).

At $\theta = \SI{5.09}{\degree}$, the system does form a Mott insulator at $\nu = 1$, signaled by a large peak in the inverse compressibility and dips in the charge and spin stiffnesses, all of which persist upon further cooling. The fact that the $\nu = 1$ insulator is the first to form as $\bar{U}$ is increased can be understood intuitively by counting the number of local fluctuations compatible with the local-moment character of a paramagnetic Mott insulator at integer filling, or equivalently by considering the degeneracy $d_{\nu}$ of the corresponding local-moment state at (integer) filling $\nu$. Specifically, in the $\mathrm{U} \left( {6} \right)$-symmetric limit, the degeneracy of a local moment is 
\begin{equation}
	\label{eqn:deg_log_mom_u6}
	d^{\mathrm{U} \left( {6} \right)}_{\nu} = {6 \choose \nu},
\end{equation}
while in the physical $\mathrm{U} \left( {2} \right)^{\otimes3}$ limit with $U' < U$ it is given by
\begin{gather}
	d^{\mathrm{U} \left( {2} \right)^{\otimes3}}_{1} = {6 \choose 1} = 6, \quad d^{\mathrm{U} \left( {2} \right)^{\otimes3}}_{2} = {6 \choose 2} - 3 = 12, \nonumber \\ 
	d^{\mathrm{U} \left( {2} \right)^{\otimes3}}_{3} = 2^3 = 8. \label{eqn:deg_log_mom_u23}
\end{gather}
Because $d_{1} < d_{2,3}$, the $\nu = 1$ Mott insulator is the first to stabilize as the interaction is increased. The same hierarchy in the critical onsite Hubbard interaction can also be derived more rigorously within a slave-rotor approach~\cite{FLO02,FLO04}, as we show in \cref{app:sec:anal}.

Another feature apparent at $\nu=2,3$ for $\theta = \SI{7.34}{\degree}$, and to some extent at $\nu=1$ for $\theta = \SI{9.43}{\degree}$, is a Pomeranchuk-like effect. In the conventional Pomeranchuk effect, liquid ${}^3$He freezes upon heating, attributed to the enhanced entropy of the solid state (due to the nuclear spins) relative to the Fermi liquid state~\cite{LEE97}. Similar behavior has also been observed in other moir\'e materials, such as twisted bilayer graphene~\cite{ROZ21,SAI21a} and TMD heterobilayers~\cite{LI21e}. In our case, at high temperatures the inverse compressibility exhibits a peak and the charge and spin stiffnesses display dips at integer fillings. Upon cooling, however, the peak in the inverse compressibility is strongly reduced and the system becomes metallic, as indicated by the charge and spin stiffnesses. This behavior can be understood intuitively as follows. For these fillings, the onsite interaction strength lies below the critical value required to stabilize a Mott insulator at zero temperature, so the ground state is metallic. At elevated temperatures, however, the Mott-insulating state is favored entropically and can therefore become stabilized by its lower free energy. In particular, the Mott insulator supports local moments and thus carries a large spin entropy, which lowers its free energy relative to that of the metallic phase.

At the smallest angle we consider, $\theta = \SI{3.89}{\degree}$, as well as in the idealized strongly correlated systems with $U=12$, the system forms correlated insulators at all integer fillings, signaled by peaks in the inverse compressibility and dips in the charge and spin stiffnesses. At $\nu = 3$, however, as the temperature is lowered below the antiferromagnetic exchange scale $J = \frac{4}{U}$, the dip in the spin stiffness $\rho_{\text{S}}$ gradually disappears, while the dip in the charge stiffness $\rho_{\text{C}}$ persists. This is consistent with the findings of Ref.~\cite{LI25}, which showed that at $\nu=3$ each chain becomes half-filled and the spins realize a one-dimensional Heisenberg antiferromagnet whose low-energy spinon excitations endow each chain with a finite spin stiffness. Further evidence for the integer-filled phases found in Ref.~\cite{LI25} is presented in \cref{sec:results:spin_dimers}.

Although all integer fillings $\nu \in \left \lbrace 1, 2, 3 \right \rbrace$ form correlated insulators in the strongly interacting regime shown in the last three columns of \cref{fig:compress_simple}, a clear hierarchy nevertheless emerges in their strength. This hierarchy can be quantified through the charge excitation gap, or equivalently through the magnitude of charge fluctuations, both of which are qualitatively reflected in the peaks of the inverse compressibility. In the $\mathrm{U} \left( {6} \right)$-symmetric case shown in the fifth column of \cref{fig:compress_simple}, $\pdv{\mu}{\nu}$ decreases with increasing $\nu$, a result that again matches the naive counting of local-moment fluctuations in these Mott-insulating states from \cref{eqn:deg_log_mom_u6}. When $U' < U$, as in the fourth and sixth columns of \cref{fig:compress_simple}, this hierarchy becomes temperature dependent. At temperatures smaller than $\bar{U}$ but still larger than $U-U'$, the system exhibits an emergent $\mathrm{U} \left( {6} \right)$ symmetry, and the inverse compressibility of the correlated insulators again decreases toward half filling. Once the temperature is lowered below $U-U'$, however, the hierarchy changes, with $\pdv{\mu}{\nu} (\nu)$ decreasing in the order $\nu = 1,3,2$, reflecting the degeneracies of the local-moment states in the $\mathrm{U} \left( {2} \right)^{\otimes3}$ limit given in \cref{eqn:deg_log_mom_u23}.

\subsection{Wigner-Mott insulators}\label{sec:results:cdw}

\begin{figure}[!t]
	\centering
	\includegraphics[width=\columnwidth]{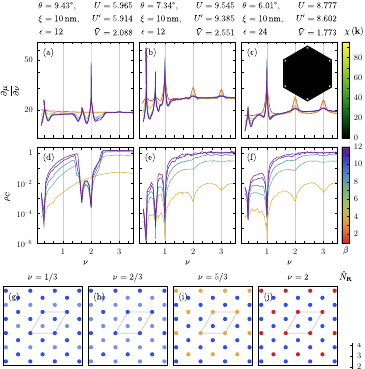}\subfloat{\label{fig:compress_CDW:a}}\subfloat{\label{fig:compress_CDW:b}}\subfloat{\label{fig:compress_CDW:c}}\subfloat{\label{fig:compress_CDW:d}}\subfloat{\label{fig:compress_CDW:e}}\subfloat{\label{fig:compress_CDW:f}}\subfloat{\label{fig:compress_CDW:g}}\subfloat{\label{fig:compress_CDW:h}}\subfloat{\label{fig:compress_CDW:i}}\subfloat{\label{fig:compress_CDW:j}}\caption{Wigner-Mott states in AA t-\ch{SnSe2}{}. (a) -- (c) show the inverse compressibility, while (d) -- (f) show the charge stiffness, for three parameter sets with significant nearest-neighbor repulsion. As in \cref{fig:compress_simple}, the parameters are listed above each column, and the traces are color-coded by the inverse temperature $\beta$ according to the colormap on the right. All results are for finite systems with $\mathcal{N} = 12$. The inset in (c) shows the momentum-resolved total charge susceptibility at $\nu = \frac13$ and $\beta = 12$, displaying a sharp peak at the $\mathrm{K}_M$ point. (g) -- (j) show the site-resolved real-space occupation of the corresponding Wigner-Mott states, with the average filling $\nu$ indicated above each panel. The gray diamond marks the associated $\sqrt{3} \times \sqrt{3}${} supercell.}
	\label{fig:compress_CDW}	
\end{figure}

Having investigated the formation of Mott insulators at integer fillings, we now turn to the regime in which the nearest-neighbor repulsion is more significant ({\it i.e.}{}, larger $\bar{V}/\bar{U}$). Three representative examples are shown in \cref{fig:compress_CDW}. At certain commensurate fillings, the charge distribution freezes into a pattern that breaks moir\'e translation symmetry down to a $\sqrt{3} \times \sqrt{3}${} supercell. This is signaled by the emergence of a strong peak in the total momentum-resolved charge susceptibility
\begin{equation}
	\chi \left( \vec{k} \right) \equiv \sum_{\eta, \eta'} \chi_{\eta \eta'} \left(\vec{k} \right),
\end{equation}
at the $\mathrm{K}_M$ point, {\it i.e.}{} at $\vec{k} = \pm \frac13 \left( \vec{b}_{M_1} + \vec{b}_{M_2} \right)$, with a much smaller signal at all other momenta. Here, $\chi_{\eta\eta'} \left( \vec{k} \right)$ is the interacting valley-resolved charge susceptibility defined in \cref{eqn:def_of_charge_susceptibility}, and $\chi \left( \vec{k} \right)$ is its valley-summed counterpart. By examining the $\ket{\alpha}$ states of the SSE samples, we also plot the real-space occupations of the Wigner-Mott states in \crefrange{fig:compress_CDW:g}{fig:compress_CDW:j}. No valley structure is resolved at the lowest temperatures we investigate. Starting from $\theta=\SI{9.43}{\degree}$, we find that the system develops three peaks in the inverse compressibility at $\nu = \frac13, \frac53, 2$, where the charge stiffness also drops (as well as a faint peak in inverse compressibility at $\nu = \frac73$). Remarkably, the system not only breaks moir\'e translation symmetry at the two fractional fillings, but also at $\nu=2$, where it does not simply form a Mott insulator with two particles on every site, but instead realizes the Wigner-Mott state shown in \cref{fig:compress_CDW:j}. Moir\'e-translation-symmetric correlated insulators can also coexist with Wigner-Mott states, as for $\theta = \SI{7.34}{\degree}$ ($\theta = \SI{6.01}{\degree}$), where the system forms Wigner-Mott states at $\nu = \frac13,\frac23$ ($\nu = \frac13$).

The formation of the Wigner-Mott states can be rationalized by solving the classical interaction Hamiltonian in \cref{eqn:int_ham} under the assumption that the orbital charge breaks moir\'e translation symmetry only down to the $\sqrt{3} \times \sqrt{3}${} supercell. This is equivalent to finding the ground states $\ket{\Psi}$ of $H_I$ with a supercell-periodic orbital charge density, $N_{\vec{R} + \Delta \vec{R},\eta} = N_{\vec{R},\eta}$, where $N_{\vec{R},\eta}$ denotes the charge of orbital $\eta$ at site $\vec{R}$, {\it i.e.}{} $\hat{N}_{\vec{R},\eta} \ket{\Psi} = N_{\vec{R},\eta} \ket{\Psi}$, and $\Delta \vec{R}$ is a supercell lattice vector. The spin degrees of freedom are not fixed by the interaction Hamiltonian and may fluctuate freely. The stability of the corresponding Wigner-Mott states can then be quantified by examining the spin degeneracy of the resulting ground states within a unit cell. 

For $\theta = \SI{9.43}{\degree}$, the corresponding results are shown in \cref{tab:large_V_CDW}. In this regime, the large nearest-neighbor repulsion requires that the sites within a supercell be filled one at a time. Whether one works in the $\mathrm{U} \left( {2} \right)^{\otimes3}$ limit or in the high-temperature emergent $\mathrm{U} \left( {6} \right)$ limit, the local-moment degeneracy is minimized at $\nu=2$, explaining the prominence of this state in the inverse compressibility shown in \cref{fig:compress_CDW:a}. The states at $\nu=\frac13,\frac53,\frac73$ also have relatively small local-moment degeneracy, which explains the other, smaller peaks in the inverse compressibility, with the $\nu = \frac73$ feature being very weak. By contrast, the remaining states have too large a local-moment degeneracy to be stabilized and are therefore not observed in the simulations.

At $\theta = \SI{7.34}{\degree}$ and $\theta = \SI{6.01}{\degree}$, the nearest-neighbor repulsion is smaller relative to the onsite one. As a result, the integer-filled phases do not break moir\'e translation symmetry. Applying the same approach as above, we find in this case that the system prefers to fill the sites of the Wigner-Mott supercell uniformly. As shown in \cref{tab:small_V_CDW}, the local-moment degeneracy then increases toward half filling, thereby stabilizing the Wigner-Mott states at smaller fillings, in agreement with what is observed in our simulations.

\begin{table}[!t]
\begin{tabular}{|c|c|c|c|c|c|}
		\hline
		\multirow{2}{*}{$\nu$} & \multicolumn{3}{c|}{$\left( N_{\vec{R},0},N_{\vec{R},1},N_{\vec{R},2} \right)$} & \multicolumn{2}{c|}{$d$} \\ \cline{2-6}
		& {\scriptsize$\vec{R} = \vec{0}$} & {\scriptsize $\vec{R}=\vec{a}_{M_1}+\vec{a}_{M_2}$} & {\scriptsize$\vec{R}=2 \left( \vec{a}_{M_1}+\vec{a}_{M_2} \right)$} & {\scriptsize $\mathrm{U} \left( {2} \right)^{\otimes3}$} & {\scriptsize $\mathrm{U} \left( {6} \right)$} \\ \hline
		1/3 & \multirow{9}{*}{$(0,0,0)$}& \multirow{6}{*}{$(0,0,0)$} & $(0,0,1)$ & 2 & 6 \\ \cline{1-1} \cline{4-6}
		2/3 & & & $(0,1,1)$ & 4 & 15 \\ \cline{1-1} \cline{4-6}
		1 & & & $(1,1,1)$ & 8 & 20 \\ \cline{1-1} \cline{4-6}
		4/3 & & & $(1,2,1)$ & 4 & 15\\ \cline{1-1} \cline{4-6}
		5/3 & & & $(1,2,2)$ & 2 & 6\\ \cline{1-1} \cline{4-6}
		2 & & & $(2,2,2)$ & 1 & 1\\ \cline{1-1} \cline{3-6}
		7/3 & & $(0,0,1)$ & \multirow{3}{*}{$(2,2,2)$} & 2 & 6 \\ \cline{1-1} \cline{3-3} \cline{5-6}
		8/3 & & $(0,1,1)$ & & 4 & 15\\ \cline{1-1} \cline{3-3} \cline{5-6}
		3 & & $(1,1,1)$ & & 8 & 20\\ \hline
	\end{tabular}\caption{Classical ground states of the Coulomb interaction Hamiltonian on the $\sqrt{3} \times \sqrt{3}${} supercell for $\theta = \SI{9.43}{\degree}$. For each filling $\nu$, we list the orbital occupations $\left( N_{\vec{R},0},N_{\vec{R},1},N_{\vec{R},2} \right)$ on the three sites of the Wigner-Mott supercell, together with the degeneracy $d$ of the corresponding local-moment manifold in both the physical $\mathrm{U} \left( {2} \right)^{\otimes3}$ limit and the $\mathrm{U} \left( {6} \right)$ limit.}
	\label{tab:large_V_CDW}
\end{table}

\begin{table}[!t]
\begin{tabular}{|c|c|c|c|c|c|}
			\hline
			\multirow{2}{*}{$\nu$} & \multicolumn{3}{c|}{$\left( N_{\vec{R},0},N_{\vec{R},1},N_{\vec{R},2} \right)$} & \multicolumn{2}{c|}{$d$} \\ \cline{2-6}
			& {\scriptsize$\vec{R} = \vec{0}$} & {\scriptsize $\vec{R}=\vec{a}_{M_1}+\vec{a}_{M_2}$} & {\scriptsize$\vec{R}=2 \left( \vec{a}_{M_1}+\vec{a}_{M_2} \right)$} & {\scriptsize $\mathrm{U} \left( {2} \right)^{\otimes3}$} & {\scriptsize $\mathrm{U} \left( {6} \right)$} \\ \hline
			1/3 & $(0,0,0)$ & $(0,0,0)$ & $(0,0,1)$ & 2 & 6 \\ \hline
			2/3 & $(0,0,0)$ & $(0,0,1)$ & $(0,0,1)$ & 4 & 36 \\ \hline
			4/3 & $(0,0,1)$ & $(0,0,1)$ & $(0,1,1)$ & 16 & 540 \\ \hline
			5/3 & $(0,0,1)$ & $(0,1,1)$ & $(0,1,1)$ & 32 & 1350 \\ \hline
			7/3 & $(0,1,1)$ & $(0,1,1)$ & $(1,1,1)$ & 128 & 4500 \\ \hline
			8/3 & $(0,1,1)$ & $(1,1,1)$ & $(1,1,1)$ & 256 & 6000 \\ \hline
		\end{tabular}\caption{Classical ground states of the Coulomb interaction Hamiltonian on the $\sqrt{3} \times \sqrt{3}${} supercell for $\theta = \SI{7.34}{\degree}$ and $\theta = \SI{6.01}{\degree}$. The layout is the same as in \cref{tab:large_V_CDW}. For each filling $\nu$, we list the orbital occupations $\left( N_{\vec{R},0},N_{\vec{R},1},N_{\vec{R},2} \right)$ on the three sites of the Wigner-Mott supercell, together with the degeneracy $d$ of the corresponding local-moment manifold in the $\mathrm{U} \left( {2} \right)^{\otimes3}$ and $\mathrm{U} \left( {6} \right)$ limits. We do not include the integer-filled solutions, since they do not break moir\'e translation symmetry.}
	\label{tab:small_V_CDW}
\end{table}

\subsection{Signatures of strong-coupling spin-dimer phases}\label{sec:results:spin_dimers}

\begin{figure*}[!t]
	\centering
	\includegraphics[width=\textwidth]{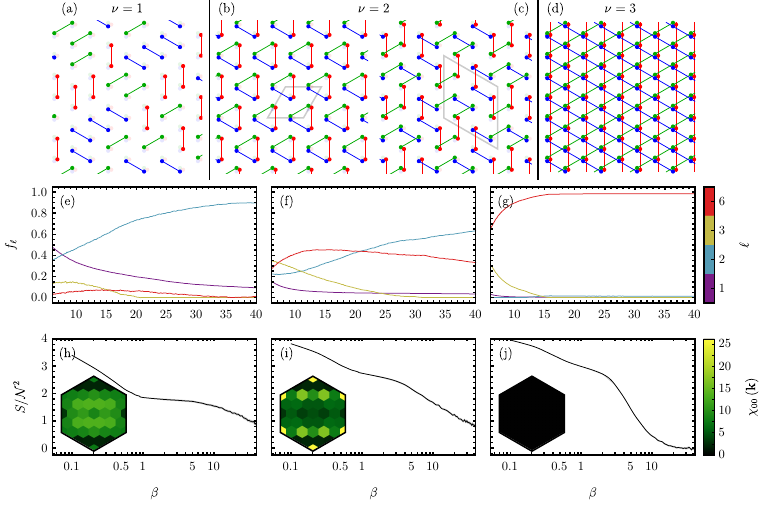}\subfloat{\label{fig:spin_corr_ins:a}}\subfloat{\label{fig:spin_corr_ins:b}}\subfloat{\label{fig:spin_corr_ins:c}}\subfloat{\label{fig:spin_corr_ins:d}}\subfloat{\label{fig:spin_corr_ins:e}}\subfloat{\label{fig:spin_corr_ins:f}}\subfloat{\label{fig:spin_corr_ins:g}}\subfloat{\label{fig:spin_corr_ins:h}}\subfloat{\label{fig:spin_corr_ins:i}}\subfloat{\label{fig:spin_corr_ins:j}}\caption{Probing the spin structure of the integer-filled correlated insulators of AA t-\ch{SnSe2}{}. (a) -- (d) show schematic representations of the strong-coupling correlated insulators found in Ref.~\cite{LI25}: (a) shows the disordered classical dimer solid at $\nu=1$; (b) and (c) show the two degenerate valence-bond solids at $\nu=2$, with the corresponding supercells indicated by the gray diamonds; and (d) shows the quantum-paramagnet phase at $\nu=3$. Colored dots represent electrons, while the colored lines indicate antiferromagnetic spin couplings. (e) -- (g) show the fraction of occupied sites belonging to spin chains of length $\ell$, denoted by $f_{\ell}$, for $\nu=1,2,3$, respectively. The errors are smaller than the plot markers. (h) -- (j) show the entropy per site as a function of inverse temperature for $\nu=1,2,3$, with the errors indicated by the gray halo. The insets display the valley-diagonal charge susceptibility $\chi_{00} \left( \vec{k} \right)$. Throughout, we use $U=12$, $U'=11.4$, and $\mathcal{N} = 6$.}
	\label{fig:spin_corr_ins}\end{figure*}

Ref.~\cite{LI25} showed that, in the strong-coupling limit at integer fillings $\nu$, the Hamiltonian of AA t-\ch{SnSe2}{} in \cref{eqn:m_pt_ham}, projected onto the low-energy subspace containing exactly $\nu$ particles on each site, takes the form of a quasi-1D Kugel-Khomskii~\cite{KUG82,KHA97} spin model,
{\small \begin{align}
	H_\nu =& \frac{J}{2} \sum_{\substack{\vec{R},\eta \\ \Delta\vec{R} = \pm \vec{e}_\eta}}
	P^\nu \left( \vec{S}_{\vec{R},\eta} \cdot \vec{S}_{\vec{R}+\Delta\vec{R},\eta} + \frac14\hat{N}_{\vec{R},\eta} \hat{N}_{\vec{R}+\Delta\vec{R},\eta} \right) P^\nu \nonumber \\
	& +\frac12 \sum_{\substack{ \vec{R},\Delta\vec{R} \\ \eta,\eta'}}
	P^\nu \delta V_{\eta \eta'}(\Delta\vec{R}) \hat{N}_{\vec{R},\eta} \hat{N}_{\vec{R}+\Delta\vec{R},\eta'} P^\nu. \label{eqn:kk_ham}
\end{align}}where $J = \frac{4}{U}$, and we remind the reader that we employ units in which $t=1$. The spin operator is defined as $\vec{S}_{\vec{R}\eta} \equiv \frac12 \sum_{s,s'} \hat{d}^\dagger_{\vec{R},\eta,s} \boldsymbol{\sigma}_{ss'} \hat{d}^\dagger_{\vec{R},\eta,s'}$, with $\boldsymbol{\sigma}$ the Pauli vector, while $\delta V_{\eta\eta'} \left( \Delta \vec{R} \right) = V_{\eta\eta'} \left( \Delta \vec{R} \right) - U \delta_{\Delta\vec{R},\vec{0}}$ denotes the density-density interaction after subtracting the onsite Hubbard repulsion. Finally, $P^{\nu}$ denotes the projector onto the subspace with exactly $\nu$ particles on each lattice site.

The onsite part, $\delta V_{\eta \eta'} \left( \vec{0} \right)= \left(U'- U \right)\left(1-\delta_{\eta \eta'} \right)$, favors a Hund's-rule-like filling pattern across the valleys of a single site, in which the three valleys are first singly occupied before double occupancy develops within any one valley. By contrast, the offsite part, $\delta V_{\eta \eta'} \left( \Delta \vec{R} \right)$ for $\Delta \vec{R} \neq 0$, favors valley polarization at integer fillings, {\it i.e.}{} a uniform \emph{equidistant} charge distribution~\cite{LI25}. In \cref{app:sec:model}, we derived a mean-field expression for the corresponding valley-polarization transition. Numerically, however, we find the associated critical temperature to be very small compared to the other energy scales of the problem. As a result, the offsite contribution to $\delta V_{\eta\eta'} \left( \Delta \vec{R} \right)$ can be neglected, and the physics of the system is governed by the first line of \cref{eqn:kk_ham}, which we now review.

At integer filling $\nu \leq 3$, exactly $\nu$ of the three orbitals in each unit cell are occupied. The charge degrees of freedom are then frozen, while the remaining spin degrees of freedom are coupled antiferromagnetically along certain directions whenever electrons occupy the same orbital. For example, two electrons separated by $\vec{e}_{\eta}$ in valley $\eta$ interact antiferromagnetically. The system can therefore be viewed as a collection of one-dimensional spin chains of varying lengths, each formed by a sequence of sites occupied by electrons from the same valley $\eta$ and aligned along the direction $\vec{e}_{\eta}$. The corresponding ground states were worked out in Ref.~\cite{LI25}. For $\nu=1,2$, the energy is minimized when the lattice is tiled by open spin chains of length two. The resulting ground states are a disordered classical dimer solid with exponential degeneracy and extensive entropy at $\nu = 1$, and a valence-bond solid at $\nu=2$ that breaks moir\'e translation symmetry and forms either a $\sqrt{3} \times \sqrt{3}${} or a $3 \times 3$ supercell. At half filling ($\nu=3$), the system reduces to a collection of one-dimensional Heisenberg chains, one along each chain direction in every valley. Schematic representations of these phases are provided in \crefrange{fig:spin_corr_ins:a}{fig:spin_corr_ins:d}.

\subsubsection{Length distribution of spin chains}\label{sec:results:spin_dimers:spin_chain}

We now investigate how these phases emerge within SSE. We consider the strongly interacting regime with $U=12$, $U'=11.4$, and vanishing nearest-neighbor repulsion, and focus on $\mathcal{N} \times \mathcal{N}$ systems with $\mathcal{N} = 6$. To improve mixing at low temperatures, we employ replica-exchange sampling~\cite{SWE86,HUK96}. 

The basic idea of the analysis is the following. In the Mott regime, the low-energy manifold can be viewed as a tiling of the two-dimensional system by one-dimensional spin chains. Spins belonging to the same chain are correlated, while spins on different chains are essentially uncorrelated. Therefore, the spin-spin correlator $\mathcal{S}(i)$ defined in \cref{eqn:def_spin_spin} contains information about the distribution of spin-chain lengths in the SSE configurations. Our goal is to express $\mathcal{S}(i)$ as a weighted sum of the spin-spin correlators of isolated one-dimensional chains of different lengths, and then invert this relation to estimate the fraction of spins belonging to chains of each length.

We first compute the spin-spin correlation functions for one-dimensional spin chains of length $\ell \geq 1$, using open boundary conditions for $\ell < \mathcal{N}$ and periodic boundary conditions for $\ell = \mathcal{N}$. The corresponding family of Hamiltonians is
\begin{equation}
	\label{eqn:spin_chain_ham}
	\mathcal{H}_{\ell} = \sum_{i=0}^{\ell-1} J \left( \vec{S}_i \cdot \vec{S}_{i+1} + \frac14 \right), 
\end{equation}
with ground state $\ket{\Psi_{\ell}}$ and energy $\mathcal{E}_{\ell}$, defined through $\mathcal{H}_{\ell} \ket{\Psi_{\ell}} = \mathcal{E}_{\ell} \ket{\Psi_{\ell}}$. Since a chain of length $\ell$ can start at any position within the two-dimensional system, the relevant spin correlator at separation $i$ is obtained by summing over all possible starting positions along the chain
\begin{equation}
	C_{\ell}(i) = \sum_{j=0}^{\ell-1} \mel**{\Psi_{\ell}}{S^{z}_{j} S^{z}_{j+i}}{\Psi_{\ell}},
\end{equation}
where we identify $\vec{S}_j = \vec{S}_{j+\mathcal{N}}$ for the periodic chain with $\ell=\mathcal{N}$.

We now return to the two-dimensional problem. Let $N_{\ell}$ denote the number of spin chains of length $\ell$ present in a given tiling of the $\mathcal{N} \times \mathcal{N}$ system. Averaging over the SSE ensemble, the full spin-spin correlator can then be written as
\begin{equation}
	\mathcal{S} \left( i \right) = \frac{1}{3\mathcal{N}^2} \sum_{\ell=1}^{\mathcal{N}} N_{\ell} C_{\ell}(i).
\end{equation}
Here, the factor $3\mathcal{N}^2$ is the total number of orbitals in the system. We are interested in extracting the chain-length distribution from the measured $\mathcal{S}(i)$. It is more convenient to parameterize this distribution in terms of the fraction of occupied sites belonging to spin chains of length $\ell$, defined as $f_{\ell} = \frac{\ell N_{\ell}}{\nu \mathcal{N}^2}$. At sufficiently low temperatures and for $\nu \leq 3$, the large intra-valley repulsion suppresses double occupancy of orbitals, so that the low-energy configurations are dominated by singly occupied orbitals arranged into spin chains. Within this approximation, the fractions $f_{\ell}$ satisfy the normalization condition $\sum_{\ell=1}^{\mathcal{N}} f_{\ell} = 1$. The spin-spin correlator becomes
\begin{equation}
	\label{eqn:chain_fitting_equation}
	\mathcal{S} \left( i \right) = \frac{\nu}{3} \sum_{\ell=1}^{\mathcal{N}} f_{\ell} \frac{C_{\ell} \left(i\right)}{\ell}. 
\end{equation}
In the full electronic model, residual charge fluctuations reduce the magnitude of the local moment relative to the ideal spin-only limit. In order to facilitate a sensible comparison of the measured correlator with the spin-chain description above, we therefore rescale $\mathcal{S}(i)$ by its onsite value, so that $\mathcal{S}(0)$ matches the spin-only value $1/4$. With this normalization, \cref{eqn:chain_fitting_equation} can be written for all fillings as
\begin{equation}
	\label{eqn:chain_fitting_equation_final}
	\frac{\mathcal{S} \left( i \right)}{4\mathcal{S} \left( 0 \right)} = \sum_{\ell=1}^{\mathcal{N}} f_{\ell} \frac{C_{\ell} \left(i\right)}{\ell},
\end{equation}
where the normalization condition on $f_{\ell}$ is automatically incorporated by the $i=0$ equation. Thus, the unknowns are the fractions $f_{\ell}$, while the constraints are provided by the measured values of $\mathcal{S}(i)$ at different separations.

Because of $C_{2x}$ symmetry, the spin correlator satisfies $\mathcal{S}(i) = \mathcal{S}(-i)$. Together with $C_{\ell}(i) = C_{\ell}(-i)$, this implies that \cref{eqn:chain_fitting_equation_final} provides only $\lfloor \mathcal{N}/2 \rfloor + 1$ independent constraints for the $\mathcal{N}$ unknown fractions $f_{\ell}$. For our simulations with $\mathcal{N}=6$, this gives four independent constraints for six possible chain lengths. We therefore solve for the chain-length distribution by minimizing the residual of \cref{eqn:chain_fitting_equation_final} as a constrained least-squares problem with $f_{\ell} \geq 0$ and $\sum_{\ell} f_{\ell}=1$. Since the problem is underdetermined, we also allow for sparse solutions by setting two of the six fractions to zero. Testing the possible choices, we find that the best description of the measured spin-spin correlator is obtained by setting $f_{4}=f_{5}=0$ and fitting the remaining four fractions.

\begin{table}[!t]
\begin{tabular}{|c|c|c|c|c|c|c|}
			\hline
			$\ell$ & 1 & 2 & 3 & 4 & 5 & 6\\ \hline
			$\mathcal{E}_{\ell}/\ell$ & $0$ & $-\frac14$ & $-\frac16$ & $-\frac{\sqrt{3}}{8}$ & $-0.185577$ & $\frac{1-\sqrt{13}}{12}$ \\ \hline
		\end{tabular}\caption{Ground-state energy per site for the spin-chain Hamiltonian in \cref{eqn:spin_chain_ham} for $\mathcal{N}=6$.}
	\label{tab:en_spins}
\end{table}

In \crefrange{fig:spin_corr_ins:d}{fig:spin_corr_ins:f}, we show the spin-chain length distribution as a function of inverse temperature down to $\beta=40$. At $\nu=1,2$, we find that most occupied sites belong to spin chains of length two, but significant chain-length fluctuations \emph{do} persist even at the lowest temperature we study. This can be understood by noting that, while local-moment fluctuations become quenched already at an inverse temperature $\beta \sim J^{-1} = 3$, fluctuations in the spin-chain length cost only a fraction of $J$, as shown in \cref{tab:en_spins}, and therefore survive down to much lower temperatures. By contrast, at $\nu = 3$, the fact that $U' < U$ already favors equal occupation of all valleys, and the system forms length-six chains already at relatively high temperature, around $\beta \approx 15$.

\subsubsection{Entropy}\label{sec:results:spin_dimers:entropy}

In addition to the spin-chain length distribution, we also estimate the entropy of the system at integer fillings as a function of temperature. To do so, we employ thermodynamic integration~\cite{FRE01}
\begin{align}
	S \left( \beta_2, \mu \right) - S \left( \beta_1, \mu \right) =\,\,& \beta_2 \left\langle H \right\rangle_{\beta_2,\mu} - \beta_1 \left\langle H \right\rangle_{\beta_1,\mu} \nonumber\\
	&- \int_{\beta_1}^{\beta_2} \dd{\beta'} \left\langle H \right\rangle_{\beta',\mu}, 
	\label{eqn:entropy}
\end{align}
where $S\left( \beta,\mu \right)$ is the entropy of the system, while $\left\langle H \right\rangle_{\beta,\mu}$ denotes the expectation value of its grand-canonical Hamiltonian at inverse temperature $\beta$ and chemical potential $\mu$. \Cref{eqn:entropy} is also derived in \cref{app:sec:sse}. At high temperatures ($\beta = 0.1$), we estimate $S \left( \beta, \mu \right)$ by perturbing in the kinetic Hamiltonian $H_0$ around the atomic limit of $H_I$, as explained in \cref{app:sec:anal}. 

The entropy per site, $S/\mathcal{N}^2$, is shown in \crefrange{fig:spin_corr_ins:h}{fig:spin_corr_ins:j} for the integer-filled correlated insulators. In all cases, a plateau appears around $\beta = 1$, corresponding to the local-moment entropy, with $S \approx \log {6 \choose \nu}$. Accordingly, the entropy increases toward charge neutrality in this regime. However, as the temperature is lowered and the spins become antiferromagnetically coupled, this hierarchy reverses, and the entropy instead increases away from half filling.

At $\nu = 3$, the entropy already vanishes around $\beta \approx 15$, at a temperature scale that coincides with the quenching of the spin-chain length fluctuations. By contrast, at $\nu=1,2$ the chain length still fluctuates, and the system therefore retains a finite entropy. We note that even at $\beta = 40$, the spin-chain length fluctuations are not yet fully quenched. As a result, $S$ has not yet reached the finite value predicted in Ref.~\cite{LI25} at $\nu = 1$, nor has it vanished at $\nu = 2$, where the system does not possess an exponentially degenerate ground state.

Finally, another signature of the phases predicted in strong coupling in Ref.~\cite{LI25} can be obtained by inspecting the momentum- and valley-resolved charge susceptibilities, shown in \crefrange{fig:spin_corr_ins:h}{fig:spin_corr_ins:j}. Although each lattice site in \crefrange{fig:spin_corr_ins:a}{fig:spin_corr_ins:d} has the same total occupation, a richer structure emerges upon projection onto a single valley. At $\nu=3$, each orbital has the same occupation, leading to a featureless $\chi_{00}\left( \vec{k} \right)$. At $\nu=2$, the charge distribution in valley $\eta=0$ reflects the breaking of moir\'e translation symmetry down to either a $\sqrt{3} \times \sqrt{3}${} or a $3 \times 3$ supercell, and correspondingly exhibits peaks at the $\mathrm{K}_M$ point and at $\frac13\vec{b}_{M_1}$, together with symmetry-related momenta. At $\nu=1$, the disordered occupation of the $\eta=0$ valley leads to the absence of peaks in the valley-resolved susceptibility $\chi_{00} \left( \vec{k} \right)$.

\section{Concluding Remarks}\label{sec:conclusions}

In this work, we have investigated the mixed-dimensional limit of AA t-\ch{SnSe2}{} in a numerically unbiased manner using QMC. To do so, we significantly extended the SSE algorithm in order to overcome the issues of inter-chain particle-number equilibration that arise in our system because of the specific form of the kinetic Hamiltonian. This can be now seen as a new numerical method: \emph{mixed-dimensional} SSE. It is worth emphasizing the differences between our work and the previous study of a mixed-dimensional Hubbard model in Ref.~\cite{XU15}. First, our model is directly experimentally motivated, unlike that of Refs.~\cite{LI14,XU15}, which is more theoretical in nature. Second, Ref.~\cite{XU15} considers a limit which, translated into our notation, is equivalent to $U \to \infty$, thereby projecting out doubly occupied states within a single orbital, and focuses on a two-orbital square-lattice model rather than our three-orbital triangular-lattice model. Beyond the fact that the $U \to \infty$ limit misses much of the physics highlighted here and in the companion paper~\cite{VAS26}, including the intermediate-coupling regime and the strong valley fluctuations, the two models also require substantially different SSE implementations.

Our findings can be probed directly in near-future experiments on AA t-\ch{SnSe2}{}. Resistivity and inverse-compressibility measurements have already been applied extensively to other moir\'e materials, such as moir\'e graphene~\cite{CAO18,CAO18a,LU19,TOM19,ZON20,PIE21,ROZ21,SAI21a,JAO22,LIU22b,ZHA25a}, and can be employed here to probe the hierarchy of correlated insulators, the Pomeranchuk effect, and the Wigner-Mott states. The twist angle provides a means to tune the interaction strength, while the distance between the sample and the screening gate can be used to tune the nearest-neighbor repulsion. In addition, scanning tunneling microscopy (STM) can probe the spatial structure of the Wigner-Mott states. Finally, the formation of spin-dimer states can be inferred from the reversal in the entropy hierarchy as a function of filling at integer fillings: while for $\beta < J$ the entropy is dominated by local moments and increases toward half filling, for $\beta > J$ it is controlled by spin-chain-length fluctuations and increases away from half filling.

The present study has focused exclusively on so-called diagonal observables, {\it i.e.}{} correlators of the charge density~\cite{SAN19}, which are readily accessible within SSE. Our algorithm can also be used to study off-diagonal observables such as fermion Green's functions~\cite{DOR01,DOR02}. The resulting spectral functions are of immediate interest for future STM experiments and for the possibility of investigating these materials with the newly developed quantum twisting microscope~\cite{INB23,XIA25,KLE26,LEE26}. Equally important are comprehensive transport simulations of the conductivity and its temperature dependence, which can be estimated directly within SSE in the mixed-dimensional limit~\cite{MOR16a} and, away from this limit, perhaps perturbatively with the help of bosonization. A related perturbative approach could also be used to study magnetotransport to leading order in the inter-chain hopping $t_{\perp}$~\cite{MEN21}. We leave these directions for future work.

Moving away from the quasi-1D limit, our present results provide an unperturbed starting point for future perturbative calculations. For example, our simulations can be used to extract the Luttinger parameters~\cite{GIA14} away from integer fillings and then investigate the stability of the corresponding crossed sliding phase as a function of $t_{\perp}$ and temperature. Of particular importance are also anti-Hund's inter-valley exchange interactions generated by electron-phonon coupling, which might promote superconductivity through a mechanism reminiscent of that occurring in fullerenes~\cite{CAP02,HAN03,NOM16,NOM15,HOS17}, especially in light of the significant inter-valley electron-phonon coupling in \ch{SnSe2}.

In closing, we discuss the physical regimes that we believe will be accurately captured by the mixed-dimensional Monte Carlo technique, thereby clarifying the range of problems to which it may be applicable. First, we note that, as an unbiased numerical method,  QMC provides an essentially exact solution of the mixed-dimensional problem itself. Importantly, arbitrarily strong inter-chain couplings that preserve the independent $\mathrm{U} \left( {1} \right)$ charge on each chain can be incorporated at the same level of accuracy with little or no additional effort. This distinguishes the mixed-dimensional approach from traditional coupled-chain techniques in Monte Carlo or DMRG, where even such couplings are typically treated perturbatively at mean-field level. However, we note that these chain mean-field techniques can include chain-$\mathrm{U} \left( {1} \right)$-breaking terms with little additional effort, and hence probe a complementary regime of parameter space~\cite{GEO00,BIE02,GIA04}. Consequently, a broad class of strongly interacting multi-chain problems that would ordinarily be analyzed using approximate coupled-wire constructions can instead be studied non-perturbatively and with fewer approximations.

We therefore focus on deviations from the idealized mixed-dimensional limit, namely those perturbations that explicitly break the chain $\mathrm{U} \left( {1} \right)$ symmetry by introducing inter-chain tunneling and are inevitably present, even if weakly, in realistic materials. As we have flagged already, accessing the single-particle Green's functions allows us to analyze related spectral and transport signatures of the mixed-dimensional fixed point. Beyond this, it can be used to extract the scaling dimensions of various chain-$U(1)$-breaking inter-chain operators, or eqivalently their associated susceptibilities. For instance, such an approach could be used to examine the relevance of single-electron tunneling operators that drive the onset of 2D coherent Fermi-liquid behaviour, or Josephson tunnelings that lead to a superconducting instability. As noted, computing Green's functions requires additional measurements within the worm algorithm that are technically involved but conceptually straightforward, and that we expect to implement in future work. Such an analysis of leading instabilities of the mixed-dimensional fixed point is likely to yield significant insight into 2D orders in doping regimes away from where we identify the more ``classical'' Mott and Wigner--Mott phases in the present work.

In cases where including inter-chain tunneling destabilizes the mixed-dimensional fixed point and seeds a flow away from it, perturbative control is lost once various generated couplings become $\mathcal{O}(1)$. Nevertheless, the perturbative analysis can still provide valuable insight into the competition between different ordering tendencies and the hierarchy of energy scales encountered {\it en route} to strong coupling. In regimes where the metallic state on each wire admits a Luttinger-liquid description, the mixed-dimensional metal may be viewed as a crossed sliding Luttinger liquid composed of incoherent one-dimensional quasiparticles. While weak chain-charge-breaking perturbations are expected to restore coherent transverse propagation below a sufficiently low crossover scale, we expect the finite-temperature physics above this scale to remain reasonably well described by the mixed-dimensional fixed point.

More broadly, because a wide class of interactions can be treated nonperturbatively and exactly by unbiased Monte Carlo sampling, there exists in principle a large manifold of mixed-dimensional fixed points within the parameter space of generic multiorbital models. Whether this manifold lies sufficiently close to realistic materials to be physically relevant is, of course, a more delicate question. For the M-point systems studied here, there are natural reasons to argue in the affirmative, since the mixed-dimensional limit emerges directly from the underlying band structure. However, other models that appear much less obviously ``mixed dimensional'' may nevertheless admit an effective description in which a similar approach proves fruitful. Exploring this possibility remains an intriguing direction for future work.

\begin{acknowledgments}

	We thank Johannes S. Hofmann for collaboration on a companion DQMC study at half-filling~\cite{VAS26}. We are grateful to Benedikt Placke, Ansgar Dorneich, Roger Melko, and Hong Yao for technical discussions on the SSE method. We also thank Kin Fai Mak, Jie Shan, Dmitri K. Efetov, Zhengchao Xia, Ming-Rui Li, Yi Jiang, Hanqi Pi, Fabian H. Essler, Eugene Demler, Shivaji L. Sondhi, Paul Fendley, Antoine Georges, Debanjan Chowdhury, and Philipp Werner for useful discussions. We acknowledge support from a UKRI Frontier Research Consolidator Grant (under the Horizon Europe Guarantee, Grant No. EP/Z002419/1, S.A.P. and D.C.). D.C. and W.K. gratefully acknowledge support from the Leverhulme Trust. K.V. acknowledges funding from Leverhulme Trust International Professorship Grant No. LIP-202-014. D.C. also thanks the Aspen Center for Physics, where part of this work was carried out. B.A.B. was supported by the Gordon and Betty Moore Foundation through Grant No. GBMF8685 towards the Princeton theory program, the Gordon and Betty Moore Foundation’s EPiQS Initiative (Grant No. GBMF11070), the Global Collaborative Network Grant at Princeton University, the Simons Investigator Grant No. 404513, the NSF-MERSEC (Grant No. MERSEC DMR 2011750), the Simons Collaboration on New Frontiers in Superconductivity (Grant No. SFI-MPS-NFS-00006741-01), Princeton Catalysis Initiative (PCI), the Schmidt Foundation at the Princeton University and the National Science Foundation through the AI Research Institutes program Award No. DMR-2433348. 
\end{acknowledgments}

\renewcommand{\addcontentsline}[3]{}

\let\addcontentsline\oldaddcontentsline

\renewcommand{\thetable}{S\arabic{table}}
\renewcommand{\thefigure}{S\arabic{figure}}
\renewcommand{\theequation}{S\arabic{section}.\arabic{equation}}
\onecolumngrid
\pagebreak
\thispagestyle{empty}
\newpage
\begin{center}
	\textbf{\large Supplementary Information for ``Mixed-dimensional quantum Monte Carlo studies of M-point moiré materials{}''}\\[.2cm]
\end{center}

\appendix
\renewcommand{\thesection}{\Roman{section}}
\tableofcontents
\let\oldaddcontentsline\addcontentsline
\newpage

\section{Model for twisted AA-stacked \ch{SnSe2}}\label{app:sec:model}

In this \siSection{}, we discuss the model studied in this work. We begin by formalizing the notation and reviewing the single-particle model for twisted AA-stacked \ch{SnSe2}  (AA t-\ch{SnSe2}{}), both at the continuum-model level~\cite{CAL25b} and in its projected Wannier representation~\cite{LI25}. We then derive the projected interaction Hamiltonian, and its truncated approximation by a nearest-neighbor density-density interaction in the Wannier basis. As explained in Ref.~\cite{LI25}, the projected interaction Hamiltonian lies close to a valley-spin $\mathrm{U} \left( {6} \right)$-symmetric limit, with small $\mathrm{U} \left( {6} \right)$ symmetry-breaking effects driving the system toward a valley-polarized state at integer fillings of the moir\'e unit cell. To accurately capture these effects within our truncated Wannier interaction Hamiltonian, we perform a mean-field calculation of the critical temperature for valley polarization and require that the truncated and the full-fledged projected interaction Hamiltonians agree in their mean-field critical temperature for valley polarization. We also briefly discuss inter-valley Hund's and anti-Hund's interaction effects, which originate from the short-wavelength components of the Coulomb interaction and from electron-phonon coupling, but which are otherwise small and therefore neglected in our treatment. Finally, we summarize the complete many-body projected Hamiltonian and its parameters, which are then used in the numerical simulations.

\subsection{Notation and single-particle Hamiltonian}\label{app:sec:model:notation}

We begin by outlining our notation, which is consistent with that of Refs.~\cite{CAL25b,LI25,VAS26}. Let $\hat{a}^\dagger_{\vec{p},s,l}$ denote the fermion operator in the plane-wave basis of monolayer $l = \pm$. The momentum $\vec{p}$ is measured from the $\Gamma$ point of the monolayer Brillouin zone (BZ), $s=\uparrow,\downarrow$ is the spin index. We define $\vec{K}^l_M$ as the M point in the BZ of layer $l$, with $\vec{K}^+_M$ and $\vec{K}^-_M$ differing by a twist angle $\theta$. For concreteness, we assume that $\vec{K}^\pm_M$ lies along the direction forming an angle $\pm \theta/2$ with the $\hat{y}$ axis. Each monolayer contains three valleys labeled by $\eta=0,1,2$, located at momenta $C^{\eta}_{3z}\vec{K}_{\pm}$, which correspond to the three single-particle–decoupled valleys of the moir\'e Hamiltonian.
We also introduce three two-dimensional momenta,
\begin{align}
	\label{app:eqn:q_vecs}
	\vec{q}_n &= C^{n}_{3z} \left( \vec{K}^{-}_{M} - \vec{K}^{+}_{M} \right),
	\qq{for} 0 \leq n \leq 2, \\
	\vec{q}_0 = k_\theta \left(1, 0\right), \qquad
	\vec{q}_1 &= k_\theta \left(-\frac{\sqrt{3}}{2}, \frac{1}{2}\right), \qquad
	\vec{q}_2 = k_\theta \left(-\frac{\sqrt{3}}{2}, -\frac{1}{2}\right),
\end{align}
whose coordinates are given in the $\left(k_x,k_y\right)$ basis. The magnitude $k_\theta=\abs{\vec{K}^{-}_{M} - \vec{K}^{+}_{M}} = 2\abs{\mathbf{K}_{+}}\sin\left(\theta/2\right)$ is set by the twist angle $\theta$. The reciprocal moir\'e lattice $\mathcal{Q} = \mathbb{Z}\vec{b}_{M_1} + \mathbb{Z}\vec{b}_{M_2}$ is generated by the reciprocal vectors,
\begin{equation}
	\vec{b}_{M_1}=2\vec{q}_0 = k_\theta \left( 2, 0 \right),\qquad  
	\vec{b}_{M_2}=-2\vec{q}_2 = k_\theta \left( \sqrt{3}, 1 \right),
\end{equation}
which also determine the associated moir\'e BZ (MBZ). The continuum M-point moir\'e Hamiltonian is defined with the aid of three additional momentum lattices,
\begin{equation}
	\label{app:eqn:three_mom_lattices_definition}
	\mathcal{Q}_{n} \equiv \mathcal{Q} + \vec{q}_{n}, \qq{for} 0 \leq n \leq 2,
\end{equation}
which together form a kagome lattice. We extend the definition of both $\mathcal{Q}_n$ and $\vec{q}_n$ to all integers by treating the index $n$ \textit{modulo}~3,
\begin{equation}
	\label{app:eqn:extension_with_modulus}
	\mathcal{Q}_{n} \equiv \mathcal{Q}_{n  \ \mathrm{mod} \ 3} \qq{and} \vec{q}_n \equiv \vec{q}_{n  \ \mathrm{mod} \ 3}, \qq{for} n \in \mathbb{Z}.
\end{equation}

The low-energy physics of the M-point moir\'e materials is governed by the moir\'e plane-wave operators,
\begin{equation}
	\label{app:eqn:low_en_ops_c}
	\hat{c}^\dagger_{\vec{k}, \vec{Q}, s, l} \equiv \hat{a}^\dagger_{C^{\eta}_{3z} \vec{K}^{l}_{M} + \vec{k} - \vec{Q}, s, l}, \qq{for} \vec{Q} \in  \mathcal{Q}_{\eta + l} \qq{and} \vec{k} \in \text{MBZ},
\end{equation}
where MBZ denotes the first moir\'e Brillouin zone. The definition of the operators $\hat{c}^\dagger_{\vec{k},\vec{Q},s,l}$ can be extended beyond the first MBZ via the embedding relation
\begin{equation}
	\hat{c}^\dagger_{\vec{k} + \vec{G}, \vec{Q}, s, l} = \hat{c}^\dagger_{\vec{k}, \vec{Q} - \vec{G}, s, l}, \qq{for} \vec{G} \in \mathcal{Q}.
\end{equation}
In the basis spanned by these moir\'e plane-wave operators, the single-particle Hamiltonian of AA t-\ch{SnSe2}{} takes the form
\begin{equation}
	\label{app:eqn:M-pt_cont}
	\hat{\mathcal{H}}_{0} = \sum_{\eta} \sum_{\substack{s,l \\ s',l'}} \sum_{\substack{\vec{Q} \in \mathcal{Q}_{\eta + l} \\ \vec{Q}' \in \mathcal{Q}_{\eta + l'}}} \left[ h_{\vec{Q},\vec{Q}'} \left( \vec{k} \right) \right]_{s l; s' l'} \hat{c}^\dagger_{\vec{k},\vec{Q},s,l} \hat{c}_{\vec{k},\vec{Q}',s',l'} \approx \sum_{s,\eta} \sum_{l,l'} \sum_{\substack{\vec{Q} \in \mathcal{Q}_{\eta + l} \\ \vec{Q}' \in \mathcal{Q}_{\eta + l'}}} \left[ h_{\vec{Q},\vec{Q}'} \left( \vec{k} \right) \right]_{l l'} \hat{c}^\dagger_{\vec{k},\vec{Q},s,l} \hat{c}_{\vec{k},\vec{Q}',s,l'},
\end{equation}
where $\left[ h_{\vec{Q},\vec{Q}'} \left( \vec{k} \right) \right]_{s l; s' l'}$ denotes the full spin-resolved moir\'e Hamiltonian matrix, and where we define
\begin{equation}
	\left[ h_{\vec{Q},\vec{Q}'} \left( \vec{k} \right) \right]_{ll'} = \frac{1}{2} \sum_{s} \left[ h_{\vec{Q},\vec{Q}'} \left( \vec{k} \right) \right]_{s l; s' l'}.
\end{equation} 
Since AA t-\ch{SnSe2}{} only weakly breaks spin $\mathrm{SU} \left( {2} \right)$ symmetry, we adopt the approximation on right-hand side of \cref{app:eqn:M-pt_cont} and neglect the small spin $\mathrm{SU} \left( {2} \right)$ symmetry-breaking effects~\cite{CAL25b}, which Ref.~\cite{CAL25b} estimates to less than $1.5\%$ in wave function for the first moir\'e conduction band.

\subsubsection{Wannier model for the first conduction band}\label{app:sec:model:notation:wannier}

As in Ref.~\cite{LI25}, we introduce the band basis for AA t-\ch{SnSe2}{},
\begin{equation}
	\sum_{l'} \sum_{\vec{Q}' \in \mathcal{Q}_{\eta + l'}}\left[ h_{\vec{Q},\vec{Q}'} \left( \vec{k} \right) \right]_{ll'} u_{\vec{Q}',l';\eta, n } \left( \vec{k} \right) =  \epsilon_{\eta,n} \left( \vec{k} \right) u_{\vec{Q},l;\eta, n }, \left( \vec{k} \right),
\end{equation}
where $u_{\vec{Q},l;\eta,n}0\left(\vec{k}\right)$ and $\epsilon_{\eta,n}0\left(\vec{k}\right)$ denote, respectively, the Bloch eigenfunction and the energy of the $n$th conduction band ($n=1,2,\dots$) in valley $\eta$ at moir\'e momentum $\vec{k}$. For the twist angles $\theta$ considered in Ref.~\cite{CAL25b} ($3.89 \leq \theta/\si{\degree} \leq 9.43$), there are either one ($6.01 \leq \theta/\si{\degree} \leq 9.43$) or two ($3.89 \leq \theta/\si{\degree} \leq 5.09$) spinful, gapped conduction bands in each valley. Since the low-energy physics of AA t-\ch{SnSe2}{} is dominated by the lowest conduction band, in this work, as in Ref.~\cite{LI25}, we focus exclusively on the first spinful conduction band in each valley and project the Hamiltonian onto this subspace. We therefore define a set of Wannier orbitals for the first conduction band in each valley, whose representation in terms of the moir\'e plane-wave fermion operators is given by
\begin{equation}
	\label{app:eqn:k_pt_wanniers_m_pt}
	\hat{f}^\dagger_{\vec{k},\eta,s} = \sum_{l} \sum_{\vec{Q} \in \mathcal{Q}_{\eta + l}} u_{\vec{Q},l;\eta, 1} \left( \vec{k} \right) \hat{c}^\dagger_{\vec{k},\vec{Q},s,l}.
\end{equation}
The first conduction bands of AA t-\ch{SnSe2}{} are topologically trivial~\cite{CAL25b} and, therefore, one can always choose a smooth gauge for the Bloch wave functions $u_{\vec{Q},l;\eta,1}\left(\vec{k}\right)$ that satisfies the appropriate embedding condition~\cite{LI25},
\begin{equation}
	u_{\vec{Q},l;\eta, 1 } \left( \vec{k} + \vec{G} \right) = u_{\vec{Q}-\vec{G},l;\eta, 1 } \left( \vec{k} \right), \qq{for any} \vec{G} \in \mathcal{Q}.
\end{equation}
As a consequence, the corresponding real-space Wannier orbitals for the first conduction band in each valley can be obtained by Fourier-transforming the operators defined in \cref{app:eqn:k_pt_wanniers_m_pt}
\begin{equation}
	\label{app:eqn:k_pt_wannier_m_pt}
	\hat{f}^\dagger_{\vec{R},\eta,s} = \frac{1}{\sqrt{N}} \sum_{\vec{k}} \hat{f}^\dagger_{\vec{k},\eta,s} e^{-i \vec{k} \cdot \vec{R}},
\end{equation}
where $N$ denotes the number of moir\'e unit cells and $\vec{R} \in \mathbb{Z}\vec{a}_{M_1} + \mathbb{Z}\vec{a}_{M_2}$ labels the sites of the moir\'e lattice, with 
\begin{equation}
	\vec{a}_{M_1} = \frac{\pi}{k_\theta} \left(\sqrt{3}, -1 \right) \qq{and}  
	\vec{a}_{M_2} = \frac{\pi}{k_\theta} \left(0, 2 \right).
\end{equation}
After defining the real-space continuum moir\'e fermionic fields
\begin{align}
	\hat{\psi}^\dagger_{\eta,s,l} \left( \vec{r} \right) &= \frac{1}{\sqrt{\Omega}} \sum_{\vec{k} \in \text{MBZ}} \sum_{\vec{Q} \in \mathcal{Q}_{\eta + l}} \hat{c}^\dagger_{\vec{k},\vec{Q},s,l} e^{-i \left( \vec{k} - \vec{Q} \right) \cdot \vec{r}}, \label{app:eqn:def_real_space_fermions_to_real}\\
	\hat{c}^\dagger_{\vec{k},\vec{Q},s,l} &= \frac{1}{\sqrt{\Omega}} \int \dd[2]{r} \hat{\psi}^\dagger_{\eta,s,l} \left( \vec{r} \right) e^{i \left( \vec{k} - \vec{Q} \right) \cdot \vec{r}}, \qq{for} \vec{Q} \in \mathcal{Q}_{\eta+l},\label{app:eqn:def_real_space_fermions_to_momentum}
\end{align}
where $\Omega$ denotes the surface area of the sample, we can also construct the real-space Wannier wave-functions for the first moir\'e conduction band. The latter are defined by\footnote{The phase factor $e^{-i\vec{q}_{\eta+l}\cdot\vec{r}}$ multiplying $\hat{\psi}^\dagger_{\eta,s,l}0\left(\vec{r}\right)$ is conventional and is introduced to ensure that Wannier wave functions centered at different lattice sites are identical and do not differ by a position-dependent phase.}
\begin{align}
	W_{\eta,l}\left(\vec{r} - \vec{R} \right) \equiv& \mel**{0}{\hat{\psi}_{\eta,s,l} \left( \vec{r} \right) e^{ i \vec{q}_{\eta+l} \cdot \vec{r}} \hat{f}^\dagger_{\vec{R},\eta,s}}{0} = \frac{1}{\sqrt{N}} \sum_{\vec{k}} \sum_{\vec{Q} \in \mathcal{Q}_{\eta + l}} u_{\vec{Q},l;\eta,1} \left( \vec{k} \right) e^{i \left( \vec{k}-\vec{Q} \right) \cdot \vec{r}} e^{ i \vec{q}_{\eta+l} \cdot \vec{r}} e^{-i \vec{k} \cdot \vec{R}} \nonumber \\
	=& \frac{1}{\sqrt{N}} \sum_{\vec{k}} \sum_{\vec{G} \in \mathcal{Q}} u_{\vec{G} + \vec{q}_{\eta + l},l;\eta,1} \left( \vec{k} \right) e^{i \left( \vec{k}-\vec{G} - \vec{q}_{\eta + l} + \vec{q}_{\eta + l} \right) \cdot \vec{r}} e^{-i \vec{k} \cdot \vec{R}} \nonumber \\
	=& \frac{1}{\sqrt{N}} \sum_{\vec{k}} \sum_{\vec{G} \in \mathcal{Q}} u_{\vec{G} + \vec{q}_{\eta + l},l;\eta,1} \left( \vec{k} \right) e^{i \left( \vec{k}-\vec{G} \right) \cdot \left(\vec{r} - \vec{R}\right)}. \label{app:eqn:f_wannier_definition}
\end{align}
In practice, we follow the gauge-smoothing procedure only in the $\eta=0$ valley and construct the Wannier wave functions in the remaining valleys using $C_{3z}$ symmetry~\cite{LI25}. Concretely, we take
\begin{equation}
	u_{\vec{Q},l,\eta,1}  \left(\vec{k} \right) = u_{C^{-\eta}_{3z}\vec{Q},l,0,1} \left( C^{-\eta}_{3z} \vec{k} \right). 
\end{equation}

The Wannier centers of the orbitals spanning the first conduction band are not located at the unit-cell origin, but instead lie close to a honeycomb Wyckoff position, namely $\vec{r}_{\text{H}} \equiv - \frac{1}{3} \vec{a}_{M_1} + \frac{1}{3} \vec{a}_{M_2}$, such that
\begin{equation}
	 \sum_{l} \int \dd[2]{r} \vec{r} \abs{W_{\eta,l} \left( \vec{r} - \vec{R} \right)}^2 = \vec{R} + C^{\eta}_{3z} \left( - \frac{1}{3} \vec{a}_{M_1} + \frac{1}{3} \vec{a}_{M_2} + \Delta \vec{r}_{0} \right) = \vec{R} + C^{\eta}_{3z} \vec{r}_{\text{H}} + \Delta \vec{r}_{\eta},
\end{equation}
where $\Delta \vec{r}_{\eta} \equiv C^{\eta}_{3z} \Delta \vec{r}_{0}$ denotes a small deviation satisfying $\abs{\vec{r}_{\eta}} \ll \abs{\vec{a}_{M_{1}}}$. As in Ref.~\cite{LI25}, we would like the Wannier orbitals from the three different valleys associated with a given unit cell $\vec{R}$ to have their Wannier centers as close as possible. To this end, we define the following \emph{translated} Wannier orbitals $\hat{d}^\dagger_{\vec{R},\eta,s}$,
\begin{equation}
	\label{app:eqn:shifted wannier orbitals}
	\hat{d}^\dagger_{\vec{R},\eta,s} = \hat{f}^\dagger_{\vec{R} + \boldsymbol{\delta}_\eta,\eta,s}, \qq{with} \boldsymbol{\delta}_{0} = \vec{0}, \, \boldsymbol{\delta}_{1} = \vec{a}_{M_2}, \, \boldsymbol{\delta}_{2} = - \vec{a}_{M_1}.
\end{equation}
By construction, the Wannier orbitals $\hat{d}^\dagger_{\vec{R},\eta,s}$ have their centers located at $\vec{R} + \vec{r}_{\text{H}} + \Delta \vec{r}_{\eta}$. Their charge density $\hat{d}^\dagger_{\vec{R},\eta,s}$ is then given by
\begin{equation}
	n_{\eta} \left( \vec{r} \right) \equiv \sum_{l} \abs{W_{\eta,l} \left( \vec{r}  + C^{\eta}_{3z}\vec{r}_{\text{H}}  \right)}^2,
\end{equation}
where the origin of the charge-density profile has been shifted to $C^{\eta}_{3z}\vec{r}_{\text{H}}$ in valley $\eta$. The Wyckoff position $\vec{r}_{\text{H}}$ is invariant under both $C_{3z}$ and $C_{2x}$ symmetries of the unit cell. Accordingly, we define the operations $C'_{3z}$ and $C'_{2x}$ as the out-of-plane three-fold and in-plane two-fold rotations about $\vec{r}_{\text{H}}$,
\begin{align}
	C'_{3z} \vec{r} &\equiv C_{3z} \left( \vec{r} - \vec{r}_{\text{H}} \right) + \vec{r}_{\text{H}} = C_{3z} \vec{r} + \frac13 \vec{a}_{M_1} + \frac23 \vec{a}_{M_2} - \frac13 \vec{a}_{M_1} + \frac13 \vec{a}_{M_2} = C_{3z} \vec{r} + \vec{a}_{M_2},  \\
	C'_{2x} \vec{r} &\equiv C_{2x} \left( \vec{r} - \vec{r}_{\text{H}} \right) + \vec{r}_{\text{H}} = C_{2x} \vec{r} + \frac13 \vec{a}_{M_1} + \frac23 \vec{a}_{M_2} - \frac13 \vec{a}_{M_1} + \frac13 \vec{a}_{M_2} = C_{2x} \vec{r} + \vec{a}_{M_2}.
\end{align}
The symmetry operations $C'_{3z}$ and $C'_{2x}$ act on the Wannier orbitals $\hat{d}^\dagger_{\vec{R},\eta,s}$ as
\begin{equation}
	C'_{3z} \hat{d}^\dagger_{\vec{R},\eta,s} \left( C'_{3z} \right)^{-1} = \hat{d}^\dagger_{C_{3z} \vec{R},\eta+1,s}, \quad
	C'_{2x} \hat{d}^\dagger_{\vec{R},\eta,s} \left( C'_{2x} \right)^{-1} = \hat{d}^\dagger_{C_{2x} \vec{R},-\eta,s}.
\end{equation}
In addition, the charge density of the Wannier orbitals is also invariant under these operations, satisfying
\begin{equation}
n_{\eta} \left( \vec{r} \right) = n_{\eta+1} \left( C_{3z} \vec{r} \right), \quad 
n_{\eta} \left( \vec{r} \right) = n_{-\eta} \left( C_{2x} \vec{r} \right).
\end{equation}
The Wannier centers for the $\hat{f}^\dagger_{\vec{R},\eta,s}$ and the $\hat{d}^\dagger_{\vec{R},\eta,s}$ fermions are represented schematically in \cref{app:fig:model:a}. 

The single-particle Hamiltonian projected into the Wannier orbitals takes the form
\begin{equation}
	\label{app:eqn:full_proj_Ham}
	H_0 = \sum_{\vec{R},\Delta \vec{R}} \sum_{\eta,s} t_{\eta} \left( \Delta \vec{R} \right) \hat{d}^\dagger_{\vec{R},\eta,s} \hat{d}_{\vec{R} + \Delta \vec{R},\eta,s},
\end{equation}
where the kinetic-energy operator obeys both $C_{3z}$ and $C_{2x}$ symmetries,
\begin{equation}
	t_{\eta} \left( \Delta \vec{R} \right) = t_{\eta+1} \left( C_{3z} \Delta \vec{R} \right), \quad 
	t_{\eta} \left( \Delta \vec{R} \right) = t_{-\eta} \left( C_{2x} \Delta \vec{R} \right),
\end{equation}
while Hermiticity and (spinless) time-reversal symmetry further require
\begin{equation}
	t_{\eta} \left( \Delta \vec{R} \right) = t^*_{\eta} \left( \Delta \vec{R} \right) = t_{\eta} \left( - \Delta \vec{R} \right).
\end{equation}
To distinguish it from the \emph{unprojected} ({\it i.e.}{}, plane-wave) moir\'e Hamiltonian $\hat{H}_0$, we henceforth denote the projected kinetic Hamiltonian of \cref{app:eqn:full_proj_Ham} by $H_0$, {\it i.e.}{}, without a hat.

The leading hopping amplitudes from \cref{app:eqn:full_proj_Ham} are defined as
\begin{equation}
	\label{app:eqn:leading_hoppings}
	t \equiv t_{0} \left( \vec{a}_{M_2} \right), \quad
	t_{\perp} \equiv t_{0} \left( 2 \vec{a}_{M_1} + \vec{a}_{M_2} \right), \quad
	t' \equiv t_{0} \left( \vec{a}_{M_1} \right),\quad
	t'' \equiv t_{0} \left( 2 \vec{a}_{M_2} \right), 
\end{equation}
where $t$ denotes the leading intra-chain hopping, $t_{\perp}$ the next-nearest inter-chain hopping, and $t'$ the nearest inter-chain hopping, which is forbidden by the momentum-space non-symmorphic $\tilde{M}_z$ symmetry~\cite{CAL25b}. The hopping amplitude are represented schematically in \cref{app:fig:model:b}. In this work, we neglect all inter-chain hopping processes and approximate the single-particle Hamiltonian by
\begin{equation}
	\label{app:eqn:single_particle_trucated}
	H_0 \approx t \sum_{\vec{R},\eta,s} \left( \hat{d}^\dagger_{\vec{R},\eta,s} \hat{d}_{\vec{R} + C^{\eta}_{3z} \vec{a}_{M_2},\eta,s} + \hat{d}^\dagger_{\vec{R},\eta,s} \hat{d}_{\vec{R} - C^{\eta}_{3z} \vec{a}_{M_2},\eta,s} \right).
\end{equation}

\subsubsection{Wannier-projected density operator}\label{app:sec:model:notation:rho_op}

To derive the interaction Hamiltonian of t-\ch{SnSe2}, we project the density operator onto the Wannier-orbital basis,
\begin{align}
	\rho \left( \vec{r} \right) &= \sum_{\eta,s,l} \hat{\psi}^\dagger_{\eta,s,l}  \left( \vec{r} \right) \hat{\psi}_{\eta,s,l} \left( \vec{r} \right) \nonumber \\
	&\approx \sum_{\eta,s,l} \sum_{\vec{R}_1,\vec{R}_2} W^*_{\eta,l} \left( \vec{r} - \vec{R}_1 \right) W_{\eta,l} \left( \vec{r} - \vec{R}_2 \right) \hat{f}^\dagger_{\vec{R}_1,\eta,s} \hat{f}_{\vec{R}_2,\eta,s} \nonumber \\
	&\approx \sum_{\eta,s,l} \sum_{\vec{R}} \abs{W_{\eta,l} \left( \vec{r}  - \vec{R} \right)}^2  \hat{f}^\dagger_{\vec{R},\eta,s} \hat{f}_{\vec{R},\eta,s} \nonumber \\
	&= \sum_{s,l} \sum_{\vec{R}} \left( \abs{W_{0,l} \left( \vec{r}  - \vec{R} \right)}^2  \hat{d}^\dagger_{\vec{R},0,s} \hat{d}_{\vec{R},0,s} + \abs{W_{1,l} \left( \vec{r}  - \vec{R} - \vec{a}_{M_2} \right)}^2  \hat{d}^\dagger_{\vec{R},1,s} \hat{d}_{\vec{R},1,s} + \abs{W_{2,l} \left( \vec{r}  - \vec{R} + \vec{a}_{M_1} \right)}^2  \hat{d}^\dagger_{\vec{R},2,s} \hat{d}_{\vec{R},2,s} \right) \nonumber \\
	&= \sum_{\eta, s,l} \sum_{\vec{R}} \abs{W_{\eta,l} \left( \vec{r}  - \vec{R} - \vec{r}_{\text{H}} + C^{\eta}_{3z} \vec{r}_{\text{H}} \right)}^2  \hat{d}^\dagger_{\vec{R},\eta,s} \hat{d}_{\vec{R},\eta,s} \nonumber \\		
	&= \sum_{\eta,s,l} \sum_{\vec{R}} n_{\eta} \left( \vec{r} - \vec{r}_{\text{H}} - \vec{R} \right) \hat{n}_{\vec{R},\eta,s}, \label{app:eqn:proj_density_operator}
\end{align}
where we have introduced the number operator $\hat{n}_{\vec{R},\eta,s} = \hat{d}^\dagger_{\vec{R},\eta,s}\hat{d}_{\vec{R},\eta,s}$. In \cref{app:eqn:proj_density_operator}, we neglect the rapidly oscillating valley-off-diagonal contributions to $\rho \left( \vec{r} \right)$. A discussion of these terms and their effects on the interaction Hamiltonian of AA t-\ch{SnSe2}{} is provided in \cref{app:sec:model:hund:ferro}. We further define the Fourier transform of the number operator as
\begin{align}
	\hat{n}_{\vec{k},\eta,s} &= \frac{1}{N} \sum_{\vec{k}} \hat{n}_{\vec{R},\eta,s} e^{i \vec{k} \cdot \vec{R}} = \frac{1}{N} \sum_{\vec{k}'} \hat{d}^\dagger_{\vec{k}'+\vec{k},\eta,s} \hat{d}_{\vec{k},\eta,s}, \label{app:eqn:orb_dens_ft_to_k} \\
	\hat{n}_{\vec{R},\eta,s} &= \sum_{\vec{k}} \hat{n}_{\vec{k},\eta,s} e^{-i \vec{k} \cdot \vec{R}}, \label{app:eqn:orb_dens_ft_to_r}
\end{align}
and, analogously, the Fourier transform of the Wannier-orbital charge-density distribution,
\begin{align}
	n_{\eta}\left( \vec{k} + \vec{G} \right) &=  \int \dd[2]{r} n_{\eta} \left( \vec{r} \right) e^{i \left( \vec{k} + \vec{G} \right) \cdot \vec{r}}, \label{app:eqn:orb_dens_func_ft_to_k}\\
	n_{\eta}\left( \vec{r} \right) &= \frac{1}{N \Omega_0}\sum_{\substack{\vec{k} \in \text{MBZ} \\ \vec{G} \in \mathcal{Q}}} n_{\eta} \left( \vec{k} + \vec{G} \right) e^{-i \left( \vec{k} + \vec{G} \right) \cdot \vec{r}}. \label{app:eqn:orb_dens_func_ft_to_r}
\end{align}
In \cref{app:eqn:orb_dens_func_ft_to_r}, $\Omega_0 = \Omega / N$ is the surface area of a moir\'e unit cell.
Using \cref{app:eqn:orb_dens_ft_to_r,app:eqn:orb_dens_func_ft_to_r}, the projected density operator can be written in momentum space as
\begin{equation}
	\rho \left( \vec{r} + \vec{r}_{\text{H}} \right) = \frac{1}{\Omega_0} \sum_{\substack{ \vec{k} \in \text{MBZ} \\ \vec{G} \in \mathcal{Q}}} \sum_{\eta,s} n_\eta \left( \vec{k} + \vec{G} \right) \hat{n}_{\vec{k},\eta,s} e^{-i \left( \vec{k} + \vec{G} \right) \cdot \vec{r}}. 
	\label{app:eqn:projected_density_oeprator_in_k_space}
\end{equation}
We will employ \cref{app:eqn:projected_density_oeprator_in_k_space} in \cref{app:sec:model:interaction} to derive the interaction Hamiltonian projected onto the Wannier basis for AA t-\ch{SnSe2}{}.

\subsection{Interaction Hamiltonian}\label{app:sec:model:interaction}

The unprojected interaction Hamiltonian for AA t-\ch{SnSe2}{} is given by
\begin{equation}
	\hat{H}_{I} = \frac{1}{2} \int \dd[2]{r_1} \dd[2]{r_2} V \left( \vec{r}_1 - \vec{r}_2 \right) \rho \left( \vec{r}_1 \right) \rho \left( \vec{r}_2 \right),
\end{equation}
where $\rho \left( \vec{r} \right)$ is the density operator introduced in the first line of \cref{app:eqn:proj_density_operator}, and $V \left( \vec{r} \right)$ is the screened gated Coulomb potential. In this work, we assume that AA t-\ch{SnSe2}{}, as in Ref.~\cite{LI25}, is placed in a dual-gated setup. In this geometry, the Coulomb potential and its Fourier transformation are given by
\begin{align}
	V \left( \vec{r} \right) &= \frac{e^2}{4 \pi \epsilon \epsilon_0 \xi } \sum_{n=-\infty}^{\infty} \frac{\left( -1 \right)^n}{\sqrt{\left( \frac{r}{\xi} \right)^2 + n^2}}= \int \frac{\dd[2]{q}}{(2 \pi)^2} V \left( \vec{q} \right) e^{- i \vec{q} \cdot \vec{r}}, \nonumber \\
	V \left( \vec{q} \right) &=  \frac{e^2}{4 \epsilon \epsilon_0}\frac{\tanh \left( q\xi/2  \right)}{q/ 2}.
\end{align}
Here $\epsilon$ denotes the dielectric constant and $\xi$ is the screening length, defined as the distance between the gates (with $e$ the electronic charge and $\epsilon_0$ the vacuum permittivity). We use $\epsilon$ as a tuning parameter to ensure that, for a given twist angle and screening length, the system remains in the intermediate-coupling regime, as discussed in more detail in \cref{app:sec:model:summary}.

To show that the density-density interaction decays exponentially with inter-orbital distance, as we discuss around \cref{app:eqn:density_density_wannier_fit}, it is also useful to determine the long-distance behavior of the dual-gated Coulomb potential. To this end, we employ the Poisson summation formula, which states that for a function $f(x)$ with Fourier transform
\begin{equation}
	\tilde{f} (k) = \int_{-\infty}^{\infty} \dd{x} f(x) e^{- i k x},
\end{equation}
one has
\begin{equation}
	\label{app:eqn:poisson_summation}
	\sum_{n=-\infty}^{\infty} f(n) e^{i k_0 n} = \sum_{n=-\infty}^{\infty} \tilde{f} \left( 2 \pi n - k_0 \right).
\end{equation}
Applying \cref{app:eqn:poisson_summation} to $f_p (x) = \frac{1}{\sqrt{p^2 + x^2}}$, whose Fourier transform is $\tilde{f}(k) = 2 K_0 \left( \abs{k} p \right)$, with $p > 0$ and $K_0$ the modified Bessel function of the second kind, we immediately obtain
\begin{equation}
	V \left( \vec{r} \right) = \frac{e^2}{4 \pi \epsilon \epsilon_0 \xi } \sum_{n=-\infty}^{\infty} 2 K_0 \left( \abs{2 n - 1} \frac{\pi r}{\xi}\right) = \frac{e^2}{\pi \epsilon \epsilon_0 \xi } \sum_{n=0}^{\infty} K_0 \left( \abs{2 n + 1} \frac{\pi r}{\xi}\right).
\end{equation}
Using the asymptotic form $K_0 (z) \sim \sqrt{\frac{\pi}{2z}} e^{-z}$ for $z \to \infty$, we find that the dual-gated Coulomb potential indeed decays exponentially at large $\abs{\vec{r}}$,
\begin{equation}
	\label{app:eqn:exp_decay_interaction}
	V \left( \vec{r} \right) \sim \frac{e^2}{\pi \epsilon \epsilon_0 \xi } \sqrt{\frac{\xi}{2 r}} e^{-\frac{\pi r}{\xi}}.
\end{equation}

\subsubsection{Wannier-projected interaction Hamiltonian}\label{app:sec:model:interaction:proj_Ham}

Using the approximate projected density operator from \cref{app:eqn:projected_density_oeprator_in_k_space}, the projected interaction Hamiltonian can be expressed as
\begin{align}
	H_{I} \approx& \frac{1}{2\Omega_0^2} \int \dd[2]{r_1} \dd[2]{r_2} \int \frac{\dd[2]{q}}{(2 \pi)^2}  V \left( \vec{q} \right) e^{- i \vec{q} \cdot \left( \vec{r}_1 - \vec{r}_2 \right)} \sum_{\substack{ \vec{k}_1,\vec{k}_2 \in \text{MBZ} \\ \vec{G}_1,\vec{G}_2 \in \mathcal{Q}}} \sum_{\substack{\eta_1,s_1 \\ \eta_2,s_2}} n_{\eta_1} \left( \vec{k}_1 + \vec{G}_1 \right) \hat{n}_{\vec{k}_1,\eta_1,s_1} e^{-i \left( \vec{k}_1 + \vec{G}_1 \right) \cdot \vec{r}_1} \nonumber \\
	&\times n_{\eta_2} \left( \vec{k}_2 + \vec{G}_2 \right) \hat{n}_{\vec{k}_2,\eta_2,s_2} e^{-i \left( \vec{k}_2 + \vec{G}_2 \right) \cdot \vec{r}_2} \nonumber \\
	=& \frac{N^2}{2N \Omega_0} \sum_{\substack{ \vec{k} \in \text{MBZ} \\ \vec{G} \in \mathcal{Q}}} \sum_{\substack{\eta_1,s_1 \\ \eta_2,s_2}}  V \left( \vec{k} + \vec{G} \right)  n^{*}_{\eta_1} \left(\vec{k} + \vec{G} \right) n_{\eta_2} \left( \vec{k} + \vec{G} \right) \hat{n}_{-\vec{k},\eta_1,s_2}  \hat{n}_{\vec{k},\eta_2,s_2} \nonumber \\
	=& \frac{N}{2} \sum_{\vec{k} \in \text{MBZ}} \sum_{\substack{\eta_1,s_1 \\ \eta_2,s_2}}  V_{\eta_1 \eta_2} \left( \vec{k} \right) \hat{n}_{-\vec{k},\eta_1,s_2}  \hat{n}_{\vec{k},\eta_2,s_2}, \label{app:eqn:proj_dens_interaction_k}
\end{align}
where we have defined the interaction in the Wannier-orbital basis as
\begin{equation}
	V_{\eta_1 \eta_2} \left( \vec{k} \right) \equiv \frac{1}{\Omega_0} \sum_{\vec{G} \in \mathcal{Q}}  V \left( \vec{k} + \vec{G} \right)  n^{*}_{\eta_1} \left(\vec{k} + \vec{G} \right) n_{\eta_2} \left( \vec{k} + \vec{G} \right).
\end{equation}
In real space, the projected interaction in \cref{app:eqn:proj_dens_interaction_k} takes the form
\begin{equation}
	\label{app:eqn:density_density_wannier_full}
	H_{I} = \frac{1}{2} \sum_{\substack{\vec{R}, \Delta \vec{R} \\ \eta_1,s_1,\eta_2,s_2}} V_{\eta_1 \eta_2} \left( \Delta \vec{R} \right) \hat{n}_{\vec{R}, \eta_1, s_1} \hat{n}_{\vec{R} + \Delta \vec{R}, \eta_2, s_2},
\end{equation}
with 
\begin{align}
	V_{\eta_1 \eta_2} \left( \Delta \vec{R} \right) &= \frac{1}{N}\sum_{\vec{k}} V_{\eta_1 \eta_2} \left( \vec{k} \right) e^{ i \vec{k} \cdot \Delta \vec{R}} \label{app:eqn:v_tensor_ft_to_r}, \\
	V_{\eta_1 \eta_2} \left( \vec{k} \right) &= \sum_{\Delta \vec{R}} V_{\eta_1 \eta_2} \left( \Delta \vec{R} \right) e^{ - i \vec{k} \cdot \Delta \vec{R}}. \label{app:eqn:v_tensor_ft_to_k}
\end{align}
We note that, in going to the second line of \cref{app:eqn:proj_dens_interaction_k}, site-off-diagonal contributions to the density operator are neglected. As discussed in the main text, this is justified by the localized nature of the Wannier orbitals. Under this approximation, the projected interaction Hamiltonian $H_I$ contains only density-density terms when written in the Wannier-orbital basis, provided one also neglects the inter-valley (anti)Hund's terms, which are discussed separately in \cref{app:sec:model:hund}.

Furthermore, owing to the exponential decay of the Coulomb interaction potential $V \left( \vec{r} \right)$ at large $\abs{\vec{r}}$ from \cref{app:eqn:exp_decay_interaction}, the interaction tensor $V_{\eta_1 \eta_2} \left( \Delta \vec{R} \right)$ likewise decays exponentially with $\abs{\Delta \vec{R}}$. This motivates a further approximation of the projected interaction Hamiltonian $H_I$, in which the interaction tensor $V_{\eta_1 \eta_2} \left( \Delta \vec{R} \right)$ is truncated to nearest neighbors. Concretely, we approximate
\begin{equation}
	\label{app:eqn:density_density_wannier_fit}
	H_I \approx \frac{1}{2} \sum_{\substack{\vec{R}, \Delta \vec{R} \\ \eta_1,s_1,\eta_2,s_2}} V^{\text{fit}}_{\eta_1 \eta_2} \left( \Delta \vec{R} \right) \hat{n}_{\vec{R}, \eta_1, s_1} \hat{n}_{\vec{R} + \Delta \vec{R}, \eta_2, s_2},
\end{equation}
where we have introduced the \emph{fitted} interaction tensor $V^{\text{fit}}_{\eta_1 \eta_2} \left( \Delta \vec{R} \right)$, defined such that
\begin{equation}
	V^{\text{fit}}_{\eta_1 \eta_2} \left( \Delta \vec{R} \right) = 0, \qq{if} \abs{\vec{R}} > \abs{\vec{a}_{M_1}}.
\end{equation}
Both the fitted and the full interaction tensors are real, positive, and symmetric under the $C_{3z}$ and $C_{2x}$ symmetries,
\begin{align}
	V_{\left( \eta_1+1 \right) \left( \eta_2+1 \right)} \left( C_{3z} \Delta \vec{R} \right) &= V_{\eta_1  \eta_2} \left( \Delta \vec{R} \right), \\
	V_{\left( -\eta_1 \right) \left( -\eta_2 \right)} \left( C_{2x} \Delta \vec{R} \right) &= V_{\eta_1  \eta_2} \left( \Delta \vec{R} \right).
\end{align}
The fitted interaction tensor contains eight independent components, which we define as
\begin{alignat}{4}
	U \equiv& V^{\text{fit}}_{0 0} \left(\vec{0}\right), &\qquad 
	U' \equiv& V^{\text{fit}}_{0 1} \left(\vec{0}\right), &\qquad 
	V_1 \equiv& V^{\text{fit}}_{1 0} \left(\vec{a}_{M_1}\right), &\qquad 
	V_2 \equiv& V^{\text{fit}}_{0 1} \left(\vec{a}_{M_2}\right), \nonumber \\ 
	V_3 \equiv& V^{\text{fit}}_{0 0} \left(\vec{a}_{M_1}\right), &\qquad 
	V_4 \equiv& V^{\text{fit}}_{0 0} \left(\vec{a}_{M_2}\right), &\qquad 
	V_5 \equiv& V^{\text{fit}}_{0 2} \left(\vec{a}_{M_2}\right), &\qquad 
	V_6 \equiv& V^{\text{fit}}_{0 1} \left(\vec{a}_{M_1}\right). \label{app:eqn:detailed_interaction_def}
\end{alignat}
Across all twist angles considered in this work, we find the hierarchy $U \gtrsim U'$, while $V_i \lesssim V_j$ for any $1 \leq i < j \leq 6$. Importantly, the interaction lies close to a $\mathrm{U} \left( {6} \right)$-symmetric limit, in which $V_{\eta_1 \eta_2} \left( \vec{R} \right)$ is independent of the valley indices. The small relative displacement of the Wannier orbitals, however, weakly breaks this $\mathrm{U} \left( {6} \right)$ symmetry down to $\mathrm{U} \left( {2} \right) \otimes \mathrm{U} \left( {2} \right) \otimes \mathrm{U} \left( {2} \right)$. Finally, we define averaged interaction scales to characterize the relative strengths of the onsite and nearest-neighbor repulsions,
\begin{equation}
	\bar{U} = \frac{1}{9} \sum_{\eta, \eta'} V^{\text{fit}}_{\eta \eta'} \left( \vec{0} \right), \quad
	\bar{V} = \frac{1}{54} \sum_{i=0}^{5} \sum_{\eta, \eta'} V^{\text{fit}}_{\eta \eta'} \left( C_{6z}^i\vec{a}_{M_1} \right).
\end{equation}

In principle, one could fit the truncated interaction to the full interaction by a least-squares minimization,
\begin{equation}
	\label{app:eqn:minimization_naive_interaction}
	\min_{U,U',V_i} \sum_{\vec{k}} \abs{V^{\text{fit}}_{\eta_1 \eta_2} \left( \vec{k} \right) - V_{\eta_1 \eta_2} \left( \vec{k} \right)}^2.
\end{equation}
We note, however, that this procedure does not guarantee that the fitted interaction captures the essential features of the full interaction Hamiltonian. In particular, Ref.~\cite{LI25} showed that $\mathrm{U} \left( {6} \right)$ symmetry breaking in the off-site interaction ($V_{\eta_1 \eta_2} \left( \Delta \vec{R} \right)$ for $\Delta \vec{R} \neq \vec{0}$) favors a valley-polarized ground state at integer fillings in the strong-coupling regime. Intuitively, since $V_{\eta_1 \eta_2} \left( \Delta \vec{R} \right)$ is a physical repulsion, valley polarization implies a more uniform charge distribution and therefore reduces the interaction energy. To ensure that the fitted interaction reproduces this behavior as well, in \cref{app:sec:model:interaction:mean_field}, we compute the relevant temperature scale (within mean field) at which the system becomes valley polarized in a Mott phase at integer fillings, assuming that the antiferromagnetic exchange coupling driving the phases of Ref.~\cite{LI25} in the $\mathrm{U} \left( {6} \right)$ limit can be neglected. We will then require that this temperature scale be reproduced between the truncated and the full-fledged interaction.

\subsubsection{Mean-field description of the valley polarization transition}\label{app:sec:model:interaction:mean_field}

We now describe the valley-polarization transition in the integer-filled Mott insulators of AA t-\ch{SnSe2}{} at the mean-field level. As discussed further in \cref{app:sec:model:summary}, the Wannier model of AA t-\ch{SnSe2}{} projected into the first six (two spins and three valleys) moir\'e conduction bands is particle-hole symmetric about half filling. We can therefore restrict attention to the integer fillings $\nu = 1$ and $\nu = 2$\footnote{Exactly at half filling ($\nu=3$), each site hosts three electrons, one in each valley. Any configuration with more than one electron occupying the same valley is suppressed by the intra-valley Hubbard repulsion $U$, which is slightly larger than the inter-valley repulsion $U'$.}. In the Mott insulator at $\nu = 1$, each site $\vec{R}$ is occupied by a single electron, which can reside in one of the three valleys $\eta_\vec{R}$; the spin state is unimportant here, since antiferromagnetic interactions are not included in this calculation. At $\nu = 2$, each site hosts two electrons. In AA t-\ch{SnSe2}{}, because the Wannier centers of the orbitals in the three valleys are close but not identical, the intra-valley Hubbard repulsion $U$ is slightly larger than the inter-valley repulsion $U'$. The two electrons therefore preferentially occupy different valleys, leaving one valley $\eta_\vec{R}$ empty. Thus, in the Mott phase, the state at each lattice site is fully specified by $\eta_\vec{R}$, which denotes the filled (empty) valley on site $\vec{R}$ at filling $\nu = 1$ ($\nu = 2$).

Neglecting antiferromagnetic interactions, the effective Hamiltonian within the subspace of integer-filled Mott states at filling $\nu$ is
\begin{equation}
	\label{app:eqn:effective_ham_mean_field_mott}
	H^{\nu}_{\text{eff}} = \frac12 \sum_{\vec{R}, \Delta\vec{R} \neq \vec{0}} f^{\nu}_{\Delta \vec{R}} \left( \eta_\vec{R}, \eta_{\vec{R} + \Delta \vec{R}} \right),
\end{equation}
where 
\begin{align}
	f^{1}_{\Delta \vec{R}} \left( \eta_1, \eta_2 \right) &= \sum_{\eta, \eta'} V_{\eta \eta'} \left( \Delta \vec{R} \right) \delta_{\eta \eta_1} \delta_{\eta' \eta_2}, \\
	f^{2}_{\Delta \vec{R}} \left( \eta_1, \eta_2 \right) &= \sum_{\eta, \eta'} V_{\eta \eta'} \left( \Delta \vec{R} \right) \left( 1 - \delta_{\eta \eta_1} \right) \left( 1 - \delta_{\eta' \eta_2} \right).
\end{align}
Our goal in what follows is to compute the ordering temperature for valley polarization at fillings $\nu = 1,2$ within a mean-field approach. To this end, we perform a mean-field decoupling of the effective Hamiltonian $H^{\nu}_{\text{eff}}$ by introducing the probability $\varrho_{\vec{R},\eta}$ that site $\vec{R}$ is in the ``state'' $\eta$. We then adopt the following mean-field ansatz and expand the probability around a uniform configuration,
\begin{equation}
	\varrho_{\vec{R},\eta} = \varrho_{\eta} + \delta \varrho_{\vec{R},\eta},
\end{equation} 
which leads to
\begin{align}
	H^{\nu}_{\text{eff}} &= \frac12  \sum_{\substack{\Delta\vec{R} \neq \vec{0} \\ \vec{R},\eta_1,\eta_2}} f^{\nu}_{\Delta \vec{R}} \left( \eta_1, \eta_2 \right) \varrho_{\vec{R},\eta_1} \varrho_{\vec{R} + \Delta \vec{R},\eta_2} \nonumber \\
	& \approx \frac12  \sum_{\substack{\Delta\vec{R} \neq \vec{0} \\ \vec{R},\eta_1,\eta_2}} f^{\nu}_{\Delta \vec{R}} \left( \eta_1, \eta_2 \right) \left( \varrho_{\eta_1} \varrho_{\eta_2} + \varrho_{\eta_1} \delta \varrho_{\vec{R} + \Delta \vec{R},\eta_2} + \delta \varrho_{\vec{R},\eta_1} \varrho_{\eta_2} \right) \nonumber \\
	& = \frac12  \sum_{\substack{\Delta\vec{R} \neq \vec{0} \\ \vec{R},\eta_1,\eta_2}} f^{\nu}_{\Delta \vec{R}} \left( \eta_1, \eta_2 \right) \left( -\varrho_{\eta_1} \varrho_{\eta_2} + \varrho_{\eta_1} \varrho_{\vec{R} + \Delta \vec{R},\eta_2} + \varrho_{\vec{R},\eta_1} \varrho_{\eta_2} \right) \nonumber \\
	& = \frac12 \left[  - N \sum_{\substack{\Delta\vec{R} \neq \vec{0} \\ \eta_1,\eta_2}} f^{\nu}_{\Delta \vec{R}} \left( \eta_1, \eta_2 \right) \varrho_{\eta_1} \varrho_{\eta_2} + \sum_{\substack{\Delta\vec{R} \neq \vec{0} \\ \vec{R},\eta}} \left( f^{\nu}_{\Delta \vec{R}} \left( \eta, \eta_{\vec{R} + \Delta \vec{R}} \right) \varrho_{\eta} + f^{\nu} \left( \eta_\vec{R}, \eta \right) \varrho_{\eta} \right) \right] \nonumber \\
	& = -\frac{N}{2} \sum_{\substack{\Delta\vec{R} \neq \vec{0} \\ \eta_1,\eta_2}} f^{\nu}_{\Delta \vec{R}} \left( \eta_1, \eta_2 \right) \varrho_{\eta_1} \varrho_{\eta_2} + \sum_{\substack{\Delta\vec{R} \neq \vec{0} \\ \vec{R},\eta}} f^{\nu}_{\Delta \vec{R}} \left( \eta_\vec{R}, \eta \right) \varrho_{\eta}, \label{app:eqn:mean_field_valley_polarization} 
\end{align}
where the approximation in the second line neglects terms that are second order in the deviation $\delta \varrho_{\vec{R},\eta}$. The mean-field partition function and free energy of the system are given by
\begin{align}
	Z_\nu &=  \Tr \left( e^{ - \beta H_{\text{eff}}^{\nu}} \right) \approx \left( Z_{\nu,\text{MF}} \left( \left\lbrace \varrho_{\eta} \right\rbrace \right) \right)^N, \nonumber \\
	Z_{\nu,\text{MF}} \left( \left\lbrace \varrho_{\eta} \right\rbrace \right) &= \exp \left( \frac{\beta}{2} \sum_{\substack{\Delta\vec{R} \neq \vec{0} \\ \eta_1,\eta_2}} f^{\nu}_{\Delta \vec{R}} \left( \eta_1, \eta_2 \right) \varrho_{\eta_1}\varrho_{\eta_2}  \right) \sum_{\eta'} \exp \left( - \beta \sum_{\eta,\Delta \vec{R} \neq \vec{0}} f^{\nu}_{\Delta \vec{R}} \left( \eta', \eta \right) \varrho_\eta \right), \\
	F &= - \frac{1}{\beta} \log Z \approx N F_{\nu,\text{MF}} \left( \left\lbrace \varrho_{\eta} \right\rbrace \right), \nonumber \\
	F_{\nu,\text{MF}} \left( \left\lbrace \varrho_{\eta} \right\rbrace \right) &= -\frac{1}{2} \sum_{\substack{\Delta\vec{R} \neq \vec{0} \\ \eta_1,\eta_2}} f^{\nu}_{\Delta \vec{R}} \left( \eta_1, \eta_2 \right) \varrho_{\eta_1}\varrho_{\eta_2} - \frac{1}{\beta}\log \left[  \sum_{\eta'} \exp \left( - \beta \sum_{\eta,\Delta \vec{R} \neq \vec{0}} f^{\nu}_{\Delta \vec{R}} \left( \eta', \eta \right) \varrho_\eta \right) \right] \nonumber \\
	&= -\frac{1}{2} \sum_{\eta_1,\eta_2} f^{\nu}_{\Delta \vec{R}} \left( \eta_1, \eta_2 \right) \varrho_{\eta_1} \varrho_{\eta_2} - \frac{1}{\beta}\log \left[ \sum_{\eta'} \exp \left( - \beta \sum_{\eta} f^{\nu} \left( \eta', \eta \right) \varrho_\eta \right) \right],
\end{align}
where we have defined
\begin{equation}
	f^{\nu} \left( \eta, \eta' \right) = \sum_{\Delta \vec{R} \neq \vec{0}} f^{\nu}_{\Delta \vec{R}} \left( \eta, \eta' \right).
\end{equation}
The ordered state in the Mott phases is obtained by minimizing the mean-field free energy $F_{\nu,\text{MF}} \left( \left\lbrace \varrho_{\eta} \right\rbrace \right)$ with respect to the mean fields $\varrho_{\eta}$, subject to the constraint $\sum_{\eta} \varrho_{\eta} = 1$. Before performing this minimization explicitly, it is useful to examine the structure of $f^{\nu} \left( \eta, \eta' \right)$ in more detail,
\begin{align}
	f^{1} \left( \eta, \eta' \right) &= \sum_{\Delta \vec{R} \neq \vec{0}} V_{\eta \eta'} \left( \Delta \vec{R} \right) = \eval{V_{\eta \eta'} \left( \vec{k} \right)}_{\vec{k}=\vec{0}} - \eval{V_{\eta \eta'} \left( \Delta \vec{R} \right)}_{\Delta \vec{R} = \vec{0}}, \nonumber \\
	f^{2} \left( \eta, \eta' \right) &= f^{1} \left( \eta, \eta' \right) + \sum_{\eta_1,\eta_2} f^{1} \left( \eta_1, \eta_2 \right) - \sum_{\eta''} \left( f^{1} \left( \eta, \eta'' \right) + f^{1} \left( \eta'', \eta' \right) \right) = f^{1} \left( \eta, \eta' \right) + \frac13 \sum_{\eta_1,\eta_2} f^{1} \left( \eta_1, \eta_2 \right),
\end{align}
where we have made use of the $C_{3z}$ symmetry of the interaction potential $V_{\eta \eta'} \left( \Delta \vec{R} \right)$. Since $f^{\nu} \left( \eta, \eta' \right)$ for $\nu = 1,2$ differ only by a constant, the corresponding mean-field free energies also differ only by a constant shift,
\begin{equation}
	F_{2,\text{MF}} \left( \left\lbrace \varrho_{\eta} \right\rbrace \right) = F_{1,\text{MF}} \left( \left\lbrace \varrho_{\eta} \right\rbrace \right) + \frac16 \sum_{\eta_1,\eta_2} f^{1} \left( \eta_1, \eta_2 \right)
\end{equation}
As a consequence, the mean-field ordering temperatures for valley polarization in the $\nu = 1$ and $\nu = 2$ Mott phases are identical, and we can therefore focus explicitly on the $\nu = 1$ case. At sufficiently high temperatures (small $\beta$), the mean-field free energy is minimized by the valley-symmetric, uniform solution $\varrho_{\eta} = \frac13$. In \cref{app:eqn:expanded_free_energy_for_vp}, we show that this solution indeed furnishes a local minimum of the free energy for sufficiently small $\beta$. Numerically, we also find it to be the global minimum. Upon lowering the temperature, the system becomes unstable toward a valley-polarized state, as argued in Ref.~\cite{LI25}. Here, we make this statement more explicit by estimating the corresponding mean-field transition temperature.

To determine the critical temperature, we expand the mean-field free energy around the valley-symmetric, uniform solution, $\varrho_{\eta}=\frac13$. Because $\sum_{\eta} \varrho_{\eta}=1$ and the problem is $C_{3z}$ symmetric, it is convenient to parametrize small fluctuations around this solution in the form
\begin{align}
	\varrho_{0} &= \frac13 + \frac{\delta_1}{\sqrt2} + \frac{\delta_2}{\sqrt6}, \nonumber \\
	\varrho_{1} &= \frac13 - \frac{\delta_1}{\sqrt2} + \frac{\delta_2}{\sqrt6}, \nonumber \\
	\varrho_{2} &= \frac13 - \delta_2 \sqrt{\frac{2}{3}}. 
\end{align}
The resulting free energy takes the form
\begin{equation}
	\label{app:eqn:expanded_free_energy_for_vp}
	F_{1,\text{MF}} \left( \delta_1, \delta_2 \right) = \frac{2 f^{1}(0,1) + f^{1}(0,0)}{6} - \frac{\log(3)}{\beta} + \left[ 3 \left( f^{1}(0,1) - f^{1}(0,0) \right)   - \beta \left( f^{1}(0,1) - f^{1}(0,0) \right)^2 \right] \frac{\delta^2_1 + \delta^2_2}{6} + \mathcal{O} \left( \delta^3 \right),
\end{equation}
where we have again used the $C_{3z}$ symmetry of the problem. The critical temperature for valley polarization can be read off directly from this expression: it is the temperature at which the symmetric critical point $\delta_1 = \delta_2 = 0$ becomes unstable,
\begin{equation}
	\beta_c = \frac{3}{f^{1}(0,1)-f^{1}(0,0)},
\end{equation}
which is positive for all systems and parameters considered here.  

In order to reproduce the correct energy scale for valley polarization in the truncated model, we therefore require that the critical temperature computed from the effective Hamiltonian in \cref{app:eqn:effective_ham_mean_field_mott} using the full interaction $V_{\eta_1 \eta_2} \left( \Delta \vec{R} \right)$ coincides with that obtained using the truncated interaction $V^{\text{fit}}_{\eta_1 \eta_2} \left( \Delta \vec{R} \right)$. This requirement fixes the interaction parameters defined in \cref{app:eqn:detailed_interaction_def} through the following constrained least-squares procedure,
\begin{align}
	\min_{U,U',V_i} \sum_{\vec{k}} \abs{V^{\text{fit}}_{\eta_1 \eta_2} \left( \vec{k} \right) - V_{\eta_1 \eta_2} \left( \vec{k} \right)}^2, &\qq{such that} \nonumber \\
	V_1 + 2 V_2 - 4 V_3 - 2 V_4 + 2 V_5 + V_6 &=\eval{\left( V_{01} \left( \vec{k} \right) - V_{00} \left( \vec{k} \right) \right)}_{\vec{k}=\vec{0}} - \eval{\left( V_{01} \left( \Delta \vec{R} \right) - V_{00} \left( \Delta \vec{R} \right) \right)}_{\Delta \vec{R} = \vec{0}}. \label{app:eqn:true_minimization_interaction}
\end{align}

\subsection{Inter-valley (anti-)Hund's interaction}\label{app:sec:model:hund}

In \cref{app:sec:model:interaction}, we derived the interaction Hamiltonian of AA t-\ch{SnSe2}{} by neglecting valley-off-diagonal contributions in the density operator from \cref{app:eqn:proj_density_operator}. In this section, we provide a rough estimate of interaction terms that break the valley-spin rotational $\mathrm{U} \left( {2} \right) \otimes \mathrm{U} \left( {2} \right) \otimes \mathrm{U} \left( {2} \right)$ symmetry of the model. Such terms can arise from two main sources:
\begin{itemize}
	\item On the one hand, a Hund's coupling originates from the Coulomb interaction. These processes involve large momentum transfer, since electrons scatter between valleys rather than within a valley, and are therefore usually neglected. As we will show, this mechanism leads to a ferromagnetic inter-valley coupling, which we also estimate.
	
	\item On the other hand, an anti-Hund's coupling can be mediated by electron-phonon interactions involving phonons that couple different valleys. This mechanism leads to an antiferromagnetic inter-valley coupling.
\end{itemize}

In twisted bilayer graphene, these two contributions are believed to be of comparable magnitude but opposite sign, making it difficult to determine which one dominates~\cite{KWA23,WAN24,LIU24e,YOU24,WAN25,WAN25a}. In this section, we outline how the (anti-)Hund's interactions arise and provide a rough estimate of the Hund's inter-valley coupling. We find that the latter is small compared to the other energy scales in the problem and is also expected to be partially cancelled by the phonon-mediated anti-Hund's interaction. For this reason, we neglect these terms in the present work and leave a more comprehensive study of their effects for future investigation.

\subsubsection{Inter-valley ferromagnetic Hund's coupling mediated by Coulomb repulsion}\label{app:sec:model:hund:ferro}

We begin by deriving the ferromagnetic Hund's coupling that originates from the short-wavelength part of the Coulomb repulsion. To this end, we start from the interaction Hamiltonian written in the Wannier-orbital basis of the monolayer,
\begin{equation}
	\label{app:eqn:full_monolayer_interaction}
	\hat{H}_{I} = \frac{1}{2} \sum_{\vec{R},\vec{R}'} V \left( \mathcal{R}_{\theta,l} \vec{R} - \mathcal{R}_{\theta_,l'} \vec{R}' \right) \hat{a}^\dagger_{\vec{R},s,l} \hat{a}_{\vec{R},s,l} \hat{a}^\dagger_{\vec{R}',s',l'}\hat{a}_{\vec{R}',s',l'},	
\end{equation}
where, in a notation similar to that used in Ref.~\cite{CAL25b}, $\hat{a}^\dagger_{\vec{R},s,l}$ creates an electron in the first conduction band in layer $l$ and monolayer unit cell $\vec{R}$, which is located at position $\mathcal{R}_{\theta,l} \vec{R}$. The rotation matrix $\mathcal{R}_{\theta,l}$ is defined as
\begin{equation}
	\label{app:eqn:notRrotation}
	\mathcal{R}_{\theta, l} =
	\begin{pmatrix}
		\cos \left(\frac{\theta  l}{2}\right) & -\sin \left(\frac{\theta  l}{2}\right) \\
		\sin \left(\frac{\theta  l}{2}\right) & \cos \left(\frac{\theta  l}{2}\right) \\	
	\end{pmatrix}.
\end{equation}
We then project the monolayer lattice fermion operators $\hat{a}^\dagger_{\vec{R},s,l}$ onto the low-energy fermions introduced in \cref{app:eqn:def_real_space_fermions_to_real}, obtaining
\begin{align}
	\hat{a}^\dagger_{\vec{R},s,l} =& \frac{1}{\sqrt{N_0}} \sum_{\vec{k}} \hat{a}^\dagger_{\vec{k},s,l} e^{-i \vec{k} \cdot \mathcal{R}_{\theta,l} \vec{R}} \nonumber \\
	\approx & \frac{1}{\sqrt{N_0}} \sum_{\eta} \sum_{\substack{\vec{k}\in \text{MBZ} \\ \vec{Q} \in \mathcal{Q}_{\eta+l}}} \hat{c}^\dagger_{\vec{k},\vec{Q},s,l} e^{-i \left( C^{\eta}_{3z} \vec{K}^l_M + \vec{k} - \vec{Q} \right) \cdot \mathcal{R}_{\theta,l} \vec{R}} \nonumber \\
	=& \sqrt{\frac{\Omega}{N_0}} \sum_{\eta} \hat{\psi}^\dagger_{\eta,s,l} \left( \mathcal{R}_{\theta,l} \vec{R} \right) e^{-i C^{\eta}_{3z} \vec{K}^l_M \cdot \mathcal{R}_{\theta,l} \vec{R}}, \label{app:eqn:monolayer_projected_operators}
\end{align}
with $N_0$ being the number of monolayer unit cells. In the first line of \cref{app:eqn:monolayer_projected_operators}, the summation over $\vec{k}$ runs over the entire \emph{monolayer} BZ, whereas in the second line we restrict to the low-energy states in the vicinity of the $C^{\eta}_{3z}\vec{K}^l_M$ points. Substituting \cref{app:eqn:monolayer_projected_operators} back into \cref{app:eqn:full_monolayer_interaction} and coarse-graining the discrete lattice sum into an integral, we obtain
\begin{equation}
	\hat{H}_{I} = \frac{1}{2} \sum_{\substack{\eta_1,\eta_2,s,l \\ \eta'_1, \eta'_2,s',l'}}\int \dd[2]{r} \dd[2]{r'} V \left( \vec{r} - \vec{r}' \right) \hat{\psi}^\dagger_{\eta_1,s,l} \left( \vec{r} \right) \hat{\psi}_{\eta_2,s,l} \left( \vec{r} \right) \hat{\psi}^\dagger_{\eta'_1,s',l'} \left( \vec{r}' \right) \hat{\psi}_{\eta'_2,s',l'} \left( \vec{r}' \right) e^{-i \Delta \vec{K} \left(\eta_1,\eta_2,l \right) \cdot \vec{r} - i \Delta \vec{K} \left(\eta'_1,\eta'_2,l' \right) \cdot \vec{r}'}, \label{app:eqn:full_monolayer_interaction_continuum}
\end{equation}
where we have defined the inter-valley momentum transfer
\begin{equation}
	\Delta\vec{K} \left(\eta_1,\eta_2,l \right) \equiv C_{3z}^{\eta_1} \vec{K}^l_M - C_{3z}^{\eta_2} \vec{K}^l_M.
\end{equation}

We now project \cref{app:eqn:full_monolayer_interaction_continuum} onto the $\hat{d}^\dagger_{\vec{R},\eta,s}$ Wannier orbitals. To this end, we first define their wave function in analogy with \cref{app:eqn:f_wannier_definition},
\begin{align}
	\tilde{W}_{\eta,l}\left(\vec{r} - \vec{R} \right) \equiv \mel**{0}{\hat{\psi}_{\eta,s,l} \left( \vec{r} \right) e^{ i \vec{q}_{\eta+l} \cdot \vec{r}} \hat{d}^\dagger_{\vec{R},\eta,s}}{0} = W_{\eta,l} \left( \vec{r} - \vec{R} - \boldsymbol{\delta}_\eta \right). \label{app:eqn:d_wannier_definition}
\end{align}
The definitions of $\tilde{W}_{\eta,l}\left(\vec{r} - \vec{R} \right)$ and $W_{\eta,l}\left(\vec{r} - \vec{R} \right)$ differ only by shifts by moir\'e lattice vectors, since the $\hat{f}^\dagger_{\vec{R},\eta,s}$ and $\hat{d}^\dagger_{\vec{R},\eta,s}$ fermions are related by a valley-dependent lattice translation, according to \cref{app:eqn:shifted wannier orbitals}. We can then project the valley-dependent density operator,
\begin{align}
	\hat{\psi}^\dagger_{\eta_1,s,l} \left( \vec{r} \right) \hat{\psi}_{\eta_2,s,l} \left( \vec{r} \right) &\approx  \sum_{\vec{R}_1,\vec{R}_2} \tilde{W}^*_{\eta_1,l} \left( \vec{r} - \vec{R}_1 \right) \tilde{W}_{\eta_2,l} \left( \vec{r} - \vec{R}_2 \right) e^{i \left( \vec{q}_{\eta_1 + l} - \vec{q}_{\eta_2 + l} \right) \cdot \vec{r}} \hat{d}^\dagger_{\vec{R}_1,\eta_1,s} \hat{d}_{\vec{R}_2,\eta_2,s} \nonumber \\
	&\approx  \sum_{\vec{R}} \tilde{W}^*_{\eta_1,l} \left( \vec{r} - \vec{R} \right) \tilde{W}_{\eta_2,l} \left( \vec{r} - \vec{R} \right) e^{i \left( \vec{q}_{\eta_1 + l} - \vec{q}_{\eta_2 + l} \right) \cdot \vec{r}} \hat{d}^\dagger_{\vec{R},\eta_1,s} \hat{d}_{\vec{R},\eta_2,s} \nonumber \\
	&= \sum_{\vec{R}} \tilde{n}_{\eta_1,\eta_2,l} \left( \vec{r} - \vec{R} \right) e^{i \left( \vec{q}_{\eta_1 + l} - \vec{q}_{\eta_2 + l} \right) \cdot \vec{r}} \hat{d}^\dagger_{\vec{R},\eta_1,s} \hat{d}_{\vec{R},\eta_2,s}, \label{app:eqn:wannier_valley_dep_density}
\end{align}
with 
\begin{equation}
	\tilde{n}_{\eta_1,\eta_2,l} \left( \vec{r} \right) = \tilde{W}^*_{\eta_1,l} \left( \vec{r} \right) \tilde{W}_{\eta_2,l} \left( \vec{r} \right).
\end{equation}
Similarly to \cref{app:eqn:f_wannier_definition}, we use the fact that the Wannier centers of the $\hat{d}^\dagger_{\vec{R},\eta,s}$ fermions lie close to $\vec{R} + \vec{r}_{\text{H}}$ independently of the valley and therefore neglect offsite overlaps in \cref{app:eqn:wannier_valley_dep_density}. We define the Fourier transform of $\tilde{n}_{\eta_1,\eta_2,l} \left( \vec{r} \right)$ as
\begin{align}
	\tilde{n}_{\eta_1, \eta_2, l} \left( \vec{k} + \vec{G} \right) &=  \int \dd[2]{r} \tilde{n}_{\eta_1, \eta_2, l} \left( \vec{r} \right) e^{i \left( \vec{k} + \vec{G} \right) \cdot \vec{r}}, \\
	\tilde{n}_{\eta_1, \eta_2, l} \left( \vec{r} \right) &= \frac{1}{N \Omega_0} \sum_{\substack{\vec{k} \in \text{MBZ} \\ \vec{G} \in \mathcal{Q}}} \tilde{n}_{\eta_1, \eta_2, l} \left( \vec{k} + \vec{G} \right) e^{-i \left( \vec{k} + \vec{G} \right) \cdot \vec{r}},
\end{align}
in analogy with \cref{app:eqn:orb_dens_func_ft_to_k,app:eqn:orb_dens_func_ft_to_r}. With these definitions, the interaction becomes
\begin{align}
	H_{I} =& \frac{1}{2} \frac{1}{\left( N \Omega_0 \right)^2} \sum_{\substack{\eta_1,\eta_2,s,l \\ \eta'_1, \eta'_2,s',l'}}\int \dd[2]{r} \dd[2]{r'} \int \frac{\dd[2]{q}}{(2 \pi)^2} V \left( \vec{q} \right) e^{-i \vec{q} \cdot \left( \vec{r} - \vec{r}' \right)} \nonumber \\
	& \times \sum_{\vec{k}, \vec{G}} \tilde{n}_{\eta_1, \eta_2, l} \left( \vec{k} + \vec{G} \right) e^{-i\left( \vec{k} + \vec{G} \right) \cdot \left( \vec{r} - \vec{R} \right)} e^{i \left( \vec{q}_{\eta_1 + l} - \vec{q}_{\eta_2 + l} \right) \cdot \vec{r}} e^{-i \Delta \vec{K}_M \left( \eta_1, \eta_2, l \right) \cdot \vec{r}} \hat{d}^\dagger_{\vec{R},\eta_1,s} \hat{d}_{\vec{R},\eta_2,s} \nonumber \\
	& \times \sum_{\vec{k}', \vec{G}'} \tilde{n}_{\eta'_1, \eta'_2, l'} \left( \vec{k}' + \vec{G}' \right) e^{-i\left( \vec{k}' + \vec{G}' \right) \cdot \left( \vec{r}' - \vec{R}' \right)} e^{i \left( \vec{q}_{\eta'_1 + l'} - \vec{q}_{\eta'_2 + l'} \right) \cdot \vec{r}'} e^{-i \Delta \vec{K}_M \left( \eta'_1, \eta'_2, l' \right) \cdot \vec{r}'} \hat{d}^\dagger_{\vec{R}',\eta'_1,s'} \hat{d}_{\vec{R}',\eta'_2,s'}.
\end{align}
The integrals over $\vec{r}$ and $\vec{r}'$ enforce the following momentum-conservation conditions,
\begin{align}
	-\vec{q} - \vec{k} - \vec{G} + \left( \vec{q}_{\eta_1 + l} - \vec{q}_{\eta_2 + l} \right) -  \Delta \vec{K}_M \left( \eta_1, \eta_2, l\right) &= 0 \nonumber \\
	\vec{q} - \vec{k}' - \vec{G}' + \left( \vec{q}_{\eta'_1 + l'} - \vec{q}_{\eta'_2 + l'} \right) -  \Delta \vec{K}_M \left( \eta'_1, \eta'_2, l'\right) &= 0
\end{align}
We now note that $\vec{k}^{(\prime)}$, $\vec{G}^{(\prime)}$, and $\vec{q}_{\eta}$ are all small compared to $\vec{K}^l_M$. Therefore, to satisfy momentum conservation, either $\abs{\Delta \vec{K}_M \left( \eta_1, \eta_2, l\right)} \sim \abs{\Delta \vec{K}_M \left( \eta'_1, \eta'_2, l'\right)} \ll \vec{K}^l_M$, which implies $\eta^{(\prime)}_1 = \eta^{(\prime)}_2$, or $\abs{\Delta \vec{K}_M \left( \eta_1, \eta_2, l\right)} \sim \abs{\Delta \vec{K}_M \left( \eta'_1, \eta'_2, l'\right)} \sim \vec{K}^l_M$. The latter case can satisfy the momentum-conservation conditions only if
\begin{equation}
	-\vec{q} \approx \Delta \vec{K}_M \left( \eta_1, \eta_2, l\right) \approx - \Delta \vec{K}_M \left( \eta'_1, \eta'_2, l'\right),
\end{equation}
which holds if and only if $\eta_1 = \eta'_2$ and $\eta_2 = \eta'_1$ with $\eta_1 \neq \eta_2$.

The condition $\eta^{(\prime)}_1 = \eta^{(\prime)}_2$ corresponds to the density-density interaction already derived in \cref{app:eqn:density_density_wannier_full}. The case $\eta_1 = \eta'_2$ and $\eta_2 = \eta'_1$ gives an additional contribution, such that the full projected interaction Hamiltonian becomes
{\small
	\begin{equation}
	H_{I} = \frac{1}{2} \sum_{\substack{\vec{R}, \Delta \vec{R} \\ \eta_1,s_1,\eta_2,s_2}} V_{\eta_1 \eta_2} \left( \Delta \vec{R} \right) \hat{n}_{\vec{R}, \eta_1, s_1} \hat{n}_{\vec{R} + \Delta \vec{R}, \eta_2, s_2} 
	+ \frac{1}{2} \sum_{\substack{\vec{R}, \Delta \vec{R} \\ \eta_1,s,\eta_2,s' \\ \eta_1 \neq \eta_2}} J^V_{\eta_1 \eta_2} \left( \Delta \vec{R} \right) \hat{d}^\dagger_{\vec{R},\eta_1,s} \hat{d}_{\vec{R},\eta_2,s} \hat{d}^\dagger_{\vec{R} + \Delta \vec{R},\eta_2,s'} \hat{d}_{\vec{R} + \Delta \vec{R},\eta_1,s'},
\end{equation}}with
\begin{align}
	J^V_{\eta_1 \eta_2} \left( \Delta \vec{R} \right) \equiv & \frac{1}{N \Omega_0} \sum_{\vec{k}, \vec{G}, l, l'} V \left( - \vec{k} - \vec{G} + \vec{q}_{\eta_1+l} - \vec{q}_{\eta_2+l} - \Delta \vec{K} \left( \eta_1, \eta_2, l \right) \right) \nonumber \\
	& \tilde{n}_{\eta_1, \eta_2, l} \left( \vec{k} + \vec{G} \right) \tilde{n}_{\eta_2, \eta_1, l'} \left( -\vec{k}  - \vec{G} + \Delta\vec{q} \left(l,l', \eta_1 \right) - \Delta\vec{q} \left(l,l', \eta_2 \right) \right) e^{-i\vec{k} \cdot \Delta \vec{R}} \nonumber \\
	= & \frac{1}{N \Omega_0} \sum_{\vec{k}, \vec{G}, l, l'} V \left( - \vec{k} - \vec{G} + \vec{q}_{\eta_1+l} - \vec{q}_{\eta_2+l} - \Delta \vec{K} \left( \eta_1, \eta_2, l \right) \right) \nonumber \\
	& \tilde{n}_{\eta_1, \eta_2, l} \left( \vec{k} + \vec{G} \right) \tilde{n}^{*}_{\eta_1, \eta_2, l'} \left( \vec{k}  + \vec{G} - \Delta\vec{q} \left(l,l', \eta_1 \right) + \Delta\vec{q} \left(l,l', \eta_2 \right) \right) e^{-i\vec{k} \cdot \Delta \vec{R}}.
\end{align}
Here we introduced the shorthand notation
\begin{equation}
	\Delta \vec{q} \left( l, l', \eta \right) = \vec{q}_{\eta+l} - \vec{q}_{\eta+l'} + \frac{l-l'}{2} \vec{q}_{\eta}.
\end{equation}
We now focus on the $\Delta\vec{R} = \vec{0}$ term and re-express the four-fermion operator using the Fierz identity,
\begin{align}
	\hat{d}^\dagger_{\vec{R},\eta_1,s} \hat{d}_{\vec{R},\eta_2,s} \hat{d}^\dagger_{\vec{R},\eta_2,s'} \hat{d}_{\vec{R},\eta_1,s'}&=\delta_{s s'} \hat{d}^\dagger_{\vec{R},\eta_1,s} \hat{d}_{\vec{R},\eta_1,s} - \hat{d}^\dagger_{\vec{R},\eta_1,s} \hat{d}_{\vec{R},\eta_1,s'} \hat{d}^\dagger_{\vec{R},\eta_2,s'} \hat{d}_{\vec{R},\eta_2,s} \nonumber \\
	&=\delta_{s s'} \hat{n}_{\vec{R},\eta_1,s} - \frac{1}{2} \sum_{s_1,s_2} \hat{n}_{\vec{R},\eta_1,s_1} \hat{n}_{\vec{R},\eta_2,s_2} - 2 \hat{\vec{S}}_{\vec{R},\eta_1} \cdot \hat{\vec{S}}_{\vec{R},\eta_2}, \label{app:eqn:fierz}
\end{align}
which holds for $\eta_1 \neq \eta_2$. In \cref{app:eqn:fierz}, we have introduced the spin oeprators
\begin{equation}
	\hat{\vec{S}}_{\vec{R},\eta} \equiv \frac12 \sum_{s s'} \hat{d}^\dagger_{\vec{R},\eta,s} \boldsymbol{\sigma}_{ss'} \hat{d}_{\vec{R},\eta,s'},
\end{equation}
where $\boldsymbol{\sigma} = \left( \sigma^x, \sigma^y, \sigma^z \right)$ is the Pauli vector. Up to chemical-potential terms, which we drop, the short-wavelength Coulomb interaction therefore generates a ferromagnetic inter-valley Hund's interaction,
\begin{equation}
	H_{I} = \frac{1}{2} \sum_{\substack{\vec{R}, \Delta \vec{R} \\ \eta_1,s_1,\eta_2,s_2}} V_{\eta_1 \eta_2} \left( \Delta \vec{R} \right) \hat{n}_{\vec{R}, \eta_1, s_1} \hat{n}_{\vec{R} + \Delta \vec{R}, \eta_2, s_2} 
	- \frac{J^V}{2} \sum_{\vec{R}, \eta_1 \neq \eta_2} \left( 2 \hat{\vec{S}}_{\vec{R},\eta_1} \cdot \hat{\vec{S}}_{\vec{R},\eta_2} + \frac{1}{2} \sum_{s_1,s_2} \hat{n}_{\vec{R},\eta_1,s_1} \hat{n}_{\vec{R},\eta_2,s_2} \right),  
\label{app:eqn:interaction hamiltonian with FM hunds}
\end{equation}
with $J^V \equiv J_{\eta \eta'} \left( \vec{0} \right)$ for any $\eta \neq \eta'$. In \cref{app:sec:model:summary}, we provide estimates of $J^V$ for the twist angles, screening lengths, and dielectric constants considered in this work.

\subsubsection{Electron-phonon coupling in monolayer \ch{SnSe2}}\label{app:sec:model:hund:eleph}

Upon electron doping, exfoliated monolayer \ch{SnSe2} becomes superconducting~\cite{ZEN18} and also exhibits a pressure-induced charge-density-wave transition~\cite{YIN18}. This has motivated \textit{ab initio}{} studies of the electronic and phononic spectra of \ch{SnSe2}, as well as of its electron-phonon coupling~\cite{KUD20,KAF20,FAN26}. Ref.~\cite{FAN26}, in particular, shows that in monolayer \ch{SnSe2} an optical phonon at the M point of the monolayer couples strongly to conduction-band electrons, suggesting the possibility of a significant phonon-mediated electron-electron interaction in AA t-\ch{SnSe2}{}. We now show that the corresponding interaction is of the antiferromagnetic anti-Hund's type.

We begin by considering the phonon and electron-phonon Hamiltonians for monolayer \ch{SnSe2},
\begin{align}
	\hat{H}^0_{\text{ph}}&= \sum_{\vec{k} \in \text{BZ}, \alpha} \omega_{\alpha} \left( \vec{k} \right) \hat{b}^\dagger_{\vec{k},\alpha} \hat{b}_{\vec{k},\alpha}, \label{app:eqn:mono_phon_ham_kin} \\
	\hat{H}^0_{\text{EPC}} &= \frac{1}{\sqrt{N_0}}\sum_{\vec{k},\vec{k}' \in \text{BZ}}\sum_{s,\alpha} F^{\alpha}\left( \vec{k}, \vec{k}' \right)\left(\hat{b}_{\vec{k}-\vec{k}',\alpha} + \hat{b}^\dagger_{\vec{k}' - \vec{k},\alpha}\right)\hat{a}^\dagger_{\vec{k},s}\hat{a}_{\vec{k}',s}. \label{app:eqn:mono_ele_phon_ham}
\end{align}
where $\text{BZ}$ is the monolayer Brillouin zone, $N_0$ denotes the number of unit cells in the monolayer, and $\hat{b}^\dagger_{\vec{k},\alpha}$ is the bosonic creation operator for a phonon with momentum $\vec{k}$ in branch (phonon band) $\alpha$, with frequency $\omega_{\alpha} \left( \vec{k} \right)$. At the same time, $\hat{a}^\dagger_{\vec{k},s}$ denotes the creation operator for the $s$-like orbital spanning the first conduction band of \ch{SnSe2}, while $F^{\alpha} \left( \vec{k}, \vec{k}' \right)$ is the real coupling strength between phonons and electrons in that band. Note that the electron-phonon Hamiltonian $\hat{H}_{\text{EPC}}$ has already been projected onto this first conduction band. Without loss of generality, we impose the following periodicity condition $\hat{b}^\dagger_{\vec{k} + \vec{g},\alpha} = \hat{b}^\dagger_{\vec{k},\alpha}$ for any reciprocal monolayer lattice vector $\vec{g}$. Since $\hat{a}^\dagger_{\vec{k},s}$ is also periodic in $\vec{k}$, it follows that $F^{\alpha} \left( \vec{k}, \vec{k}' \right)$ is periodic in both $\vec{k}$ and $\vec{k}'$.

We next show explicitly that the crystalline symmetries of the problem allow a phonon to scatter electrons between two different M valleys. Under a symmetry operation $g$, the electron and phonon operators transform as
\begin{align}
	g \hat{b}^\dagger_{\vec{k},\alpha} g^{-1} &=  \sum_{\alpha'} D^{b}_{\alpha' \alpha} \left( g, \vec{k} \right) \hat{b}^\dagger_{g \vec{k}, \alpha'}, \\
	g \hat{a}^\dagger_{\vec{k},s} g^{-1} &=  \sum_{s'} D^{c}_{s' s} (g) \hat{a}^\dagger_{g \vec{k}, s'},
\end{align} 
where $D^{b}{\alpha' \alpha} \left( g, \vec{k} \right)$ [$D^{c}{s' s} (g)$] is the unitary momentum-dependent (momentum-independent) representation matrix of the symmetry operation $g$ acting on the phonon (electron) operator. Requiring that $\commutator{g}{\hat{H}{\text{EPC}}} = 0$ immediately implies that the electron-phonon coupling tensor obeys
\begin{equation}
	\sum_{\alpha} D^{b}_{\alpha' \alpha} \left( g, \vec{k}' - \vec{k} \right) \left[ F^{\alpha} \left( \vec{k}, \vec{k}' \right)\right]^{(*)} = F^{\alpha'} \left( g \vec{k}, g \vec{k}' \right),
\end{equation}
with ${}^{(*)}$ indicating complex conjugation when $g$ is antiunitary. For inter-M-valley scattering, $\vec{k}$, $\vec{k}'$, and $\vec{k}'-\vec{k}$ are all distinct M points, up to reciprocal lattice vectors. If $g$ belongs to the little group of the M point, it can force $F^{\alpha} \left( \vec{k}, \vec{k}' \right)$ to vanish whenever the representation $D^{b}_{\alpha' \alpha} \left( g, \vec{k} \right)$ is nontrivial. Either by consulting the \href{https://www.topologicalquantumchemistry.fr/topophonons/#/?MPID=mp-665}{Topological Phonon Database}~\cite{BRA17,XU24a}, or by inducing the band representation of the phonon spectrum of \ch{SnSe2} using the tables from the \href{https://cryst.ehu.es}{Bilbao Crystallographic Sever}~\cite{PET24}, one obtains the band representation of the full phonon spectrum of \ch{SnSe2},
\begin{equation}
	\left( E_u \oplus A_{2u} \right) @ 1a \uparrow \mathcal{G} \oplus \left( A_1 + E \right) @ 2c \uparrow \mathcal{G} = \Gamma_1^+ \oplus 2 \Gamma_2^- \oplus \Gamma_3^+ \oplus 2 \Gamma_3^-, K_1 \oplus 2 K_2 \oplus 3 K_3, 2 M_1^+ \oplus 2 M_1^- \oplus M_2^+ \oplus 4 M_2^-,
\end{equation}
with the analysis being performed in layer group $\mathcal{G} = P \bar{3}m11'$. There are exactly two phonon bands transforming as the trivial irreducible representation $M_1^+$ at the M point, showing that two phonon bands can scatter electrons between different M valleys. This matches the \textit{ab initio}{} results of Ref.~\cite{FAN26}.

To proceed, we adapt \cref{app:eqn:mono_phon_ham_kin,app:eqn:mono_ele_phon_ham} to the twisted bilayer setting by appropriately rotating the momenta in the two layers and, in addition, assuming that the presence of two layers does not modify either the phonon spectrum or the electron-phonon Hamiltonian,
\begin{align}
	\hat{H}_{\text{ph}} &= \sum_{l,\alpha} \sum_{\vec{k} \in \text{BZ}_l} \omega_{\alpha} \left( \mathcal{R}^{-1}_{\theta,l} \vec{k} \right) \hat{b}^\dagger_{\vec{k},l,\alpha} \hat{b}_{\vec{k},l,\alpha}, \label{app:eqn:EPC:phonon Hamiltonian} \\
	\hat{H}_{\text{EPC}} &= \frac{1}{\sqrt{N_0}}\sum_{s,\alpha,l} \sum_{\vec{k},\vec{k}' \in \text{BZ}_l} F^{\alpha}\left( \mathcal{R}^{-1}_{\theta,l} \vec{k}, \mathcal{R}^{-1}_{\theta,l} \vec{k}' \right)\left(\hat{b}_{\vec{k}-\vec{k}',l,\alpha} + \hat{b}^\dagger_{\vec{k}' - \vec{k},l,\alpha}\right) \hat{a}^\dagger_{\vec{k},s,l}\hat{a}_{\vec{k}',s,l}, \label{app:eqn:EPC: generic monolayer basis expression}
\end{align}
where $\text{BZ}_l$ is the monolayer Brillouin zone of layer $l$, $\mathcal{R}_{\theta,l}$ is the rotation matrix defined in \cref{app:eqn:notRrotation}, and $N_0$ still denotes the number of monolayer unit cells in the sample.

\subsubsection{Inter-valley antiferromagnetic anti-Hund's coupling mediated by phonons}\label{app:sec:model:hund:antiferro}

We now restrict \cref{app:eqn:EPC: generic monolayer basis expression} to \emph{inter-valley} processes in the vicinity of the M points, taking $\vec{k}^{(\prime)} = C^{\eta}{3z} \vec{K}_M^{l} + \delta \vec{k}^{(\prime)}$, which yields
\begin{equation}
	\label{app:eqn:EPC: generic monolayer basis expression 2}
    \hat{H}_{\text{EPC}} \approx \frac{1}{\sqrt{N_0}} \sum_{\delta \vec{k}, \delta \vec{k}'} \sum_{\eta \neq \eta'}\sum_{s,\alpha,l} F^{\alpha}_{\eta \eta'} \left(\hat{b}_{C^{\eta+\eta'}_{3z} \vec{K}^l_M + \delta \vec{k} - \delta \vec{k}',l,\alpha} + \hat{b}^\dagger_{-C^{\eta+\eta'}_{3z} \vec{K}^l_M - \delta \vec{k} + \delta \vec{k}',l,\alpha}\right) \hat{a}^\dagger_{C^{\eta}_{3z} \vec{K}_M^l + \delta \vec{k}, s, l} \hat{a}_{C^{\eta'}_{3z} \vec{K}_M^l + \delta \vec{k}', s, l},
\end{equation}
where we have introduced the compact notation
\begin{equation}
	F_{\eta \eta'}^\alpha = F^{\alpha} \left( C_{3z}^\eta \vec{K}_M, C_{3z}^\eta \vec{K}_M \right), \qq{with} \vec{K}_M \equiv \mathcal{R}^{-1}_{\theta,l}  \vec{K}^l_M,
\end{equation}
and made the approximation $F^{\alpha} \left( \mathcal{R}^{-1}_{\theta,l} \left( C_{3z}^\eta \vec{K}^l_M + \delta \vec{k} \right), \mathcal{R}^{-1}_{\theta,l} \left( C_{3z}^{\eta'} \vec{K}^l_M + \delta \vec{k}' \right) \right) \approx F_{\eta \eta'}^\alpha$. We have also used the fact that $C^{\eta}_{3z} \vec{K}_M^l -C^{\eta'}_{3z} \vec{K}_M^{l} = C_{3z}^{\eta + \eta'} \vec{K}_M^{l}$, up to a reciprocal monolayer lattice vector, for $\eta \neq \eta'$. In this calculation, we focus on \emph{inter-valley} phonon processes, characterized by $\eta \neq \eta'$. The discarded \emph{intra-valley} processes are associated with $\Gamma$ phonons and ultimately lead to a renormalization of the valley-preserving density-density interaction. 

Next, we rewrite \cref{app:eqn:EPC: generic monolayer basis expression 2} in terms of the moir'e plane-wave operators from \cref{app:eqn:low_en_ops_c} and decompose the sum $\sum_{\delta \vec{k}}$ into $\sum_{\vec{k} \in \text{MBZ}} \sum_{\vec{Q} \in \mathcal{Q}{\eta+l}}$, 
\begin{align}
        \hat{H}_{\text{EPC}}=& \frac{1}{\sqrt{N_0}}\sum_{\vec{k},\vec{k}' \in \text{MBZ}}  \sum_{\eta \neq \eta'}
        \sum_{s,\alpha,l} \sum_{ \substack{ \vec{Q} \in \mathcal{Q}_{\eta+l} \\ \vec{Q}' \in \mathcal{Q}_{\eta'+l}}} F^{\alpha}_{\eta \eta'} \hat{c}^\dagger_{\vec{k},\vec{Q},s,l} \hat{c}_{\vec{k}',\vec{Q}',s,l} \nonumber \\
        &\times \left( \hat{b}_{C^{\eta+\eta'}_{3z} \vec{K}^l_M + \vec{k} - \vec{k}' - \left( \vec{Q} - \vec{Q}' \right),l,\alpha} + 
        \hat{b}^\dagger_{- C^{\eta+\eta'}_{3z} \vec{K}^l_M + \vec{k}' - \vec{k} - \left( \vec{Q}'-\vec{Q} \right), l, \alpha} \right). \label{app:eqn:EPC: generic monolayer basis expression 3}
\end{align}

We proceed by projecting onto the lowest-energy bands of the moir\'e model by ``inverting'' \cref{app:eqn:k_pt_wanniers_m_pt}:
\begin{equation}
	\label{app:eqn:EPC: projection inversion}
    \hat{c}^\dagger_{\vec{k},\vec{Q},s,l} \approx u_{\vec{Q},l;\eta, 1} \left( \vec{k} \right)\hat{f}^\dagger_{\vec{k},\eta,s}\,,
\end{equation}
where the valley $\eta$ is uniquely determined by the layer $l$ and the lattice on which $\vec{Q}$ lies. Substituting \cref{app:eqn:EPC: projection inversion} into \cref{app:eqn:EPC: generic monolayer basis expression 3}, the projected electron-phonon coupling term becomes
\begin{align}
      H_{\text{EPC}}=& \frac{1}{\sqrt{N_0}}\sum_{\vec{k},\vec{k}' \in \text{MBZ}}  \sum_{\substack{\eta \neq \eta' \\ s,\alpha,l}} \sum_{\vec{Q}'' \in \mathcal{Q}_{\eta + \eta' + l}} g^{l,\alpha}_{\eta\eta'}\left( \vec{k}, \vec{k}', \vec{Q}'' \right) \nonumber \\
      &\times \hat{f}^\dagger_{\vec{k},\eta,s} \hat{f}_{\vec{k}',\eta',s} 
      \left( \hat{b}_{C^{\eta+\eta'}_{3z} \vec{K}^l_M + \vec{k} - \vec{k}' - \vec{Q}'',l,\alpha} + 
      \hat{b}^\dagger_{- C^{\eta+\eta'}_{3z} \vec{K}^l_M + \vec{k}' - \vec{k} + \vec{Q}'', l, \alpha} \right), \label{app:eqn:EPC: wannier basis expression 1} 
\end{align}
with the electron-phonon vertex in the Wannier basis being defined by
\begin{equation}
	\label{app:eqn:EPC: vertex redefinition}
	g^{l,\alpha}_{\eta\eta'} \left( \vec{k}, \vec{k}', \vec{Q}'' \right) \equiv \sum_{\vec{Q} \in \mathcal{Q}_{\eta+l}} u^*_{\vec{Q},l;\eta,1} \left( \vec{k} \right) F^{\alpha}_{\eta\eta'} u_{\vec{Q} -\vec{Q}'',l;\eta', 1} \left( \vec{k}' \right).
\end{equation}
In going from \cref{app:eqn:EPC: generic monolayer basis expression 3} to \cref{app:eqn:EPC: wannier basis expression 1}, we have reorganized the sum over $\vec{Q}' \in \mathcal{Q}_{\eta' + l}$ into a sum over $\vec{Q}'' = \vec{Q}-\vec{Q}' \in \mathcal{Q}_{\eta + \eta' + l}$. We have also dropped the ``hat'' over $H_{\text{EPC}}$ to indicate that \cref{app:eqn:EPC: wannier basis expression 1} has been projected into the first moir\'e conduction band of AA t-\ch{SnSe2}{} and into the phonon modes near the M points. We also note that the electron-phonon vertex in the Wannier basis obeys the following Hermiticity condition
\begin{equation}
	\label{app:eqn:EPC: vertex hermiticity}
	g^{l,\alpha}_{\eta \eta'} \left( \vec{k}, \vec{k}', \vec{Q}'' \right) =  \left[g^{l,\alpha}_{\eta' \eta} \left( \vec{k}', \vec{k}, -\vec{Q}'' \right)\right]^*.
\end{equation}
The expression for $g^{l,\alpha}_{\eta \eta'} \left( \vec{k}, \vec{k}', \vec{Q}'' \right)$ can be simplified further using the Fourier transform
\begin{equation}
	u_{\vec{G} + \vec{q}_{\eta+l},l;\eta,1} \left( \vec{k} \right) = \frac{1}{\Omega_0 \sqrt{N}} \int \dd[2]{r} \tilde{W}_{\eta,l} \left( \vec{r} \right) e^{- i \vec{k} \cdot \left( \vec{r} - \boldsymbol{\delta}_{\eta} \right)} e^{i \vec{G} \cdot \vec{r}},
\end{equation}
which follows from \cref{app:eqn:f_wannier_definition,app:eqn:d_wannier_definition}. The electron-phonon vertex is then given by
\begin{align}
	g^{l,\alpha}_{\eta\eta'} \left( \vec{k}, \vec{k}', \vec{Q}'' \right) =& \sum_{\vec{G} \in \mathcal{Q}}  \frac{F^{\alpha}_{\eta \eta'}}{N \Omega^2_0} \int \dd[2]{r} \dd[2]{r'} \tilde{W}^*_{\eta,l} \left( \vec{r} \right) \tilde{W}_{\eta',l} \left( \vec{r}' \right) e^{ i \vec{k} \cdot \left( \vec{r} - \boldsymbol{\delta}_\eta \right)} e^{- i\vec{G} \cdot \vec{r}} e^{ - i \vec{k}' \cdot \left( \vec{r}' - \boldsymbol{\delta}_{\eta'} \right)} e^{i \left( \vec{G} + \vec{q}_{\eta + l} - \vec{Q}'' - \vec{q}_{\eta' + l} \right) \cdot \vec{r}'} \nonumber \\
	= & \frac{F^{\alpha}_{\eta \eta'}}{N \Omega_0} \int \dd[2]{r} \dd[2]{r'} \tilde{W}^*_{\eta,l} \left( \vec{r} \right) \tilde{W}_{\eta',l} \left( \vec{r}' \right) e^{ i \vec{k} \cdot \left( \vec{r} - \boldsymbol{\delta}_\eta \right)} e^{ - i \vec{k}' \cdot \left( \vec{r}' - \boldsymbol{\delta}_{\eta'} \right)} e^{i \left( \vec{q}_{\eta + l} - \vec{Q}'' - \vec{q}_{\eta' + l} \right) \cdot \vec{r}'} \sum_{\vec{R}} \delta \left( \vec{r} - \vec{r}' - \vec{R} \right) \nonumber \\
	\approx & \frac{F^{\alpha}_{\eta \eta'}}{N \Omega_0} \int \dd[2]{r} \tilde{W}^*_{\eta,l} \left( \vec{r} \right) \tilde{W}_{\eta',l} \left( \vec{r} \right) e^{ i \vec{k} \cdot \left( \vec{r} - \boldsymbol{\delta}_\eta \right)} e^{ - i \vec{k}' \cdot \left( \vec{r} - \boldsymbol{\delta}_{\eta'} \right)} e^{i \left( \vec{q}_{\eta + l} - \vec{Q}'' - \vec{q}_{\eta' + l} \right) \cdot \vec{r}} \nonumber \\
	\approx & \frac{F^{\alpha}_{\eta \eta'}}{N \Omega_0} \tilde{n}_{\eta,\eta',l} \left( \vec{k}- \vec{k}' + \vec{q}_{\eta+l} - \vec{Q}'' - \vec{q}_{\eta'+l} \right) e^{ - i \vec{k} \cdot \boldsymbol{\delta}_\eta} e^{ i \vec{k}' \cdot \boldsymbol{\delta}_{\eta'}}, \label{app:eqn:epc_moire_approx}
\end{align}
where, in the second line, we have neglected the offsite contributions, which are vanishingly small because of the exponential decay of the Wannier orbitals.

The kinetic phonon Hamiltonian can likewise be projected onto the phonon modes near the M point,
\begin{equation}
	H_{\text{ph}} = \sum_{l,\alpha,\eta} \sum_{\delta \vec{k}} \omega_{\text{M},\alpha} \hat{b}^\dagger_{C^{\eta}_{3z}\vec{K}_M^l + \delta \vec{k},l,\alpha} \hat{b}_{C^{\eta}_{3z}\vec{K}_M^l + \delta \vec{k},l,\alpha},
\end{equation}
where, for simplicity, we have neglected the phonon dispersion around the M point and defined $\omega_{M,\alpha} = \omega_{\alpha} \left( \vec{K}_M \right)$. We can now write down the Hamiltonian of AA t-\ch{SnSe2}{} projected onto the low-energy degrees of freedom,
\begin{equation}
    H = H_0 + H_I + H_{\text{ph}} + H_{\text{EPC}},
\end{equation}
where $H_0 = \sum_{\vec{k},\eta,s} \epsilon_{\eta,1} \left( \vec{k} \right) \hat{f}^\dagger_{\vec{k},\eta,s} \hat{f}_{\vec{k},\eta,s}$ is the kinetic Hamiltonian of AA t-\ch{SnSe2}{} projected into the first moir\'e conduction band, while $H_I$ is the corresponding Coulomb-originated interaction term. The terms $H_{\text{ph}}$ and $H_{\text{EPC}}$ are, respectively, the quadratic phonon and electron-phonon Hamiltonians defined in \cref{app:eqn:EPC:phonon Hamiltonian,app:eqn:EPC: wannier basis expression 1}.

We wish to understand the effect of phonons on the electron spectrum. At lowest order, this effect arises from the emission and absorption of phonons, leading to a \emph{phonon-mediated} electron-electron interaction that involves two vertices and is therefore second order in the electron-phonon coupling tensor from \cref{app:eqn:EPC: vertex redefinition}. To isolate such a process, we perform a Schrieffer-Wolff transformation~\cite{SCH66,LIU24e}
\begin{align}
        \tilde{H} =& e^{-\hat{S}} H e^{+\hat{S}} = \left(H_0 + H_\text{ph} +H_{I}\right) + \left(H_{\text{EPC}}+ \commutator{H_0 + H_\text{ph}}{\hat{S}^{(1)}}\right) \\
        +&\left(\commutator{H_{\text{EPC}}}{\hat{S}^{(1)}} + \commutator{H_0 + H_\text{ph}}{\hat{S}^{(2)}} + \commutator{\commutator{ H_0 + H_\text{ph} }{ \hat{S}^{(1)}}}{\hat{S}^{(1)}}\right) + \dots \label{app:eqn:EPC:SW-1}
\end{align}
where we expand $\hat{S} = \hat{S}^{(1)} + \hat{S}^{(2)}+\dots$ in powers of the vertex $g$. We also neglect any terms generated by $\commutator{ H_I }{\hat{S}}$. Our goal is to find an $\hat{S}^{(1)}$ such that
\begin{equation}
\label{eqn:app:EPC:SW-2}
    H_{\text{EPC}}+ \commutator{H_0 + H_\text{ph}}{ \hat{S}^{(1)} }=0
\end{equation}
which reduces the effective Hamiltonian of \cref{app:eqn:EPC:SW-1}, projected onto the \emph{low-energy subspace} of the bosonic vacuum via the projector $P$, to
\begin{equation}
\label{eqn:app:EPC:SW-3}
    H_{\text{eff}} = P \left(H_0 + H_\text{ph} +H_{I}\right) P + \frac12 P \commutator{H_{\text{EPC}}}{\hat{S}^{(1)}} P + \dots.
\end{equation}
One can check that the operator $\hat{S}^{(1)}$ satisfying \cref{eqn:app:EPC:SW-2} is
\begin{align}
	\hat{S}^{(1)} =& \frac{1}{\sqrt{N_0}}\sum_{\vec{k},\vec{k}' \in \text{MBZ}}  \sum_{\substack{\eta \neq \eta' \\ s,\alpha,l}} \sum_{\vec{Q}'' \in \mathcal{Q}_{\eta + \eta' + l}} g^{l,\alpha}_{\eta\eta'}\left( \vec{k}, \vec{k}', \vec{Q}'' \right) \hat{f}^\dagger_{\vec{k},\eta,s} \hat{f}_{\vec{k}',\eta',s} \nonumber \\
	& \times \left(\frac{1}{\omega_{\text{M},\alpha} - \left( \epsilon_{\eta} \left( \vec{k} \right) - \epsilon_{\eta'} \left( \vec{k}' \right) \right)} \hat{b}_{C^{\eta+\eta'}_{3z} \vec{K}^l_M + \vec{k} - \vec{k}' - \vec{Q}'',l,\alpha}  \right. \nonumber\\
	& \quad -\left.\frac{1}{\omega_{\text{M},\alpha} + \left( \epsilon_{\eta} \left( \vec{k} \right) - \epsilon_{\eta'} \left( \vec{k}' \right) \right)} \hat{b}^\dagger_{- C^{\eta+\eta'}_{3z} \vec{K}^l_M + \vec{k}' - \vec{k} + \vec{Q}'', l, \alpha} \right) \nonumber \\
	\approx& \frac{1}{\sqrt{N_0}}\sum_{\vec{k},\vec{k}' \in \text{MBZ}}  \sum_{\substack{\eta \neq \eta' \\ s,\alpha,l}} \sum_{\vec{Q}'' \in \mathcal{Q}_{\eta + \eta' + l}} \frac{g^{l,\alpha}_{\eta\eta'}\left( \vec{k}, \vec{k}', \vec{Q}'' \right)}{\omega_{\text{M},\alpha}} \hat{f}^\dagger_{\vec{k},\eta,s} \hat{f}_{\vec{k}',\eta',s} \nonumber \\
	&\times \left(\hat{b}_{C^{\eta+\eta'}_{3z} \vec{K}^l_M + \vec{k} - \vec{k}' - \vec{Q}'',l,\alpha} - \hat{b}^\dagger_{- C^{\eta+\eta'}_{3z} \vec{K}^l_M + \vec{k}' - \vec{k} + \vec{Q}'', l, \alpha} \right), 
	\label{eqn:app:EPC:SW-4}
\end{align}
where, in going from the first to the second expression, we have used the fact that the phonon energy is much larger than the characteristic energy scale of the first moir\'e conduction band and have therefore neglected the term $\epsilon_{\eta} \left( \vec{k} \right) - \epsilon_{\eta'} \left( \vec{k}' \right)$ in the denominators. Evaluating $\left[H_{\text{EPC}},\hat{S}^{(1)}\right]$ and projecting onto the low-energy subspace of the bosonic vacuum, the resulting four-fermion term\footnote{A one-body term arises as well, but is ignored for the purpose of this section, which is just to show the emergence of an antiferromagnetic coupling.} takes the form
\begin{equation}
    H_{\text{eff}} = - \sum_{\substack{\vec{k}_1,\vec{k}_2, \vec{q} \in \text{MBZ}}} \sum_{\substack{\eta_1 \neq \eta_1',\\ \eta_2 \neq \eta_2' \\ \eta_1 + \eta'_1 = \eta_2 + \eta_2'}} \sum_{s_1,s_2} \mathcal{V}_{\eta_1 \eta_1'; \eta_2 \eta_2'} \left( \vec{k}_1, \vec{k}_2, \vec{q} \right)  \hat{f}^\dagger_{\vec{k}_1 + \vec{q},\eta_1,s_1} \hat{f}_{\vec{k}_1,\eta_1',s_1} \hat{f}^\dagger_{\vec{k}_2,\eta_2,s_2} \hat{f}_{\vec{k}_2 + \vec{q},\eta_2',s_2},
    \label{app:eqn:EPC:phonon-mediated interaction k space}
\end{equation}
with interaction vertex
\begin{align}
    \mathcal{V}_{\eta_1 \eta_1'; \eta_2 \eta_2'} \left( \vec{k}_1, \vec{k}_2, \vec{q} \right) =& 
    \frac{1}{N_0} \sum_{\alpha,l} \sum_{\vec{Q}'' \in \mathcal{Q}_{\eta_1 + \eta'_1 +l }} \frac{1}{\omega_{\text{M},\alpha}} g^{l,\alpha}_{\eta_1 \eta_1'} \left( \vec{k}_1 + \vec{q}, \vec{k}_1, \vec{Q}'' \right) g^{l,\alpha}_{\eta_2 \eta_2'} \left( \vec{k}_2,\vec{k}_2 + \vec{q}, -\vec{Q}'' \right) \nonumber \\
    \approx& \frac{1}{N_0 \left( N \Omega_0 \right)^2} \sum_{\alpha,l} \sum_{\vec{Q}'' \in \mathcal{Q}_{\eta_1 + \eta'_1 +l }} \frac{F^{\alpha}_{\eta_1 \eta_1'} F^{\alpha}_{\eta_2 \eta_2'}}{\omega_{M,\alpha}} \tilde{n}_{\eta_1,\eta_1',l} \left( \vec{q} + \vec{q}_{\eta_1+l} - \vec{Q}'' - \vec{q}_{\eta_1'+l} \right) e^{ - i \left( \vec{k}_1 + \vec{q} \right) \cdot \boldsymbol{\delta}_{\eta_1}} e^{ i \vec{k}_1 \cdot \boldsymbol{\delta}_{\eta_1'}} \nonumber \\
    &\times \tilde{n}_{\eta_2,\eta_2',l} \left( -\vec{q} + \vec{q}_{\eta_2+l} + \vec{Q}'' - \vec{q}_{\eta_2'+l} \right) e^{ - i \vec{k}_2 \cdot \boldsymbol{\delta}_{\eta_2}} e^{ i \left( \vec{k}_2 + \vec{q} \right) \cdot \boldsymbol{\delta}_{\eta_2'}},
    \label{app:eqn:EPC:phonon-mediated vertex k space}
\end{align}
where, in the final approximation, we have used \cref{app:eqn:epc_moire_approx}. Fourier transforming to real space, we obtain
\begin{equation}
	H_{\text{eff}} \approx - \sum_{\vec{R}_1,\vec{R}_2} \sum_{\substack{\eta_1 \neq \eta_1',\\ \eta_2 \neq \eta_2' \\ \eta_1 + \eta'_1 = \eta_2 + \eta_2'}} \sum_{s_1,s_2} \mathcal{V}_{\eta_1 \eta_1'; \eta_2 \eta_2'} \left( \vec{R}_2 - \vec{R}_1 \right)  \hat{d}^\dagger_{\vec{R}_1,\eta_1,s_1} \hat{d}_{\vec{R}_1,\eta_1',s_1} \hat{d}^\dagger_{\vec{R}_2,\eta_2,s_2}, \hat{d}_{\vec{R}_2,\eta_2',s_2}, \label{app:eqn:total_phonon_mediated_interaction_real}
\end{equation} 
where 
\begin{align}
	\mathcal{V}_{\eta_1 \eta_1'; \eta_2 \eta_2'} \left( \Delta \vec{R} \right) = & \frac{1}{N_0 \left( N \Omega_0 \right)^2} \sum_{\alpha,l} \sum_{\vec{q} \in \text{MBZ}} \sum_{\vec{Q}'' \in \mathcal{Q}_{\eta_1 + \eta'_1 +l }} \frac{F^{\alpha}_{\eta_1 \eta_1'} F^{\alpha}_{\eta_2 \eta_2'}}{\omega_{\text{M},\alpha}} \tilde{n}_{\eta_1,\eta_1',l} \left( \vec{q} + \vec{q}_{\eta_1+l} - \vec{Q}'' - \vec{q}_{\eta_1'+l} \right) \nonumber \\
	&\times \tilde{n}_{\eta_2,\eta_2',l} \left( -\vec{q} + \vec{q}_{\eta_2+l} + \vec{Q}'' - \vec{q}_{\eta_2'+l} \right) e^{ - i \vec{q} \cdot \Delta \vec{R}} \nonumber \\
	 = & \frac{1}{N_0 \left( N \Omega_0 \right)^2} \sum_{\vec{q} \in \text{MBZ}} \sum_{\alpha,l,\vec{G} \in \mathcal{Q}}  \frac{F^{\alpha}_{\eta_1 \eta_1'} F^{\alpha}_{\eta_2 \eta_2'}}{\omega_{\text{M},\alpha}} \tilde{n}_{\eta_1,\eta_1',l} \left( \vec{q} + \vec{G} + 2 \vec{q}_{\eta_1 + l} \right) \nonumber \\
	 & \times \tilde{n}^*_{\eta_2',\eta_2,l} \left( \vec{q} + \vec{G} + 2 \vec{q}_{\eta_2'+l} \right) e^{ - i \vec{q} \cdot \Delta \vec{R}}, 
\end{align}
with $\eta_{1,2} \neq \eta_{1,2}'$ and $\eta_1 + \eta_1' = \eta_2 + \eta_2'$. The phonon-mediated interaction in \cref{app:eqn:total_phonon_mediated_interaction_real} contains two distinct processes. On the one hand, the terms with $\eta^{(\prime)}_1 = \eta^{(\prime)}_2$ break valley $\mathrm{U} \left( {1} \right)$ symmetry and are akin to pair-hopping terms, {\it i.e.}{} processes in which two particles simultaneously hop from one valley to another. Such terms arise \emph{only} in M-point moir\'e systems through an Umklapp process, because two electrons can tunnel into a different valley by emitting two M phonons in a process that conserves momentum only up to reciprocal monolayer lattice vectors~\cite{LIU24e}. On the other hand, the terms with $\eta_1 = \eta_2' \neq \eta_1' = \eta_2$ preserve valley $\mathrm{U} \left( {1} \right)$ symmetry and give rise to an anti-Hund's process. In what follows, we focus on this latter contribution.

Dropping the valley-$\mathrm{U} \left( {1} \right)$-breaking term, as well as the offsite contributions in \cref{app:eqn:total_phonon_mediated_interaction_real}, we immediately obtain
\begin{align}
	H_{\text{eff}} \approx& - \frac{J^{\text{ph}}}{2} \sum_{\substack{\vec{R} \\ \eta_1,s,\eta_2,s' \\ \eta_1 \neq \eta_2}} \hat{d}^\dagger_{\vec{R},\eta_1,s} \hat{d}_{\vec{R},\eta_2,s} \hat{d}^\dagger_{\vec{R},\eta_2,s'} \hat{d}_{\vec{R},\eta_1,s'} \nonumber \\
	=& \frac{J^{\text{ph}}}{2} \sum_{\vec{R}, \eta_1 \neq \eta_2} \left( 2 \hat{\vec{S}}_{\vec{R},\eta_1} \cdot \hat{\vec{S}}_{\vec{R},\eta_2} + \frac{1}{2} \sum_{s_1,s_2} \hat{n}_{\vec{R},\eta_1,s_1} \hat{n}_{\vec{R},\eta_2,s_2} \right),
\end{align}
where the final equality follows from the Fierz identity in \cref{app:eqn:fierz} after dropping the chemical-potential terms, and where
\begin{equation}
	J^{\text{ph}} = 2 \mathcal{V}_{\eta_1 \eta_2; \eta_2 \eta_1} \left( \vec{0} \right) = \frac{2}{N_0 \left( N \Omega_0 \right)^2} \sum_{\vec{q} \in \text{MBZ}} \sum_{\alpha,l,\vec{G} \in \mathcal{Q}} \frac{\abs{F^{\alpha}_{\eta_1 \eta_2} \tilde{n}_{\eta_1,\eta_2,l} \left( \vec{q} + \vec{G} + 2 \vec{q}_{\eta_1 + l} \right) }^2}{\omega_{\text{M},\alpha}} > 0.
\end{equation}
Thus, the electron-phonon coupling generates an anti-Hund's antiferromagnetic interaction in AA t-\ch{SnSe2}{}. While the exact value is small (and a quantitative estimation is beyond the scope of this work), we stress that this effect further diminishes any Hund's ferromagnetic interaction arising from the short-range Coulomb repulsion. 

\subsection{Summary of the model}\label{app:sec:model:summary}

\begin{figure}[!t]
	\centering
	\includegraphics[width=\textwidth]{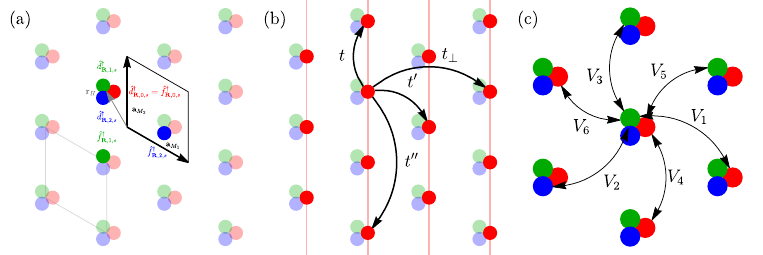}
	\subfloat{\label{app:fig:model:a}}\subfloat{\label{app:fig:model:b}}\subfloat{\label{app:fig:model:c}}\caption{Wannier model for the first conduction band of AA t-\ch{SnSe2}{}. (a) shows the moir\'e lattice vectors and the conventional unit cell employed in Refs.~\cite{CAL25b,LI25}. The Wannier orbitals $\hat{f}^\dagger_{\vec{R},\eta,s}$ and the translated orbitals $\hat{d}^\dagger_{\vec{R},\eta,s}$ are illustrated schematically, with their Wannier centers indicated by colored dots, where the color labels the valley: red ($\eta=0$), green ($\eta=1$), and blue ($\eta=2$). In what follows, we adopt a different unit-cell convention where the origin is chosen close to the Wannier centers of the three orbitals, indicated by the gray diamond. (b) illustrates the dominant hopping processes in valley $\eta=0$. We restrict to the leading hopping amplitude $t$, which gives rise to quasi-one-dimensional motion of electrons along the red ``chains'' in valley $\eta=0$. (c) shows the six interaction parameters $V_i$ ($1 \leq i \leq 6$), which parameterize the nearest-neighbor repulsion between the Wannier orbitals.}
	\label{app:fig:model}
\end{figure}

The Wannier model for the first conduction band of AA t-\ch{SnSe2}{} is illustrated schematically in \cref{app:fig:model}. In this work, we study the following model using Quantum Monte Carlo (QMC),
\begin{align}
	H &= H_{0} + H_{I}, \label{app:eqn:final_qmc_hamiltonia}\\
	H_{0} &= -t \sum_{\vec{R},\eta,s} \left( \hat{d}^\dagger_{\vec{R},\eta,s} \hat{d}_{\vec{R} + C^{\eta}_{3z} \vec{a}_{M_2},\eta,s} + \hat{d}^\dagger_{\vec{R},\eta,s} \hat{d}_{\vec{R} - C^{\eta}_{3z} \vec{a}_{M_2},\eta,s} \right) - \mu \sum_{\vec{R},\eta,s} \hat{n}_{\vec{R},\eta,s}, \label{app:eqn:final_single_particle}\\
	H_{I} &= \frac12 \sum_{\substack{\vec{R}, \Delta \vec{R} \\ \eta_1,s_1,\eta_2,s_2}} V^{\text{fit}}_{\eta_1 \eta_2} \left( \Delta \vec{R} \right) \left( \hat{n}_{\vec{R}, \eta_1, s_1} - \frac12 \right) \left( \hat{n}_{\vec{R} + \Delta \vec{R}, \eta_2, s_2} - \frac12 \right).\label{app:eqn:final_interaction}
\end{align}
Compared to \cref{app:eqn:single_particle_trucated}, the kinetic Hamiltonian in \cref{app:eqn:final_single_particle} is obtained by performing the following unitary transformation, which flips the sign of the hopping amplitude $t$ and shifts the momentum in each valley, while otherwise leaving the model unchanged,
\begin{equation}
	\label{app:eqn:unitary_flip_for_SSE_t}
	\hat{d}^\dagger_{\vec{R},\eta,s} \to \hat{d}^\dagger_{\vec{R},\eta,s} e^{\frac{i}{2} C^{\eta}_{3z} \vec{b}_{M_1}  \cdot \vec{R}}.
\end{equation}
This transformation is introduced because the QMC method employed here, reviewed in \cref{app:sec:sse}, requires all off-diagonal matrix elements of the Hamiltonian to be strictly negative (any diagonal terms can always be rendered negative by adding a constant). A chemical potential term is also included in \cref{app:eqn:final_single_particle}, as we will be working in the grand canonical ensemble.

For the interaction Hamiltonian, we normal order with respect to half filling of each orbital of the projected Wannier model by adding a constant and an additional chemical potential term, both of which leave the physics unchanged,
\begin{equation}
	H_{I} = \frac12\sum_{\substack{\vec{R}, \Delta \vec{R} \\ \eta_1,s_1,\eta_2,s_2}} V^{\text{fit}}_{\eta_1 \eta_2} \left( \Delta \vec{R} \right) \hat{n}_{\vec{R}, \eta_1, s_1} \hat{n}_{\vec{R} + \Delta \vec{R}, \eta_2, s_2} - 2 \sum_{\substack{\vec{R}, \Delta \vec{R} \\ \eta_1,\eta_2,s}}  V^{\text{fit}}_{\eta_1 \eta_2} \left( \Delta \vec{R} \right) \hat{n}_{\vec{R}, \eta_2, s} + N \sum_{\Delta \vec{R},\eta_1,\eta_2} V^{\text{fit}}_{\eta_1 \eta_2} \left( \Delta \vec{R} \right).
\end{equation}
This choice ensures that the grand-canonical Hamiltonian $H$ is explicitly symmetric under a simultaneous flip of the chemical potential $\mu$ and a particle-hole transformation $\mathcal{P}$, 
\begin{equation}
	\mathcal{P} \hat{d}^\dagger_{\vec{R},\eta,s} \mathcal{P}^{-1} = (-1)^{C^{\eta}_{3z} \vec{b}_{M_2}\cdot \vec{R}} \hat{d}_{\vec{R},\eta,s}.
\end{equation}
For the remainder of this work, we restrict to the truncated fitted interaction potential $V^{\text{fit}}_{\eta_1 \eta_2} \left( \Delta \vec{R} \right)$ and drop the superscript ${}^{\text{fit}}$ for notational simplicity. In addition, we shift the origin of the unit cell to $\vec{r}_{\text{H}} = -\frac13 \vec{a}_{M_1} + \frac13 \vec{a}_{M_2}$, which is both a $C_{3z}$- and $C_{2x}$-invariant point. From this point onward, the operators $C_{3z}$ and $C_{2x}$ denote rotations taken with respect to this new choice of unit-cell origin. Up to the unitary transformation in \cref{app:eqn:unitary_flip_for_SSE_t} and the fitting procedure used for the nearest-neighbor interaction tensor, the model in \cref{app:eqn:final_qmc_hamiltonia} is identical to that of Ref.~\cite{LI25}. For the reasons explained in \cref{sec:model:int}, we further neglect the small (anti)Hund's terms in the Hamiltonian.

\newcolumntype{C}[1]{>{\columncolor{#1}[\tabcolsep][\tabcolsep]}c}
\newcolumntype{L}[1]{>{\columncolor{#1}[\tabcolsep][\tabcolsep]}l}

\begin{table}[t]
	\centering
\begin{tabular}{|l|c|C{green!20}|C{blue!20}|C{red!20}|C{blue!20}|*{2}{c|}*{8}{L{green!20}|}l|l|L{orange!20}|}
			\hline
			No. & $\theta$ & $t/\si{\milli\electronvolt}$ & $t_{\perp}/t$ & $t'/t$ & $t''/t$ & $\epsilon$ & $\xi/\si{\nano\meter}$ & $U/t$ & $U'/t$ & $V_1/t$ & $V_2/t$ & $V_3/t$ & $V_4/t$ & $V_5/t$ & $V_6/t$ & $\bar{U}/t$ & $\bar{V}/t$ & $J^V/t$ \\
			\hline
			$1$ &  &  &  &  &  &  & $2.5$ & $4.083$ & $4.037$ & $0.625$ & $0.710$ & $0.801$ & $0.797$ & $0.887$ & $1.013$ & $4.052$ & $0.803$ & $0.218$\\
\cline{1-1}\cline{8-19}
$2$ &  &  &  &  &  &  & $5$ & $5.234$ & $5.184$ & $1.249$ & $1.363$ & $1.478$ & $1.480$ & $1.595$ & $1.749$ & $5.201$ & $1.483$ & $0.218$\\
\cline{1-1}\cline{8-19}
$3$ & \multirow{-3}{*}{\SI{9.43}{\degree}} & \multirow{-3}{*}{$19.983$} & \multirow{-3}{*}{$0.469$} & \multirow{-3}{*}{$-0.086$} & \multirow{-3}{*}{$0.233$} & \multirow{-3}{*}{$12$} & $10$ & $5.965$ & $5.914$ & $1.835$ & $1.959$ & $2.082$ & $2.087$ & $2.210$ & $2.372$ & $5.931$ & $2.088$ & $0.218$\\
\hline
$4$ &  &  &  &  &  &  & $2.5$ & $6.324$ & $6.186$ & $0.450$ & $0.606$ & $0.677$ & $0.801$ & $0.869$ & $0.992$ & $6.232$ & $0.727$ & $0.341$\\
\cline{1-1}\cline{8-19}
$5$ &  &  &  &  &  &  & $5$ & $8.277$ & $8.121$ & $1.202$ & $1.430$ & $1.549$ & $1.705$ & $1.821$ & $2.000$ & $8.173$ & $1.612$ & $0.341$\\
\cline{1-1}\cline{8-19}
$6$ & \multirow{-3}{*}{\SI{7.34}{\degree}} & \multirow{-3}{*}{$11.388$} & \multirow{-3}{*}{$0.254$} & \multirow{-3}{*}{$-0.083$} & \multirow{-3}{*}{$0.202$} & \multirow{-3}{*}{$12$} & $10$ & $9.545$ & $9.385$ & $2.086$ & $2.343$ & $2.485$ & $2.650$ & $2.789$ & $2.992$ & $9.438$ & $2.551$ & $0.341$\\
\hline
$7$ &  &  &  &  &  &  & $2.5$ & $5.686$ & $5.536$ & $0.186$ & $0.287$ & $0.306$ & $0.427$ & $0.446$ & $0.493$ & $5.586$ & $0.354$ & $0.312$\\
\cline{1-1}\cline{8-19}
$8$ &  &  &  &  &  &  & $5$ & $7.549$ & $7.380$ & $0.687$ & $0.855$ & $0.913$ & $1.072$ & $1.130$ & $1.224$ & $7.436$ & $0.975$ & $0.312$\\
\cline{1-1}\cline{8-19}
$9$ & \multirow{-3}{*}{\SI{6.01}{\degree}} & \multirow{-3}{*}{$5.871$} & \multirow{-3}{*}{$0.112$} & \multirow{-3}{*}{$-0.007$} & \multirow{-3}{*}{$0.164$} & \multirow{-3}{*}{$24$} & $10$ & $8.777$ & $8.602$ & $1.422$ & $1.624$ & $1.706$ & $1.877$ & $1.960$ & $2.081$ & $8.660$ & $1.773$ & $0.312$\\
\hline
$10$ &  &  &  &  &  &  & $2.5$ & $7.282$ & $7.056$ & $0.119$ & $0.206$ & $0.208$ & $0.335$ & $0.336$ & $0.358$ & $7.131$ & $0.257$ & $0.387$\\
\cline{1-1}\cline{8-19}
$11$ &  &  &  &  &  &  & $5$ & $9.833$ & $9.574$ & $0.599$ & $0.765$ & $0.809$ & $0.990$ & $1.033$ & $1.107$ & $9.660$ & $0.879$ & $0.387$\\
\cline{1-1}\cline{8-19}
$12$ & \multirow{-3}{*}{\SI{5.09}{\degree}} & \multirow{-3}{*}{$4.152$} & \multirow{-3}{*}{$0.057$} & \multirow{-3}{*}{$-0.047$} & \multirow{-3}{*}{$0.131$} & \multirow{-3}{*}{$24$} & $10$ & $11.55$ & $11.28$ & $1.474$ & $1.690$ & $1.769$ & $1.969$ & $2.048$ & $2.162$ & $11.37$ & $1.847$ & $0.387$\\
\hline
$13$ &  &  &  &  &  &  & $2.5$ & $6.470$ & $6.231$ & $0.046$ & $0.084$ & $0.091$ & $0.146$ & $0.155$ & $0.177$ & $6.311$ & $0.114$ & $0.337$\\
\cline{1-1}\cline{8-19}
$14$ &  &  &  &  &  &  & $5$ & $8.827$ & $8.549$ & $0.339$ & $0.434$ & $0.479$ & $0.568$ & $0.616$ & $0.688$ & $8.642$ & $0.517$ & $0.337$\\
\cline{1-1}\cline{8-19}
$15$ & \multirow{-3}{*}{\SI{4.41}{\degree}} & \multirow{-3}{*}{$2.214$} & \multirow{-3}{*}{$0.024$} & \multirow{-3}{*}{$-0.132$} & \multirow{-3}{*}{$0.088$} & \multirow{-3}{*}{$48$} & $10$ & $10.43$ & $10.14$ & $1.017$ & $1.158$ & $1.243$ & $1.345$ & $1.432$ & $1.548$ & $10.23$ & $1.286$ & $0.337$\\
\hline
$16$ &  &  &  &  &  &  & $2.5$ & $6.323$ & $6.047$ & $0.023$ & $0.044$ & $0.047$ & $0.079$ & $0.084$ & $0.096$ & $6.139$ & $0.061$ & $0.314$\\
\cline{1-1}\cline{8-19}
$17$ &  &  &  &  &  &  & $5$ & $8.765$ & $8.443$ & $0.230$ & $0.296$ & $0.331$ & $0.391$ & $0.428$ & $0.481$ & $8.550$ & $0.357$ & $0.314$\\
\cline{1-1}\cline{8-19}
$18$ & \multirow{-3}{*}{\SI{3.89}{\degree}} & \multirow{-3}{*}{$1.375$} & \multirow{-3}{*}{$0.008$} & \multirow{-3}{*}{$-0.191$} & \multirow{-3}{*}{$0.063$} & \multirow{-3}{*}{$72$} & $10$ & $10.46$ & $10.13$ & $0.829$ & $0.942$ & $1.016$ & $1.091$ & $1.168$ & $1.266$ & $10.24$ & $1.049$ & $0.314$ \\
			\hline
			19 & N/A & 1 & \multicolumn{3}{c|}{0} & N/A & N/A & 12.00 & 12.00 & \multicolumn{6}{C{green!20}|}{0} & 12.00 & 0 & 0 \\
			\hline
			20 & N/A & 1 & \multicolumn{3}{c|}{0} & N/A & N/A & 12.00 & 11.40 & \multicolumn{6}{C{green!20}|}{0} & 11.60 & 0 & 0 \\
			\hline
		\end{tabular}\caption{Parameters of the projected AA t-\ch{SnSe2}{} Wannier Hamiltonian studied in this work. For each twist angle $\theta$, we list the leading kinetic hopping amplitudes from \cref{app:eqn:leading_hoppings}, as well as the interaction parameters defined in \cref{app:eqn:detailed_interaction_def} for different dielectric constants $\epsilon$ and screening lengths $\xi$. The column colors corresponding to different parameters are explained in the text.}
	\label{app:tab:parameters_master}
\end{table}

For our QMC simulations, we consider 18 different parameter sets for AA t-\ch{SnSe2}{}, corresponding to various twist angles $\theta$, dielectric constants $\epsilon$, and screening lengths $\xi$, as summarized in \cref{app:tab:parameters_master} and in \cref{fig:parameters}. Specifically, we focus on the same commensurate twist angles studied in Ref.~\cite{CAL25b}. For each angle, we consider three different screening lengths, $\xi/\si{\nano\meter} \in \left\lbrace 2.5, 5, 10 \right\rbrace$. Varying $\xi$ primarily tunes the relative strength of onsite and offsite interactions, with larger screening lengths enhancing the relative importance of offsite terms. 

In our QMC calculations, we retain all parameters highlighted in green in \cref{app:tab:parameters_master}, as these are the largest in magnitude. The parameters $t_{\perp}$ and $t''$, highlighted in blue, correspond to subleading longer-range $\tilde{M}_z$-preserving hopping processes, as explained in \cref{app:eqn:leading_hoppings,app:fig:model}. Besides being smaller than $t$, these hoppings become relatively less important at smaller angles, as the orbitals of AA t-\ch{SnSe2}{} become more localized. By contrast, the $\tilde{M}_z$-breaking $t'$ term, highlighted in red, is also small, but its relative importance increases with decreasing twist angle. As explained in Ref.~\cite{CAL25b}, it arises from in-plane relaxation effects. Because relaxation is harder to quantify numerically, we likewise drop this term in our simulations, in a spirit similar to the chiral approximation in twisted bilayer graphene~\cite{TAR19}. Finally, we neglect the ferromagnetic Hund's interaction $J^{V}$, highlighted in orange, both because it is small in magnitude and because it is further reduced by the anti-Hund's antiferromagnetic interaction originating from electron-phonon coupling. In other words, our estimate of the Hund's interaction is likely an upper bound. We leave a comprehensive study of the perturbative effects of all these terms to future work.

For each twist angle, we additionally adjust the dielectric constant $\epsilon$ so as to sample the phase diagram broadly while placing particular emphasis on the intermediate-coupling regime, which we define here by $4 \lesssim \bar{U}/t \lesssim 10$. This choice is motivated by several complementary considerations:
\begin{itemize}
	\item Our goal is to obtain a representative view of the weak-, intermediate-, and strong-coupling regimes, while giving increased resolution to the intermediate-coupling regime, where neither weak-coupling nor strong-coupling approaches are expected to be quantitatively controlled. By contrast, the weak-coupling regime is reached primarily at the largest twist angles, where the moir\'e unit cell is relatively small and the electron densities corresponding to integer fillings of the moir\'e unit cell become more difficult to access experimentally through electrostatic doping. At the opposite end, the strong-coupling regime at integer fillings is already partly constrained by the available analytical description of Ref.~\cite{LI25}, and our QMC calculations are used in part to benchmark and extend that picture. 
	
	\item A further motivation for increasing $\epsilon$ beyond a nominal value such as $\epsilon = 12$ is that our screened Coulomb interaction does not explicitly include screening from remote bands. Ref.~\cite{LI25} showed that, in untwisted \ch{SnSe2}, the onsite Hubbard interaction of the first conduction band is reduced by approximately a factor of three once such remote-band screening effects are taken into account. For the moir\'e conduction band considered here, where the Wannier orbitals are even more spatially extended, these screening effects are expected to be at least as important. We therefore mimic this missing screening by working with a correspondingly larger effective dielectric constant.
	
	\item Finally, our parameter choices should be viewed as a broad sampling of the phase diagram of the Hamiltonian in \cref{app:eqn:final_qmc_hamiltonia}, with deliberate additional emphasis on the physically relevant and particularly rich intermediate-coupling regime. The values listed in \cref{app:eqn:final_interaction} are therefore chosen not to exclude the weak- or strong-coupling limits, but rather to complement them with finer coverage in the regime that is least accessible analytically.
\end{itemize}
Within our numerical simulations, the resulting 18 parameter sets lie predominantly in the intermediate-coupling regime, while still providing access to its crossover toward weaker and stronger interactions. To gain additional insight into the strong-coupling limit, we also include two idealized interaction models with no nearest-neighbor repulsion and $U/t = 12$, both with and without explicit $\mathrm{U} \left( {6} \right)$ symmetry breaking.

\section{Stochastic Series Expansion Method}\label{app:sec:sse}

In this \siSection{}, we discuss the Stochastic Series Expansion (SSE) QMC method~\cite{SAN91,SAN92,SAN97a,SAN97,SAN99,DOR01,DOR02,SEN02,SYL02,SAN03,MEL07,GRO09,XU15,MOR16a,MAJ16,LI19b,SAN19} as applied to AA t-\ch{SnSe2}{}. We begin by mapping the many-body fermionic Hamiltonian from \cref{app:eqn:final_qmc_hamiltonia} to an equivalent hard-core bosonic one via a Jordan-Wigner (JW) transformation. This mapping is carried out here in order to make the absence of a sign problem fully explicit. Since the transformation is exact, the bosonic and fermionic formulations are fully equivalent, and observables of the original fermionic model can be related directly to corresponding observables in the bosonic representation. In practice, the SSE implementation is most naturally formulated in terms of the resulting hard-core bosonic Hamiltonian, and this is the Hamiltonian we actually employ in our simulations. We next briefly review the SSE method and specialize it to our specific model discussed in \cref{app:sec:model:summary}. The QMC updates relevant for our system are discussed separately in \cref{app:sec:updates}. Finally, we explain how various observables and operator correlation functions can be estimated within SSE. In particular, we describe how to compute charge and spin correlators. We also derive expressions for the charge and spin stiffness of the model and show how these quantities can be estimated using SSE. Lastly, we outline a procedure for estimating the entropy of the system from the numerical simulations.

\subsection{Jordan-Wigner transformation and absence of sign problem}\label{app:sec:sse:no_sign_problem}

\begin{figure}[!t]
	\centering
	\includegraphics[width=0.6666667\textwidth]{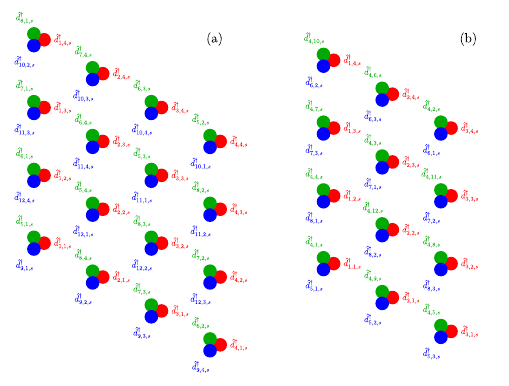}\subfloat{\label{app:fig:chains:a}}\subfloat{\label{app:fig:chains:b}}\caption{Indexing of the fermionic operators in AA t-\ch{SnSe2}{} along one-dimensional chains. (a) shows a $4\times 4$ supercell in which the fermionic operators indexed according to \cref{app:eqn:fermion_operators} are explicitly indicated. In this case, there are 12 one-dimensional chains, each containing four orbitals distributed equally among the three valleys. In contrast, for the $3\times 4$ system shown in (b), the chains are unevenly distributed: there are three chains with four orbitals in valley $\eta=0$ (red), four chains with three orbitals in valley $\eta=2$ (blue), and a single chain with 12 orbitals in valley $\eta=1$ (green). The $\eta=1$ chain wraps around the sample multiple times when periodic boundary conditions are imposed. To avoid such complications, throughout this work we restrict to $\mathcal{N}\times \mathcal{N}$ supercells.}
	\label{app:fig:chains}
\end{figure}                                                                          

To show explicitly that the system does not suffer from a sign problem within SSE, we first perform a Jordan-Wigner (JW) transformation that maps the fermionic operators $\hat{d}^\dagger_{\vec{R},\eta,s}$ to hardcore bosons $\hat{b}^\dagger_{\vec{R},\eta,s}$. We restrict throughout to systems of size $\mathcal{N} \times \mathcal{N}$. Since the JW transformation exploits the quasi-one-dimensional structure of the problem, we introduce the following indexing convention for operators carrying lattice-site, valley, and spin labels. In particular, the fermionic operators $\hat{d}^\dagger_{\vec{R},\eta,s}$ are reindexed along the $3\mathcal{N}$ quasi-one-dimensional chains as
\begin{equation}
	\hat{d}^\dagger_{\alpha,i,s} \equiv \begin{cases}
		\hat{d}^\dagger_{(\alpha-1) \vec{a}_{M_1} + (i-1) \vec{a}_{M_2},0,s}, &\quad 1 \leq \alpha \leq \mathcal{N} \\
		\hat{d}^\dagger_{(\alpha-\mathcal{N}-1) \vec{a}_{M_2} + (i-1) \left( - \vec{a}_{M_1} - \vec{a}_{M_2} \right),1,s}, &\quad \mathcal{N} + 1 \leq \alpha \leq 2\mathcal{N} \\
		\hat{d}^\dagger_{(\alpha-2\mathcal{N}-1) \left( - \vec{a}_{M_1} - \vec{a}_{M_2} \right) + (i-1) \vec{a}_{M_1},2,s}, &\quad 2\mathcal{N} + 1 \leq \alpha \leq 3\mathcal{N} \\
	\end{cases}. \label{app:eqn:fermion_operators}
\end{equation}
Specifically, $\hat{d}^\dagger_{\alpha,i,s}$ creates a spin-$s$ electron at site $i$ of the $\alpha$-th chain. This convention is illustrated in \cref{app:fig:chains:a} for the case $\mathcal{N}=4$. In the same spirit, we define the number operator along the chains as $\hat{n}_{\alpha,i,s} = \hat{d}^\dagger_{\alpha,i,s} \hat{d}_{\alpha,i,s}$. The JW transformation then maps the fermions to hardcore bosons according to 
\begin{align}
	\hat{d}^\dagger_{\alpha,i,s} &= S^{\dagger}_{\alpha,i,s} \hat{b}^\dagger_{\alpha,i,s} = \hat{b}^\dagger_{\alpha,i,s} S^{\dagger}_{\alpha,i,s}, \label{app:eqn:JW_f_cre} \\
	\hat{d}_{\alpha,i,s} &= S_{\alpha,i,s} \hat{b}_{\alpha,i,s} = \hat{b}_{\alpha,i,s} S_{\alpha,i,s}, \label{app:eqn:JW_f_des} \\
	\hat{n}_{\alpha,i,s} &= \hat{d}^\dagger_{\alpha,i,s} \hat{d}_{\alpha,i,s} = \hat{b}^\dagger_{\alpha,i,s} \hat{b}_{\alpha,i,s}, \label{app:eqn:JW_num_op} 
\end{align}
The corresponding JW string operators are defined as
\begin{equation}
	S_{\alpha,i,s} = \exp\left[-i \pi \left( \sum_{\alpha'=1}^{\alpha-1} \sum_{s'} \hat{N}_{\alpha',s'} \right) \right] \exp \left[-i \pi \left( \delta_{s \downarrow} \hat{N}_{\alpha,\uparrow} + \sum_{i'=1}^{i-1} \hat{n}_{\alpha,i',s} \right)  \right],
\end{equation}
where 
\begin{equation}
	\hat{N}_{\alpha,s} =  \sum_{i=1}^{\mathcal{N}} \hat{n}_{\alpha,i,s},
\end{equation}
denotes the total number of particles of spin $s$ along the $\alpha$-th chain. The JW ordering is chosen such that, for a fixed chain, the string first traverses the spin-$\uparrow$ fermions and then the spin-$\downarrow$ fermions, before proceeding to the remaining parallel chains within the same valley and finally to the other two valleys. The essential feature of this construction is the relative ordering of the two spin species within a given chain, since this ensures that the JW-string contributions cancel in the hopping term and thereby renders the resulting bosonic Hamiltonian sign-problem free. By contrast, the ordering among different chains is inconsequential.

It follows directly that if the fermionic operators $\hat{d}^\dagger_{\alpha,i,s}$ satisfy standard anticommutation relations, the operators $\hat{b}^\dagger_{\alpha,i,s}$ obey bosonic commutation relations on different sites,
\begin{equation}
	\commutator{\hat{b}^\dagger_{\alpha,i,s}}{\hat{b}^\dagger_{\alpha',i',s'}} = 
	\commutator{\hat{b}_{\alpha,i,s}}{\hat{b}_{\alpha',i',s'}} = 
	\commutator{\hat{b}_{\alpha,i,s}}{\hat{b}^\dagger_{\alpha',i',s'}} = 0, \qq{for} (\alpha,i,s) \neq (\alpha',i',s')
\end{equation}
together with the onsite hardcore constraints
\begin{equation}
	\left( \hat{b}^\dagger_{\alpha,i,s} \right)^2=
	\left( \hat{b}_{\alpha,i,s} \right)^2=0, \quad
	\anticommutator{\hat{b}^\dagger_{\alpha,i,s}}{\hat{b}_{\alpha,i,s}} = 1.
\end{equation}

We now consider the kinetic energy operator from \cref{app:eqn:final_single_particle} and rewrite it for an $\mathcal{N} \times \mathcal{N}$ system using the chain notation introduced above
\begin{equation}
	H_{0} = -t \sum_{s} \sum_{\alpha=1}^{3\mathcal{N}} \left[ \sum_{i=1}^{\mathcal{N}-1} \left( \hat{d}^\dagger_{\alpha,i,s} \hat{d}_{\alpha,i+1,s} + \hat{d}^\dagger_{\alpha,i+1,s} \hat{d}_{\alpha,i,s} \right) + \hat{\varphi}_{\alpha,s} \left( \hat{d}^\dagger_{\alpha,\mathcal{N},s} \hat{d}_{\alpha,1,s} + \hat{d}^\dagger_{\alpha,1,s} \hat{d}_{\alpha,\mathcal{N},s} \right) - \mu \sum_{i=1}^{\mathcal{N}} \hat{n}_{\alpha,i,s} \right],
\end{equation}
where, in anticipation of the discussion below, we have assumed periodic \emph{twisted} boundary conditions along each chain, parametrized by a phase $\hat{\varphi}_{\alpha,s}$ that depends on the spin sector and on the particle configuration of the $\alpha$-th chain. Under the JW transformation of \cref{app:eqn:JW_f_cre,app:eqn:JW_f_des}, the individual hopping terms transform as
\begin{align}
	\hat{d}^\dagger_{\alpha,i,s} \hat{d}_{\alpha,i+1,s} = \hat{b}^\dagger_{\alpha,i,s} S^{\dagger}_{\alpha,i,s} S_{\alpha,i+1,s} \hat{b}_{\alpha,i+1,s} = \hat{b}^\dagger_{\alpha,i,s} e^{-i \pi \hat{n}_{\alpha,i,s}} \hat{b}_{\alpha,i+1,s} = \hat{b}^\dagger_{\alpha,i,s} \hat{b}_{\alpha,i+1,s}, \nonumber \\
	\hat{d}^\dagger_{\alpha,\mathcal{N},s} \hat{d}_{\alpha,1,s} = \hat{b}^\dagger_{\alpha,\mathcal{N},s} S^{\dagger}_{\alpha,\mathcal{N},s} S_{\alpha,1,s} \hat{b}_{\alpha,1,s} = \hat{b}^\dagger_{\alpha,\mathcal{N},s} e^{i \pi \left( \hat{N}_{\alpha,s} - \hat{n}_{\alpha,\mathcal{N},s} \right)} \hat{b}_{\alpha,1,s} = - e^{i \pi \hat{N}_{\alpha,s}} \hat{b}^\dagger_{\alpha,\mathcal{N},s} \hat{b}_{\alpha,1,s}.
\end{align}
This implies that the kinetic Hamiltonian takes the following form in the bosonic representation
\begin{equation}
	H_{0} = -t \sum_{s} \sum_{\alpha=1}^{3\mathcal{N}} \left[ \sum_{i=1}^{\mathcal{N}-1} \left( \hat{b}^\dagger_{\alpha,i,s} \hat{b}_{\alpha,i+1,s} + \hat{b}^\dagger_{\alpha,i+1,s} \hat{b}_{\alpha,i,s} \right) - \hat{\varphi}_{\alpha,s} e^{i \pi \hat{N}_{\alpha,s}} \left( \hat{b}^\dagger_{\alpha,\mathcal{N},s} \hat{b}_{\alpha,1,s} + \hat{b}^\dagger_{\alpha,1,s} \hat{b}_{\alpha,\mathcal{N},s} \right) - \mu \sum_{i=1}^{\mathcal{N}} \hat{n}_{\alpha,i,s} \right] \label{app:eqn:boundary_condition_ham}.
\end{equation}
In the bosonic number basis, all off-diagonal matrix elements are non-positive ({\it i.e.}{} the kinetic Hamiltonian is stoquastic) only if
\begin{equation}
	\label{app:eqn:boundary_condition_twisted}
	\hat{\varphi}_{\alpha,s} = -e^{i \pi \hat{N}_{\alpha,s}},
\end{equation}
which is simply a parity-dependent sign. Therefore, adopting periodic boundary conditions for the hardcore bosons while maintaining the absence of a sign problem in SSE requires parity-dependent boundary conditions in the original fermionic model. We adopt this choice and consider fermionic systems on an $\mathcal{N} \times \mathcal{N}$ lattice in which the boundary condition along each chain and for each spin sector is antiperiodic (periodic) when the corresponding particle number is even (odd). Under this prescription, the fermionic Hamiltonian maps to a hardcore bosonic model with genuine periodic boundary conditions and remains sign-problem free within SSE. 

Thus, for each fixed particle-number sector on a given chain, the fermionic model maps to a hard-core bosonic one with periodic boundary conditions and non-positive off-diagonal matrix elements. In the SSE simulation, if an update changes the particle-number parity of a chain, the corresponding fermionic boundary condition is updated as well: antiperiodic for even particle number and periodic for odd particle number. This is a sector-dependent boundary prescription rather than a modification of the local Hamiltonian, and, given that the system is topologically trivial, its effect is confined to finite-size boundary terms, so the thermodynamic-limit physics is unchanged.

From this point onward, we impose the aforementioned parity-dependent boundary conditions along each chain and take the kinetic Hamiltonian to be
\begin{equation}
	H_{0} = -t \sum_{s} \sum_{\alpha=1}^{3\mathcal{N}}  \sum_{i=1}^{\mathcal{N}} \left( \hat{b}^\dagger_{\alpha,i,s} \hat{b}_{\alpha,i+1,s} + \hat{b}^\dagger_{\alpha,i+1,s} \hat{b}_{\alpha,i,s} - \mu \hat{n}_{\alpha,i,s} \right),
\end{equation}
with periodic boundary conditions $\hat{b}^\dagger_{\alpha,i+\mathcal{N},s} = \hat{b}^\dagger_{\alpha,i,s}$. In the bosonic representation, the interaction term $H_{I}$ retains identical form as in \cref{app:eqn:final_interaction}. Finally, we note that our choice to restrict to systems of size $\mathcal{N} \times \mathcal{N}$ guarantees that all quasi-one-dimensional chains have equal length, a consequence of translation symmetry together with $C_{3z}$ rotational symmetry. For other aspect ratios, such as a $3 \times 4$ system, the chains generally differ in length and may wind around the torus multiple times, as illustrated in \cref{app:fig:chains:b}.

\subsection{Review of the SSE method}\label{app:sec:sse:review}
The SSE algorithm has been employed extensively in previous studies of fermionic~\cite{SAN92,SEN02,XU15,MOR16a}, bosonic~\cite{DOR01,DOR02,MAJ16}, and spin systems~\cite{SAN91,SAN97a,SAN97,SAN99,DOR01,SYL02,SAN03,MEL07,GRO09}. We now briefly outline its application to the present model. Up to a constant, the grand-canonical Hamiltonian of the first conduction band of AA t-\ch{SnSe2}{} can be decomposed into a sum of nearest-neighbor bond operators,
\begin{equation}
	\label{app:eqn:SSE_hamiltonian_decomposition}
	H = - \sum_{b=1}^{3\mathcal{N}^2} \left( H_{1,b} + H_{2,b} + H_{3,b} + H_{4,b} \right),
\end{equation}
where the index $b$ labels a nearest-neighbor ``bond''. The corresponding bond Hamiltonian terms are
\begin{align}
	H_{1,b} =& \mathcal{C} - \sum_{\substack{ \eta_1,s_1 \\ \eta_2,s_2}} V_{\eta_1 \eta_2} \left( \vec{R}^b_1 - \vec{R}^b_2 \right) \left( \hat{n}_{\vec{R}^b_1, \eta_1, s_1} - \frac12 \right) \left( \hat{n}_{\vec{R}^b_2, \eta_2, s_2} - \frac12 \right) + 
	\frac{\mu}{2g}  \sum_{\eta,s} \left( \hat{n}_{\vec{R}^b_1,\eta,s} + \hat{n}_{\vec{R}^b_2,\eta,s} \right)  \nonumber \\
	&- \frac{1}{4g} \sum_{\substack{ \eta_1,s_1 \\ \eta_2,s_2}} V_{\eta_1 \eta_2} \left( \vec{0} \right) \left[  \left( \hat{n}_{\vec{R}^b_1, \eta_1, s_1} - \frac12 \right) \left( \hat{n}_{\vec{R}^b_1, \eta_2, s_2} - \frac12 \right) + \left( \hat{n}_{\vec{R}^b_2, \eta_1, s_1} - \frac12 \right) \left( \hat{n}_{\vec{R}^b_2, \eta_2, s_2} - \frac12 \right) \right], \label{app:eqn:sse_hams_1b} \\
	H_{2,b} =& t \left( \hat{b}^\dagger_{\vec{R}^b_1,\eta^b,\uparrow}\hat{b}_{\vec{R}^b_2,\eta^b,\uparrow} + \hat{b}^\dagger_{\vec{R}^b_2,\eta^b,\uparrow}\hat{b}_{\vec{R}^b_1,\eta^b,\uparrow} \right), \label{app:eqn:sse_hams_2b} \\
	H_{3,b} =& t \left( \hat{b}^\dagger_{\vec{R}^b_1,\eta^b,\downarrow}\hat{b}_{\vec{R}^b_2,\eta^b,\downarrow} + \hat{b}^\dagger_{\vec{R}^b_2,\eta^b,\downarrow}\hat{b}_{\vec{R}^b_1,\eta^b,\downarrow} \right), \label{app:eqn:sse_hams_3b}\\
	H_{4,b} =& t. \label{app:eqn:sse_hams_4b}
\end{align}
In \cref{app:eqn:sse_hams_1b} denotes half the coordination number of a triangular-lattice site ($g=3$) or, equivalently, the number of unique -- according to \cref{app:eqn:unique_bonds} -- nearest-neighbor bonds associated with a given site. An $\mathcal{N}\times \mathcal{N}$ system therefore contains $g \mathcal{N}^{2}=3\mathcal{N}^{2}$ unique nearest-neighbor bonds, each specified by the triplet $(\vec{R}^{b}_{1},\vec{R}^{b}_{2},\eta^{b})$ defined through
\begin{equation}
	\label{app:eqn:unique_bonds}
	\vec{R}^b_1 = x \vec{a}_{M_1} + y \vec{a}_{M_2}, \quad
	\vec{R}^b_2 = x \vec{a}_{M_1} + y \vec{a}_{M_2} + C_{3z}^n \vec{a}_{M_2}, \qq{and}
	\eta^b = n
\end{equation}
which corresponds to the bond index $b=n \mathcal{N}^{2} + x \mathcal{N} + y + 1$, with $0\le n\le2$ and $0\le x,y \le \mathcal{N}-1$. In \crefrange{app:eqn:sse_hams_1b}{app:eqn:sse_hams_4b}, the constant $\mathcal{C}$ is chosen such that the diagonal operator $H_{1,b}$ has only non-negative matrix elements. This property allows the partition function to be expanded as a sum of strictly non-negative contributions. Finally, the operators $H_{4,b}$ are introduced specifically for the SSE method as they will be exchanged with $H_{2,b}$ and $H_{3,b}$ operators during the off-diagonal updates, as we explain in more detail in \cref{app:sec:updates:offdiagonal}. 

To construct the expansion of the grand canonical partition function, we introduce the Fock basis\footnote{Strictly speaking, the SSE implementation is carried out in the hard-core bosonic representation obtained via the JW transformation. Since this mapping is exact, we keep the fermionic language and denote the local states as $0$, $\uparrow$, $\downarrow$, and $\uparrow\downarrow$, referring to them as empty, spin-up, spin-down, and doubly occupied electronic states, respectively.}
\begin{equation}
	\ket{\alpha} = \ket{\left \lbrace \xi_{\vec{R},\eta} \right\rbrace},
\end{equation}
in which $\xi_{\vec{R},\eta}\in\{0,\uparrow,\downarrow,\uparrow\downarrow\}$ denotes the local occupation state. The partition function can then be expressed through a Taylor expansion,
\begin{equation}
	Z=\Tr \left( e^{-\beta H} \right)= \sum_{\alpha}\sum_{n=0}^{\infty} \frac{\beta^n}{n!}
	\mel**{\alpha}{\left( -H \right)^n}{\alpha} =\sum_{\alpha}\sum_{n=0}^{\infty} \sum_{S_n} \frac{\beta^n}{n!}
	 \mel**{\alpha}{\prod_{p=1}^n H_{a_{p},b_{p}}}{\alpha},
\end{equation}
where $S_{n}$ denotes an ordered sequence of index pairs defining the operator string $\prod_{p=1}^n H_{a_{p},b_{p}}$
\begin{equation}
	\label{app:eqn:operator_sequences}
	S_n=[a,b]_1[a,b]_2\dots [a,b]_n,
\end{equation}
with the shorthand $[a,b]_p = \left[a_p,b_p \right]$ (for $1\le a\le4$ and $1\le b\le3\mathcal{N}^{2}$). It is also convenient to define the normalized intermediate states generated by the first $p$ operators of the string $S_n$
\begin{equation}
	\bra{\alpha (p)} \sim \bra{\alpha} \prod_{j=1}^p H_{a_{j},b_{j}}, \qq{for} 0 \leq p \leq n,
\end{equation}
in which $\ket{\alpha(0)}=\ket{\alpha(n)}=\ket{\alpha}$. Because of the structure of the bond operators in \crefrange{app:eqn:sse_hams_1b}{app:eqn:sse_hams_4b}, all intermediate states $\ket{\alpha(p)}$ remain Slater-determinant states expressed in the same Fock basis. The configurations contributing to the partition function are sampled within Monte Carlo according to their statistical weights, as will be detailed in \cref{app:sec:updates}. Crucially, the absence of a sign problem follows from the fact that all matrix elements of the bond operators are non-negative. In addition, note that the $H_{1,b}$ and $H_{4,b}$ operators are diagonal in the Fock basis, whereas $H_{2,b}$ and $H_{3,b}$ are off-diagonal, and we will refer to them accordingly in what follows.

In practice, it is more convenient to truncate the Taylor series at a fixed power $L$, chosen sufficiently large that the resulting error is exponentially small. To see this explicitly, we bound the truncation error as
\begin{equation}
	Z - Z_{\text{truncated}} = \sum_{n=L+1}^{\infty} \frac{\beta^n}{n!} \Tr \left( \left(- H \right)^n \right) \leq \dim(H) \sum_{n=L+1}^{\infty} \frac{\left( \beta \norm{-H} \right)^n}{n!},
\end{equation}
where $\dim \left( H \right)$ is the Hilbert-space dimension of $H$, and $\norm{-H}$ denotes the operator norm of $-H$, {\it i.e.}{} its largest eigenvalue. The remaining infinite sum can in turn be bounded as
\begin{align}
	\sum_{n=L+1}^{\infty} \frac{x^n}{n!} =& \frac{x^{L+1}}{(L+1)!} \left( 1 + \frac{x}{L+2} + \frac{x^2}{(L+2)(L+3)} + \dots \right) \leq
	\frac{x^{L+1}}{(L+1)!} \sum_{n=0}^{\infty} \left( \frac{x}{L+2} \right)^n = \frac{x^{L+1}}{(L+1)!} \frac{1}{1- \frac{x}{L+2}} \leq \nonumber \\
	\leq &  \frac{2 x^{L+1}}{(L+1)!} \sim \frac{x^{L+1}}{\sqrt{2 \pi (L+1)} \left( \frac{L+1}{e} \right)^{L+1}} =  \frac{1}{\sqrt{2 \pi (L+1)}} \left(  \frac{e x}{L+1} \right)^{L+1}, \qq{for} 0 \leq \frac{x}{L+2} \leq \frac12,
\end{align}
which shows that, for sufficiently large $L$, the truncation error is indeed exponentially small,
\begin{equation}
	Z - Z_{\text{truncated}} \sim \frac{\dim \left( H \right)}{\sqrt{2 \pi (L+1)}} \left(  \frac{e \beta \norm{-H}}{L+1} \right)^{L+1}.
\end{equation}
In what follows, we use $Z$ to denote either the full or the truncated partition functions -- the difference between them is vanishingly small.

Numerically, $L$ is determined self-consistently during the equilibration (warm-up) stage by setting $L=\frac54 n_{\max}$, where $n_{\max}$ denotes the largest expansion order encountered during equilibration\footnote{An initial $L=64$ is chosen. SSE configurations are then generated using the updates detailed in \cref{app:sec:updates}. The quantity $n_{\max}$ keeps track of the maximal value of $n$ encountered among the generated configurations. Every time $n_{\max}$ increases, so does $L$. After the steady state distribution has been reached, $L$ does not change anymore.}.
As a consistency check, we note that the configurations sampled during the measurement stage, {\it i.e.}{} after the equilibration stage in which $L$ is determined, satisfy $n \leq 0.9 L$\footnote{As explained in \cref{app:sec:updates:pt:prob}, this can be understood from the fact that at sufficiently low temperatures the expansion order is approximately Poisson distributed, up to energy fluctuations, with mean $\left\langle n \right\rangle = \left\langle -\beta H \right\rangle$. Over a finite set of generated QMC samples, one expects $n_{\max} = \left\langle n \right\rangle + z \sqrt{\left\langle n \right\rangle}$, where $z$ is an order-one constant that increases slowly with the number of samples. Since we estimate $n_{\max}$ during a warm-up stage whose length is comparable to that of the measurement stage, the choice $L=\frac54 n_{\max}$ provides a conservative cutoff and is safely larger than the expansion orders expected to occur in practice.}. 
For bookkeeping purposes, we extend each operator sequence $S_n$ to length $L$ by inserting $L-n$ identity operators, denoted by $H_{0,0}$. This defines the sequence $\tilde{S}_L$. Mathematically, the partition function can then be written, up to exponentially small corrections, as a sum over fixed-length operator sequences $\tilde{S}_L$ that may contain identity operators.

In this formulation, the explicit summation over $n$ is replaced by a summation over $\tilde{S}_L$. Since identity operators can be inserted in ${L \choose n}$ different ways, each contribution must be divided by this combinatorial factor. The partition function thus reads
\begin{equation}
	\label{app:eqn:part_function_initial_SSE}
	Z=\Tr \left( e^{-\beta H} \right)= \sum_{\alpha} \sum_{\tilde{S}_L} \frac{\beta^{n} \left( L-n \right)!}{L!}
	\mel**{\alpha}{\prod_{p=1}^{L} H_{\tilde{a}_{p},\tilde{b}_{p}}}{\alpha},
\end{equation}
where $n$ denotes the number of non-identity operators in $\tilde{S}_L$, and the operator-index pairs $[\tilde{a},\tilde{b}]_p$ may now also take the value $[0,0]$ corresponding to identity operators. The partition function is therefore expressed as a sum over SSE configurations, which we define as the tuples $\left(\alpha, \tilde{S}_L \right)$ formed by an initial state $\alpha$ and an operator sequence $\tilde{S}_L$. We also define $\ket{\tilde{\alpha}(p)}$ as the normalized state obtained after acting with the first $p$ operators of the sequence $\tilde{S}_L$,
\begin{equation}
	\bra{\tilde{\alpha} (p)} \sim \bra{\alpha} \prod_{j=1}^p H_{\tilde{a}_{j},\tilde{b}_{j}}, \qq{for} 0 \leq p \leq L,
\end{equation}
with $\ket{\tilde{\alpha}(0)}=\ket{\tilde{\alpha}(L)}=\ket{\alpha}$. It is important to distinguish between the two types of sequences: strings with a tilde ($\tilde{S}_L$) include identity operators, while those without a tilde ($S_n$) contain only non-identity operators. Similarly, indices with a tilde label positions within the $\tilde{S}_L$ sequence (indexed by the propagation index $1 \leq p \leq L$), whereas indices without a tilde refer to positions within the $S_n$ sequence.

\subsection{Measuring observables}\label{app:sec:sse:observables}

Before discussing how the terms in the partition function in \cref{app:eqn:part_function_initial_SSE} ({\it i.e.}{}, the SSE configurations) are sampled, we first explain how the various observables are measured. In this work, we focus exclusively on so-called \emph{diagonal} observables ({\it i.e.}{}, expectation values of operators that are diagonal in the Fock basis)~\cite{SAN97a,SEN02,SAN19}, as well as on observables built from operators that can be written as linear combinations of the bond Hamiltonian terms in \crefrange{app:eqn:sse_hams_1b}{app:eqn:sse_hams_4b} (such as the energy of the system). Methods for sampling \emph{off-diagonal} observables have also been developed~\cite{DOR01,DOR02,SAN19} and will be employed in future work

We begin by introducing the weight functions
\begin{align}
	W \left( \alpha, S_n \right) &= \frac{1}{Z} \frac{\beta^n}{n!} \mel**{\alpha}{\prod_{p=1}^{n} H_{a_{p},b_{p}}}{\alpha} = \sum_{\tilde{S}_L \supset S_n} \tilde{W} \left( \alpha, \tilde{S}_L \right), \label{app:eqn:weight_SSE}\\
	\tilde{W} \left( \alpha, \tilde{S}_L \right) &= \frac{1}{Z} \frac{\beta^n \left(L-n\right)!}{L!} \mel**{\alpha}{\prod_{p=1}^{L} H_{\tilde{a}_{p},\tilde{b}_{p}}}{\alpha}, \label{app:eqn:weight_tilde_SSE}
\end{align}
where $\tilde{S}_L \supset S_n$ denotes the augmented sequences obtained from $S_n$ by inserting $(L-n)$ identity operators. We use $\left\langle \dots \right\rangle$ to denote thermal expectation values in the grand canonical ensemble at inverse temperature $\beta$ and chemical potential $\mu$, and $\left\langle \dots \right\rangle_{\text{MC}}$ for expectation values evaluated with respect to the Monte Carlo weight $\tilde{W} \left( \alpha, \tilde{S}_L \right)$. Specifically,
\begin{align}
	\left\langle \mathcal{O} \right\rangle &= \frac1Z \Tr\left( e^{- \beta H} \mathcal{O} \right), \\
	\left\langle \mathcal{F} \right\rangle_{\text{MC}} &= \sum_{\left( \alpha, \tilde{S}_L \right)} \mathcal{F} \left( \alpha, \tilde{S}_L \right)  \tilde{W} \left( \alpha, \tilde{S}_L \right),
\end{align}
where in the first line case $\mathcal{O}$ is an operator and in the second line $\mathcal{F}$ is a function of the SSE configuration $\left( \alpha, \tilde{S}_L \right)$ ({\it e.g.}{}, one such function would be the number of non-identity operators $n$ in the operator string $\tilde{S}_L$).

In what follows, we consider thermal expectation values involving one or two operators. We focus on diagonal operators $D$ (which correspond to the diagonal observables of interest) and on the bond operators from \crefrange{app:eqn:sse_hams_1b}{app:eqn:sse_hams_4b}. We denote the eigenvalues of $D$ acting on the states at propagation index $p$ in the operator sequences by
\begin{equation}
	\label{app:eqn:SSE_def_diag_obs}
	D \ket{\alpha(p)} = d_p \ket{\alpha(p)} \qq{and} D \ket{\tilde{\alpha}(p)} = \tilde{d}_p \ket{\tilde{\alpha}(p)}.
\end{equation}
Finally, for an operator $\mathcal{O}$, we define its imaginary-time evolution as
\begin{equation}
	\mathcal{O} (\tau) \equiv e^{-\tau H} \mathcal{O} e^{\tau H}. 
\end{equation}

\subsubsection{Single-operator expectation values}\label{app:sec:sse:observables:single}

We begin by considering the thermal expectation value of one of the bond operators introduced in \crefrange{app:eqn:sse_hams_1b}{app:eqn:sse_hams_4b}
\begin{align}
	\left\langle H_{i,j} \right\rangle =& \frac{1}{Z} \Tr \left( H_{i,j} e^{-\beta H} \right) = \frac{1}{Z} \sum_{\alpha}\sum_{n=0}^{\infty} \sum_{S_n} \frac{\beta^n}{n!}
	\mel**{\alpha}{H_{i,j} \prod_{p=1}^n H_{a_{p},b_{p}}}{\alpha} \nonumber \\
	=& \frac{1}{Z} \sum_{\alpha}\sum_{n=0}^{\infty} \sum_{S_{n+1}} \frac{\beta^n}{n!} \delta_{[i,j],[a,b]_1}
	\mel**{\alpha}{\prod_{p=1}^{n+1} H_{a_{p},b_{p}}}{\alpha} \nonumber \\
	=& \frac{1}{Z} \sum_{\alpha}\sum_{n=0}^{\infty} \sum_{S_{n+1}} \frac{\beta^n}{n!} \left( \frac{1}{n+1} \sum_{p'=1}^{n+1} \delta_{[i,j],[a,b]_{p'}} \right)
	\mel**{\alpha}{\prod_{p=1}^{n+1} H_{a_{p},b_{p}}}{\alpha}\nonumber \\
	=& \sum_{\alpha} \sum_{\tilde{S}_L} \frac{1}{\beta} N_{H_{i,j}} \tilde{W} \left( \alpha, \tilde{S}_L \right) \nonumber \\
	=& \frac{1}{\beta} \left\langle N_{H_{i,j}} \right\rangle_{\text{MC}},
\end{align}
where $N_{H_{i,j}}$ denotes the number of occurrences of the operator $H_{i,j}$ in a given augmented operator sequence $\tilde{S}_L$. This immediately yields an estimator for the grand canonical energy of the system, by summing over ${i,j}$
\begin{equation}
	\left\langle H \right\rangle = -\frac{1}{\beta} \left\langle n \right\rangle_{\text{MC}}. 
\end{equation}

We now turn to diagonal operators. The expectation value of a generic diagonal operator $D$ can be written as
\begin{align}
	\left\langle D \right\rangle =& \frac{1}{Z} \Tr \left( D e^{-\beta H} \right) = \frac{1}{Z} \sum_{\alpha}\sum_{n=0}^{\infty} \sum_{S_n} \frac{\beta^n}{n!}
	\mel**{\alpha}{D \prod_{p=1}^n H_{a_{p},b_{p}}}{\alpha} \nonumber \\
	=& \frac{1}{Z} \sum_{\alpha}\sum_{n=0}^{\infty} \sum_{S_n} \frac{\beta^n}{n!} d_{1}
	\mel**{\alpha}{\prod_{p=1}^{n} H_{a_{p},b_{p}}}{\alpha} \nonumber \\
	=& \frac{1}{Z} \sum_{\alpha}\sum_{n=0}^{\infty} \sum_{S_n} \frac{\beta^n}{n!} \left( \frac{1}{n} \sum_{p'=1}^{n} d_{p'} \right)
	\mel**{\alpha}{\prod_{p=1}^{n} H_{a_{p},b_{p}}}{\alpha}\nonumber \\
	=& \sum_{\alpha} \sum_{\tilde{S}_L} \left( \frac{1}{n} \sum_{p=1}^{n} d_{p} \right) \tilde{W} \left( \alpha, \tilde{S}_L \right) \nonumber \\
	=& \left\langle \left( \frac{1}{n} \sum_{p=1}^{n} d_{p} \right) \right\rangle_{\text{MC}},
\end{align}
where $d_p$ was defined in \cref{app:eqn:SSE_def_diag_obs}.

\subsubsection{Double-operator expectation values}\label{app:sec:sse:observables:double}

We now consider imaginary-time correlation functions. We first derive the estimator for correlations of the bond operators introduced in \crefrange{app:eqn:sse_hams_1b}{app:eqn:sse_hams_4b}
\begin{align}
	\left\langle H_{i',j'} (\tau) H_{i,j} (0) \right\rangle =& \frac{1}{Z} \sum_{\alpha}\sum_{n,m=0}^{\infty} \frac{(\beta-\tau)^n \tau^m}{n! m!} \mel**{\alpha}{(-H)^n H_{i',j'} (-H)^m H_{i,j}}{\alpha} \nonumber \\
	=& \frac{1}{Z} \sum_{\alpha}\sum_{n,m=0}^{\infty} \sum_{S_{n+m+2}} \frac{\beta^{n+m}}{n! m!} \left(1 - \frac{\tau}{\beta} \right)^n \left(\frac{\tau}{\beta} \right)^m \delta_{[i',j'],[a,b]_{n+1}} \delta_{[i,j],[a,b]_{n+m+2}} 
	\mel**{\alpha}{\prod_{p=1}^{n+m+2} H_{a_{p},b_{p}}}{\alpha} \nonumber \\
	=& \frac{1}{Z} \sum_{\alpha}\sum_{n=0}^{\infty} \sum_{m=0}^{n} \sum_{S_{n+2}} \frac{\beta^{n}}{m! (n-m)!} \left(1 - \frac{\tau}{\beta} \right)^{n-m} \left(\frac{\tau}{\beta} \right)^m \delta_{[i',j'],[a,b]_{n-m+1}} \delta_{[i,j],[a,b]_{n+2}} 
	\mel**{\alpha}{\prod_{p=1}^{n+2} H_{a_{p},b_{p}}}{\alpha} \nonumber \\
	=& \sum_{\alpha}\sum_{n=0}^{\infty} \sum_{m=0}^{n} \sum_{S_{n+2}} \frac{(n+2)!}{\beta^2 m! (n-m)!} \left(1 - \frac{\tau}{\beta} \right)^{n-m} \left(\frac{\tau}{\beta} \right)^m \delta_{[i',j'],[a,b]_{n-m+1}} \delta_{[i,j],[a,b]_{n+2}} 
	W \left( \alpha, S_{n+2} \right) \nonumber \\
	=& \sum_{\alpha}\sum_{n=2}^{\infty} \sum_{m=0}^{n-2} \sum_{S_n} \frac{n!}{\beta^2 m! (n-2-m)!} \left(1 - \frac{\tau}{\beta} \right)^{n-2-m} \left(\frac{\tau}{\beta} \right)^m \delta_{[i',j'],[a,b]_{n-m-1}} \delta_{[i,j],[a,b]_n} 
	W \left( \alpha, S_n \right) \nonumber \\
	=& \sum_{\alpha} \sum_{\tilde{S}_L} \sum_{m=0}^{n-2} \frac{n!}{\beta^2 m! (n-2-m)!} \left(1 - \frac{\tau}{\beta} \right)^{n-2-m} \left(\frac{\tau}{\beta} \right)^m \frac{1}{n}\sum_{p=1}^{n}\delta_{[i',j'],[a,b]_{p-m-1}} \delta_{[i,j],[a,b]_p} 
	\tilde{W} \left( \alpha, \tilde{S}_L \right) \nonumber \\
	=& \left\langle \sum_{m=1}^{n-1} \frac{(n-1)!}{\beta^2 (m-1)! (n-1-m)!} \left(1 - \frac{\tau}{\beta} \right)^{n-1-m} \left(\frac{\tau}{\beta} \right)^{m-1} \sum_{p=1}^{n}\delta_{[i',j'],[a,b]_{p-m}} \delta_{[i,j],[a,b]_p} \right\rangle_{\text{MC}},
\end{align}
where $0 \leq \tau \leq \beta$, and the cyclic convention $[a,b]_{p}$ for $p<1$ or $p>n$ is understood as $[a,b]_{[(p-1)\mod n] +1}$. For two diagonal operators $D$ and $D'$, an analogous derivation yields
\begin{align}
	\left\langle D' (\tau) D (0) \right\rangle=& \frac{1}{Z} \sum_{\alpha}\sum_{n,m=0}^{\infty} \frac{(\beta-\tau)^n \tau^m}{n! m!} \mel**{\alpha}{(-H)^n D' (-H)^m D}{\alpha} \nonumber \\
	=& \frac{1}{Z} \sum_{\alpha}\sum_{n,m=0}^{\infty} \sum_{S_{n+m}} \frac{\beta^{n+m}}{n! m!} \left(1 - \frac{\tau}{\beta} \right)^n \left(\frac{\tau}{\beta} \right)^m d'_n d_{n+m} \mel**{\alpha}{\prod_{p=1}^{n+m} H_{a_{p},b_{p}}}{\alpha} \nonumber \\
	=& \frac{1}{Z} \sum_{\alpha}\sum_{n=0}^{\infty} \sum_{S_n} \sum_{m=0}^n \frac{\beta^n}{(n-m)! m!} \left(1 - \frac{\tau}{\beta} \right)^{n-m} \left(\frac{\tau}{\beta} \right)^m \frac{1}{n}\sum_{p=1}^n d'_{p-m} d_p \mel**{\alpha}{\prod_{p=1}^n H_{a_{p},b_{p}}}{\alpha} \nonumber \\
	=& \sum_{\alpha} \sum_{\tilde{S}_L} \sum_{m=0}^n \frac{n!}{(n-m)! m!} \left(1 - \frac{\tau}{\beta} \right)^{n-m} \left(\frac{\tau}{\beta} \right)^m \frac{1}{n}\sum_{p=1}^n d'_{p-m} d_p \tilde{W} \left( \alpha, \tilde{S}_L\right) \nonumber \\
	=& \left\langle \sum_{m=0}^n \frac{(n-1)!}{(n-m)! m!} \left(1 - \frac{\tau}{\beta} \right)^{n-m} \left(\frac{\tau}{\beta} \right)^m \sum_{p=1}^n d'_{p-m} d_p \right\rangle_{\text{MC}}.
\end{align}
Equal-time correlators follow directly from these expressions and take the form
\begin{align}
	\left\langle H_{i',j'} H_{i,j} \right\rangle &=  \left\langle  \frac{n-1}{\beta} \sum_{p=1}^{n}\delta_{[i',j'],[a,b]_{p-1}} \delta_{[i,j],[a,b]_p} \right\rangle_{\text{MC}}, \\
	\left\langle D' D \right\rangle &= \left\langle \frac{1}{n} \sum_{p=1}^{n} d'_p d_p \right\rangle_{\text{MC}}.
\end{align}
It is also useful to obtain estimators for the zero-frequency correlators. Using the identity
\begin{equation}
	\int_{0}^{\beta} \dd{\tau} \left(1 - \frac{\tau}{\beta} \right)^{n-m} \left(\frac{\tau}{\beta} \right)^m = \beta \frac{(n-m)! m!}{(n+1)!},
\end{equation}
one finds
\begin{align}
	\int_{0}^{\beta} \dd{\tau} \left\langle H_{i',j'} (\tau) H_{i,j}(0) \right\rangle &= \frac{1}{\beta} \left\langle \sum_{m=1}^{n-1} \sum_{p=1}^{n} \delta_{[i',j'],[a,b]_{p-m}} \delta_{[i,j],[a,b]_p} \right\rangle_{\text{MC}} \nonumber \\
	&= \frac{1}{\beta} \left\langle N_{H_{i,j}} N_{H_{i',j'}} - \delta_{[i,j],[i'j']} N_{H_{i,j}}  \right\rangle_{\text{MC}}, \label{app:eqn:zero_freq_ham_terms} \\
	\int_{0}^{\beta} \dd{\tau} \left\langle D' (\tau) D (0) \right\rangle &= \left\langle \frac{1}{n(n+1)}\left[ \sum_{p=1}^n d_p d'_p + \left( \sum_{p=1}^n d_p \right) \left( \sum_{p=1}^n d'_p \right) \right] \right\rangle_{\text{MC}}. \label{app:eqn:zero_freq_diag_ops}
\end{align}

\subsubsection{Charge and spin correlators}\label{app:sec:sse:observables:charge}

Among the diagonal observables we focus on in this work are the valley-resolved and valley-averaged connected charge correlators and susceptibilities, defined as
\begin{alignat}{2}
	\mathcal{C}_{\eta_1 \eta_2} \left( \vec{R} \right) &= \left\langle \hat{N}_{\vec{R}', \eta_1} \hat{N}_{\vec{R}' + \vec{R},\eta_2} \right\rangle_c, &\quad
	\mathcal{C} \left( \vec{R} \right) &= \left\langle \hat{N}_{\vec{R}'}  \hat{N}_{\vec{R}' + \vec{R}} \right\rangle_c, \\
	\chi_{\eta_1 \eta_2} \left( \vec{R} \right) &= \int_{0}^{\beta} \dd{\tau} \left\langle \hat{N}_{\vec{R}',\eta_1} (\tau) \hat{N}_{\vec{R}' + \vec{R},\eta_2} (0) \right\rangle_c, &\quad
	\chi \left( \vec{R} \right) &= \int_{0}^{\beta} \dd{\tau} \left\langle \hat{N}_{\vec{R}'} (\tau) \hat{N}_{\vec{R}' + \vec{R}} (0) \right\rangle_c,
\end{alignat}
which, by translation invariance, are independent of the choice of $\vec{R}'$. Here, the valley-resolved and total charge densities are defined by
\begin{equation}
	\hat{N}_{\vec{R},\eta} = \sum_{s} \hat{n}_{\vec{R},\eta,s} \qq{and}
	\hat{N}_{\vec{R}} = \sum_{\eta} \hat{N}_{\vec{R},\eta}.
\end{equation}
In the equations above, we have defined the connected correlator between two quantities $\mathcal{O}_1$ and $\mathcal{O}_2$ to be 
\begin{equation}
	\left\langle \mathcal{O}_1 (\tau_1)\mathcal{O}_2 (\tau_2) \right\rangle_c \equiv \left\langle \left( \mathcal{O}_1 (\tau_1) - \left\langle \mathcal{O}_1 (\tau_1) \right\rangle \right) \left( \mathcal{O}_2 (\tau_2) - \left\langle \mathcal{O}_2 (\tau_2) \right\rangle \right) \right\rangle.
\end{equation}
The Fourier transforms of the correlators and susceptibilities are given by
\begin{equation}
	{\chi,C}_{(\eta_1 \eta_2)} \left( \vec{k} \right) = \frac{1}{\mathcal{N}^2} \sum_{\vec{R}} {\chi,C}_{(\eta_1 \eta_2)} \left( \vec{R} \right) e^{i \vec{k} \cdot \vec{R}}.
\end{equation}
Because each chain can be independently rotated by an arbitrary $\mathrm{SU} \left( {2} \right)$ transformation, spins on different chains are uncorrelated. As a result, the spin-spin correlations are fully characterized by the correlation function along each chain. For this purpose, we define
\begin{equation}
	\label{app:eqn:def_spin_spin}
	\mathcal{S}(i) = \left\langle \hat{S}^z_{\vec{R},\eta} \hat{S}^z_{\vec{R} + i C^{\eta}_{3z} \vec{a}_{M_2},\eta} \right\rangle, \qq{for any} \vec{R}, \eta, 0 \leq i \leq \mathcal{N}-1
\end{equation} 
where the spin operators are given by
\begin{equation}
	\hat{S}^z_{\vec{R},\eta} \equiv \frac{1}{2} \left( \hat{n}_{\vec{R},\eta,\uparrow} - \hat{n}_{\vec{R},\eta,\downarrow} \right).
\end{equation}

\subsubsection{Charge and spin stiffnesses}\label{app:sec:sse:observables:stiffness}

To obtain the charge and spin stiffnesses of the system, we first thread a flux through the system and modify the kinetic part of the Hamiltonian accordingly,
\begin{equation}
	H_0 \left( \phi_{\text{C},\text{S}} \right) = -t \sum_{\vec{R},\eta,s} \left( e^{i\phi_{\text{C},\text{S}} \left(C^{\eta}_{3z} \vec{a}_{M_2} \right) \cdot \vec{\hat{x}} f^{\text{C},\text{S}}_{s}}\hat{b}^\dagger_{\vec{R} + C^{\eta}_{3z} \vec{a}_{M_2},\eta,s} \hat{b}_{\vec{R},\eta,s} + \text{h.c.} \right).
\end{equation}
Here $\phi_{\text{C}}$ denotes a flux coupled to the charge sector, while $\phi_{\text{S}}$ denotes a flux coupled to the spin sector. The factor $f^{\text{C},\text{S}}_{s}$ specifies whether the flux couples to the charge or spin degrees of freedom,
\begin{equation}
	f^{\text{C}}_{s} = \delta_{s,\uparrow} + \delta_{s,\downarrow}, \quad
	f^{\text{S}}_{s} = \delta_{s,\uparrow} - \delta_{s,\downarrow}.
\end{equation}
Expanding the kinetic Hamiltonian to second order in $\phi$ yields
\begin{align}
	H_0 (\phi) &= H_0 + i \phi A  - \frac{\phi^2}{2} B  + \mathcal{O} \left( \phi^3 \right) \\
	A &= -\sum_{\vec{R},\eta,s} f_s \left(C^{\eta}_{3z} \vec{a}_{M_2} \right) \cdot \vec{\hat{x}} \left( J^+_{s,\eta} \left( \vec{R} \right) - J^-_{s,\eta} \left( \vec{R} \right) \right) \\
	B &= -\sum_{\vec{R},\eta,s} \left[ \left(C^{\eta}_{3z} \vec{a}_{M_2} \right) \cdot \vec{\hat{x}} \right]^2 \left( J^+_{s,\eta} \left( \vec{R} \right) + J^-_{s,\eta} \left( \vec{R} \right) \right),
\end{align}
where the ``current'' operators are defined by
\begin{align}
	J^+_{s,\eta} \left( \vec{R} \right) &\equiv t \hat{b}^\dagger_{\vec{R} + C^{\eta}_{3z} \vec{a}_{M_2},\eta,s} \hat{b}_{\vec{R},\eta,s}, \\
	J^-_{s,\eta} \left( \vec{R} \right) &\equiv t \hat{b}^\dagger_{\vec{R},\eta,s} \hat{b}_{\vec{R} + C^{\eta}_{3z} \vec{a}_{M_2},\eta,s}.
\end{align}
We note in passing that $A$ is the paramagnetic current operator, whereas $B$ can be related to the diamagnetic contribution to the current in presence of a flux. If we define the grand potential in the presence of a flux as $\Phi \left( \phi \right)$, then the charge and spin stiffness are given by
\begin{equation}
	\rho_{\text{C},\text{S}} \equiv \frac{1}{\mathcal{N}^2} \eval{\pdv[2]{\Phi \left( \phi_{\text{C},\text{S}} \right)}{\phi_{\text{C},\text{S}}}}_{\phi_{\text{C},\text{S}} = 0}.
\end{equation}
The grand canonical partition function in the presence of a flux can be expanded up to second order in $\phi$ as
\begin{align}
	\frac{Z (\phi)}{Z(0)} &= 1 - i \phi \int_{0}^{\beta} \dd{\tau} \left\langle A (\tau) \right\rangle - \frac{\phi^2}{2} \int_{0}^{\beta} \dd{\tau_1} \int_{0}^{\beta} \dd{\tau_2} \left\langle A \left( \tau_1 \right) A \left( \tau_2 \right) \right\rangle + \frac{\phi^2}{2} \int_{0}^{\beta} \dd{\tau} \left\langle B (\tau) \right\rangle + \mathcal{O} \left( \phi^3 \right) \nonumber \\
	&= 1 - i \beta \phi \left\langle A \right\rangle  - \frac{\beta \phi^2}{2} \int_{0}^{\beta} \dd{\tau} \left\langle A (\tau) A(0) \right\rangle + \frac{\beta \phi^2}{2} \left\langle B \right\rangle + \mathcal{O} \left( \phi^3 \right),
\end{align}
with all expectation values evaluated with respect to the $\phi=0$ grand-canonical Hamiltonian. By $C_{2x}$ symmetry, $\left\langle A \right\rangle = 0$, and therefore the grand potential satisfies
\begin{equation}
	\Phi \left( \phi \right) - \Phi \left( 0 \right) = \frac{\phi^2}{2} \left( \int_{0}^{\beta} \dd{\tau} \left\langle A(\tau) A(0) \right\rangle - \left\langle B \right\rangle \right) + \mathcal{O} \left( \phi^3 \right),
\end{equation}
which immediately gives the following expression for the stiffness,
\begin{equation}
	\rho = \frac{1}{\mathcal{N}^2} \left( \int_{0}^{\beta} \dd{\tau} \left\langle A(\tau) A(0) \right\rangle - \left\langle B \right\rangle \right).
\end{equation}
The second term can be simplified using the $C_{3z}$ symmetry of the problem,
\begin{align}
	\label{app:eqn:stiff_b_correlator}
	\left\langle B \right\rangle =& - \left[ \left(C^{\eta}_{3z} \vec{a}_{M_2} \right) \cdot \vec{\hat{x}} \right]^2 \sum_{\vec{R},\eta,s}  \left\langle J^+_{s,\eta} \left( \vec{R} \right) + J^-_{s,\eta} \left( \vec{R} \right) \right\rangle \nonumber \\
	=& \frac{1}{3} \left\langle H_0 \right\rangle \sum_{\eta} \left[ \left(C^{\eta}_{3z} \vec{a}_{M_2} \right) \cdot \vec{\hat{x}} \right]^2,
\end{align}
while the first term can be evaluated as
\begin{align}
	\int_{0}^{\beta} \dd{\tau} \left\langle A(\tau) A(0) \right\rangle  =& \sum_{\substack{ \vec{R}_1,\vec{R}_2 \\ \eta_1,\eta_2,s_1,s_2}} \left[ \left(C^{\eta_1}_{3z} \vec{a}_{M_2} \right) \cdot \vec{\hat{x}} \right] \left[ \left(C^{\eta_2}_{3z} \vec{a}_{M_2} \right) \cdot \vec{\hat{x}} \right] f_{s_1} f_{s_2} \nonumber \\
	&\times \left( \left\langle J^+_{s_1,\eta_1} \left( \vec{R}_1,\tau \right) J^+_{s_2,\eta_2} \left( \vec{R}_2,0 \right) \right\rangle + \left\langle J^-_{s_1,\eta_1} \left( \vec{R}_1,\tau \right) J^-_{s_2,\eta_2} \left( \vec{R}_2,0 \right) \right\rangle \right. \nonumber \\
	&\left. \quad - \left\langle J^+_{s_1,\eta_1} \left( \vec{R}_1,\tau \right) J^-_{s_2,\eta_2} \left( \vec{R}_2,0 \right) \right\rangle - \left\langle J^-_{s_1,\eta_1} \left( \vec{R}_1,\tau \right) J^+_{s_2,\eta_2} \left( \vec{R}_2,0 \right) \right\rangle \right)\nonumber \\
	=& \frac{1}{\beta}\sum_{\substack{ \vec{R}_1,\vec{R}_2 \\ \eta_1,\eta_2,s_1,s_2}} \left[ \left(C^{\eta_1}_{3z} \vec{a}_{M_2} \right) \cdot \vec{\hat{x}} \right] \left[ \left(C^{\eta_2}_{3z} \vec{a}_{M_2} \right) \cdot \vec{\hat{x}} \right] f_{s_1} f_{s_2} \nonumber \\
	&\times \left[ \left\langle N_{J^+_{s_1,\eta_1} \left( \vec{R}_1 \right)} N_{J^+_{s_2,\eta_2} \left( \vec{R}_2 \right)} - \delta_{s_1 s_2} \delta_{\eta_1 \eta_2} \delta_{\vec{R}_1 \vec{R}_2} N_{J^+_{s_1,\eta_1} \left( \vec{R}_1 \right)} \right\rangle_{\text{MC}} \right. \nonumber \\
	&\quad +\left\langle N_{J^-_{s_1,\eta_1} \left( \vec{R}_1 \right)} N_{J^-_{s_2,\eta_2} \left( \vec{R}_2 \right)} - \delta_{s_1 s_2} \delta_{\eta_1 \eta_2} \delta_{\vec{R}_1 \vec{R}_2} N_{J^-_{s_1,\eta_1} \left( \vec{R}_1 \right)} \right\rangle_{\text{MC}} \nonumber \\
	&\left. \quad -\left\langle N_{J^+_{s_1,\eta_1} \left( \vec{R}_1 \right)} N_{J^-_{s_2,\eta_2} \left( \vec{R}_2 \right)} \right\rangle_{\text{MC}} -\left\langle N_{J^-_{s_1,\eta_1} \left( \vec{R}_1 \right)} N_{J^+_{s_2,\eta_2} \left( \vec{R}_2 \right)} \right\rangle_{\text{MC}} \right] \nonumber \\
	=& \frac{1}{\beta}\sum_{\substack{\eta_1,\eta_2 \\ s_1,s_2}} \left[ \left(C^{\eta_1}_{3z} \vec{a}_{M_2} \right) \cdot \vec{\hat{x}} \right] \left[ \left(C^{\eta_2}_{3z} \vec{a}_{M_2} \right) \cdot \vec{\hat{x}} \right] f_{s_1} f_{s_2}  \left\langle \left( N_{J^+_{s_1,\eta_1}} - N_{J^-_{s_1,\eta_1}}\right) \left( N_{J^+_{s_2,\eta_2}} - N_{J^-_{s_2,\eta_2}}\right) \right\rangle_{\text{MC}} \nonumber \\
	&-\frac{1}{\beta}\sum_{\vec{R},\eta,s} \left[ \left(C^{\eta}_{3z} \vec{a}_{M_2} \right) \cdot \vec{\hat{x}} \right]^2 \left\langle N_{J^+_{s,\eta} \left( \vec{R} \right)} + N_{J^-_{s,\eta} \left( \vec{R} \right)} \right\rangle_{\text{MC}} \nonumber \\
	=& \frac{1}{\beta}\sum_{\substack{\eta_1,\eta_2 \\ s_1,s_2}} \left[ \left(C^{\eta_1}_{3z} \vec{a}_{M_2} \right) \cdot \vec{\hat{x}} \right] \left[ \left(C^{\eta_2}_{3z} \vec{a}_{M_2} \right) \cdot \vec{\hat{x}} \right] f_{s_1} f_{s_2}  \left\langle \left( N_{J^+_{s_1,\eta_1}} - N_{J^-_{s_1,\eta_1}}\right) \left( N_{J^+_{s_2,\eta_2}} - N_{J^-_{s_2,\eta_2}}\right) \right\rangle_{\text{MC}} \nonumber \\
	&+\frac{1}{3} \left\langle H_0 \right\rangle \sum_{\eta} \left[ \left(C^{\eta}_{3z} \vec{a}_{M_2} \right) \cdot \vec{\hat{x}} \right]^2. \label{app:eqn:stiff_aa_correlator}
\end{align}
To evaluate the zero-frequency correlators between the ``current'' operators in the first equality of \cref{app:eqn:stiff_aa_correlator}, we use the fact that the operators in \cref{app:eqn:sse_hams_2b,app:eqn:sse_hams_3b} can be written as
\begin{align}
	H_{2,b} &= J^+_{\uparrow,\eta^b} \left( \vec{R}_1^b \right) + J^-_{\uparrow,\eta^b} \left( \vec{R}_1^b \right), \\
	H_{3,b} &= J^+_{\downarrow,\eta^b} \left( \vec{R}_1^b \right) + J^-_{\downarrow,\eta^b} \left( \vec{R}_1^b \right).
\end{align}
Thus, when acting on a state in the Fock basis, $H_{a,b}$ (for $a=2,3$) reduces to a \emph{single} current operator $J^{\pm}_{s,\eta} \left( \vec{R} \right)$. This allows the zero-frequency correlators in the first line of \cref{app:eqn:stiff_aa_correlator} to be estimated from the Monte Carlo simulations using the prescription in \cref{app:eqn:zero_freq_ham_terms}, simply by counting the number of $J^{\pm}_{s,\eta} \left( \vec{R} \right)$ operators in the operator string $\tilde{S}_{L}$, which we denote by $N_{J^{\pm}_{s,\eta} \left( \vec{R} \right)}$. The final result depends only on the total number of current operators in a given operator string that move particles in a given direction,
\begin{equation}
	N_{J^{\pm}_{s,\eta}} = \sum_{\vec{R}} N_{J^{\pm}_{s,\eta} \left( \vec{R} \right)},
\end{equation}
and is therefore given by
\begin{align}
	\rho_{\text{C}} &= \frac{1}{\beta \mathcal{N}^2}\sum_{\substack{\eta_1,\eta_2 \\ s_1,s_2}} \left[ \left(C^{\eta_1}_{3z} \vec{a}_{M_2} \right) \cdot \vec{\hat{x}} \right] \left[ \left(C^{\eta_2}_{3z} \vec{a}_{M_2} \right) \cdot \vec{\hat{x}} \right] \left\langle \left( N_{J^+_{s_1,\eta_1}} - N_{J^-_{s_1,\eta_1}}\right) \left( N_{J^+_{s_2,\eta_2}} - N_{J^-_{s_2,\eta_2}}\right) \right\rangle_{\text{MC}}, \label{app:sec:chg_stiff}\\
	\rho_{\text{S}} &= \frac{1}{\beta \mathcal{N}^2}\sum_{\substack{\eta_1,\eta_2 \\ s_1,s_2}} \left[ \left(C^{\eta_1}_{3z} \vec{a}_{M_2} \right) \cdot \vec{\hat{x}} \right] \left[ \left(C^{\eta_2}_{3z} \vec{a}_{M_2} \right) \cdot \vec{\hat{x}} \right] s_1 s_2 \left\langle \left( N_{J^+_{s_1,\eta_1}} - N_{J^-_{s_1,\eta_1}}\right) \left( N_{J^+_{s_2,\eta_2}} - N_{J^-_{s_2,\eta_2}}\right) \right\rangle_{\text{MC}}. \label{app:sec:spin_stiff}
\end{align}
In \cref{app:sec:spin_stiff}, $\uparrow$ ($\downarrow$) should be interpreted as $+1$ ($-1$). 

\subsubsection{Entropy}\label{app:sec:sse:observables:entropy}

Finally, we explain how the entropy can be estimated numerically in our SSE simulations using thermodynamic integration~\cite{FRE01}. In this subsection, we explicitly indicate the natural variables of all thermodynamic quantities involved. We begin by noting that the grand-canonical energy of the system is related to the partition function through
\begin{equation}
	\pdv{\beta} \log Z \left( \beta, \mu \right) = - \left\langle H \right\rangle_{\beta,\mu}.
\end{equation}
This relation allows us to integrate from infinite temperature down to a given inverse temperature $\beta$ and thus obtain the partition function explicitly,
\begin{equation}
	\log Z \left( \beta, \mu \right) = \log Z \left(0, \mu \right) - \int_{0}^{\beta} \dd{\beta'} \left\langle H \right\rangle_{\beta',\mu}.
\end{equation}
At infinite temperature, the partition function is simply given by the dimension of the Hilbert space of the problem, $\dim (\mathbb{H})$, which in turn yields the grand potential
\begin{equation}
	\Phi \left( \beta, \mu \right) =  \frac{1}{\beta}\left( \int_{0}^{\beta} \dd{\beta'} \left\langle H \right\rangle_{\beta',\mu} - \log\dim (\mathbb{H}) \right).
\end{equation}
On the other hand, the grand-canonical potential can also be expressed in terms of the entropy and the grand-canonical internal energy of the system,
\begin{equation}
	\Phi \left( \beta, \mu \right) =  \left\langle H \right\rangle_{\beta,\mu} - \frac{1}{\beta}S \left( \beta, \mu \right).
\end{equation}
Combining these expressions, we directly obtain the entropy of the system,
\begin{equation}
	\label{app:eqn:entropy_from_zero}
	S \left( \beta, \mu \right) = \beta \left\langle H \right\rangle_{\beta,\mu} - \int_{0}^{\beta} \dd{\beta'} \left\langle H \right\rangle_{\beta',\mu} + \log\dim (\mathbb{H}) = \log\dim (\mathbb{H}) - \int_{0}^{\beta} \dd{\beta'} \left( \left\langle H \right\rangle_{\beta',\mu} - \left\langle H \right\rangle_{\beta,\mu} \right).
\end{equation}
In order to directly employ \cref{app:eqn:entropy_from_zero}, one needs to carry out simulations at very large temperatures. While this is not necessarily a problem, in practice, we find it much easier to start from a small but nonzero inverse temperature $\beta_1$, where other methods (such as perturbation theory) can be reliably employed\footnote{At small $\beta$, the energy varies rapidly with $\beta$, so obtaining an accurate thermodynamic integral directly from SSE would require a dense sampling of temperatures in that regime. Instead, we start the integration from a sufficiently small but finite $\beta$ where the entropy can be estimated reliably using perturbation theory around the atomic limit, and then use the numerical data only in the regime where the variation with $\beta$ is milder.}. We therefore find
\begin{equation}
	\label{app:eqn:entropy_diff_thermo_integration}
	S \left( \beta_2, \mu \right) - S \left( \beta_1, \mu \right) = \beta_2 \left\langle H \right\rangle_{\beta_2,\mu} - \beta_1 \left\langle H \right\rangle_{\beta_1,\mu} - \int_{\beta_1}^{\beta_2} \dd{\beta'} \left\langle H \right\rangle_{\beta',\mu}. 
\end{equation}

\section{SSE updates}\label{app:sec:updates}

We use a Markov-chain Monte Carlo procedure to sample the SSE configurations $\left(\alpha, \tilde{S}_L\right)$ in \cref{app:eqn:part_function_initial_SSE} according to their weights in the grand-canonical partition function $Z$. In the first part of this \siSection{}, we review the moves used in conventional SSE algorithms. Specifically, there are two main types of standard SSE updates: diagonal updates -- in which diagonal operators are added to or removed from the SSE configuration $\left(\alpha, \tilde{S}_L\right)$ -- and off-diagonal updates -- in which diagonal and off-diagonal operators, defined in \cref{app:sec:sse:review}, are exchanged with one another~\cite{SAN19}.

In addition to the conventional SSE updates, we also devise and employ a new type of update designed specifically to ensure ergodicity at low temperature in classically-coupled quantum chains ({\it i.e.}{} one-dimensional Hubbard chains interacting with one another via density-density interactions) -- the type of systems studied here -- which we term \emph{inter-chain updates}. These enable particle transfer between different chains at temperatures where overall charge fluctuations are suppressed. Finally, we also review replica-exchange QMC~\cite{SWE86,HUK96}, as applied to SSE, which improves tunneling between low-energy configurations at low temperatures. The crux of this method is to run multiple replicas in parallel at different temperature and then exchange their configurations in a way that preserves detailed balance.

A full SSE sweep (also referred to as a full QMC updating cycle or a Monte Carlo step) consists of a diagonal update, in which all eligible diagonal and identity operators are attempted to be swapped, followed by a number of off-diagonal updates. The number of such updates is determined self-consistently, as explained in detail in this \siSection{}. Interspersed with the diagonal updates are a number of inter-chain updates, chosen such that the number of inter-chain updates is on average one quarter of the number of off-diagonal updates. Whenever replica-exchange QMC is employed ({\it i.e.}{}, for the data shown in \cref{fig:compress_simple}), each SSE sweep is followed by an attempt to exchange between the replicas that are simulated in parallel.

\subsection{Diagonal updates and graphical representation}\label{app:sec:updates:diagonal}

We begin by discussing the conventional diagonal SSE updates. Starting from a given configuration $\left(\alpha, \tilde{S}_L\right)$, a diagonal update consists of local insertions and removals of diagonal operators, namely $H_{1,b}$ and $H_{4,b}$ defined in \cref{app:eqn:sse_hams_1b,app:eqn:sse_hams_4b}, respectively. Concretely, for fixed $\alpha$, we sweep through the operator sequence $\tilde{S}_L$ and attempt to replace identity operators ($H_{0,0}$) by diagonal ones ($H_{1,b}$ and $H_{4,b}$), and vice versa. Each local move therefore attempts a substitution of the form $[0,0] \leftrightarrow [\tilde{a},\tilde{b}]_p$ for $1 \leq p \leq L$ and $\tilde{a}=1,4$. These moves are accepted with the following Metropolis probabilities
\begin{align}
	P \left( [0,0] \rightarrow [\tilde{a},\tilde{b}]_p \right) =& \min \left(1, \frac{N_{\text{Ops}} \beta M_{\tilde{a},\tilde{b}}(p)}{L-n} \right), \\
	P \left( [\tilde{a},\tilde{b}]_p \rightarrow [0,0] \right) =& \min \left(1, \frac{L-n+1}{N_{\text{Ops}} \beta M_{\tilde{a},\tilde{b}}(p)} \right), 
\end{align}
where $\tilde{a}=1,4$ denotes the type of diagonal operator being inserted (or removed), while $N_{\text{Ops}} = 2 \times 3 \times \mathcal{N}^2$ is the total number of diagonal operators available in the system. In addition, $M_{\tilde{a},\tilde{b}}(p)$ denotes the matrix element of the diagonal operator to be inserted or removed,
\begin{equation}
	M_{\tilde{a},\tilde{b}}(p) = \mel**{\tilde{\alpha}(p-1)}{H_{\tilde{a},\tilde{b}}}{\tilde{\alpha}(p)}.
\end{equation}
The operators $H_{\tilde{a},\tilde{b}}$ are supported only on a small number of orbitals. Concretely, the operator $H_{1,\tilde{b}}$ acts on the lattice sites $\vec{R}^{b}_1$ and $\vec{R}^{b}_2$ and on all three valleys, and is therefore supported on six orbitals. By convention, the operator $H_{4,\tilde{b}}$ acts only on the two lattice sites $\vec{R}^{b}_1$ and $\vec{R}^{b}_2$ in valley $\eta^b$, and is therefore supported on only two orbitals. The operator $H_{4,\tilde{b}}$ is introduced so that it can be exchanged with the off-diagonal hopping operators $H_{2,\tilde{b}}$ and $H_{3,\tilde{b}}$ during the off-diagonal updates, which will be discussed in \cref{app:sec:updates:offdiagonal}.

During the diagonal update (which consists of $L$ local moves), we start from the initial state $\ket{\alpha}$ and then sweep through the operator string $\tilde{S}_L$ (in the order of increasing propagation index $p$). At any given stage, we store only the state $\ket{\tilde{\alpha}(p)}$ for the current value of $p$. If a non-diagonal operator is encountered, such as $H_{\tilde{a},\tilde{b}}$ with $\tilde{a}=2,3$, the state is simply propagated forward ({\it i.e.}{}, we obtain $\ket{\tilde{\alpha}(p+1)} \propto H_{\tilde{a},\tilde{b}} \ket{\tilde{\alpha}(p)}$). If instead a diagonal operator or an identity operator is encountered, a swap is attempted according to the above prescription. A diagonal update is completed once all eligible operators in $\tilde{S}_L$ have been considered for swapping.

\begin{figure}[!t]
	\centering
	\includegraphics[width=\textwidth]{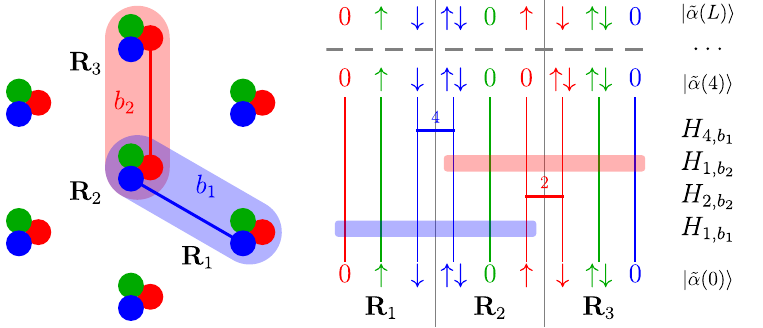}
	\caption{Schematic representation of an SSE configuration $\left(\alpha, \tilde{S}_L\right)$. The left panel shows a portion of the lattice in which three sites, $\vec{R}_1$, $\vec{R}_2$, and $\vec{R}_3$, and two bonds, $b_1$ and $b_2$, are highlighted. The right panel graphically represents the corresponding SSE configuration. Vertical lines denote the orbital world lines, while horizontal vertices represent operators in the string $\tilde{S}_L$. The vertical direction corresponds the propagation index $p$. Diagonal operators with support on six orbitals are shown as extended colored rectangles (blue and red, corresponding respectively to $b_1$ and $b_2$), while operators with support on only two orbitals appear as short horizontal vertices. One off-diagonal operator acting on $b_2$ is also shown explicitly. For operators acting on only two orbitals, the number above indicates whether the corresponding Hamiltonian term $H_{a,b}$ represents a spin-$\uparrow$ hopping ($a=2$), a spin-$\downarrow$ hopping ($a=3$), or a diagonal term ($a=4$). The basis state $\ket{\tilde{\alpha}(0)}=\ket{\alpha}$ is shown at the bottom, and one propagated state $\ket{\tilde{\alpha}(4)}$ is also indicated within the operator string. The real-space support of the operators displayed on the right is highlighted on the left panel.}
	\label{app:fig:sse_schematic}
\end{figure}

At this stage, it is useful to introduce a graphical representation of the operator string, such as that shown in \cref{app:fig:sse_schematic}. The operators $H_{\tilde{a},\tilde{b}}$ are stacked vertically in the same order as in $\tilde{S}_L$ and are connected by ``world lines'' that track the state, {\it i.e.}{} the occupation and spin state, of each orbital as the operator string is traversed along the propagation index $p$. In this representation, the propagation index -- akin, though not identical, to imaginary time -- increases from bottom to top, starting from the basis state $\ket{\tilde{\alpha}(0)}=\ket{\alpha}$. Acting with the operator at position $p$ transforms the state $\ket{\tilde{\alpha}(p+1)} \propto H_{\tilde{a},\tilde{b}} \ket{\tilde{\alpha}(p)}$, and this change is represented by a vertex connecting the world lines of the orbitals on which the operator has support.

Each operator $H_{\tilde{a},\tilde{b}}$ therefore appears as a vertex with a number of legs equal to twice the number of orbitals on which it acts (at a given propagation index). For example, the diagonal operator $H_{1,\tilde{b}}$ connects six world lines, corresponding to the three valleys on each of the two lattice sites $\vec{R}_1^b$ and $\vec{R}_2^b$, while the operator $H_{4,\tilde{b}}$ connects only the two world lines associated with valley $\eta^b$ on those sites. Off-diagonal operators, such as $H_{2,\tilde{b}}$ and $H_{3,\tilde{b}}$, also appear as vertices, but in this case they change the local configuration of the connected world lines, corresponding to hopping processes between the two lattice sites of bond $b$. As a result, the operator $H_{1,\tilde{b}}$ has 12 legs (six incoming and six outgoing), while the operators $H_{\tilde{a},\tilde{b}}$ with $2 \leq \tilde{a} \leq 4$ have four legs (two incoming from smaller $p$ and two outgoing towards larger $p$).

\subsection{Off-diagonal updates}\label{app:sec:updates:offdiagonal}

To exchange off-diagonal operators, we employ off-diagonal loop updates~\cite{SAN99,SEN02,SAN19}. For completeness, we briefly review how these updates are implemented. We begin by introducing a shorthand notation in which $s$ denotes an SSE configuration $\left(\alpha, \tilde{S}_L \right)$ and $\tilde{W}(s)$ denotes its weight in the partition function, as defined in \cref{app:eqn:weight_tilde_SSE}. The detailed-balance condition for a single off-diagonal update $s \to s'$ then reads
\begin{equation}
	\label{app:eqn:detailed_balance_off_diagonal}
	P \left( s \to s' \right) \tilde{W} (s) = P \left( s' \to s \right) \tilde{W} \left( s' \right),
\end{equation} 
where $P \left( s \to s' \right)$ denotes the probability of transitioning (updating) from configuration $s$ to $s'$. 

A single off-diagonal update is carried out as follows. The state on a world line in the SSE configuration is first locally ({\it i.e.}{}, at a fixed propagation index $p$) changed to one of the three other allowed states, chosen at random ({\it e.g.}{}, the orbital state may change from $\uparrow$ to $\uparrow\downarrow$). This creates a worm-like defect in the SSE configuration consisting of two discontinuities. One end of the defect, termed the worm tail, is kept fixed, while the other end, termed the worm head, is propagated through the SSE configuration, thereby extending the defect. As the worm head propagates, both the local states and the operator vertices ({\it i.e.}{} the operator types) are updated so that the vertices remain allowed. Note that the worm head can propagate in the direction of both increasing and decreasing propagation index $p$. Eventually, the worm head meets the tail and the loop closes, resulting in a new SSE configuration. For this reason, off-diagonal updates are also termed \emph{loop updates}. During a given SSE sweep, multiple such loop updates are performed. In practice, their number is chosen such that the worm head visits, on average, a total of $2L$ vertices~\cite{SEN02}. 

\subsubsection{Vertex scattering rules}\label{app:sec:updates:offdiagonal:scattering}

\begin{figure}[!t]
	\centering
	\includegraphics[width=\textwidth]{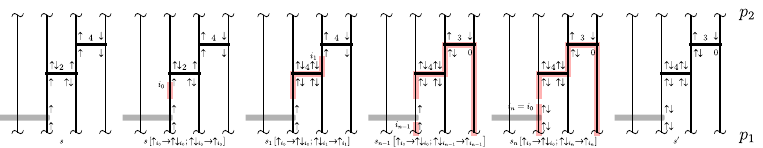}
	\caption{Loop off-diagonal update. We illustrate a fragment (both spatially and in terms of the propagation index $p$) of the SSE configuration $s$ undergoing a single-loop update to $s'$. The first panel shows a portion of the configuration $s$ (between propagation indices $0 < p_1 < p_2 < L$) containing three vertices. The input and output states of each operator are shown explicitly: one 12-legged vertex (drawn as the thick grey vertex) and two four-legged vertices. In the second panel a worm defect is introduced at the leg $i_0$, where the local state along the world line is changed from $\uparrow$ to $\uparrow\downarrow$. In the third panel the worm head propagates through the first vertex, transforming it from $H_{2,b}$ to $H_{4,b}$ and moving the worm head to the leg $i_1$. The worm head then continues to propagate through the operator string until it reaches the position $i_{n-1}$, as illustrated in the fourth panel. During this process the vertices encountered by the worm are updated accordingly, including the 12-legged vertex shown in the fifth panel. At a given vertex, the worm head can move sideways, turn, or continue straight, as illustrated respectively in the third, fourth, and fifth panels. Finally, when the worm head reaches the original position $i_n=i_0$, the two discontinuities annihilate and a new valid configuration $s'$ is obtained (last panel). The reverse process $s' \to s$ corresponds to traversing the same sequence of updates in the opposite direction, {\it i.e.}{} by reading the panels from right to left.}
	\label{app:fig:illustration_loop}
\end{figure}

The way the worm head ``propagates'' through the operator vertices is determined by requiring that the detailed-balance condition from \cref{app:eqn:detailed_balance_off_diagonal} holds. This leads to a set of ``vertex scattering rules'' whose derivation we will now review~\cite{SEN02}. To do so, we consider introducing a worm at a certain vertex leg $i_0$ (here and in what follows, the indices $i_0$, $i_1$, {\it etc.}{} encode both the propagation index $p$, the unit cell, and the orbital -- a certain point along a given world line). We assume that locally at $i_0$ the state along the world line is changed from $\xi_{i_0}$ to $\xi'_{i_0}$ according to $\xi_{i_0} \to \xi'_{i_0}$. This produces a ``faulty'' SSE configuration with two discontinuities: along the world line passing through $i_0$, the state changes from $\xi_{i_0}$ to $\xi'_{i_0}$ and then back to $\xi_{i_0}$ immediately before the vertex containing the leg $i_0$. We denote this configuration by $s\left[ \xi_{i_0} \to \xi'_{i_0}; \xi'_{i_0} \to \xi_{i_0} \right]$. It is worth noting that $s\left[ \xi_{i_0} \to \xi'_{i_0}; \xi'_{i_0} \to \xi_{i_0} \right]$ still contains the same operator vertices as $s$: it only differs from $s$ by the presence of a short worm defect on the world line passing through $i_0$. This is exemplified by the second panel of \cref{app:fig:illustration_loop}.

To grow this line defect, the vertex adjacent to $i_0$ can be changed to another allowed vertex by updating both the state at $i_0$ according to $\xi_{i_0} \to \xi'_{i_0}$, the state on another leg of that vertex (termed the exit leg and which we denote by $o_0$) according to $\xi_{o_0} \to \xi'_{o_0}$, and potentially the operator itself. The choice of the exit leg is in general non-deterministic: given a vertex and the states along the world lines connected to it, there can be several allowed ways for the worm head to \emph{scatter} ({\it i.e.}{}, proceed through the vertex), as shown in \cref{app:fig:illustration_loop}. The corresponding scattering rules form the object of this subsection, and will be derived below. For now we note that, because of spin and charge conservation on the vertices, at a given exit leg there is at most one possible update of the electron state that can lead to an allowed vertex. The $o_0$ leg is connected to an inward leg of another vertex, $i_1 = o_0$. After changing the first vertex, the configuration becomes $s_1\left[ \xi_{i_0} \to \xi'_{i_0}; \xi'_{i_1} \to \xi_{i_1} \right]$, which again contains two discontinuities located at $i_0$ and $i_1$. Growing the worm further yields a configuration $s_2\left[ \xi_{i_0} \to \xi'_{i_0}; \xi'_{i_2} \to \xi_{i_2} \right]$ with discontinuities at $i_0$ and $i_2$. All these steps are shown schematically in \cref{app:fig:illustration_loop}.

The process continues in this manner, propagating the worm head through the SSE configuration until the worm head reaches a leg $i_n$ that coincides with the position of the worm tail $i_0$. The resulting configuration is $s_n\left[ \xi_{i_0} \to \xi'_{i_0}; \xi'_{i_n} \to \xi_{i_n} \right] = s'$, which no longer contains any defects because the discontinuities associated with the worm head and worm tail annihilate each other. In other words, the worm has closed into a loop and thereby generates a new SSE configuration. We stress, however, that such a closure is not guaranteed in general; a more detailed discussion of this issue is provided at the end of \cref{app:sec:updates:offdiagonal:particularize}.

The transition probability $P \left( s \to s' \right)$ corresponding to a single (successful) loop update can therefore be obtained by summing over all possible choices of the initial discontinuity and over all possible paths by which the worm head propagates through the SSE configuration before the loop closes
\begin{align}
	P \left( s \to s' \right) = & 
	\sum_{n} \sum_{\substack{i_0,i_1, \dots, i_n \\ \xi'_{i_0}, \xi'_{i_1}, \dots, \xi'_{i_n} \\ s_0, s_1, \dots, s_n}} \delta_{s,s_0} \delta_{s',s_n} \delta_{i_0,i_n} P \left( i_0 \right) P \left( \xi_{i_0} \to \xi'_{i_0} \right) \nonumber \\
	& \times \prod_{m=0}^{n-1} P \left( s_m \left[ \xi_{i_0} \to \xi'_{i_0}; \xi'_{i_m} \to \xi_{i_m} \right] \to s_{m+1}\left[ \xi_{i_0} \to \xi'_{i_0}; \xi'_{i_{m+1}} \to \xi_{i_{m+1}} \right] \right), \label{app:eqn:loop_process_direct}
\end{align}
where $P \left( i_0 \right)$ denotes the probability that the entrance leg of the loop is chosen at $i_0$, while $P \left( \xi_{i_0} \to \xi'_{i_0} \right)$ denotes the probability of performing the initial state change, both of which will be specified below \cref{app:eqn:detailed_balance_interm_1}. The product over $m$ describes the successive propagation steps of the worm head. Each factor corresponds to a local update in which a single vertex is modified while the worm head moves from $i_m$ to $i_{m+1}$. Consequently, the probability $P \left( s_m \left[ \xi_{i_0} \to \xi'_{i_0}; \xi'_{i_m} \to \xi_{i_m} \right] \to s_{m+1}\left[ \xi_{i_0} \to \xi'_{i_0}; \xi'_{i_{m+1}} \to \xi_{i_{m+1}} \right] \right)$ is only taken to be non-vanishing if the configurations $s_m$ and $s_{m+1}$ differ by exactly one vertex.

The reverse process can be obtained by traversing the same loop in the opposite direction, which corresponds to reversing every propagation step of the worm head.
\begin{align}
	P \left( s' \to s \right) = & 
	\sum_{n} \sum_{\substack{i_0,i_1, \dots, i_n \\ \xi_{i_0}, \xi_{i_1}, \dots, \xi_{i_n} \\ s_0, s_1, \dots, s_n}} \delta_{s',s_n} \delta_{s,s_0} \delta_{i_0,i_n} P \left( i_n \right) P \left( \xi'_{i_n} \to \xi_{i_n} \right) \nonumber \\
	& \times \prod_{m=0}^{n-1} P \left( s_{m+1}\left[ \xi_{i_0} \to \xi'_{i_0}; \xi'_{i_{m+1}} \to \xi_{i_{m+1}} \right] \to s_m \left[ \xi_{i_0} \to \xi'_{i_0}; \xi'_{i_m} \to \xi_{i_m} \right] \right), \label{app:eqn:loop_process_inverse}
\end{align}
An illustration of a loop connecting $s$ to $s'$ -- corresponding to a single term in the forward transition probability -- together with its reverse process is shown in \cref{app:fig:illustration_loop}. 

One way to satisfy the detailed-balance condition from \cref{app:eqn:detailed_balance_off_diagonal} for arbitrary configurations $s$ and $s'$ is to impose it \emph{individually} for every pair of forward and reverse paths.
\begin{align}
	&\tilde{W} \left( s_0 \right) P \left( i_0 \right) P \left( \xi_{i_0} \to \xi'_{i_0} \right) 
	\prod_{m=0}^{n-1} P \left( s_m \left[ \xi_{i_0} \to \xi'_{i_0}; \xi'_{i_m} \to \xi_{i_m} \right] \to s_{m+1}\left[ \xi_{i_0} \to \xi'_{i_0}; \xi'_{i_{m+1}} \to \xi_{i_{m+1}} \right] \right) \nonumber \\
	=& \tilde{W} \left( s_n \right) P \left( i_0 \right) P \left( \xi'_{i_n} \to \xi_{i_n} \right) 
	\prod_{m=0}^{n-1} P \left( s_{m+1}\left[ \xi_{i_0} \to \xi'_{i_0}; \xi'_{i_{m+1}} \to \xi_{i_{m+1}} \right] \to s_m \left[ \xi_{i_0} \to \xi'_{i_0}; \xi'_{i_m} \to \xi_{i_m} \right] \right). \label{app:eqn:detailed_balance_interm_1}
\end{align}
Canceling the factor $P\left( i_0 \right)$ and choosing a loop-update procedure in which the initial state change is selected uniformly (such that $P \left( \xi_{i_0} \to \xi'_{i_0} \right)$ is independent of the states), the detailed-balance condition becomes
\begin{equation}
	\prod_{m=0}^{n-1} \frac{ P \left( s_m \left[ \xi_{i_0} \to \xi'_{i_0}; \xi'_{i_m} \to \xi_{i_m} \right] \to s_{m+1}\left[ \xi_{i_0} \to \xi'_{i_0}; \xi'_{i_{m+1}} \to \xi_{i_{m+1}} \right] \right)}{P \left( s_{m+1}\left[ \xi_{i_0} \to \xi'_{i_0}; \xi'_{i_{m+1}} \to \xi_{i_{m+1}} \right] \to s_m \left[ \xi_{i_0} \to \xi'_{i_0}; \xi'_{i_m} \to \xi_{i_m} \right] \right)} = \frac{\tilde{W} \left( s_n \right)}{\tilde{W} \left( s_0 \right)}. \label{app:eqn:detailed_balance_interm_2}
\end{equation}
This condition can be satisfied for arbitrary paths if we impose the local relation
\begin{equation}
	\frac{ P \left( s_m \left[ \xi_{i_0} \to \xi'_{i_0}; \xi'_{i_m} \to \xi_{i_m} \right] \to s_{m+1}\left[ \xi_{i_0} \to \xi'_{i_0}; \xi'_{i_{m+1}} \to \xi_{i_{m+1}} \right] \right)}{P \left( s_{m+1}\left[ \xi_{i_0} \to \xi'_{i_0}; \xi'_{i_{m+1}} \to \xi_{i_{m+1}} \right] \to s_m \left[ \xi_{i_0} \to \xi'_{i_0}; \xi'_{i_m} \to \xi_{i_m} \right] \right)} = \frac{\tilde{W} \left( s_{m+1} \right)}{\tilde{W} \left( s_m \right)}. \label{app:eqn:detailed_balance_interm_3}
\end{equation}
Remember that the configurations $s_m$ and $s_{m+1}$ differ by only a single vertex located at a certain position in the SSE operator string. We denote this vertex by $v$ in the configuration $s_m$ and by $v'$ in the configuration $s_{m+1}$. The symbol $v$ therefore encodes both the type of operator and the local states on which it acts. We denote the corresponding matrix element of the vertex by $M_v$.

In this notation, the detailed-balance condition from \cref{app:eqn:detailed_balance_interm_3} reduces to a purely local relation between the probabilities of transforming one vertex into another while propagating the worm through the vertex
\begin{equation}
	\frac{ P \left( v \left[ \dots; \xi'_i \to \xi_i \right] \to v' \left[ \dots; \xi'_o \to \xi_o \right] \right)}{P \left( v' \left[ \dots; \xi'_o \to \xi_o \right] \to v \left[ \dots; \xi'_i \to \xi_i \right] \right)} = \frac{M_{v'}}{M_{v}}. \label{app:eqn:detailed_balance_vertex}
\end{equation}
In \cref{app:eqn:detailed_balance_vertex}, $v \left[ \dots; \xi'_i \to \xi_i \right]$ denotes the state of the vertex with the head of the worm at one of its legs $i$. Once the worm head discontinuity is propagated through the worm, the vertex changes to $v' \left[ \dots; \xi'_o \to \xi_o \right]$. In addition to \cref{app:eqn:detailed_balance_vertex}, the vertex probabilities need to satisfy the following normalization condition
\begin{equation}
	\sum_{\left \lbrace v' \left[ \dots; \xi'_o \to \xi_o \right] \right \rbrace}P \left( v \left[ \dots; \xi'_i \to \xi_i \right] \to v' \left[ \dots; \xi'_o \to \xi_o \right] \right) = 1, \label{app:eqn:normalization_worm_vertex}
\end{equation}
which indicates that the worm head \emph{must} propagate in one way or another through the vertex with unit probability. We note that the process in which the worm head bounces $v \left[ \dots; \xi'_i \to \xi_i \right] \to v \left[ \dots; \xi'_i \to \xi_i \right]$ is allowed and in fact is generically needed in order to satisfy the detailed-balance and normalization conditions.

\subsubsection{The vertex scattering rules for AA t-\ch{SnSe2}{}}\label{app:sec:updates:offdiagonal:particularize}

\Cref{app:eqn:detailed_balance_vertex,app:eqn:normalization_worm_vertex} fully specify how the worm head propagates through the SSE configuration. Particularizing these expressions to the operator vertices that appear in our problem is straightforward. There are two types of vertices in our model: 12-legged and four-legged vertices. The 12-legged vertices correspond to the $H_{1,b}$ operators, which describe density-density interactions. As a result, charge and spin are individually conserved along each of the six world lines passing through the vertex. There are then only two possibilities: either the worm head passes straight through the vertex or it bounces. The propagation probability of the worm through a 12-legged vertex therefore takes the standard Metropolis form
\begin{equation}
	P \left( v \left[ \dots; \xi'_i \to \xi_i \right] \to v' \left[ \dots; \xi'_o \to \xi_o \right] \right) = \min \left( 1, \frac{M_{v'}}{M_{v}} \right), \quad \text{for } v \neq v' \text{ 12-legged vertices.}
\end{equation}

\begin{figure}[!t]
	\centering
	\includegraphics[width=0.5\textwidth]{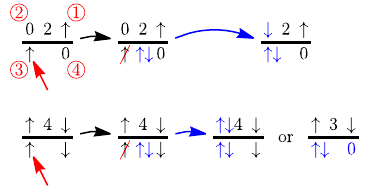}
	\caption{Two types of scattering processes at four-legged vertices. Each row illustrates a different scattering process. The state indicated by the red arrow is changed according to $\uparrow \to \uparrow\downarrow$. In the first row, after this change, the worm can exit only through the top-left leg, so there is a single possible outcome, chosen with probability $1$, meaning that In the second row, there are two possible outcomes, each chosen with probability $\frac12$. In the final states, the changed legs of the vertices are shown in blue.}
	\label{app:fig:four_leg_worm_prop}
\end{figure}

For the four-legged vertices, corresponding to operators $H_{a,b}$ with $2 \leq a \leq 4$, we can use the fact that they all have the same weight $M_v = t$. In 
this case, the detailed-balance condition from \cref{app:eqn:detailed_balance_vertex} reduces to
\begin{equation}
	P \left( v \left[ \dots; \xi'_i \to \xi_i \right] \to v' \left[ \dots; \xi'_o \to \xi_o \right] \right) = P \left( v' \left[ \dots; \xi'_o \to \xi_o \right] \to v \left[ \dots; \xi'_i \to \xi_i \right] \right),
\end{equation}
for four-legged vertices $v$ and $v'$. This simplification makes it possible to satisfy detailed balance \emph{without including any bounces} at the four-legged vertices~\cite{SYL02}, {\it i.e.}{} $P \left( v \left[ \dots; \xi'_i \to \xi_i \right] \to v \left[ \dots; \xi'_i \to \xi_i \right] \right) = 0$. Specifically, direct enumeration shows that only two cases can occur: for some four-legged vertices, the worm head can propagate in two different ways, each with probability $\frac12$, while in the remaining cases it can propagate in only one way, with probability $1$. Both cases are exemplified in \cref{app:fig:four_leg_worm_prop}. In the notation of \cref{app:eqn:detailed_balance_vertex}, the corresponding scattering probabilities are given by
\begin{align}
	P \left( \mel**{\uparrow,0}{H_{2,\tilde{b}}}{0,\uparrow} \left[ \dots; {\uparrow\downarrow}_{3} \to {\uparrow}_{3} \right] \right.&\left. \to \mel**{\uparrow\downarrow,0}{H_{2,\tilde{b}}}{\downarrow,\uparrow} \left[ \dots; {\downarrow}_{2} \to {0}_{2} \right] \right) = 1, \\
	P \left( \mel**{\uparrow,\downarrow}{H_{4,\tilde{b}}}{\uparrow,\downarrow} \left[ \dots; {\uparrow\downarrow}_{3} \to {\uparrow}_{3} \right] \right.&\left. \to \mel**{\uparrow\downarrow,\downarrow}{H_{4,\tilde{b}}}{\uparrow\downarrow,\downarrow} \left[ \dots; {\uparrow\downarrow}_{2} \to {\uparrow}_{2} \right] \right) = 1/2, \\
	P \left( \mel**{\uparrow,\downarrow}{H_{4,\tilde{b}}}{\uparrow,\downarrow} \left[ \dots; {\uparrow\downarrow}_{3} \to {\uparrow}_{3} \right] \right.&\left. \to \mel**{\uparrow\downarrow,0}{H_{3,\tilde{b}}}{\uparrow,\downarrow} \left[ \dots; {0}_{4} \to {\downarrow}_{4} \right] \right) = 1/2,
\end{align}
where the numbering of the vertex legs is given in \cref{app:fig:four_leg_worm_prop}. 

We note that our specific decomposition of the Hamiltonian into bond terms, as given in \cref{app:eqn:SSE_hamiltonian_decomposition}, has three main advantages:
\begin{itemize}
	\item The ``scattering'' rules at each vertex are sufficiently simple that long lookup tables, which can lead to frequent cache misses, can be avoided.
	\item The probability that the worm bounces at a vertex is thereby minimized, making the worms more effective and increasing the ergodicity of the off-diagonal loop updates.
	\item This decomposition also allows us to implement a new type of update in which particles can move between different chains in regimes where thermal particle-number fluctuations are suppressed, as we detail in \cref{app:sec:updates:inter-chain}.
\end{itemize}

In addition to the above rules, we note that at high temperatures there are typically some sites without associated vertices in the system ({\it i.e.}{}, there are orbitals whose world lines do not pass through any vertices). The states on such sites can be changed at random, since they do not affect the configuration weight. Finally, we note that in certain parameter regimes the length of a loop can sometimes become extremely large before it closes -- in practice, it may even fail to close altogether. It is therefore necessary to impose a maximum length beyond which the loop construction is terminated and a new starting point is chosen (we use $450L$ for this cut-off length). Following Ref.~\cite{SEN02}, after an incomplete loop update we perform a diagonal update before attempting the next loop. This reduces the likelihood that the next loop starts from essentially the same local environment and therefore again exceeds the termination length. Loop termination does not violate detailed balance and does not introduce any systematic errors in the results. In most cases, incomplete loop termination occurs sufficiently rarely that it does not adversely affect the simulation.

Following Ref.~\cite{SEN02}, after such an incomplete loop update we perform a diagonal update before attempting the next loop. This additional diagonal update is itself a standard SSE move satisfying detailed balance, and its role is only to decorrelate the operator string, reducing the likelihood that the next loop starts from essentially the same local environment and again exceeds the termination length. The loop termination therefore changes only the update dynamics, not the sampled equilibrium distribution, and does not introduce any systematic errors in the results. In most cases, incomplete loop termination occurs sufficiently rarely that it does not adversely affect the simulation.

\subsection{Inter-chain updates}\label{app:sec:updates:inter-chain}

The main advantage of SSE is that, in general, off-diagonal loop updates defined in \cref{app:sec:updates:offdiagonal} provide a global QMC update mechanism: a loop can ``touch'' and modify vertices throughout the operator string. At the same time, in our model for the mixed-dimensional limit of AA t-\ch{SnSe2}{}, the $\mathrm{U} \left( {2} \right)$ symmetry along each chain restricts an off-diagonal loop to a single chain. This hampers inter-chain equilibration, particularly at low temperatures. 

To see this, consider the case in which the system is in a Mott phase, so that each site is occupied on average by one particle. At a given site, however, that particle can belong to any of the three valleys. With only diagonal and off-diagonal updates, changing the valley occupation of a particle at a given site requires a sequence in which one off-diagonal loop update removes or creates a particle in one valley, followed by another off-diagonal loop update that creates or removes a particle in a different valley. In the Mott phase, however, charge fluctuations are strongly suppressed, and such processes almost never occur in practice.

To get around this limitation, we design an update in which the diagonal 12-legged vertices adjacent to a given site are temporarily ignored, such that local charge fluctuations in the corresponding chains are no longer suppressed. Two successive loop updates starting from that site are then performed, creating or destroying particles in different chains containing the site. Finally, once the diagonal 12-legged vertices are reintroduced, the move is accepted or rejected through a final Metropolis step. In this way, particles can move between chains even when overall charge fluctuations are strongly suppressed.

We now explain in detail the construction of the inter-chain update. For a given SSE configuration $s = \left( \alpha, \tilde{S}_L \right)$ and a specific site $\vec{R}$, we split the weight from \cref{app:eqn:weight_tilde_SSE} as
\begin{equation}
	\tilde{W} (s) = \tilde{W}_d (s) \tilde{W}_o (s),
\end{equation}
where we define the local diagonal weight and the remainder of the weight, respectively, as
\begin{align}
	\tilde{W}_d \left( \alpha, \tilde{S}_L \right) &= \prod_{\substack{p=1 \\ \vec{R} \in \left\lbrace \vec{R}_1^{\tilde{b}_p}, \vec{R}_2^{\tilde{b}_p} \right\rbrace  \text{ and } \tilde{a}_p = 1}}^{L} \mel**{\tilde{\alpha} (p-1)}{ H_{\tilde{a}_{p},\tilde{b}_{p}}}{\tilde{\alpha} (p)}, \label{app:eqn:op_string_mat_elems_diag}\\
	\tilde{W}_o \left( \alpha, \tilde{S}_L \right) &= \frac{1}{Z} \frac{\beta^n \left(L-n\right)!}{L!} \prod_{\substack{p=1 \\ \vec{R} \not\in \left\lbrace \vec{R}_1^{\tilde{b}_p}, \vec{R}_2^{\tilde{b}_p} \right\rbrace  \text{ or } \tilde{a}_p \neq 1}}^{L} \mel**{\tilde{\alpha} (p-1)}{ H_{\tilde{a}_{p},\tilde{b}_{p}}}{\tilde{\alpha} (p)}. \label{app:eqn:op_string_mat_elems_rest}
\end{align}
In other words, $\tilde{W}_d (s)$ is obtained from $\tilde{W} (s)$ by removing all diagonal 12-legged vertices supported on world lines that contain the site of interest $\vec{R}$. Because diagonal operators do not change the state, the product of matrix elements in \cref{app:eqn:op_string_mat_elems_rest} constitutes a \textit{bona fide} operator string, in contrast to the expression in \cref{app:eqn:op_string_mat_elems_diag}. We can therefore rewrite the two factors as
\begin{align}
	\tilde{W}_d \left( \alpha, \tilde{S}_L \right) &= \prod_{\substack{p=1 \\ \vec{R} \in \left\lbrace \vec{R}_1^{\tilde{b}_p}, \vec{R}_2^{\tilde{b}_p} \right\rbrace  \text{ and } \tilde{a}_p = 1}}^{L} M_{\tilde{a},\tilde{b}} (p), \label{app:eqn:op_string_mat_elems_diag_simple}\\
	\tilde{W}_o \left( \alpha, \tilde{S}_L \right) &= \frac{1}{Z} \frac{\beta^n \left(L-n\right)!}{L!} \bigg\langle \alpha \bigg\vert \prod_{\substack{p=1 \\ \vec{R} \not\in \left\lbrace \vec{R}_1^{\tilde{b}_p}, \vec{R}_2^{\tilde{b}_p} \right\rbrace  \text{ or } \tilde{a}_p \neq 1}}^{L} H_{\tilde{a}_{p},\tilde{b}_{p}} \bigg\vert \alpha \bigg\rangle. \label{app:eqn:op_string_mat_elems_rest_simple}
\end{align}

\begin{table}[t]
	\centering
\begin{tabular}{|c|c|c|c|}
			\hline
			$\mel**{\tilde{\alpha}(p)}{\left(  \hat{N}_{\vec{R},\eta}, \hat{N}_{\vec{R},\eta'} \right)}{\tilde{\alpha}(p)}$ &
			$\mel**{\tilde{\alpha}'(p)}{\left(  \hat{N}_{\vec{R},\eta}, \hat{N}_{\vec{R},\eta'} \right)}{\tilde{\alpha}'(p)}$ &
			$p^{\text{worm}}_{s \to s'}$ & $p^\text{worm}_{s' \to s}$ \\
			\hline
			$(0,0)$ & N/A & N/A & N/A \\ \hline
			$(0,1)$ or $(1,0)$ & $(1,0)$ or $(0,1)$ & $1/2$ & $1/2$ \\ \hline
			$(0,2)$ or $(2,0)$ & $(1,1)$ & $1/4$ & $1/2$ \\ \hline 
			$(1,1)$ & $(0,2)$ or $(2,0)$ & $1/2$ & $1/4$ \\ \hline 
			$(1,2)$ or $(2,1)$ & $(2,1)$ or $(1,2)$ & $1/2$ & $1/2$ \\ \hline 
			$(2,2)$ & N/A & N/A & N/A \\ \hline 
		\end{tabular}\caption{Probabilities of choosing worms that implement inter-chain particle transfers between two chains. For each initial occupation configuration at the two orbitals $\left( \vec{R},\eta \right)$ and $\left( \vec{R},\eta' \right)$, we list the corresponding allowed final occupation configuration, together with the probability of choosing the worms that realize the direct update $s \to s'$ and the reverse update $s' \to s$.}
	\label{app:tab:probabilities_loops_diag_updates}
\end{table}

Let us now describe the inter-chain update between configurations $s$ and $s'$. At the selected site $\vec{R}$ (chosen at random), there are three orbitals. Two of them, denoted by $\eta$ and $\eta'$, are chosen at random, and an attempt is made to move one fermion from $\left( \vec{R}, \eta \right)$ to $\left( \vec{R}, \eta' \right)$ at a randomly chosen propagation index $1 \leq p \leq L$ along the corresponding world lines. Depending on the number of particles present in the state $\ket{\tilde{\alpha}(p)}$, one then attempts two consecutive loops that effectively transfer a single particle between $\left( \vec{R}, \eta \right)$ and $\left( \vec{R}, \eta' \right)$ at position $p$. Assuming that all possible particle moves ({\it i.e.}{} all possible worm choices for the two loop updates that result in a particle transfer) are chosen with uniform probability, the probability of selecting the two loop updates that move particles in this way is summarized in \cref{app:tab:probabilities_loops_diag_updates}. Specifically, we list the probability of choosing the two worms for the two loops that produce the forward update $s \to s'$ or the reverse one $s' \to s$, which we denote by $p^{\text{worm}}_{s \to s'}$ and $p^{\text{worm}}_{s' \to s}$, respectively. We restrict throughout to worms that move exactly one particle between the two orbitals, and all such worms are chosen with equal probability. 

The detailed-balance condition for the inter-chain update is exactly the same as in \cref{app:eqn:detailed_balance_off_diagonal}. The probability $P \left( s \to s' \right)$ now admits the formal decomposition
\begin{equation}
	P \left( s \to s' \right) = p^{\text{accept}}_{s \to s'} p^{\text{proposal}}_{s \to s'} p^{\text{worm}}_{s \to s'},
\end{equation}
where $p^{\text{proposal}}_{s \to s'}$ is the proposal probability for the two successive loop updates that change the SSE configuration from $s$ to $s'$, $p^{\text{worm}}_{s \to s'}$ is the probability of choosing the corresponding worms, while $p^{\text{accept}}_{s \to s'}$ denotes the probability that the proposed update is finally accepted. The key point is that the loops that effect the change $s \to s'$ are propagated (non-deterministically) according to the weights $\tilde{W}_o (s)$, which ignore the diagonal 12-legged vertices adjacent to the site $\vec{R}$. As a result, the proposal probabilities for the forward and reverse processes satisfy the detailed-balance condition
\begin{equation}
	\tilde{W}_o (s) p^{\text{proposal}}_{s \to s'} = \tilde{W}_o \left( s' \right) p^{\text{proposal}}_{s' \to s}.
\end{equation}
It follows that the detailed-balance condition for the full inter-chain update is equivalent to
\begin{equation}
	p^{\text{accept}}_{s \to s'} p^{\text{worm}}_{s \to s'} \tilde{W}_d (s) = p^{\text{accept}}_{s' \to s} p^{\text{worm}}_{s' \to s} \tilde{W}_d \left( s' \right),
\end{equation}
which can be readily satisfied by the standard Metropolis filter for the update $s \to s'$,
\begin{equation}
	p^{\text{accept}}_{s \to s'} = \min \left(1, \frac{p^{\text{worm}}_{s' \to s} \tilde{W}_d \left( s' \right)}{p^{\text{worm}}_{s \to s'} \tilde{W}_d (s)} \right).
\end{equation}
One expects that, as a result of the two consecutive loops, a fermion at position $\vec{R}$ has only changed its valley quantum number. Consequently, the ratio $\frac{\tilde{W}_d \left( s' \right)}{\tilde{W}_d (s)}$ is not exponentially small, thereby providing a mechanism for electrons to propagate between chains in regimes where charge fluctuations are suppressed.

\subsection{Replica-exchange Monte Carlo}\label{app:sec:updates:pt}

As derived in Ref.~\cite{LI25} and reviewed in the main text, at integer fillings in the strong-coupling limit of AA t-\ch{SnSe2}{}, the low-lying states of AA t-\ch{SnSe2}{} in the mixed-dimensional limit are given by tilings of the lattice by one-dimensional antiferromagnetic spin chains. At very low temperatures, much lower than the corresponding antiferromagnetic exchange interaction, the system can therefore become trapped for long times in one of many nearly degenerate low-energy configurations. Equivalently, the free-energy landscape contains many valleys separated by barriers that are difficult to cross using local updates alone. This leads to long autocorrelation times and makes it hard to sample the full low-energy manifold ergodically. To better explore this regime, we employ replica-exchange QMC, or parallel tempering (PT)~\cite{SWE86,HUK96}. In PT, a number of independent replicas are simulated in parallel at different temperatures. After each SSE sweep, we attempt to exchange replicas at adjacent temperatures in a way that preserves detailed balance for the full ensemble. The basic idea is that cold replicas can move up the temperature ladder, where thermal fluctuations allow them to escape a given free-energy valley and explore the low-energy manifold more efficiently, before returning to lower temperatures.

\subsubsection{Replica-exchange probability}\label{app:sec:updates:pt:prob}

In this subsection, we outline the PT update. For this purpose, we attach an inverse-temperature index to all weights in order to keep track of which replica belongs to which temperature. At the same time, we explicitly indicate the padded operator-string length for all configurations. In a single PT update, two replicas at inverse temperatures $\beta$ and $\beta'$, whose SSE configurations are denoted by $s_{L} = \left( \alpha, \tilde{S}_{L}  \right)$ and $s'_{L'} = \left( \alpha', \tilde{S'}_{L'} \right)$, are attempted to be exchanged,
\begin{equation}
	\label{app:eqn:pt_process}
	\begin{pmatrix}
		s_{L} = \left( \alpha, \tilde{S}_{L}  \right) @ \beta \\
		s'_{L'} = \left( \alpha', \tilde{S'}_{L'} \right) @ \beta'
	\end{pmatrix} \leftrightarrow
	\begin{pmatrix}
		s_{L'} = \left( \alpha, \tilde{S}_{L'}  \right) @ \beta' \\
		s'_{L} = \left( \alpha', \tilde{S'}_{L} \right) @ \beta
	\end{pmatrix}.
\end{equation}
In such a move, the operator strings of the two replicas are kept unchanged, while their inverse temperatures $\beta$ and padded lengths $L$ are exchanged (we assume that $n,n' \leq L,L'$ so that the move is well defined). The detailed-balance condition for the move in \cref{app:eqn:pt_process} is
\begin{equation}
	\label{app:eqn:detailed_balance_PT}
	P \left[ \left( s_{L} @ \beta, s'_{L'} @ \beta' \right) \to \left( s'_{L} @ \beta, s_{L'} @ \beta' \right) \right] \tilde{W}_{\beta} \left( s_{L} \right) \tilde{W}_{\beta'} \left( s'_{L'} \right)  = P \left[ \left( s'_{L} @ \beta, s_{L'} @ \beta' \right) \to \left( s_{L} @ \beta, s'_{L'} @ \beta' \right) \right] \tilde{W}_{\beta} \left( s'_{L} \right) \tilde{W}_{\beta'} \left( s_{L'} \right),
\end{equation}
where $P \left[ \left( s_{L} @ \beta, s'_{L'} @ \beta' \right) \to \left( s'_{L} @ \beta, s_{L'} @ \beta' \right) \right]$ denotes the probability of swapping the temperatures of the two configurations, while $P \left[ \left( s'_{L} @ \beta, s_{L'} @ \beta' \right) \to \left( s_{L} @ \beta, s'_{L'} @ \beta' \right) \right]$ denotes the probability of the reverse process. These probabilities can in turn be decomposed into proposal and acceptance probabilities,
\begin{equation}
	P \left[ \left( s_{L} @ \beta, s'_{L'} @ \beta' \right) \to \left( s'_{L} @ \beta, s_{L'} @ \beta' \right) \right] = p^{\text{accept}}_{\left( s_{L} @ \beta, s'_{L'} @ \beta' \right) \to \left( s'_{L} @ \beta, s_{L'} @ \beta' \right)} p^{\text{proposal}}_{\left( s_{L} @ \beta, s'_{L'} @ \beta' \right) \to \left( s'_{L} @ \beta, s_{L'} @ \beta' \right)}.
\end{equation}
For simplicity, in what follows we take $\beta < \beta'$ without loss of generality. This implies $L < L'$. When changing the padded length of $\tilde{S}_L$ from $L$ to $L'$, one must insert $L'-L$ identity operators into $\tilde{S}_L$, chosen uniformly at random. Conversely, shrinking the padded length of $\tilde{S'}_{L'}$ is achieved by removing $L'-L$ identity operators, chosen at random among the $L'-n'$ identity operators present in $\tilde{S'}_{L'}$. This gives the proposal probability for the direct process move,
\begin{equation}
	p^{\text{proposal}}_{\left( s_{L} @ \beta, s'_{L'} @ \beta' \right) \to \left( s'_{L} @ \beta, s_{L'} @ \beta' \right)} = {L' \choose L' - L} {L'- n' \choose L' - L}. 
\end{equation}
Similarly, for the reverse one, the proposal probability is
\begin{equation}
	p^{\text{proposal}}_{\left( s'_{L} @ \beta, s_{L'} @ \beta' \right) \to \left( s_{L} @ \beta, s'_{L'} @ \beta' \right)} = {L' \choose L' - L} {L'- n \choose L' - L}.
\end{equation}
Using these expressions in the detailed-balance condition \cref{app:eqn:detailed_balance_PT}, we obtain
\begin{align}
	\frac{p^{\text{accept}}_{\left( s_{L} @ \beta, s'_{L'} @ \beta' \right) \to \left( s'_{L} @ \beta, s_{L'} @ \beta' \right)}}{p^{\text{accept}}_{\left( s'_{L} @ \beta, s_{L'} @ \beta' \right) \to \left( s_{L} @ \beta, s'_{L'} @ \beta' \right)}} &= \frac{ \tilde{W}_{\beta} \left( s'_{L} \right) \tilde{W}_{\beta'} \left( s_{L'} \right) }{ \tilde{W}_{\beta} \left( s_{L} \right) \tilde{W}_{\beta'} \left( s'_{L'} \right) } \frac{ p^{\text{proposal}}_{\left( s'_{L} @ \beta, s_{L'} @ \beta' \right) \to \left( s_{L} @ \beta, s'_{L'} @ \beta' \right)} }{ p^{\text{proposal}}_{\left( s_{L} @ \beta, s'_{L'} @ \beta' \right) \to \left( s'_{L} @ \beta, s_{L'} @ \beta' \right)} } \nonumber \\
	&= \frac{ {\beta}^{n'} \left( L - n' \right)! {\beta'}^{n} \left( L' - n \right)! }{ {\beta}^{n} \left( L - n \right)! {\beta'}^{n'} \left( L' - n' \right)! } \frac{ {L'- n' \choose L' - L} }{ {L'- n \choose L' - L} } \nonumber \\
	&= \left( \frac{\beta'}{\beta} \right)^{n-n'}\frac{ \left( L - n' \right)!  \left( L' - n \right)! }{ \left( L - n \right)! \left( L' - n' \right)! } \frac{\left(L' - n' \right)!}{\left( L' - L \right)! \left( L - n' \right)! } \frac{ \left( L' - L \right)! \left( L - n \right)!}{ \left( L'- n \right)! }  \nonumber \\
	&= \left( \frac{\beta'}{\beta} \right)^{n-n'}.
\end{align}
As expected, the result does not depend on $L$ or $L'$, which enter the weights only for bookkeeping purposes. The detailed balance condition can then be satisfied by the standard Metropolis filter for the PT swap update,
\begin{equation}
	p^{\text{accept}}_{\left( s_{L} @ \beta, s'_{L'} @ \beta' \right) \to \left( s'_{L} @ \beta, s_{L'} @ \beta' \right)} = \min \left(1, \left( \frac{\beta'}{\beta} \right)^{n-n'} \right).
\end{equation}
For PT to be efficient, adjacent replicas must be close enough in temperature that the swap probability remains non-vanishing. In what follows, we derive a low-temperature analytical estimate for the PT swap probability in SSE simulations.

\subsubsection{Estimating the swap probability}\label{app:sec:updates:pt:swap_probability}

To estimate the PT swap probability between two temperatures $\beta$ and $\beta'$, we make the following two observations:
\begin{itemize}
	\item The expansion order $n$ follows Poisson distribution,
	\begin{equation}
		n \sim \text{Po} \left( -\beta \left\langle H \right\rangle \right),
	\end{equation}
	under the assumption that the energy-fluctuations are ignored, which is justified in finite systems at low temperatures.
	\item At sufficiently low temperatures, $\left\langle H \right\rangle$ can be approximated by the ground-state grand-canonical energy,
	\begin{equation}
		\left\langle H \right\rangle \approx E_0.
	\end{equation}
\end{itemize}
Approximating the Poisson distribution by a normal distribution at low enough temperatures, we assume that the SSE expansion order is distributed as 
\begin{equation}
	n \sim \mathcal{N} \left( \mu_{\beta}, \sigma_{\beta}^2 \right),
\end{equation}
with $\mu_{\beta} = \sigma_{\beta}^2 = - \beta E_0$. Within this approximation, the expected PT swap probability can be computed as
\begin{align}
	\left\langle p^{\text{accept}}_{\left( s_{L} @ \beta, s'_{L'} @ \beta' \right) \to \left( s'_{L} @ \beta, s_{L'} @ \beta' \right)} \right\rangle \approx& \int_{-\infty}^{\infty} \frac{\dd{n}}{\sqrt{2 \pi \sigma^2_{\beta}}} \int_{-\infty}^{\infty} \frac{\dd{n'}}{\sqrt{2 \pi \sigma^2-_{\beta'}}} \left[ \Theta \left(n-n'\right) \right. \nonumber \\
	&\left. + \Theta \left(n'-n\right) \exp \left( \left( n - n'\right) \log \left( \frac{\beta'}{\beta} \right) - \frac{\left( n-\mu_\beta \right)^2}{2 \sigma^2_{\beta}} - \frac{\left( n'-\mu_{\beta'} \right)^2}{2 \sigma^2_{\beta'}} \right) \right] \nonumber \\
	=& \int_{-\infty}^{\infty} \frac{\dd{\Delta n}}{\sqrt{2 \pi \Sigma^2}} \left[ \Theta \left( \Delta n \right) + \Theta \left( -\Delta n\right) e^{t \Delta n} \right] e^{- \frac{\left( \Delta n - M \right)^2}{2 \Sigma^2}} \nonumber \\
	=& \Phi \left( \frac{M}{\Sigma} \right) + e^{M t + \frac{\Sigma^2 t^2}{2}} \Phi \left( \frac{-M - \Sigma^2 t}{\Sigma} \right),
\end{align}
where $\Sigma = \sqrt{\sigma^2_{\beta} + \sigma^2_{\beta'}}$, $M = \mu_{\beta} - \mu_{\beta'}$, $\Phi \left( x \right)$ is the cumulative distribution function of the $\mathcal{N} \left(0, 1 \right)$ normal distribution, and $t = \log \left( \frac{\beta'}{\beta} \right)$. Defining 
\begin{equation}
	\Delta \beta = \beta' - \beta \qq{and} \bar{\beta} = \frac{\beta + \beta'}{2}, 
\end{equation}
and working in the regime of small temperature differences, $\Delta \beta \ll \bar{\beta}$, we find that the expected replica-swap probability can be approximated by
\begin{equation}
	\label{app:eqn:expected_replica_swap_rate}
	\left\langle p^{\text{accept}}_{\left( s_{L} @ \beta, s'_{L'} @ \beta' \right) \to \left( s'_{L} @ \beta, s_{L'} @ \beta' \right)} \right\rangle \approx \operatorname{erfc} \left(  \frac{\Delta \beta }{ \bar{\beta}} \sqrt{\frac{-\bar{\beta} E_0}{4}}\right).
\end{equation}

\subsubsection{Constructing efficient parallel-tempering ladders}\label{app:sec:updates:pt:construct}

\Cref{app:eqn:expected_replica_swap_rate} can be used to infer how to construct most efficiently a temperature ladder between two extremal inverse temperatures, $\beta_{\min}$ (the inverse temperature above which the system is ergodic) and $\beta_{\max}$ (the inverse temperature of interest). We assume that the parallel-tempering ladder consists of $N$ replicas at inverse temperatures $\beta_i$ (with $1 \leq i \leq N$, $\beta_1 = \beta_{\min}$, and $\beta_N = \beta_{\max}$). The ladder is most efficient when all neighboring replicas are exchanged with equal probability. This is achieved when the ratio $\frac{\beta_{i+1} - \beta_i}{\sqrt{\frac{\beta_{i+1} + \beta_i}{2}}}$ is constant for all neighboring pairs. In particular, assuming that the inverse-temperature differences are small, one convenient choice is
\begin{equation}
	\sqrt{\beta_i} = \sqrt{\beta_{\min}} + \frac{i-1}{N-1} \left( \sqrt{\beta_{\max}} - \sqrt{\beta_{\min}} \right), \qq{for} 1 \leq i \leq N,
\end{equation} 
in which case the expected replica-swap rate is
\begin{equation}
	\label{app:eqn:expected_replica_swap}
	p^{\text{swap}} \equiv \left\langle p^{\text{accept}}_{\left( s_{L} @ \beta, s'_{L'} @ \beta' \right) \to \left( s'_{L} @ \beta, s_{L'} @ \beta' \right)} \right\rangle \approx \operatorname{erfc} \left(\frac{\sqrt{\beta_{\max}} - \sqrt{\beta_{\min}}}{N-1} \sqrt{\frac{-E_0}{4}} \right).
\end{equation}
It is clear that the replica-swap rate can be increased arbitrarily by increasing the number of replicas $N$. However, the relevant figure of merit is \emph{not} a large swap rate, but rather a short \emph{roundtrip time} -- the characteristic time for a replica to travel from the cold end of the ladder to the hot end and back. In practice, we employ a checkerboard decomposition of the replicas: after every SSE sweep, we first attempt swaps between pairs $(i,i+1)$ with odd $1 \leq i < N$, and then between pairs with even $i$. After each SSE sweep, a replica changes its index by $\left\langle \left( \Delta i \right)^2 \right\rangle = 4 p^2 + 2p(1-p) \approx 2 p$. As a result, the characteristic roundtrip time scales as
\begin{equation}
	\tau^{\text{roundtrip}} \sim \frac{N^2}{p^{\text{swap}}} \approx \frac{N^2}{\operatorname{erfc} \left(\frac{\sqrt{\beta_{\max}} - \sqrt{\beta_{\min}}}{N-1} \sqrt{\frac{-E_0}{4}} \right)}.
\end{equation}
Given $\beta_{\max}$, $\beta_{\min}$, and $E_0$, the optimal number of replicas can then be obtained by minimizing the roundtrip time $\tau^{\text{roundtrip}}$.

\section{Analytical results}\label{app:sec:anal}

In this \siSection{}, we present two analytical results employed in this work. We begin by obtaining the grand potential of the AA t-\ch{SnSe2}{} Hamiltonian in the strong-coupling limit by perturbing around the atomic limit. This result is used in the main text to obtain the entropy of the system in the high-temperature paramagnetic phase. This entropy then serves as the starting point for the thermodynamic integration of numerically obtained SSE data, as outlined in \cref{app:sec:sse:observables:entropy}, allowing us to determine the entropy of the system at lower temperatures. We then provide a more rigorous explanation of the hierarchy of correlated insulators and their inverse compressibility using the parton mean-field approach~\cite{FLO02,FLO04}, which is presented in greater detail in our companion work~\cite{VAS26}. Consequently, we only briefly review this approach here and present its conclusions.

\subsection{Perturbation theory around the atomic limit} \label{app:sec:anal:perturbation}

In this section, we perform perturbation theory around the atomic limit, treating the kinetic energy term perturbatively. The goal is to obtain the grand potential to leading order in the kinetic term, and thereby determine the entropy of the system in the high-temperature paramagnetic phase. 

We start by performing a mean-field decoupling of the offsite repulsion terms in the interacting Hamiltonian of AA t-\ch{SnSe2}{} in the mixed-dimensional limit,
\begin{equation}
	\label{app:eqn:interaction_hamiltonia_with_atomic}
	H_I = \sum_{\vec{R}} H_{\text{At}} \left( \vec{R} \right) + \frac12 \sum_{\substack{\Delta \vec{R} \neq \vec{0} \\ \vec{R}, \eta_1, \eta_2}} V_{\eta_1 \eta_2} \left( \Delta \vec{R} \right) :\mathrel{\hat{N}_{\eta_1,\vec{R} + \Delta \vec{R}}}: :\mathrel{\hat{N}_{\eta_2,\vec{R}}}: + \mu \sum_{\vec{R},\eta} \hat{N}_{\eta,\vec{R}},
\end{equation}
where the normal ordering is defined as $:\mathrel{\hat{N}_{\eta,\vec{R}}}: = \hat{N}_{\eta,\vec{R}} - 1$. In \cref{app:eqn:interaction_hamiltonia_with_atomic}, we have also introduced the atomic Hamiltonian\footnote{In \cref{app:eqn:final_single_particle,app:eqn:final_interaction}, the chemical potential was placed in the kinetic term. In this and the following section, it is more convenient to include it in the interacting atomic contribution. The last term in \cref{app:eqn:interaction_hamiltonia_with_atomic} simply reflects the fact that $H_I$ is defined without the chemical potential, whereas the atomic Hamiltonian $H_{\text{At}} \left( \vec{R} \right)$ defined here \emph{does} include the chemical potential term.}
\begin{equation}
	\label{app:eqn:atomic_hamiltonian}
	H_{\text{At}} \left( \vec{R} \right) = \frac12 \sum_{\eta_1,\eta_2} V_{\eta_1 \eta_2} \left( \vec{0} \right) :\mathrel{\hat{N}_{\eta_1,\vec{R}}}: :\mathrel{\hat{N}_{\eta_2,\vec{R}}}: - \mu \sum_{\eta} \hat{N}_{\eta,\vec{R}},
\end{equation}
Taking $:\mathrel{\hat{N}_{\vec{R},\eta}}: = \bar{N} + \delta \hat{N}_{\vec{R},\eta}$, where $\bar{N} = \frac{\nu-3}{3}$ is the average filling of a single orbital, measured relative to the offset such that $-1 \leq \bar{N} \leq 1$, the offsite interaction can be decoupled as
\begin{align}
	&\frac12 \sum_{\substack{\Delta \vec{R} \neq \vec{0} \\ \vec{R}, \eta_1, \eta_2}} V_{\eta_1 \eta_2} \left( \Delta \vec{R} \right) :\mathrel{\hat{N}_{\eta_1,\vec{R} + \Delta \vec{R}}}: :\mathrel{\hat{N}_{\eta_2,\vec{R}}}: = \frac12 \sum_{\substack{\Delta \vec{R} \neq \vec{0} \\ \vec{R}, \eta_1, \eta_2}} V_{\eta_1 \eta_2} \left( \Delta \vec{R} \right) \left( \bar{N} + \delta \hat{N}_{\eta_1,\vec{R} + \Delta \vec{R}} \right) \left( \bar{N} + \delta \hat{N}_{\eta_2,\vec{R}} \right)  \nonumber \\
	\approx & \frac12 \sum_{\substack{\Delta \vec{R} \neq \vec{0} \\ \vec{R}, \eta_1, \eta_2}} V_{\eta_1 \eta_2} \left( \Delta \vec{R} \right) \left[ \bar{N}^2 + \bar{N} \left(:\mathrel{\hat{N}_{\eta_1,\vec{R} + \Delta \vec{R}}}: -  \bar{N} \right) + \left( :\mathrel{\hat{N}_{\eta_2,\vec{R}}}: - \bar{N} \right) \bar{N} \right] \nonumber \\
	= & \frac12 \sum_{\substack{\Delta \vec{R} \neq \vec{0} \\ \vec{R}, \eta_1, \eta_2}} V_{\eta_1 \eta_2} \left( \Delta \vec{R} \right) \left( 2 \bar{N} :\mathrel{\hat{N}_{\eta_1,\vec{R}}}: -\bar{N}^2 \right) \nonumber \\
	= & \frac12 \sum_{\substack{\Delta \vec{R} \neq \vec{0} \\ \vec{R}, \eta, \Delta \eta}} V_{\eta (\eta + \Delta \eta )} \left( \Delta \vec{R} \right) \left( 2 \bar{N} :\mathrel{\hat{N}_{\eta,\vec{R}}}: -\bar{N}^2 \right) \nonumber \\
	= & 18 \bar{V} \sum_{\vec{R},\eta} \left( \bar{N} :\mathrel{\hat{N}_{\eta,\vec{R}}}: - \frac{\bar{N}^2}{2} \right) = \sum_{\vec{R}} H_{V} \left( \vec{R}, \bar{N} \right),
\end{align}
where $H_{V} \left( \vec{R}, \bar{N} \right) = 18 \bar{V} \sum_{\eta} \left( \bar{N} :\mathrel{\hat{N}_{\eta,\vec{R}}}: - \frac{\bar{N}^2}{2} \right)$. The mean-field decoupled interaction Hamiltonian can then be written as a sum of single-site terms, the last of which depends on the average orbital occupation and must be determined self-consistently
\begin{equation}
	H_{I} \approx \sum_{\vec{R}} H_{\text{At}} \left( \vec{R} \right) + H_{V} \left( \vec{R}, \bar{N} \right) + \mu \sum_{\vec{R},\eta} \hat{N}_{\eta,\vec{R}}.
\end{equation}
Note, however, that $H_{V}$ takes the form of a chemical-potential term whose coefficient depends on $\bar{N}$, which must itself be determined self-consistently. For a given chemical potential $\mu$ and inverse temperature $\beta$, the interaction Hamiltonian $H_I$ is then trivially exactly solvable.

The goal of this section is to perform perturbation theory in the kinetic Hamiltonian $H_{0}$ and obtain the grand potential perturbatively in the kinetic terms. To this end, we define the unperturbed atomic action $S_{\text{At}{}}$, the kinetic perturbation $\delta \mathcal{S}$, and the corresponding unperturbed and perturbed partition functions
\begin{align}
	\mathcal{S}_{\text{At}{}} \left[ \hat{c}^\dagger_{}, \hat{c}_{} \right] &= \int_{0}^{\beta} \dd{\tau} \left[ \sum_{\vec{R},\eta,s} \hat{c}^\dagger_{\vec{R},\eta,s} (\tau) \partial_{\tau} \hat{c}_{\vec{R},\eta,s} (\tau) + \sum_{\vec{R}} \left( H_{\text{At}} \left( \vec{R}, \tau \right) + H_{V} \left( \vec{R}, \bar{N}, \tau \right) \right) \right] \\
	\delta \mathcal{S} \left[ \hat{c}^\dagger_{}, \hat{c}_{} \right] &= \int_{0}^{\beta} \dd{\tau} \left( H_0 (\tau) + \mu \sum_{\vec{R},\eta} \hat{N}_{\eta,\vec{R}} (\tau) \right) \\
	Z_{\text{At}{}} &= \int \mathcal{D} \left[ \hat{c}^\dagger_{}, \hat{c}_{} \right] \exp \left( - \mathcal{S}_{\text{At}{}} \left[ \hat{c}^\dagger_{}, \hat{c}_{} \right] \right) \\
	Z &= \int \mathcal{D} \left[ \hat{c}^\dagger_{}, \hat{c}_{} \right] \exp \left( - \mathcal{S}_{\text{At}{}} \left[ \hat{c}^\dagger_{}, \hat{c}_{} \right] - \delta \mathcal{S} \left[ \hat{c}^\dagger_{}, \hat{c}_{} \right] \right) = Z_{\text{At}{}} \left\langle e^{- \delta \mathcal{S} \left[ \hat{c}^\dagger_{}, \hat{c}_{} \right]} \right\rangle_{\text{At}{}}
\end{align}
where $\left\langle \dots \right\rangle_{\text{At}{}}$ denotes the expectation value in the atomic limit, while the imaginary-time argument $\tau$ indicates imaginary-time evolution. The average orbital filling $\bar{N}$ is determined self-consistently for each chemical potential $\mu$.

Up to the first nontrivial order in the kinetic term, the perturbed partition function is 
\begin{equation}
	Z = Z_{\text{At}} \left( 1 + \frac12 \left\langle  \left( \delta \mathcal{S} \left[ \hat{c}^\dagger_{}, \hat{c}_{} \right] \right)^2 \right\rangle_{\text{At}} + \dots \right),
\end{equation}
where the ellipses denote higher order terms. The corresponding grand potential is, to the same order,
\begin{equation}
	\Phi \left( \beta, \mu \right) = \Phi^{\text{At}{}} \left( \beta, \mu \right) - \frac{1}{2 \beta} \left\langle  \left( \delta \mathcal{S} \left[ \hat{c}^\dagger_{}, \hat{c}_{} \right] \right)^2 \right\rangle_{\text{At}}.
\end{equation}
We now proceed to evaluate the nontrivial correction
\begin{align}
	\left\langle  \left( \delta \mathcal{S} \left[ \hat{c}^\dagger_{}, \hat{c}_{} \right] \right)^2 \right\rangle_{\text{At}} =& \int_0^{\beta} \dd{\tau_1} \int_0^{\beta} \dd{\tau_2} \sum_{\substack{\vec{R}_1, \vec{R}'_1, \eta_1, s_1 \\ \vec{R}_2, \vec{R}'_2, \eta_2, s_2}} t_{\eta_1} \left( \vec{R}'_1 - \vec{R}_1 \right) t_{\eta_2} \left( \vec{R}'_2 - \vec{R}_2 \right) \nonumber \\
	& \quad \times \left\langle  \left( \hat{c}^\dagger_{\vec{R}_1,\eta_1,s_1} \left( \tau_1 \right) \hat{c}_{\vec{R}'_1,\eta_1,s_1} \left( \tau_1 \right) \right) \left( \hat{c}^\dagger_{\vec{R}'_2,\eta_2,s_2} \left( \tau_2 \right) \hat{c}_{\vec{R}_s,\eta_2,s_2} \left( \tau_2 \right) \right)  \right\rangle_{\text{At}} \nonumber \\
	=& \beta \int_{0}^{\beta} \dd{\tau} \sum_{\substack{\vec{R}, \vec{R}' \\ \eta, s}} t^2_{\eta} \left( \vec{R}' - \vec{R} \right) \left\langle  \hat{c}^\dagger_{\vec{R},\eta,s} (\tau) \hat{c}_{\vec{R}',\eta,s} (\tau) \hat{c}^\dagger_{\vec{R}',\eta,s} (0) \hat{c}_{\vec{R},\eta,s} (0)  \right\rangle_{\text{At}}  \nonumber \\
	=& \beta N \int_{0}^{\beta} \dd{\tau} \sum_{\Delta \vec{R}, \eta, s} t^2_{\eta} \left( \Delta \vec{R} \right) \left\langle  \hat{c}^\dagger_{\vec{R},\eta,s} (\tau) \hat{c}_{\vec{R},\eta,s} (0) \right\rangle_{\text{At}} \left\langle  \hat{c}_{\vec{R}',\eta,s} (\tau) \hat{c}^\dagger_{\vec{R}',\eta,s} (0)  \right\rangle_{\text{At}} \nonumber \\
	=& \beta N \int_{0}^{\beta} \dd{\tau} \sum_{\Delta \vec{R}, \eta, s} t^2_{\eta} \left( \Delta \vec{R} \right) \left\langle  \hat{c}^\dagger_{\vec{R},\eta,s} (\tau) \hat{c}_{\vec{R},\eta,s} (0) \right\rangle_{\text{At}} \left\langle  \hat{c}_{\vec{R}',\eta,s} (\tau) \hat{c}^\dagger_{\vec{R}',\eta,s} (0)  \right\rangle_{\text{At}}\nonumber \\
	=& - \beta N  \sum_{\Delta \vec{R}, \eta, s} t^2_{\eta} \left( \Delta \vec{R} \right) \int_{0}^{\beta} \dd{\tau} G_{\text{At}} (-\tau) G_{\text{At}} (\tau) \nonumber \\
	=& - 2 \times 2 \times 3 \beta N  t^2 \int_{0}^{\beta} \dd{\tau} G_{\text{At}} (-\tau) G_{\text{At}} (\tau), \label{app:eqn:second_order_grand_potential_correction}
\end{align}
where $G_{\text{At}} (\tau)$ is the site-, valley-, and spin-independent, site-diagonal atomic Green's function 
\begin{equation}
	G_{\text{At}} (\tau) \equiv - \left\langle  \mathcal{T}_{\tau} \hat{c}_{\vec{R},\eta,s} (\tau) \hat{c}^\dagger_{\vec{R},\eta,s} (0) \right\rangle_{\text{At}},
\end{equation} 
with $\mathcal{T}_{\tau}$ denoting the imaginary-time-ordering operator. The decoupling of the four-fermion correlation function into a product of two two-fermion Green's functions in the third equality of \cref{app:eqn:second_order_grand_potential_correction} does not follow from Wick's theorem, since the atomic action $\mathcal{S}_{\text{At}{}} \left[ \hat{c}^\dagger_{}, \hat{c}_{} \right]$ is interacting. Instead, it follows from the fact that the atomic action conserves charge separately for each lattice site, valley, and spin. Up to second order in the hopping amplitude, the grand potential of the system is therefore
\begin{equation}
	\label{app:eqn:grand_potential_perturbation}
	\Phi \left( \beta, \mu \right) = \Phi^{\text{At}} \left( \beta, \mu \right) + 6 N t^2 \int_{0}^{\beta} \dd{\tau} G_{\text{At}} (-\tau) G_{\text{At}} (\tau).
\end{equation}
\Cref{app:eqn:grand_potential_perturbation} gives the high-temperature entropy used as the starting point for the thermodynamic integration in \cref{app:eqn:entropy_diff_thermo_integration}. Combined with the energy extracted from SSE simulations, this allows us to obtain the entropy as a function of temperature, as shown in \cref{fig:spin_corr_ins}. Specifically, the high-temperature entropy can be extracted from the grand potential via 
\begin{equation}
	S \left( \beta, \mu \right) = \beta^2 \pdv{ \Phi \left( \beta, \mu \right)}{\beta}.
\end{equation}

\subsection{Parton mean-field description of metal-to-insulator transitions in AA t-\ch{SnSe2}{}}\label{app:sec:anal:parton}

In this section, we provide a more rigorous explanation of the hierarchy of correlated-insulator strengths in AA t-\ch{SnSe2}{} at integer fillings, as observed in the main text. The SSE data reveal two notable features:~(1) the system remains metallic up to a larger \textit{critical} interaction strength $U_c$ than, for example, the single-orbital Hubbard model~\cite{ROZ92,GEO92a,GEO96,FLO02}; and~(2) both the charge gap of the insulating states and the value of $U_c$ depend strongly on the integer filling $\nu$ and on the interaction anisotropy $U - U '$. In the main text, we argue that these observations can already be partially understood in the atomic limit, where they arise from the atomic level degeneracies and the resulting valley fluctuations. We strengthen this explanation here by performing a parton decomposition of the electronic degrees of freedom and treating the resulting problem at the mean-field level. A more detailed presentation of this construction is given in our companion paper~\cite{VAS26}; here, we summarize only the essential ingredients and refer the reader to Ref.~\cite{VAS26} for a more complete discussion.

Our approach is a multi-orbital ({\it i.e.}{}, multi-valley) extension of the rotor parton decomposition of Florens and Georges~\cite{FLO02,FLO04}, originally formulated for the $\mathrm{U} \left( {N} \right)$ Hubbard model. The central idea of the rotor approach is to decompose each electron into a charge-carrying quantum rotor and a charge-neutral spinon, $\hat{d}^\dagger_{} \rightarrow \hat{\gamma}^\dagger_{} e^{i \hat{\varphi}}$, and to study the metal-to-insulator transition by tracking the fluctuations of the rotor angular momentum $\hat{L}$ conjugate to $\hat{\varphi}$, which encode the physical charge fluctuations. Multi-orbital generalizations in the literature have often employed auxiliary spin/boson fields instead~\cite{DE05,HAS10,YU12,CRI23}, since pair-hopping and Hund's couplings are ubiquitous in atomic-scale problems and make a rotor description less natural. In AA t-\ch{SnSe2}{}, however, the moir\'e scale naturally suppresses these terms, allowing for an intuitive interpretation of the Mott transition in the rotor language.

\subsubsection{Atomic contribution}\label{app:sec:anal:parton:atomic}
We begin by rewriting the atomic Hamiltonian from \cref{app:eqn:atomic_hamiltonian} in a symmetric basis that separates the total from the valley-imbalance charge operators (which we will henceforth call a \emph{diagonal} basis)
\begin{align}
    H_{\text{At}} = & - \mu \sum_{\eta} \hat{N}_\eta +\frac{U}{2}\sum_\eta (\hat{N}_\eta-1)^2 + U' \sum_{\eta<\eta'}(\hat{N}_\eta-1)(\hat{N}_{\eta'}-1) \nonumber \\
    =&-\mu\sum_{\eta}\hat{N}_\eta + \frac{U+2U'}{6}\left(\sum_\eta\hat{N}_{\eta}-3 \right)^2 + \frac{U-U'}{12}\left(\hat{N}_0 + \hat{N}_1 - 2\hat{N}_2\right)^2 + \frac{U-U'}{4}\left(\hat{N}_0 - \hat{N}_1 \right)^2 \nonumber \\
    =&-\mu\hat{Q}_{\text{ch}} + \frac{\bar{U}}{6}\left(\hat{Q}_{\text{ch}}-3 \right)^2 + \frac{U_+}{2}\hat{Q}_+^2 + \frac{U_-}{2}\hat{Q}_-^2, \label{app:eqn:Atomic Hamiltonian changed basis}
\end{align}
where we have temporarily suppressed the lattice index $\vec{R}$ for notational simplicity. In the last line of \cref{app:eqn:Atomic Hamiltonian changed basis}, we introduced the charge and valley-imbalance modes
\begin{equation}
    \hat Q_\text{ch}=\sum_\eta \hat{N}_\eta , \qquad
    \hat Q_+ = \hat{N}_0+\hat{N}_1-2\hat{N}_2, \qquad
    \hat Q_- = \hat{N}_0-\hat{N}_1 .
\end{equation}
with stiffnesses
\begin{equation}
    \bar{U}=\tfrac{U+2U'}{3} , \qquad
    U_+ = \tfrac{U-U'}{6} , \qquad
    U_- = \tfrac{U-U'}{2} . 
\end{equation}
The neutral-mode stiffnesses $U_\pm$ vanish as $U - U' \to 0^+$; these soft valley fluctuations are central to the physics. By inspection, the spectrum of \cref{app:eqn:Atomic Hamiltonian changed basis} can be reproduced exactly by replacing the charge and valley-imbalance operators with rotor angular momenta $\hat{L}_{\text{ch}}$, $\hat{L}_+$, $\hat{L}_-$. This is because both sets of operators have spectra in the integers $\mathbb{Z}$. We will exploit this observation below.

At this point, we introduce the quantum rotor variables $\hat{\ell}_{\eta}$ and $\hat{\varphi}_{\eta}$, with commutation relations
\begin{alignat}{2}
	\commutator{\hat{\ell}_{\eta}}{\hat{\varphi}_{\eta'}}
	&= -i\delta_{\eta\eta'},
	&\qquad
	\commutator{\hat{\ell}_{\eta}}{e^{i\hat{\varphi}_{\eta'}}}
	&= \delta_{\eta\eta'} e^{i\hat{\varphi}_{\eta}}, \nonumber \\
	\commutator{\hat{\ell}_{\eta}}{\hat{\ell}_{\eta'}}
	&= 0,
	&\qquad
	\commutator{\hat{\varphi}_{\eta}}{\hat{\varphi}_{\eta'}}
	&= 0.\label{app:eqn:rotor operators relations 1}
\end{alignat}
as well as their diagonal-basis counterparts,
\begin{align}
	\left(\hat{L}_{\text{ch}},\hat{L}_{+},\hat{L}_{-}\right)^T&=A\left(\hat{\ell}_{0},\hat{\ell}_{1},\hat{\ell}_{2}\right)^T\\
	\left(\hat{\vartheta}_{\text{ch}},\hat{\vartheta}_{+},\hat{\vartheta}_{-}\right)^T&=(A^{-1})^T\left(\hat{\varphi}_{0},\hat{\varphi}_{1},\hat{\varphi}_{2}\right)^T
\end{align}
where 
\begin{equation}
    A = \begin{pmatrix}
        +1 & +1 & +1\\
        +1 & +1 & -2 \\
        +1 & -1 & 0 \\
    \end{pmatrix} .
\end{equation}
The operators $\hat{L}_{\kappa}$ and $\hat{\vartheta}_{\kappa}$ obey commutation relations analogous to those in \cref{app:eqn:rotor operators relations 1}. The auxiliary-rotor decomposition is
\begin{equation}
	\label{app:eqn:rotor decomposition}
    \hat{d}^\dagger_{\eta,s} = \hat{\gamma}^\dagger_{\eta,s}\,e^{i\hat\varphi_\eta} ,
\end{equation}
where $\hat{\gamma}^\dagger_{\eta,s}$ are charge-neutral spinons and $\hat{\varphi}_{\eta}$ are the rotor angles. By itself, this decomposition does not rewrite the interactions in terms of rotor angular momenta, since any electron density operator is mapped to a spinon density operator. The rotor description is instead enabled by the operator constraint, imposed separately in each valley,
\begin{equation}
	\label{app:eqn:Atomic Hamiltonian rotor description constraints}
    \hat\ell_\eta = \sum_s \left( \hat{\gamma}^\dagger_{\eta,s} \hat{\gamma}_{\eta,s} - \frac{1}{2} \right),
\end{equation}
which identifies $\hat{N}_\eta - 1 \equiv \hat\ell_\eta$ and therefore allows the replacement $(\hat{N}_\eta-1)^2 \to \hat\ell_\eta^2$. The atomic Hamiltonian then takes the form
\begin{equation}
	\label{app:eqn:Atomic Hamiltonian rotor description}
	H_{\text{At}} =-\mu\sum_{\eta,s}\hat{\gamma}^\dagger_{\eta,s}\hat{\gamma}_{\eta,s} + \frac{\bar{U}}{6}\hat{L}_{\text{ch}}^2 + \frac{U_+}{2}\hat{L}_+^2 + \frac{U_-}{2}\hat{L}_-^2  .
\end{equation}
For the Hamiltonian in \cref{app:eqn:Atomic Hamiltonian rotor description} to have the same spectrum as the original Hamiltonian in \cref{app:eqn:Atomic Hamiltonian changed basis}, the constraints in \cref{app:eqn:Atomic Hamiltonian rotor description constraints} must be imposed. The advantage of this approach is that the constraints can be imposed only on average, rather than exactly, while still reproducing the original model at sufficiently low temperatures. In this low-temperature regime, the unphysical part of the rotor spectrum, corresponding for example to values of $\hat{\ell}_\eta$ that cannot arise from an electron occupancy ({\it e.g.}{}, $\left\langle \hat{\ell}_{\eta} \right\rangle = 3$), is strongly suppressed~\cite{VAS26}.

\subsubsection{The lattice problem}\label{app:sec:anal:parton:lattice}

\begin{figure}[!t]
	\centering
	\includegraphics[width=\textwidth]{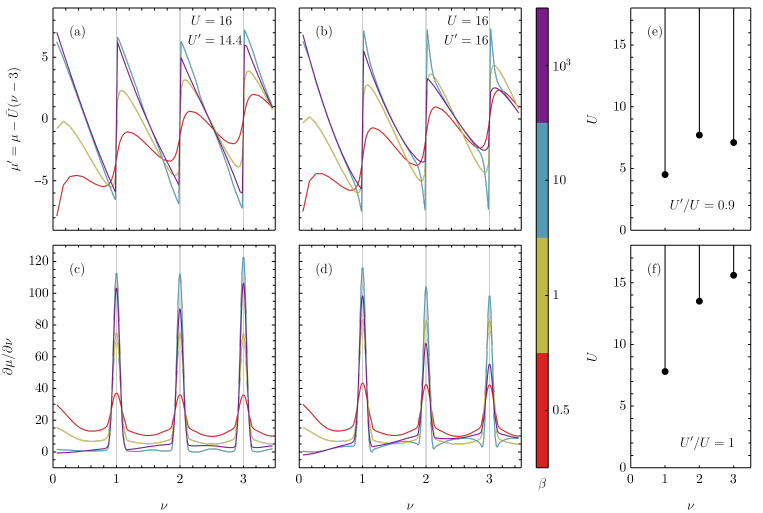}\subfloat{\label{app:fig:PartonMF:a}}\subfloat{\label{app:fig:PartonMF:b}}\subfloat{\label{app:fig:PartonMF:c}}\subfloat{\label{app:fig:PartonMF:d}}\subfloat{\label{app:fig:PartonMF:e}}\subfloat{\label{app:fig:PartonMF:f}}\subfloat{\label{app:fig:PartonMF:g}}\caption{Parton mean-field description of AA t-\ch{SnSe2}{}. We show the filling and inverse-compressibility results for a idealized strong-coupling system with only onsite Hubbard interactions, with $U = 16$ and either $U' = 14.4$ or $U' = 16$. (a) and (b) show the electrochemical potential $\mu' \equiv \mu - \bar{U} (\nu - 3)$, while (c) and (d) show the inverse compressibility, both as functions of filling. The curves are colored according to the inverse temperature $\beta$, as indicated by the colour map. The parameters used in each column are identical and are shown in the top panels: (a) and (c) correspond to the physical $\mathrm{U} \left( {2} \right)^{\otimes 3}$-symmetric local interaction, while (b) and (d) correspond to the idealized $\mathrm{U} \left( {6} \right)$-symmetric interaction. The inverse compressibility can be used as a proxy for the relative strength, {\it i.e.}{} the charge gap, of the correlated insulators. Panels (e) and (f) show the \textit{critical} interaction strength $U_c$ (black dots) at which the system transitions from a metal to an insulator, and its dependence on the filling and local interaction anisotropy $U'/U$.}
	\label{app:fig:PartonMF}
\end{figure}

The above discussion can be extended straightforwardly to the lattice by decomposing the electron operators into spinons and rotors on each site. The kinetic terms then become
\begin{equation}
    \hat{d}^\dagger_{\vec{R},\eta,s} \hat{d}_{\vec{R} + \Delta \vec{R},\eta,s} = \hat{\gamma}^\dagger_{\vec{R},\eta,s}\hat{\gamma}_{\vec{R} + \Delta \vec{R},\eta,s} e^{i(\hat\varphi_{\vec{R},\eta}-\hat\varphi_{\vec{R} + \Delta \vec{R},\eta})} ,
\end{equation}
which leads to a Hamiltonian that couples the spinon and rotor degrees of freedom. We decouple them at the mean-field level
\begin{align}
    \hat{\gamma}^\dagger_{\vec{R},\eta,s} \hat{\gamma}_{\vec{R} + \Delta \vec{R},\eta,s} e^{i \left( \hat\varphi_{\vec{R},\eta} - \hat\varphi_{\vec{R} + \Delta \vec{R},\eta} \right)} \approx &  \hat{\gamma}^\dagger_{\vec{R},\eta,s} \hat{\gamma}_{\vec{R} + \Delta \vec{R},\eta,s} \left\langle  e^{i \left( \hat\varphi_{\vec{R},\eta}-\hat\varphi_{\vec{R} + \Delta \vec{R},\eta} \right)}  \right\rangle_{\theta} \nonumber \\ 
	& + \left\langle \hat{\gamma}^\dagger_{\vec{R},\eta,s} \hat{\gamma}_{\vec{R} + \Delta \vec{R},\eta,s}  \right\rangle_\gamma
	e^{i \left( \hat\varphi_{\vec{R},\eta} - \hat\varphi_{\vec{R}+\Delta \vec{R},\eta} \right)} + \dots . \label{app:eqn:Lattice Hamiltonian rotor description}
\end{align}
In \cref{app:eqn:Lattice Hamiltonian rotor description}, $\left\langle \dots \right\rangle_{\gamma(\theta)}$ denotes an expectation value with respect to the now-decoupled spinon (rotor) Hamiltonian. At this stage, the rotor Hamiltonian is an XY model on the lattice, with couplings defined on the bonds. We further simplify the problem by treating this rotor model at the single-site mean-field level, which reduces it to an effective single-site rotor problem. With these approximations, and assuming no spontaneous valley or charge-current order, the mean-field Hamiltonian takes the form~\cite{VAS26}
\begin{align}
    \hat H_\gamma & = \sum_{\vec{k},\eta,s} \left[ Q^2 \epsilon_{\eta,1} \left( \vec{k} \right) - \left( \mu + h \right) \right]
    \hat{\gamma}^\dagger_{\vec{k},\eta,s} \hat{\gamma}_{\vec{k},\eta,s}, \nonumber \\    
    \hat H_\theta &= h\hat L_\text{ch} + \sum_{x\in\{\text{ch},+,-\}} \frac{U_x}{2} \hat L_x^2 + K \sum_{\eta \in \{0,1,2\}} \cos\hat\varphi_\eta .
    \label{app:eqn:Lattice Hamiltonian rotor description final}
\end{align}
In \cref{app:eqn:Lattice Hamiltonian rotor description final}, $\epsilon_{\eta,1} \left( \vec{k} \right)$ is the bare band dispersion, $Q \equiv \left\langle \cos\hat\varphi_\eta \right\rangle_\theta$ is the square root of the quasiparticle weight, $Z = Q^2$, and the parameters $h$ and $K$ are determined self-consistently from
\begin{align}
    \left\langle \hat L_\text{ch} \right\rangle_\theta &= 6 \int  \dd{\epsilon} D(\epsilon) n_{\text{F}} \left[ Q^2 \epsilon - \left( \mu + h \right) \right]-3, \nonumber \\
    K &= 4Q \int \dd{\epsilon} D(\epsilon) \epsilon n_{\text{F}} \left[ Q^2 \epsilon - \left( \mu + h \right) \right].  \label{app:eqn:Lattice Hamiltonian rotor description final constraints}
\end{align}
Here, $n_{\text{F}}(\epsilon) \equiv \left(1 + e^{\beta \epsilon} \right)^{-1}$ is the Fermi-Dirac distribution function, and $D(\epsilon)$ is the bare density of states for a single valley, which is the same for all valleys by $C_{3z}$ symmetry. The first equation in \cref{app:eqn:Lattice Hamiltonian rotor description final constraints} is the lattice-averaged version of the original constraint in \cref{app:eqn:Atomic Hamiltonian rotor description constraints}. The $K$ term couples the charge and valley-imbalance rotors and originates from the hopping terms. Finally, we note that our approach here works strictly for $U'<U$. In the case of $U'=U$, one must revert to the full $\mathrm{U} \left( {6} \right)$ formulation of Ref.~\cite{FLO04}; \cref{app:eqn:Lattice Hamiltonian rotor description final constraints} is the equivalent of Eq. (38) in Ref.~\cite{FLO04}.

The mechanism behind the enhancement of metallicity as $\left(U - U' \right) \to 0^+$, and the resulting strong dependence of $U_c$ on the interaction anisotropy, is transparent in this language. Although the stiffness of the total charge operator $\bar{U}$ remains large, the hopping term $K \cos\hat\varphi_\eta$ is written in the valley basis and therefore couples charge fluctuations to the neutral channels through $\hat \varphi = [A^T]^{-1} \hat \vartheta$. These neutral channels soften as the $\mathrm{U} \left( {6} \right)$ point is approached, since $U_\pm \propto \left( U - U' \right)$, thereby stabilizing the metallic phase up to large values of $\bar{U}$.

The results are summarized in \cref{app:fig:PartonMF}. The rotor description qualitatively captures the trends seen in the SSE data for the integer correlated insulators. In particular, it finds that the $\nu=1$ Mott insulator requires the smallest $U_c$ to form and has the largest charge gap compared to the $\nu=2$ and $\nu=3$ Mott insulators. It also captures another key feature of the data, namely the dependence of the insulating strength on the interaction anisotropy. For the idealized $\mathrm{U} \left( {6} \right)$-symmetric Hubbard interaction, the charge gap of the insulators decreases towards half-filling, while the critical interaction strength increases in the same direction, as shown in \cref{app:fig:PartonMF:c,app:fig:PartonMF:f}, respectively. In contrast, in the physical $\mathrm{U} \left( {2} \right)^{\otimes 3}$-symmetric case with $U' < U$, the weakest insulator occurs at $\nu = 2$, both in terms of the charge gap and the critical interaction strength. A similar conclusion was reached in \cref{sec:results:ci} by counting the atomic-level degeneracies in each case.

Finally, at sufficiently high temperatures, $U-U'\lesssim T\lesssim U$, the $\mathrm{U} \left( {6} \right)$ degeneracy-breaking terms remain incoherent, and the correlated insulators therefore follow the original $\mathrm{U} \left( {6} \right)$ hierarchy. It should also be noted that the rotor description assumes a disordered spinon channel and therefore does not capture the magnetic order associated with the Mott insulating phases. This at least partially explains why the rotor description overestimates $U_c$ compared to the SSE results. For example, \cref{app:fig:PartonMF:f} suggests that a system with onsite interactions $U = U' = 12$ would form a correlated insulator only at $\nu=1$, whereas our SSE simulations show insulating behaviour at all integer fillings down to the lowest temperatures studied, $\beta=12$.

\section{Additional numerical results for AA t-\ch{SnSe2}{}}\label{app:sec:numerical_results}

In this \siSection{}, we present comprehensive numerical results obtained using SSE on the AA t-\ch{SnSe2}{} Hamiltonian from \cref{app:sec:model:summary} for the 20 different parameters listed in \cref{app:tab:parameters_master}. We focus on diagonal observables such as the inverse compressibility, electrochemical potential, charge and spin stiffnesses, and the momentum-resolved charge susceptibility at high symmetry points of the MBZ. We begin by briefly discussing the details of the numerical simulations and then we define the main observables computed here. We then present comprehensive plots of the observables as a function of temperature and fillings.

\subsection{Overview}\label{app:sec:numerical_results:overview}

For each parameter set listed in \cref{app:tab:parameters_master}, we consider about 70 different values of the chemical potential, chosen to cover the filling range $0 \lesssim \nu \leq 3$ over a range of temperatures. For ease of comparison, throughout this \siSection{} we use energy units in which the leading intra-chain hopping is set to $t = 1$.

Specifically, we simulate $\beta \in \left\{ 1, 2, 4, 6, 9, 12 \right\}$ and consider $\mathcal{N} \times \mathcal{N}$ systems with $\mathcal{N} = 12$, which correspond to 864 fermionic degrees of freedom. For each combination of $(\mu, \beta)$, we perform $3 \times 2^{17}$ warm-up QMC sweeps and $2^{18}$ measuring sweeps. The simulations at a certain chemical potential $\mu$ are ``chained'' together in order of decreasing temperature, with the final configuration of the run at smaller $\beta$ serving as the initial condition for the run at larger $\beta$. The first $2^{18}$ warm-up sweeps are annealing steps, during which the temperature is decreased exponentially from the inverse temperature of the initial condition to the target inverse temperature. For the initial inverse temperature $\beta = 1$, we use random initial conditions and an initial inverse temperature $\beta = 0.1$.

Among the first observables that we consider in this \siSection{} is the charge susceptibility of the system at high-symmetry momenta. To further characterize the breaking of discrete crystalline symmetries, we resolve the irreducible representations of the little group at the corresponding high-symmetry momenta. Specifically, we define
\begin{align}
	\chi_{\Gamma_A} &= \sum_{\eta_1, \eta_2} u^*_{A; \eta_1} \eval{\chi_{\eta_1 \eta_2} \left( \vec{k} \right)}_{\vec{k} = \vec{0}} u_{A; \eta_2}, \label{app:eqn:def_susc_hsm_gamma_a} \\ 
	\chi_{\Gamma_E} &= \sum_{\substack{\eta_1, \eta_2 \\ i,j}} u^*_{E, i; \eta_1} \eval{\chi_{\eta_1 \eta_2} \left( \vec{k} \right)}_{\vec{k} = \vec{0}} u_{E, j; \eta_2}, \label{app:eqn:def_susc_hsm_gamma_e} \\
	\chi_{\mathrm{K}_A} &= \sum_{\eta_1, \eta_2} u^*_{A; \eta_1} \eval{\chi_{\eta_1 \eta_2} \left( \vec{k} \right)}_{\vec{k} = \frac13\left( \vec{b}_{M_1} + \vec{b}_{M_2} \right)} u_{A; \eta_2}, \label{app:eqn:def_susc_hsm_k_a} \\
	\chi_{\mathrm{K}_E} &= \sum_{\substack{\eta_1, \eta_2 \\ i,j}} u^*_{E, i; \eta_1} \eval{\chi_{\eta_1 \eta_2} \left( \vec{k} \right)}_{\vec{k} = \frac13\left( \vec{b}_{M_1} + \vec{b}_{M_2} \right)} u_{E, j; \eta_2}, \label{app:eqn:def_susc_hsm_k_e} \\
	\chi_{\mathrm{M}^-} &= \sum_{\eta_1, \eta_2} u^*_{-; \eta_1} \eval{\chi_{\eta_1 \eta_2} \left( \vec{k} \right)}_{\vec{k} = \frac12 \vec{b}_{M_1}} u_{-; \eta_2}, \label{app:eqn:def_susc_hsm_m_-}\\
	\chi_{\mathrm{M}^+} &= \sum_{\substack{\eta_1, \eta_2 \\ i,j}} u^*_{+, i; \eta_1} \eval{\chi_{\eta_1 \eta_2} \left( \vec{k} \right)}_{\vec{k} = \frac12 \vec{b}_{M_1}} u_{+; j,\eta_2}, \label{app:eqn:def_susc_hsm_m_+}
\end{align} 
where 
\begin{alignat}{4}
	u_{A} &= && \frac{1}{\sqrt{3}} \begin{pmatrix} 1 & 1 & 1 \end{pmatrix} &\quad
	u_{-} &= && \frac{1}{\sqrt{2}} \begin{pmatrix} 0 & 1 & -1 \end{pmatrix}, \nonumber \\
	u_{E,1} &= && \frac{1}{\sqrt{3}} \begin{pmatrix} 1 & e^{i \frac{2\pi}{3}} & e^{-i \frac{2\pi}{3}} \end{pmatrix} &\quad
	u_{+,1} &= && \begin{pmatrix} 1 & 0 & 0 \end{pmatrix}, \nonumber \\
	u_{E,2} &= && \frac{1}{\sqrt{3}} \begin{pmatrix} 1 & e^{-i \frac{2\pi}{3}} & e^{i \frac{2\pi}{3}} \end{pmatrix} &\quad
	u_{+,2} &= && \frac{1}{\sqrt{2}} \begin{pmatrix} 0 & 1 & 1 \end{pmatrix}. 
\end{alignat}
The symmetry-resolved components of the charge-charge susceptibility at high-symmetry momenta have the following interpretation. $\chi_{\Gamma_A}$ is related to the total charge-charge susceptibility and, \emph{in the thermodynamic limit}, can be related to the compressibility,
\begin{equation}
	\label{app:eqn:relation_compressibility_susc}
	\chi_{\Gamma_A} = \frac{1}{3} \frac{\beta}{\mathcal{N}^2} \sum_{\vec{R}} \left\langle \hat{N}_{\vec{R}'} \hat{N}_{\vec{R}' + \vec{R}} \right\rangle_c = \frac{1}{3} \pdv{\nu}{\mu},
\end{equation}
where the last equality holds in the $\mathcal{N} \to \infty$ limit. On the other hand, $\chi_{\Gamma_E}$ measures the tendency toward valley polarization without translation-symmetry breaking. The remaining momentum-resolved components of the susceptibility can be associated with various translation-breaking orders, including charge-density-wave (CDW) and Wigner-Mott (WM) states. For example, $\chi_{\mathrm{K}_A}$ and $\chi_{\mathrm{K}_E}$ probe the tendency of the system to develop a $\sqrt{3}\times \sqrt{3}$ CDW or WM state that either preserves or breaks the $C_{3z}$ symmetry. Similarly, $\chi_{\mathrm{M}^+}$ and $\chi_{\mathrm{M}^-}$ probe the formation of a $2 \times 1$ CDW or WM state that either preserves or breaks the $C_{2x}$ symmetry.

Beyond the above-mentioned components of the charge-charge susceptibility, we also compute the charge and spin stiffnesses $\rho_{\text{C}}$ and $\rho_{\text{S}}$, which indicate whether the system can transport charge and spin, respectively.

The inverse compressibility of the system, $\pdv{\mu}{\nu}$, can be estimated in two ways: either through so-called correlated sampling using \cref{app:eqn:relation_compressibility_susc}, or through direct differentiation of the numerically estimated filling $\nu \left( \mu, \beta \right)$. In principle, correlated sampling is almost always the better option for computing observables in QMC. In the present case, however, \cref{app:eqn:relation_compressibility_susc} holds strictly only in the thermodynamic limit. At low temperature ({\it i.e.}{} below the charge gap of the system) and for the finite systems considered in this work, the inverse compressibility estimated in this way can diverge even in metallic phases, where the charge gap decreases only relatively slowly with system size, as $1/\mathcal{N}$. For this reason, and also because $\nu \left( \mu, \beta \right)$ can be estimated essentially without error (the measured uncertainty is very small), in this work we instead take the numerical derivative of the chemical potential with respect to the filling $\nu$ after smoothing with a narrow Gaussian kernel. At higher temperatures, both approaches yield identical results.

In addition to the inverse compressibility, we also plot the electrochemical potential of the system ({\it i.e.}{} the chemical potential from which we subtract the uniform charging energy of the corresponding gated device)~\cite{RAI23a}. Experimental works~\cite{ROZ21,SAI21a,ZHA25a} typically report the electrochemical potential $\mu'$, which in our system is related to the usual chemical potential $\mu$ via
\begin{equation}
	\mu' = \mu - \left(\nu - 3 \right) \left( \bar{U} + 6 \bar{V} \right).
\end{equation}
Finally, we also plot the spin-spin correlation function $\mathcal{S} (i)$ along a chain, as defined in \cref{app:eqn:def_spin_spin}, at the lowest temperature simulated (namely $\beta = 12$). For all quantities computed in this work, the errors are estimated using the binning method outlined in Ref.~\cite{FLY89}. 

\begin{figure}[H]
\centering
\includegraphics[width=\textwidth]{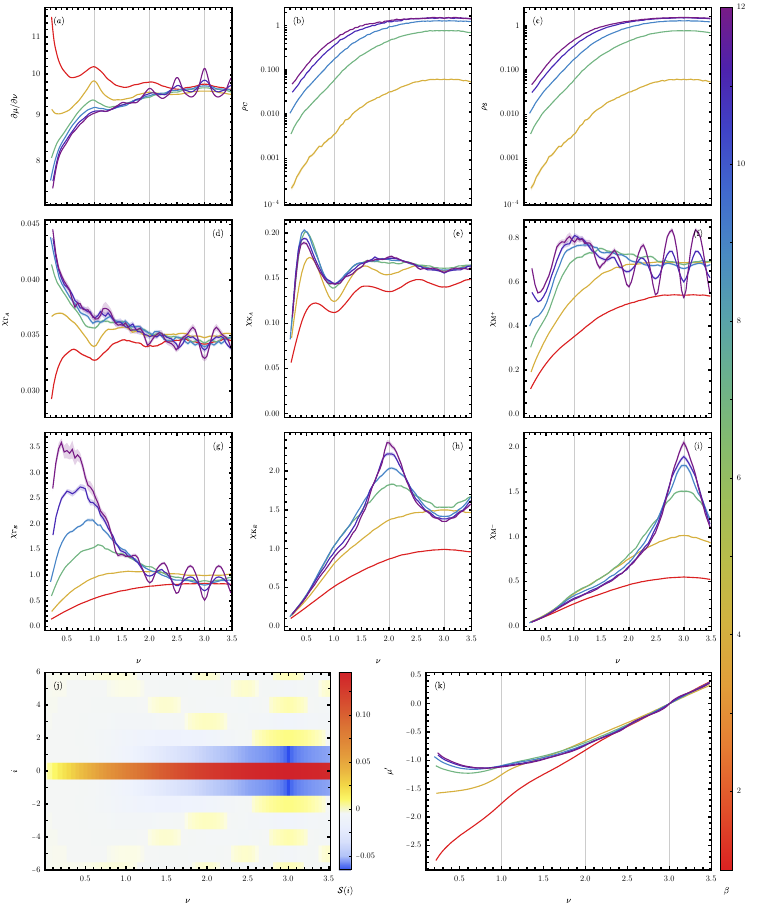}
\subfloat{\label{app:fig:allPlotSummary_1:a}}\subfloat{\label{app:fig:allPlotSummary_1:b}}\subfloat{\label{app:fig:allPlotSummary_1:c}}\subfloat{\label{app:fig:allPlotSummary_1:d}}\subfloat{\label{app:fig:allPlotSummary_1:e}}\subfloat{\label{app:fig:allPlotSummary_1:f}}\subfloat{\label{app:fig:allPlotSummary_1:g}}\subfloat{\label{app:fig:allPlotSummary_1:h}}\subfloat{\label{app:fig:allPlotSummary_1:i}}\subfloat{\label{app:fig:allPlotSummary_1:j}}\subfloat{\label{app:fig:allPlotSummary_1:k}}\caption{Numerical results for AA t-\ch{SnSe2}{} at $\theta = 9.43 \degree$, $\epsilon = 12$, and $\xi = \SI{2.5}{\nano\meter}$, corresponding to $\bar{U}=4.052$ and $\bar{V} = 0.803$. (a) shows the inverse compressibility $\pdv{\mu}{\nu}$. The charge and spin stiffnesses, $\rho_{\text{C}}$ and $\rho_{\text{S}}$, are displayed in (b) and (c), respectively. The various components of the charge-charge susceptibility defined in \crefrange{app:eqn:def_susc_hsm_gamma_a}{app:eqn:def_susc_hsm_m_+} are shown in (d)-(i). In (j), we plot the spin-spin correlation function along a chain, $\mathcal{S}(i)$, at $\beta=12$ as a function of $\nu$, while (k) shows the electrochemical potential $\mu'$. We consider $\beta \in \left\{1, 2, 4, 6, 9, 12\right\}$, except for the charge- and spin-stiffness plots, where we restrict to $\beta \geq 2$.}
\label{app:fig:allPlotSummary_1}
\end{figure}

To facilitate comparison between different parameter sets, all figures presented in \cref{app:sec:numerical_results:all_results} share the same format. Each figure corresponds to a single parameter set from \cref{app:tab:parameters_master} and contains 11 panels. Taking \cref{app:fig:allPlotSummary_1} as an example, \cref{app:fig:allPlotSummary_1:a} shows the inverse compressibility $\pdv{\mu}{\nu}$, while \cref{app:fig:allPlotSummary_1:b,app:fig:allPlotSummary_1:c} show, respectively, the charge and spin stiffnesses of the system. The six components of the charge-charge susceptibility defined in \crefrange{app:eqn:def_susc_hsm_gamma_a}{app:eqn:def_susc_hsm_m_+} are shown in \crefrange{app:fig:allPlotSummary_1:d}{app:fig:allPlotSummary_1:i}. Each trace is plotted as a function of filling and is colored according to the inverse temperature $\beta$, using the color map shown on the right. A narrow Gaussian filter is applied to the susceptibility plots to reduce finite-size effects at low temperatures. For each trace, a halo of the same color indicates the associated error. Whenever this halo is not visible, the error is smaller than the line width. \Cref{app:fig:allPlotSummary_1:j} shows the spin-spin correlation function along a chain, $\mathcal{S}(i)$, as a function of filling at the lowest temperature simulated, $\beta=12$. Finally, \cref{app:fig:allPlotSummary_1:k} shows the electrochemical potential of the system as a function of filling for different inverse temperatures. The reader is reminded that we use units in which $t=1$.

\clearpage

\subsection{Results}\label{app:sec:numerical_results:all_results}
\begin{figure}[H]
\centering
\includegraphics[width=\textwidth]{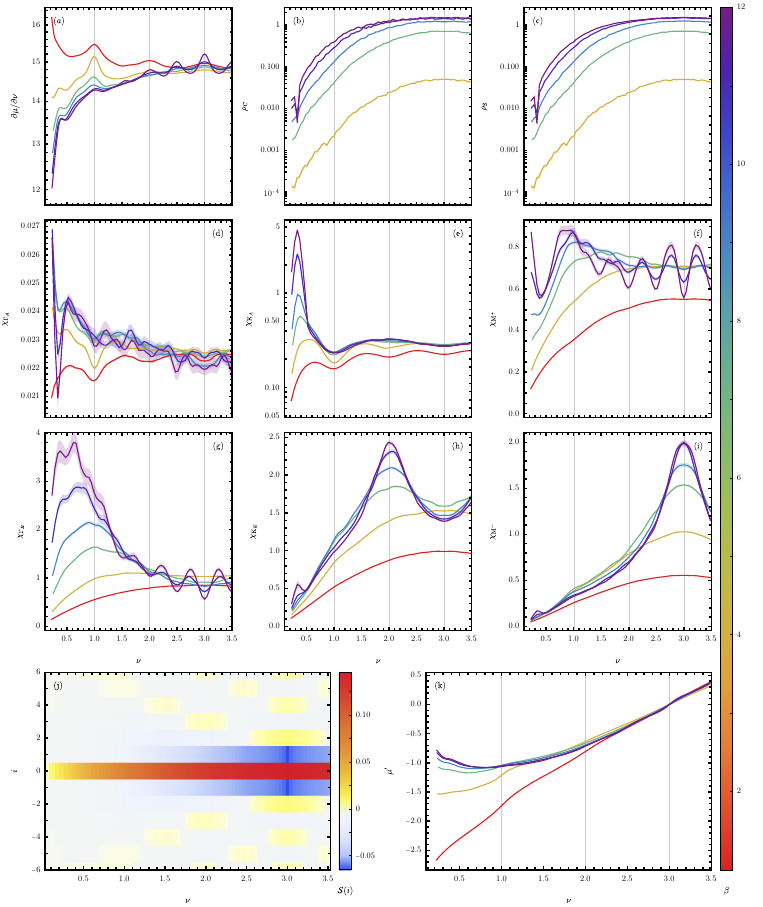}
\subfloat{\label{app:fig:allPlotSummary_2:a}}\subfloat{\label{app:fig:allPlotSummary_2:b}}\subfloat{\label{app:fig:allPlotSummary_2:c}}\subfloat{\label{app:fig:allPlotSummary_2:d}}\subfloat{\label{app:fig:allPlotSummary_2:e}}\subfloat{\label{app:fig:allPlotSummary_2:f}}\subfloat{\label{app:fig:allPlotSummary_2:g}}\subfloat{\label{app:fig:allPlotSummary_2:h}}\subfloat{\label{app:fig:allPlotSummary_2:i}}\subfloat{\label{app:fig:allPlotSummary_2:j}}\subfloat{\label{app:fig:allPlotSummary_2:k}}\caption{Numerical results for AA t-\ch{SnSe2}{} at $\theta = 9.43 \degree$, $\epsilon = 12$, and $\xi = \SI{5}{\nano\meter}$, corresponding to $\bar{U}=5.201$ and $\bar{V} = 1.483$. (a) shows the inverse compressibility $\pdv{\mu}{\nu}$. The charge and spin stiffnesses, $\rho_{\text{C}}$ and $\rho_{\text{S}}$, are displayed in (b) and (c), respectively. The various components of the charge-charge susceptibility defined in \crefrange{app:eqn:def_susc_hsm_gamma_a}{app:eqn:def_susc_hsm_m_+} are shown in (d)-(i). In (j), we plot the spin-spin correlation function along a chain, $\mathcal{S}(i)$, at $\beta=12$ as a function of $\nu$, while (k) shows the electrochemical potential $\mu'$. We consider $\beta \in \left\{1, 2, 4, 6, 9, 12\right\}$, except for the charge- and spin-stiffness plots, where we restrict to $\beta \geq 2$.}
\label{app:fig:allPlotSummary_2}
\end{figure}
\begin{figure}[H]
\centering
\includegraphics[width=\textwidth]{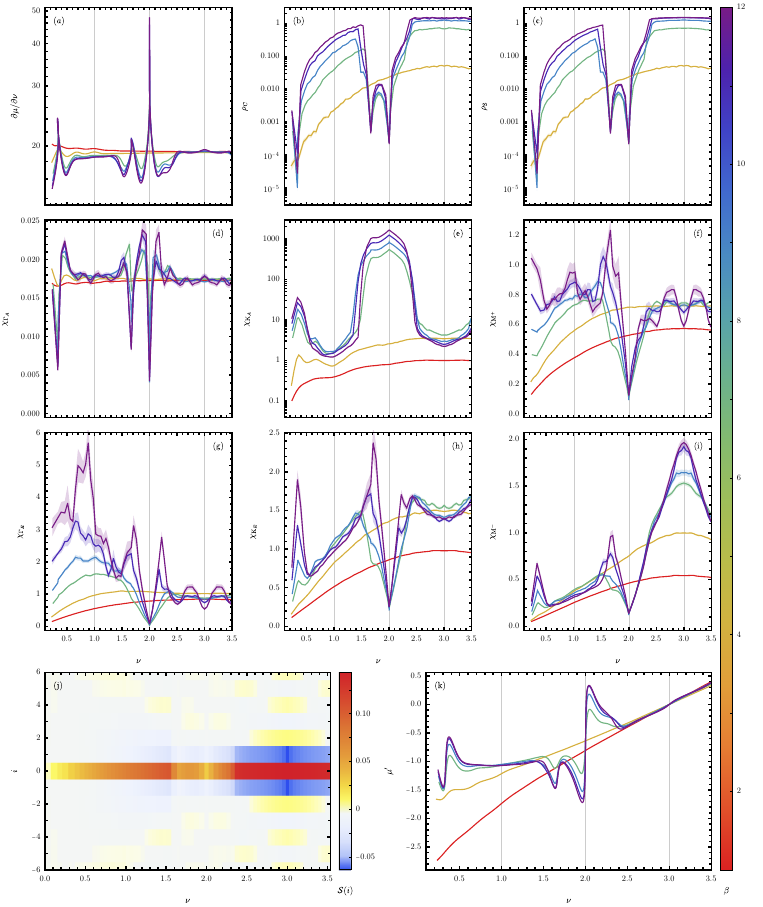}
\subfloat{\label{app:fig:allPlotSummary_3:a}}\subfloat{\label{app:fig:allPlotSummary_3:b}}\subfloat{\label{app:fig:allPlotSummary_3:c}}\subfloat{\label{app:fig:allPlotSummary_3:d}}\subfloat{\label{app:fig:allPlotSummary_3:e}}\subfloat{\label{app:fig:allPlotSummary_3:f}}\subfloat{\label{app:fig:allPlotSummary_3:g}}\subfloat{\label{app:fig:allPlotSummary_3:h}}\subfloat{\label{app:fig:allPlotSummary_3:i}}\subfloat{\label{app:fig:allPlotSummary_3:j}}\subfloat{\label{app:fig:allPlotSummary_3:k}}\caption{Numerical results for AA t-\ch{SnSe2}{} at $\theta = 9.43 \degree$, $\epsilon = 12$, and $\xi = \SI{10}{\nano\meter}$, corresponding to $\bar{U}=5.931$ and $\bar{V} = 2.088$. (a) shows the inverse compressibility $\pdv{\mu}{\nu}$. The charge and spin stiffnesses, $\rho_{\text{C}}$ and $\rho_{\text{S}}$, are displayed in (b) and (c), respectively. The various components of the charge-charge susceptibility defined in \crefrange{app:eqn:def_susc_hsm_gamma_a}{app:eqn:def_susc_hsm_m_+} are shown in (d)-(i). In (j), we plot the spin-spin correlation function along a chain, $\mathcal{S}(i)$, at $\beta=12$ as a function of $\nu$, while (k) shows the electrochemical potential $\mu'$. We consider $\beta \in \left\{1, 2, 4, 6, 9, 12\right\}$, except for the charge- and spin-stiffness plots, where we restrict to $\beta \geq 2$.}
\label{app:fig:allPlotSummary_3}
\end{figure}
\begin{figure}[H]
\centering
\includegraphics[width=\textwidth]{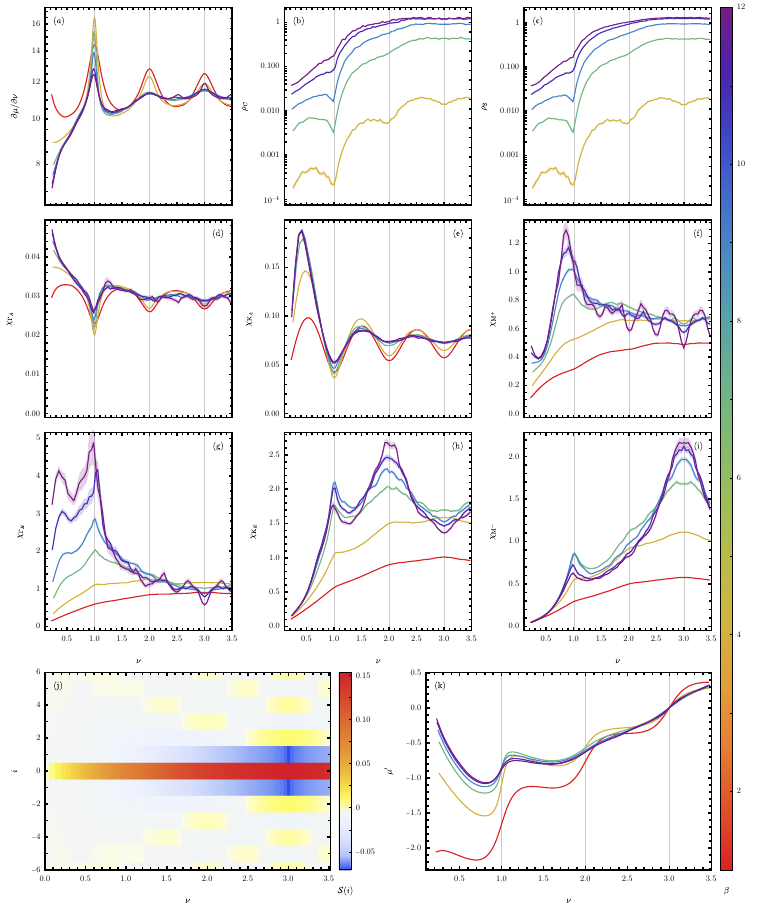}
\subfloat{\label{app:fig:allPlotSummary_4:a}}\subfloat{\label{app:fig:allPlotSummary_4:b}}\subfloat{\label{app:fig:allPlotSummary_4:c}}\subfloat{\label{app:fig:allPlotSummary_4:d}}\subfloat{\label{app:fig:allPlotSummary_4:e}}\subfloat{\label{app:fig:allPlotSummary_4:f}}\subfloat{\label{app:fig:allPlotSummary_4:g}}\subfloat{\label{app:fig:allPlotSummary_4:h}}\subfloat{\label{app:fig:allPlotSummary_4:i}}\subfloat{\label{app:fig:allPlotSummary_4:j}}\subfloat{\label{app:fig:allPlotSummary_4:k}}\caption{Numerical results for AA t-\ch{SnSe2}{} at $\theta = 7.34 \degree$, $\epsilon = 12$, and $\xi = \SI{2.5}{\nano\meter}$, corresponding to $\bar{U}=6.232$ and $\bar{V} = 0.727$. (a) shows the inverse compressibility $\pdv{\mu}{\nu}$. The charge and spin stiffnesses, $\rho_{\text{C}}$ and $\rho_{\text{S}}$, are displayed in (b) and (c), respectively. The various components of the charge-charge susceptibility defined in \crefrange{app:eqn:def_susc_hsm_gamma_a}{app:eqn:def_susc_hsm_m_+} are shown in (d)-(i). In (j), we plot the spin-spin correlation function along a chain, $\mathcal{S}(i)$, at $\beta=12$ as a function of $\nu$, while (k) shows the electrochemical potential $\mu'$. We consider $\beta \in \left\{1, 2, 4, 6, 9, 12\right\}$, except for the charge- and spin-stiffness plots, where we restrict to $\beta \geq 2$.}
\label{app:fig:allPlotSummary_4}
\end{figure}
\begin{figure}[H]
\centering
\includegraphics[width=\textwidth]{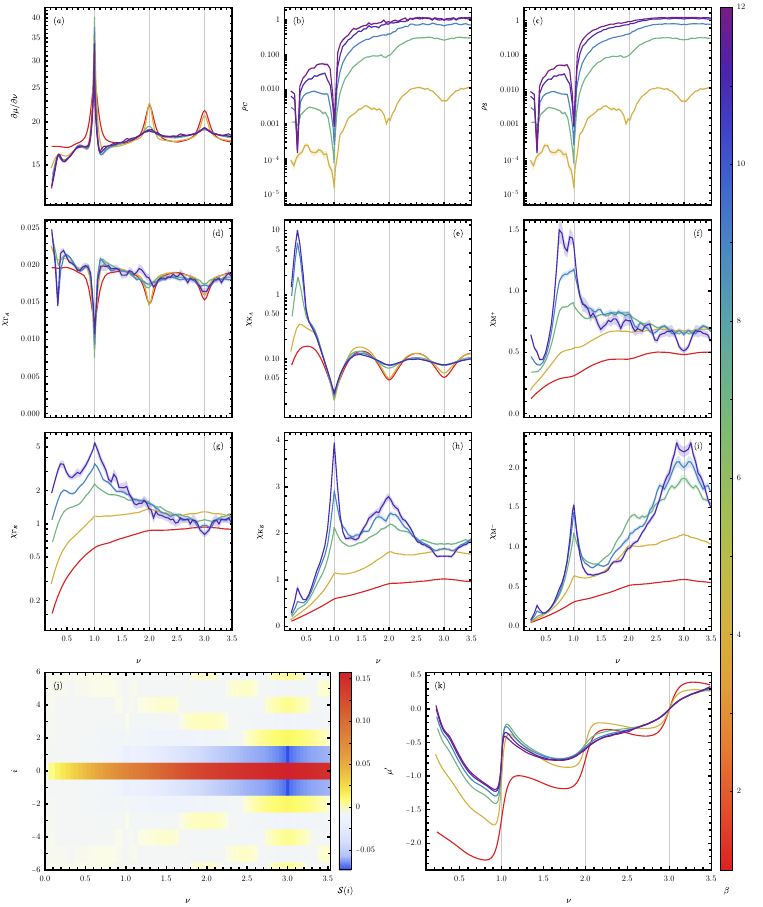}
\subfloat{\label{app:fig:allPlotSummary_5:a}}\subfloat{\label{app:fig:allPlotSummary_5:b}}\subfloat{\label{app:fig:allPlotSummary_5:c}}\subfloat{\label{app:fig:allPlotSummary_5:d}}\subfloat{\label{app:fig:allPlotSummary_5:e}}\subfloat{\label{app:fig:allPlotSummary_5:f}}\subfloat{\label{app:fig:allPlotSummary_5:g}}\subfloat{\label{app:fig:allPlotSummary_5:h}}\subfloat{\label{app:fig:allPlotSummary_5:i}}\subfloat{\label{app:fig:allPlotSummary_5:j}}\subfloat{\label{app:fig:allPlotSummary_5:k}}\caption{Numerical results for AA t-\ch{SnSe2}{} at $\theta = 7.34 \degree$, $\epsilon = 12$, and $\xi = \SI{5}{\nano\meter}$, corresponding to $\bar{U}=8.173$ and $\bar{V} = 1.612$. (a) shows the inverse compressibility $\pdv{\mu}{\nu}$. The charge and spin stiffnesses, $\rho_{\text{C}}$ and $\rho_{\text{S}}$, are displayed in (b) and (c), respectively. The various components of the charge-charge susceptibility defined in \crefrange{app:eqn:def_susc_hsm_gamma_a}{app:eqn:def_susc_hsm_m_+} are shown in (d)-(i). In (j), we plot the spin-spin correlation function along a chain, $\mathcal{S}(i)$, at $\beta=12$ as a function of $\nu$, while (k) shows the electrochemical potential $\mu'$. We consider $\beta \in \left\{1, 2, 4, 6, 9, 12\right\}$, except for the charge- and spin-stiffness plots, where we restrict to $\beta \geq 2$ and for the susceptibility plots, where we restrict to $\beta \leq 9$.}
\label{app:fig:allPlotSummary_5}
\end{figure}
\begin{figure}[H]
\centering
\includegraphics[width=\textwidth]{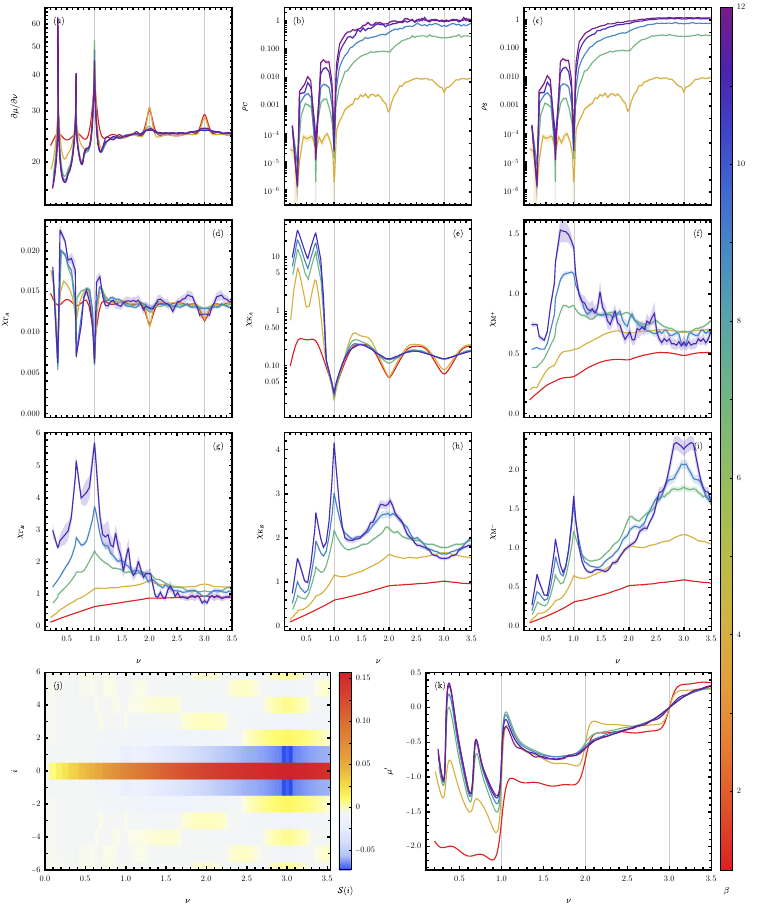}
\subfloat{\label{app:fig:allPlotSummary_6:a}}\subfloat{\label{app:fig:allPlotSummary_6:b}}\subfloat{\label{app:fig:allPlotSummary_6:c}}\subfloat{\label{app:fig:allPlotSummary_6:d}}\subfloat{\label{app:fig:allPlotSummary_6:e}}\subfloat{\label{app:fig:allPlotSummary_6:f}}\subfloat{\label{app:fig:allPlotSummary_6:g}}\subfloat{\label{app:fig:allPlotSummary_6:h}}\subfloat{\label{app:fig:allPlotSummary_6:i}}\subfloat{\label{app:fig:allPlotSummary_6:j}}\subfloat{\label{app:fig:allPlotSummary_6:k}}\caption{Numerical results for AA t-\ch{SnSe2}{} at $\theta = 7.34 \degree$, $\epsilon = 12$, and $\xi = \SI{10}{\nano\meter}$, corresponding to $\bar{U}=9.438$ and $\bar{V} = 2.551$. (a) shows the inverse compressibility $\pdv{\mu}{\nu}$. The charge and spin stiffnesses, $\rho_{\text{C}}$ and $\rho_{\text{S}}$, are displayed in (b) and (c), respectively. The various components of the charge-charge susceptibility defined in \crefrange{app:eqn:def_susc_hsm_gamma_a}{app:eqn:def_susc_hsm_m_+} are shown in (d)-(i). In (j), we plot the spin-spin correlation function along a chain, $\mathcal{S}(i)$, at $\beta=12$ as a function of $\nu$, while (k) shows the electrochemical potential $\mu'$. We consider $\beta \in \left\{1, 2, 4, 6, 9, 12\right\}$, except for the charge- and spin-stiffness plots, where we restrict to $\beta \geq 2$ and for the susceptibility plots, where we restrict to $\beta \leq 9$.}
\label{app:fig:allPlotSummary_6}
\end{figure}
\begin{figure}[H]
\centering
\includegraphics[width=\textwidth]{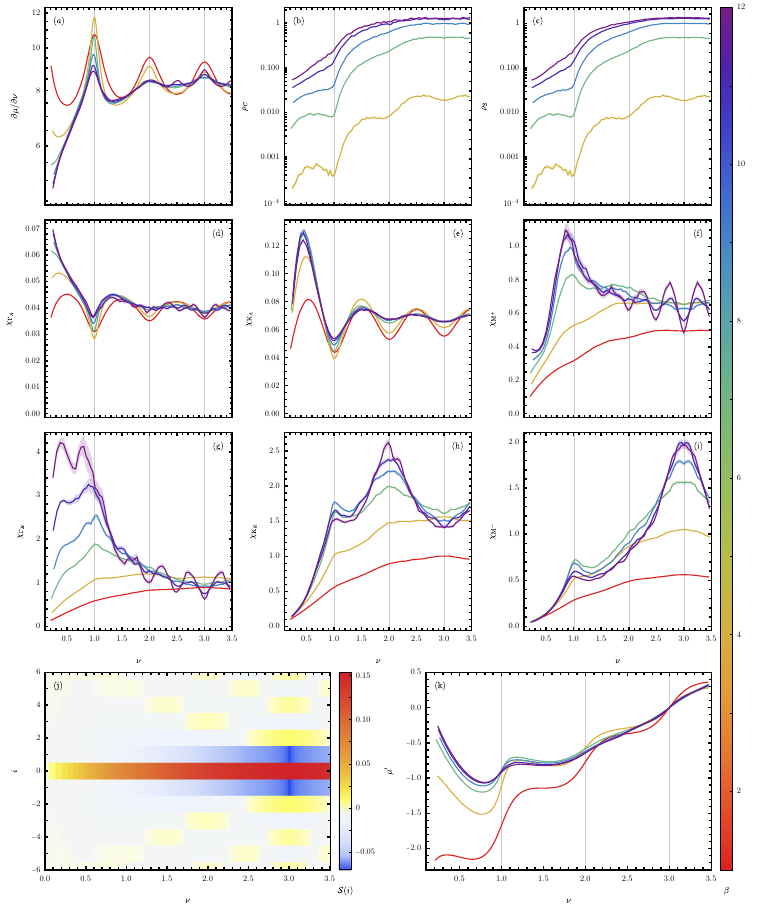}
\subfloat{\label{app:fig:allPlotSummary_7:a}}\subfloat{\label{app:fig:allPlotSummary_7:b}}\subfloat{\label{app:fig:allPlotSummary_7:c}}\subfloat{\label{app:fig:allPlotSummary_7:d}}\subfloat{\label{app:fig:allPlotSummary_7:e}}\subfloat{\label{app:fig:allPlotSummary_7:f}}\subfloat{\label{app:fig:allPlotSummary_7:g}}\subfloat{\label{app:fig:allPlotSummary_7:h}}\subfloat{\label{app:fig:allPlotSummary_7:i}}\subfloat{\label{app:fig:allPlotSummary_7:j}}\subfloat{\label{app:fig:allPlotSummary_7:k}}\caption{Numerical results for AA t-\ch{SnSe2}{} at $\theta = 6.01 \degree$, $\epsilon = 24$, and $\xi = \SI{2.5}{\nano\meter}$, corresponding to $\bar{U}=5.586$ and $\bar{V} = 0.354$. (a) shows the inverse compressibility $\pdv{\mu}{\nu}$. The charge and spin stiffnesses, $\rho_{\text{C}}$ and $\rho_{\text{S}}$, are displayed in (b) and (c), respectively. The various components of the charge-charge susceptibility defined in \crefrange{app:eqn:def_susc_hsm_gamma_a}{app:eqn:def_susc_hsm_m_+} are shown in (d)-(i). In (j), we plot the spin-spin correlation function along a chain, $\mathcal{S}(i)$, at $\beta=12$ as a function of $\nu$, while (k) shows the electrochemical potential $\mu'$. We consider $\beta \in \left\{1, 2, 4, 6, 9, 12\right\}$, except for the charge- and spin-stiffness plots, where we restrict to $\beta \geq 2$.}
\label{app:fig:allPlotSummary_7}
\end{figure}
\begin{figure}[H]
\centering
\includegraphics[width=\textwidth]{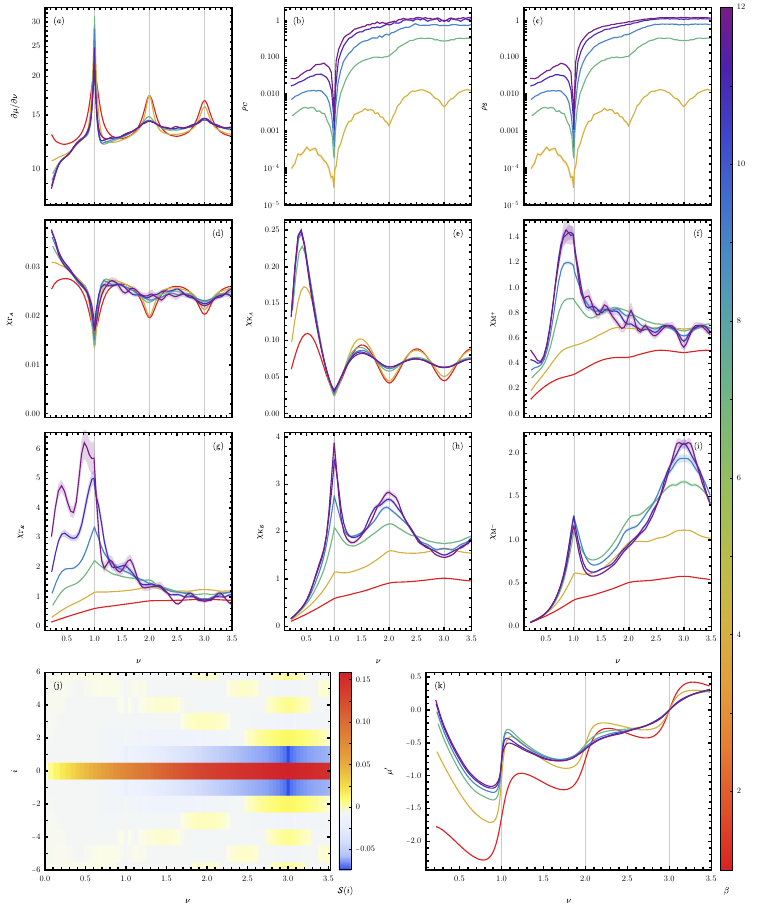}
\subfloat{\label{app:fig:allPlotSummary_8:a}}\subfloat{\label{app:fig:allPlotSummary_8:b}}\subfloat{\label{app:fig:allPlotSummary_8:c}}\subfloat{\label{app:fig:allPlotSummary_8:d}}\subfloat{\label{app:fig:allPlotSummary_8:e}}\subfloat{\label{app:fig:allPlotSummary_8:f}}\subfloat{\label{app:fig:allPlotSummary_8:g}}\subfloat{\label{app:fig:allPlotSummary_8:h}}\subfloat{\label{app:fig:allPlotSummary_8:i}}\subfloat{\label{app:fig:allPlotSummary_8:j}}\subfloat{\label{app:fig:allPlotSummary_8:k}}\caption{Numerical results for AA t-\ch{SnSe2}{} at $\theta = 6.01 \degree$, $\epsilon = 24$, and $\xi = \SI{5}{\nano\meter}$, corresponding to $\bar{U}=7.436$ and $\bar{V} = 0.975$. (a) shows the inverse compressibility $\pdv{\mu}{\nu}$. The charge and spin stiffnesses, $\rho_{\text{C}}$ and $\rho_{\text{S}}$, are displayed in (b) and (c), respectively. The various components of the charge-charge susceptibility defined in \crefrange{app:eqn:def_susc_hsm_gamma_a}{app:eqn:def_susc_hsm_m_+} are shown in (d)-(i). In (j), we plot the spin-spin correlation function along a chain, $\mathcal{S}(i)$, at $\beta=12$ as a function of $\nu$, while (k) shows the electrochemical potential $\mu'$. We consider $\beta \in \left\{1, 2, 4, 6, 9, 12\right\}$, except for the charge- and spin-stiffness plots, where we restrict to $\beta \geq 2$.}
\label{app:fig:allPlotSummary_8}
\end{figure}
\begin{figure}[H]
\centering
\includegraphics[width=\textwidth]{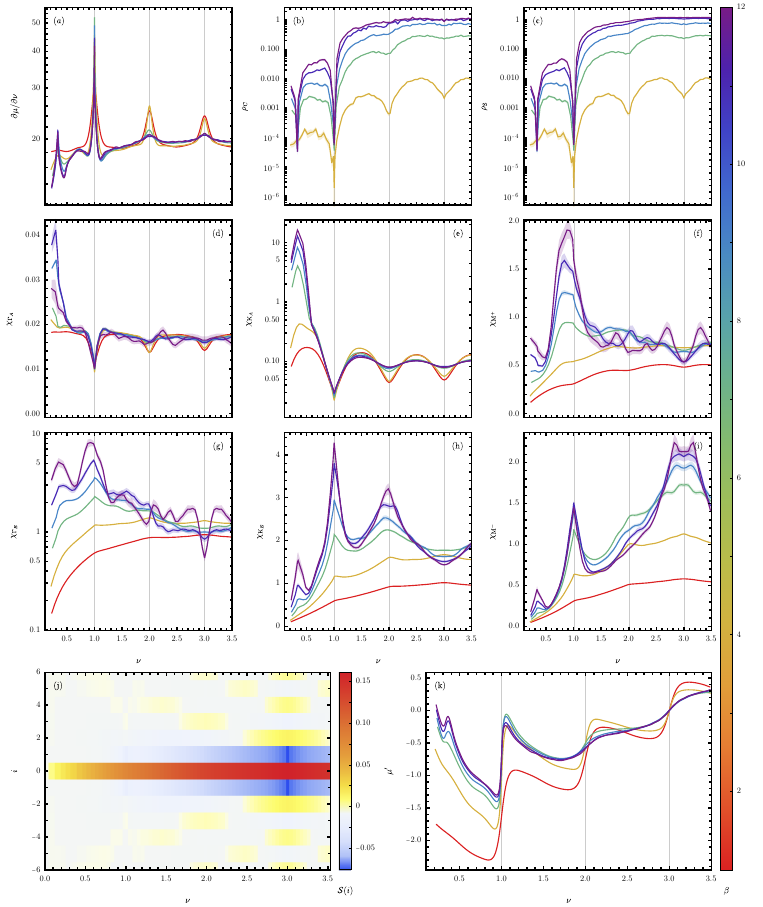}
\subfloat{\label{app:fig:allPlotSummary_9:a}}\subfloat{\label{app:fig:allPlotSummary_9:b}}\subfloat{\label{app:fig:allPlotSummary_9:c}}\subfloat{\label{app:fig:allPlotSummary_9:d}}\subfloat{\label{app:fig:allPlotSummary_9:e}}\subfloat{\label{app:fig:allPlotSummary_9:f}}\subfloat{\label{app:fig:allPlotSummary_9:g}}\subfloat{\label{app:fig:allPlotSummary_9:h}}\subfloat{\label{app:fig:allPlotSummary_9:i}}\subfloat{\label{app:fig:allPlotSummary_9:j}}\subfloat{\label{app:fig:allPlotSummary_9:k}}\caption{Numerical results for AA t-\ch{SnSe2}{} at $\theta = 6.01 \degree$, $\epsilon = 24$, and $\xi = \SI{10}{\nano\meter}$, corresponding to $\bar{U}=8.66$ and $\bar{V} = 1.773$. (a) shows the inverse compressibility $\pdv{\mu}{\nu}$. The charge and spin stiffnesses, $\rho_{\text{C}}$ and $\rho_{\text{S}}$, are displayed in (b) and (c), respectively. The various components of the charge-charge susceptibility defined in \crefrange{app:eqn:def_susc_hsm_gamma_a}{app:eqn:def_susc_hsm_m_+} are shown in (d)-(i). In (j), we plot the spin-spin correlation function along a chain, $\mathcal{S}(i)$, at $\beta=12$ as a function of $\nu$, while (k) shows the electrochemical potential $\mu'$. We consider $\beta \in \left\{1, 2, 4, 6, 9, 12\right\}$, except for the charge- and spin-stiffness plots, where we restrict to $\beta \geq 2$.}
\label{app:fig:allPlotSummary_9}
\end{figure}
\begin{figure}[H]
\centering
\includegraphics[width=\textwidth]{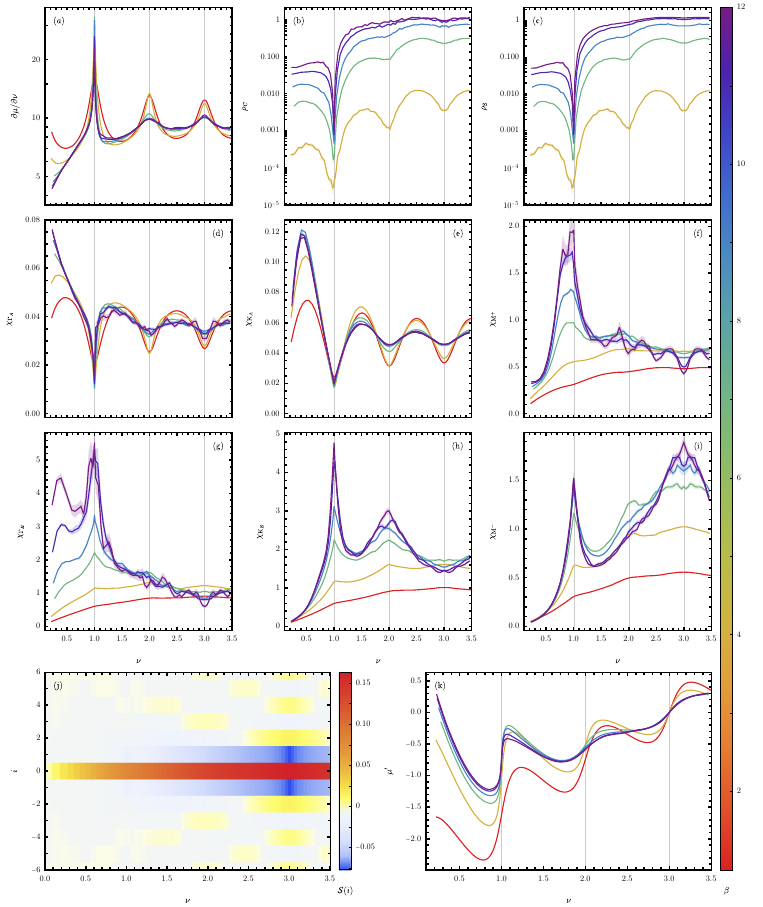}
\subfloat{\label{app:fig:allPlotSummary_10:a}}\subfloat{\label{app:fig:allPlotSummary_10:b}}\subfloat{\label{app:fig:allPlotSummary_10:c}}\subfloat{\label{app:fig:allPlotSummary_10:d}}\subfloat{\label{app:fig:allPlotSummary_10:e}}\subfloat{\label{app:fig:allPlotSummary_10:f}}\subfloat{\label{app:fig:allPlotSummary_10:g}}\subfloat{\label{app:fig:allPlotSummary_10:h}}\subfloat{\label{app:fig:allPlotSummary_10:i}}\subfloat{\label{app:fig:allPlotSummary_10:j}}\subfloat{\label{app:fig:allPlotSummary_10:k}}\caption{Numerical results for AA t-\ch{SnSe2}{} at $\theta = 5.09 \degree$, $\epsilon = 24$, and $\xi = \SI{2.5}{\nano\meter}$, corresponding to $\bar{U}=7.131$ and $\bar{V} = 0.257$. (a) shows the inverse compressibility $\pdv{\mu}{\nu}$. The charge and spin stiffnesses, $\rho_{\text{C}}$ and $\rho_{\text{S}}$, are displayed in (b) and (c), respectively. The various components of the charge-charge susceptibility defined in \crefrange{app:eqn:def_susc_hsm_gamma_a}{app:eqn:def_susc_hsm_m_+} are shown in (d)-(i). In (j), we plot the spin-spin correlation function along a chain, $\mathcal{S}(i)$, at $\beta=12$ as a function of $\nu$, while (k) shows the electrochemical potential $\mu'$. We consider $\beta \in \left\{1, 2, 4, 6, 9, 12\right\}$, except for the charge- and spin-stiffness plots, where we restrict to $\beta \geq 2$.}
\label{app:fig:allPlotSummary_10}
\end{figure}
\begin{figure}[H]
\centering
\includegraphics[width=\textwidth]{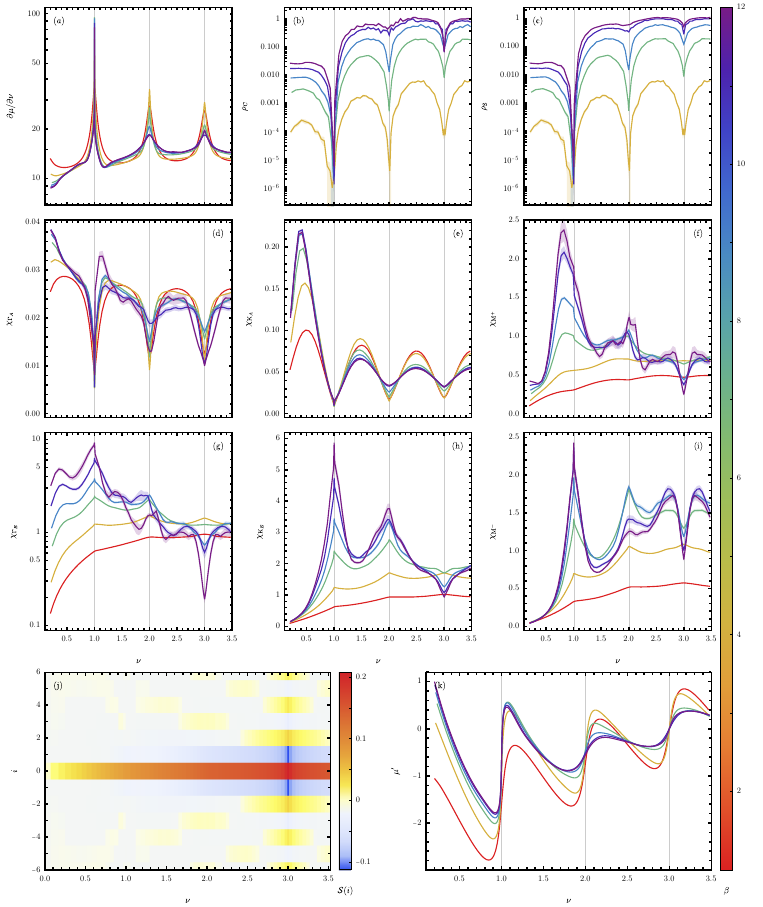}
\subfloat{\label{app:fig:allPlotSummary_11:a}}\subfloat{\label{app:fig:allPlotSummary_11:b}}\subfloat{\label{app:fig:allPlotSummary_11:c}}\subfloat{\label{app:fig:allPlotSummary_11:d}}\subfloat{\label{app:fig:allPlotSummary_11:e}}\subfloat{\label{app:fig:allPlotSummary_11:f}}\subfloat{\label{app:fig:allPlotSummary_11:g}}\subfloat{\label{app:fig:allPlotSummary_11:h}}\subfloat{\label{app:fig:allPlotSummary_11:i}}\subfloat{\label{app:fig:allPlotSummary_11:j}}\subfloat{\label{app:fig:allPlotSummary_11:k}}\caption{Numerical results for AA t-\ch{SnSe2}{} at $\theta = 5.09 \degree$, $\epsilon = 24$, and $\xi = \SI{5}{\nano\meter}$, corresponding to $\bar{U}=9.66$ and $\bar{V} = 0.879$. (a) shows the inverse compressibility $\pdv{\mu}{\nu}$. The charge and spin stiffnesses, $\rho_{\text{C}}$ and $\rho_{\text{S}}$, are displayed in (b) and (c), respectively. The various components of the charge-charge susceptibility defined in \crefrange{app:eqn:def_susc_hsm_gamma_a}{app:eqn:def_susc_hsm_m_+} are shown in (d)-(i). In (j), we plot the spin-spin correlation function along a chain, $\mathcal{S}(i)$, at $\beta=12$ as a function of $\nu$, while (k) shows the electrochemical potential $\mu'$. We consider $\beta \in \left\{1, 2, 4, 6, 9, 12\right\}$, except for the charge- and spin-stiffness plots, where we restrict to $\beta \geq 2$.}
\label{app:fig:allPlotSummary_11}
\end{figure}
\begin{figure}[H]
\centering
\includegraphics[width=\textwidth]{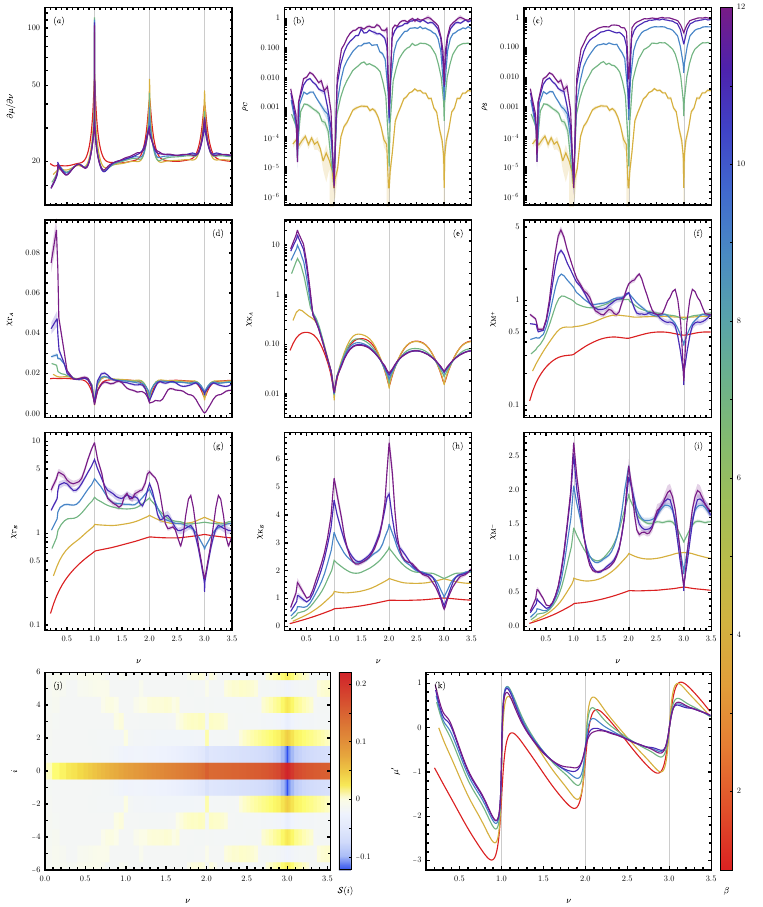}
\subfloat{\label{app:fig:allPlotSummary_12:a}}\subfloat{\label{app:fig:allPlotSummary_12:b}}\subfloat{\label{app:fig:allPlotSummary_12:c}}\subfloat{\label{app:fig:allPlotSummary_12:d}}\subfloat{\label{app:fig:allPlotSummary_12:e}}\subfloat{\label{app:fig:allPlotSummary_12:f}}\subfloat{\label{app:fig:allPlotSummary_12:g}}\subfloat{\label{app:fig:allPlotSummary_12:h}}\subfloat{\label{app:fig:allPlotSummary_12:i}}\subfloat{\label{app:fig:allPlotSummary_12:j}}\subfloat{\label{app:fig:allPlotSummary_12:k}}\caption{Numerical results for AA t-\ch{SnSe2}{} at $\theta = 5.09 \degree$, $\epsilon = 24$, and $\xi = \SI{10}{\nano\meter}$, corresponding to $\bar{U}=11.372$ and $\bar{V} = 1.847$. (a) shows the inverse compressibility $\pdv{\mu}{\nu}$. The charge and spin stiffnesses, $\rho_{\text{C}}$ and $\rho_{\text{S}}$, are displayed in (b) and (c), respectively. The various components of the charge-charge susceptibility defined in \crefrange{app:eqn:def_susc_hsm_gamma_a}{app:eqn:def_susc_hsm_m_+} are shown in (d)-(i). In (j), we plot the spin-spin correlation function along a chain, $\mathcal{S}(i)$, at $\beta=12$ as a function of $\nu$, while (k) shows the electrochemical potential $\mu'$. We consider $\beta \in \left\{1, 2, 4, 6, 9, 12\right\}$, except for the charge- and spin-stiffness plots, where we restrict to $\beta \geq 2$.}
\label{app:fig:allPlotSummary_12}
\end{figure}
\begin{figure}[H]
\centering
\includegraphics[width=\textwidth]{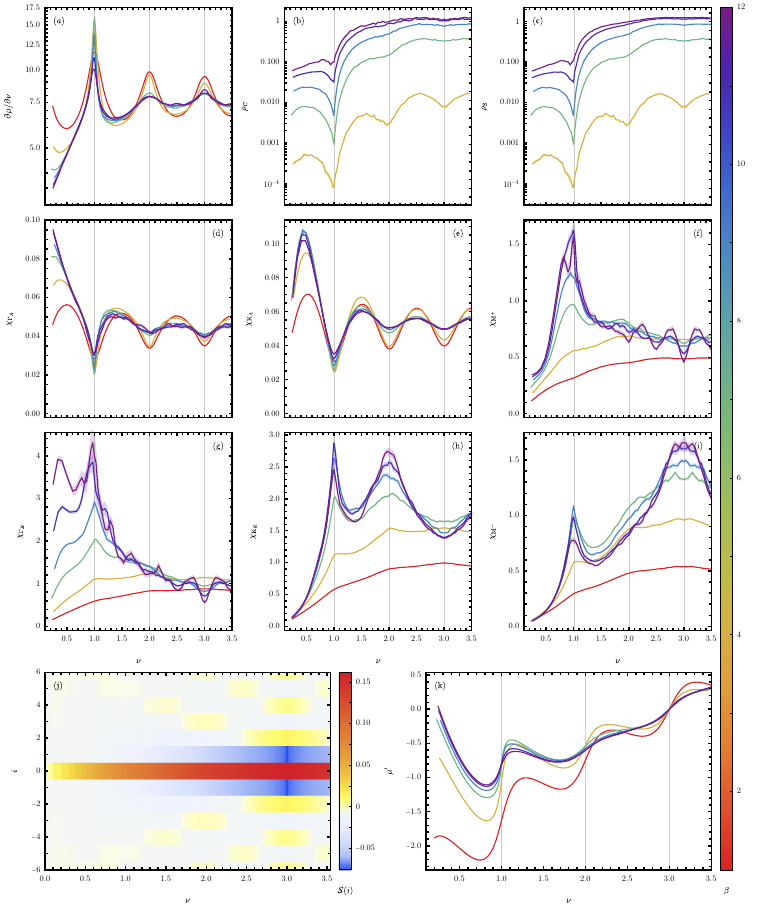}
\subfloat{\label{app:fig:allPlotSummary_13:a}}\subfloat{\label{app:fig:allPlotSummary_13:b}}\subfloat{\label{app:fig:allPlotSummary_13:c}}\subfloat{\label{app:fig:allPlotSummary_13:d}}\subfloat{\label{app:fig:allPlotSummary_13:e}}\subfloat{\label{app:fig:allPlotSummary_13:f}}\subfloat{\label{app:fig:allPlotSummary_13:g}}\subfloat{\label{app:fig:allPlotSummary_13:h}}\subfloat{\label{app:fig:allPlotSummary_13:i}}\subfloat{\label{app:fig:allPlotSummary_13:j}}\subfloat{\label{app:fig:allPlotSummary_13:k}}\caption{Numerical results for AA t-\ch{SnSe2}{} at $\theta = 4.41 \degree$, $\epsilon = 48$, and $\xi = \SI{2.5}{\nano\meter}$, corresponding to $\bar{U}=6.311$ and $\bar{V} = 0.114$. (a) shows the inverse compressibility $\pdv{\mu}{\nu}$. The charge and spin stiffnesses, $\rho_{\text{C}}$ and $\rho_{\text{S}}$, are displayed in (b) and (c), respectively. The various components of the charge-charge susceptibility defined in \crefrange{app:eqn:def_susc_hsm_gamma_a}{app:eqn:def_susc_hsm_m_+} are shown in (d)-(i). In (j), we plot the spin-spin correlation function along a chain, $\mathcal{S}(i)$, at $\beta=12$ as a function of $\nu$, while (k) shows the electrochemical potential $\mu'$. We consider $\beta \in \left\{1, 2, 4, 6, 9, 12\right\}$, except for the charge- and spin-stiffness plots, where we restrict to $\beta \geq 2$.}
\label{app:fig:allPlotSummary_13}
\end{figure}
\begin{figure}[H]
\centering
\includegraphics[width=\textwidth]{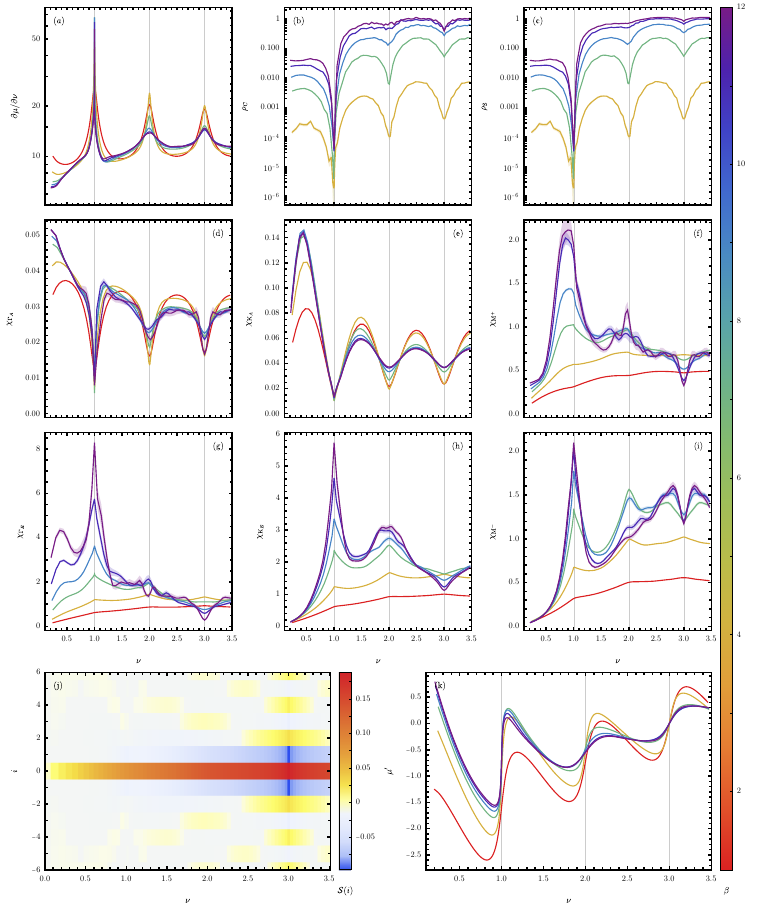}
\subfloat{\label{app:fig:allPlotSummary_14:a}}\subfloat{\label{app:fig:allPlotSummary_14:b}}\subfloat{\label{app:fig:allPlotSummary_14:c}}\subfloat{\label{app:fig:allPlotSummary_14:d}}\subfloat{\label{app:fig:allPlotSummary_14:e}}\subfloat{\label{app:fig:allPlotSummary_14:f}}\subfloat{\label{app:fig:allPlotSummary_14:g}}\subfloat{\label{app:fig:allPlotSummary_14:h}}\subfloat{\label{app:fig:allPlotSummary_14:i}}\subfloat{\label{app:fig:allPlotSummary_14:j}}\subfloat{\label{app:fig:allPlotSummary_14:k}}\caption{Numerical results for AA t-\ch{SnSe2}{} at $\theta = 4.41 \degree$, $\epsilon = 48$, and $\xi = \SI{5}{\nano\meter}$, corresponding to $\bar{U}=8.642$ and $\bar{V} = 0.517$. (a) shows the inverse compressibility $\pdv{\mu}{\nu}$. The charge and spin stiffnesses, $\rho_{\text{C}}$ and $\rho_{\text{S}}$, are displayed in (b) and (c), respectively. The various components of the charge-charge susceptibility defined in \crefrange{app:eqn:def_susc_hsm_gamma_a}{app:eqn:def_susc_hsm_m_+} are shown in (d)-(i). In (j), we plot the spin-spin correlation function along a chain, $\mathcal{S}(i)$, at $\beta=12$ as a function of $\nu$, while (k) shows the electrochemical potential $\mu'$. We consider $\beta \in \left\{1, 2, 4, 6, 9, 12\right\}$, except for the charge- and spin-stiffness plots, where we restrict to $\beta \geq 2$.}
\label{app:fig:allPlotSummary_14}
\end{figure}
\begin{figure}[H]
\centering
\includegraphics[width=\textwidth]{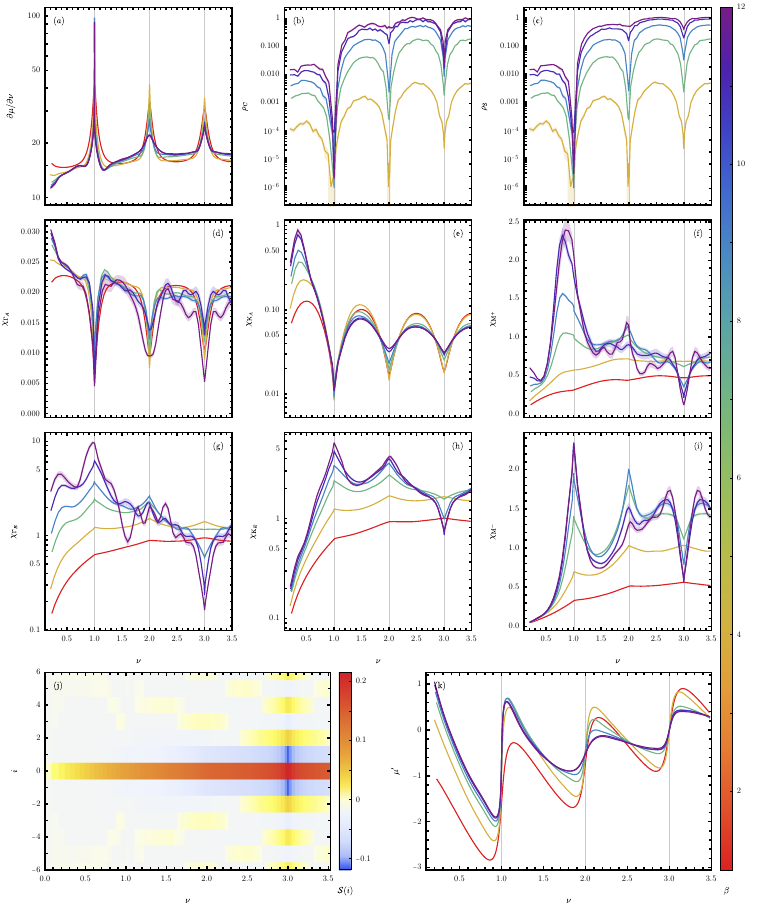}
\subfloat{\label{app:fig:allPlotSummary_15:a}}\subfloat{\label{app:fig:allPlotSummary_15:b}}\subfloat{\label{app:fig:allPlotSummary_15:c}}\subfloat{\label{app:fig:allPlotSummary_15:d}}\subfloat{\label{app:fig:allPlotSummary_15:e}}\subfloat{\label{app:fig:allPlotSummary_15:f}}\subfloat{\label{app:fig:allPlotSummary_15:g}}\subfloat{\label{app:fig:allPlotSummary_15:h}}\subfloat{\label{app:fig:allPlotSummary_15:i}}\subfloat{\label{app:fig:allPlotSummary_15:j}}\subfloat{\label{app:fig:allPlotSummary_15:k}}\caption{Numerical results for AA t-\ch{SnSe2}{} at $\theta = 4.41 \degree$, $\epsilon = 48$, and $\xi = \SI{10}{\nano\meter}$, corresponding to $\bar{U}=10.239$ and $\bar{V} = 1.286$. (a) shows the inverse compressibility $\pdv{\mu}{\nu}$. The charge and spin stiffnesses, $\rho_{\text{C}}$ and $\rho_{\text{S}}$, are displayed in (b) and (c), respectively. The various components of the charge-charge susceptibility defined in \crefrange{app:eqn:def_susc_hsm_gamma_a}{app:eqn:def_susc_hsm_m_+} are shown in (d)-(i). In (j), we plot the spin-spin correlation function along a chain, $\mathcal{S}(i)$, at $\beta=12$ as a function of $\nu$, while (k) shows the electrochemical potential $\mu'$. We consider $\beta \in \left\{1, 2, 4, 6, 9, 12\right\}$, except for the charge- and spin-stiffness plots, where we restrict to $\beta \geq 2$.}
\label{app:fig:allPlotSummary_15}
\end{figure}
\begin{figure}[H]
\centering
\includegraphics[width=\textwidth]{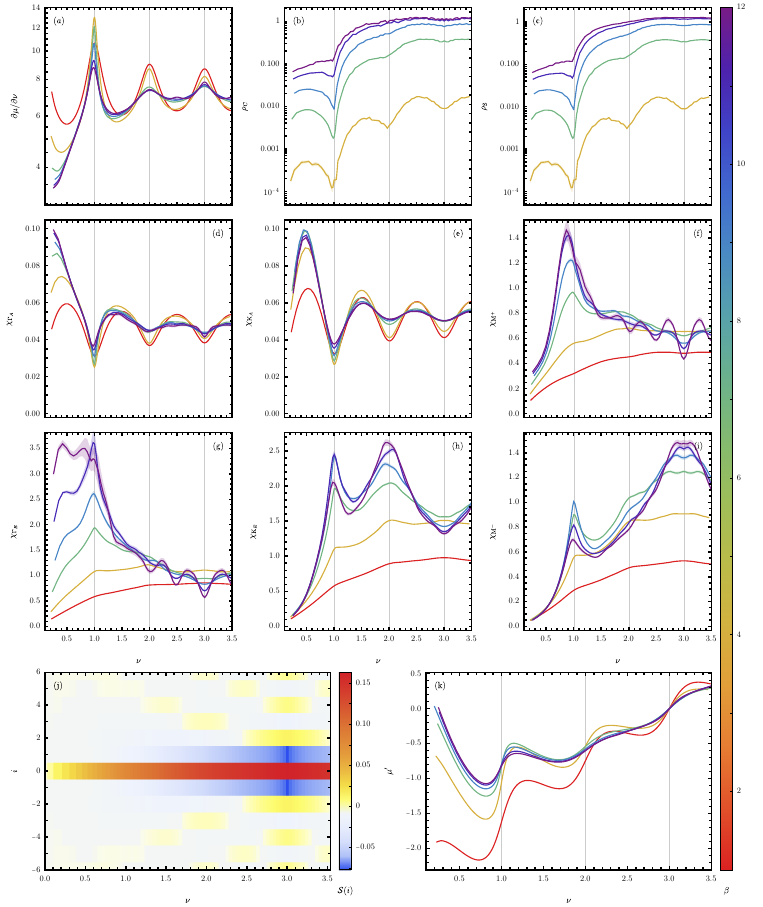}
\subfloat{\label{app:fig:allPlotSummary_16:a}}\subfloat{\label{app:fig:allPlotSummary_16:b}}\subfloat{\label{app:fig:allPlotSummary_16:c}}\subfloat{\label{app:fig:allPlotSummary_16:d}}\subfloat{\label{app:fig:allPlotSummary_16:e}}\subfloat{\label{app:fig:allPlotSummary_16:f}}\subfloat{\label{app:fig:allPlotSummary_16:g}}\subfloat{\label{app:fig:allPlotSummary_16:h}}\subfloat{\label{app:fig:allPlotSummary_16:i}}\subfloat{\label{app:fig:allPlotSummary_16:j}}\subfloat{\label{app:fig:allPlotSummary_16:k}}\caption{Numerical results for AA t-\ch{SnSe2}{} at $\theta = 3.89 \degree$, $\epsilon = 72$, and $\xi = \SI{2.5}{\nano\meter}$, corresponding to $\bar{U}=6.139$ and $\bar{V} = 0.061$. (a) shows the inverse compressibility $\pdv{\mu}{\nu}$. The charge and spin stiffnesses, $\rho_{\text{C}}$ and $\rho_{\text{S}}$, are displayed in (b) and (c), respectively. The various components of the charge-charge susceptibility defined in \crefrange{app:eqn:def_susc_hsm_gamma_a}{app:eqn:def_susc_hsm_m_+} are shown in (d)-(i). In (j), we plot the spin-spin correlation function along a chain, $\mathcal{S}(i)$, at $\beta=12$ as a function of $\nu$, while (k) shows the electrochemical potential $\mu'$. We consider $\beta \in \left\{1, 2, 4, 6, 9, 12\right\}$, except for the charge- and spin-stiffness plots, where we restrict to $\beta \geq 2$.}
\label{app:fig:allPlotSummary_16}
\end{figure}
\begin{figure}[H]
\centering
\includegraphics[width=\textwidth]{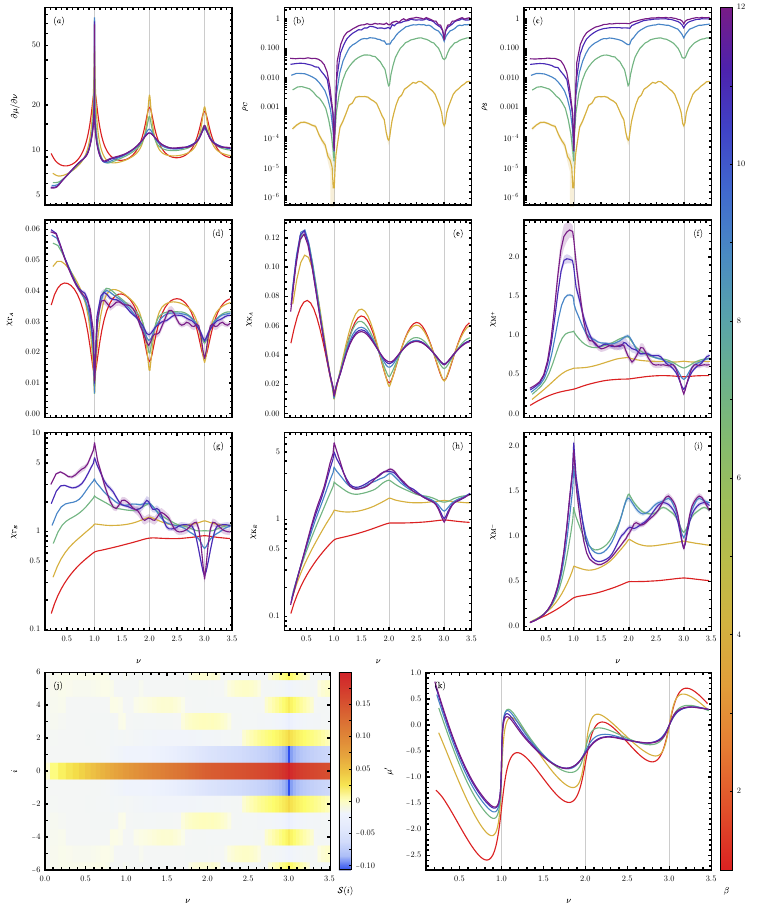}
\subfloat{\label{app:fig:allPlotSummary_17:a}}\subfloat{\label{app:fig:allPlotSummary_17:b}}\subfloat{\label{app:fig:allPlotSummary_17:c}}\subfloat{\label{app:fig:allPlotSummary_17:d}}\subfloat{\label{app:fig:allPlotSummary_17:e}}\subfloat{\label{app:fig:allPlotSummary_17:f}}\subfloat{\label{app:fig:allPlotSummary_17:g}}\subfloat{\label{app:fig:allPlotSummary_17:h}}\subfloat{\label{app:fig:allPlotSummary_17:i}}\subfloat{\label{app:fig:allPlotSummary_17:j}}\subfloat{\label{app:fig:allPlotSummary_17:k}}\caption{Numerical results for AA t-\ch{SnSe2}{} at $\theta = 3.89 \degree$, $\epsilon = 72$, and $\xi = \SI{5}{\nano\meter}$, corresponding to $\bar{U}=8.55$ and $\bar{V} = 0.357$. (a) shows the inverse compressibility $\pdv{\mu}{\nu}$. The charge and spin stiffnesses, $\rho_{\text{C}}$ and $\rho_{\text{S}}$, are displayed in (b) and (c), respectively. The various components of the charge-charge susceptibility defined in \crefrange{app:eqn:def_susc_hsm_gamma_a}{app:eqn:def_susc_hsm_m_+} are shown in (d)-(i). In (j), we plot the spin-spin correlation function along a chain, $\mathcal{S}(i)$, at $\beta=12$ as a function of $\nu$, while (k) shows the electrochemical potential $\mu'$. We consider $\beta \in \left\{1, 2, 4, 6, 9, 12\right\}$, except for the charge- and spin-stiffness plots, where we restrict to $\beta \geq 2$.}
\label{app:fig:allPlotSummary_17}
\end{figure}
\begin{figure}[H]
\centering
\includegraphics[width=\textwidth]{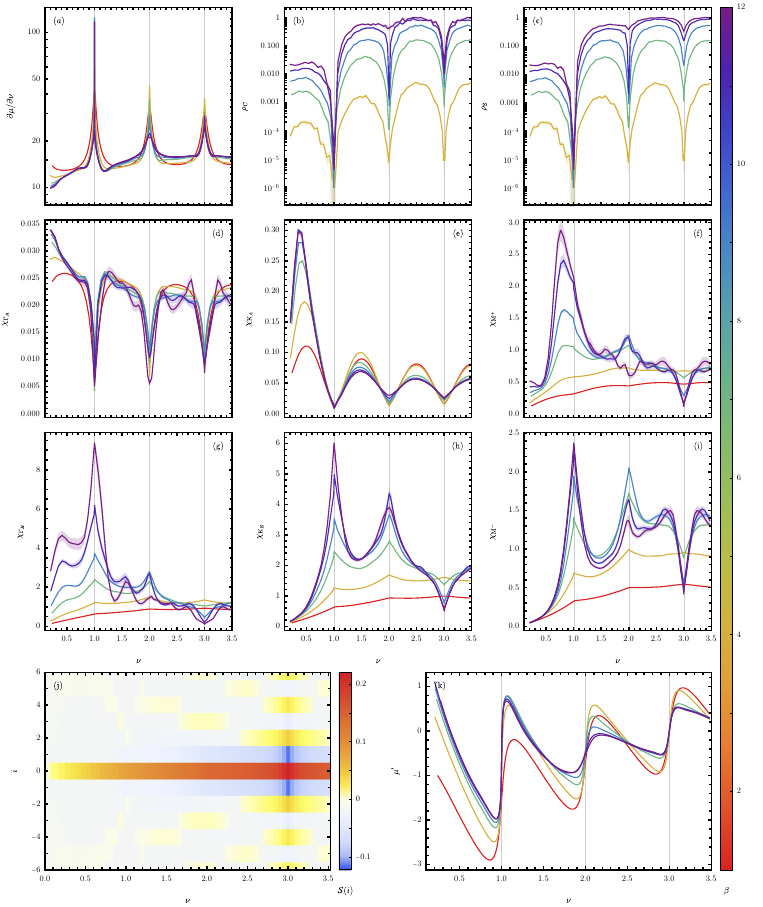}
\subfloat{\label{app:fig:allPlotSummary_18:a}}\subfloat{\label{app:fig:allPlotSummary_18:b}}\subfloat{\label{app:fig:allPlotSummary_18:c}}\subfloat{\label{app:fig:allPlotSummary_18:d}}\subfloat{\label{app:fig:allPlotSummary_18:e}}\subfloat{\label{app:fig:allPlotSummary_18:f}}\subfloat{\label{app:fig:allPlotSummary_18:g}}\subfloat{\label{app:fig:allPlotSummary_18:h}}\subfloat{\label{app:fig:allPlotSummary_18:i}}\subfloat{\label{app:fig:allPlotSummary_18:j}}\subfloat{\label{app:fig:allPlotSummary_18:k}}\caption{Numerical results for AA t-\ch{SnSe2}{} at $\theta = 3.89 \degree$, $\epsilon = 72$, and $\xi = \SI{10}{\nano\meter}$, corresponding to $\bar{U}=10.244$ and $\bar{V} = 1.049$. (a) shows the inverse compressibility $\pdv{\mu}{\nu}$. The charge and spin stiffnesses, $\rho_{\text{C}}$ and $\rho_{\text{S}}$, are displayed in (b) and (c), respectively. The various components of the charge-charge susceptibility defined in \crefrange{app:eqn:def_susc_hsm_gamma_a}{app:eqn:def_susc_hsm_m_+} are shown in (d)-(i). In (j), we plot the spin-spin correlation function along a chain, $\mathcal{S}(i)$, at $\beta=12$ as a function of $\nu$, while (k) shows the electrochemical potential $\mu'$. We consider $\beta \in \left\{1, 2, 4, 6, 9, 12\right\}$, except for the charge- and spin-stiffness plots, where we restrict to $\beta \geq 2$.}
\label{app:fig:allPlotSummary_18}
\end{figure}
\begin{figure}[H]
\centering
\includegraphics[width=\textwidth]{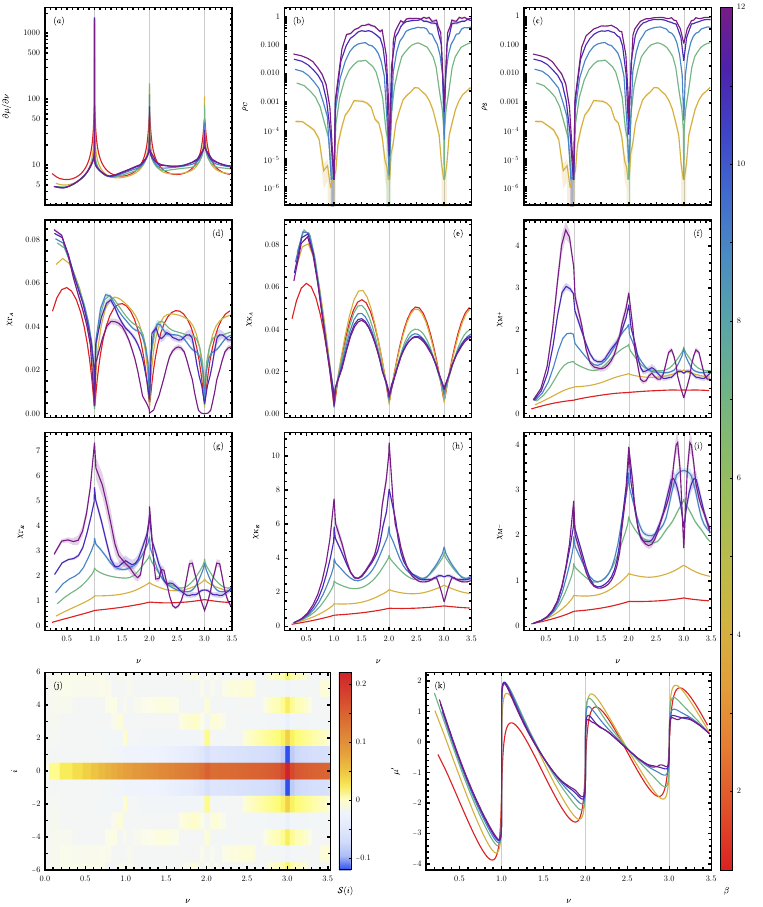}
\subfloat{\label{app:fig:allPlotSummary_19:a}}\subfloat{\label{app:fig:allPlotSummary_19:b}}\subfloat{\label{app:fig:allPlotSummary_19:c}}\subfloat{\label{app:fig:allPlotSummary_19:d}}\subfloat{\label{app:fig:allPlotSummary_19:e}}\subfloat{\label{app:fig:allPlotSummary_19:f}}\subfloat{\label{app:fig:allPlotSummary_19:g}}\subfloat{\label{app:fig:allPlotSummary_19:h}}\subfloat{\label{app:fig:allPlotSummary_19:i}}\subfloat{\label{app:fig:allPlotSummary_19:j}}\subfloat{\label{app:fig:allPlotSummary_19:k}}\caption{Numerical results for the AA t-\ch{SnSe2}{} Hamiltonian with only on-site interaction and $U = U' = 12$. (a) shows the inverse compressibility $\pdv{\mu}{\nu}$. The charge and spin stiffnesses, $\rho_{\text{C}}$ and $\rho_{\text{S}}$, are displayed in (b) and (c), respectively. The various components of the charge-charge susceptibility defined in \crefrange{app:eqn:def_susc_hsm_gamma_a}{app:eqn:def_susc_hsm_m_+} are shown in (d)-(i). In (j), we plot the spin-spin correlation function along a chain, $\mathcal{S}(i)$, at $\beta=12$ as a function of $\nu$, while (k) shows the electrochemical potential $\mu'$. We consider $\beta \in \left\{1, 2, 4, 6, 9, 12\right\}$, except for the charge- and spin-stiffness plots, where we restrict to $\beta \geq 2$.}
\label{app:fig:allPlotSummary_19}
\end{figure}
\begin{figure}[H]
\centering
\includegraphics[width=\textwidth]{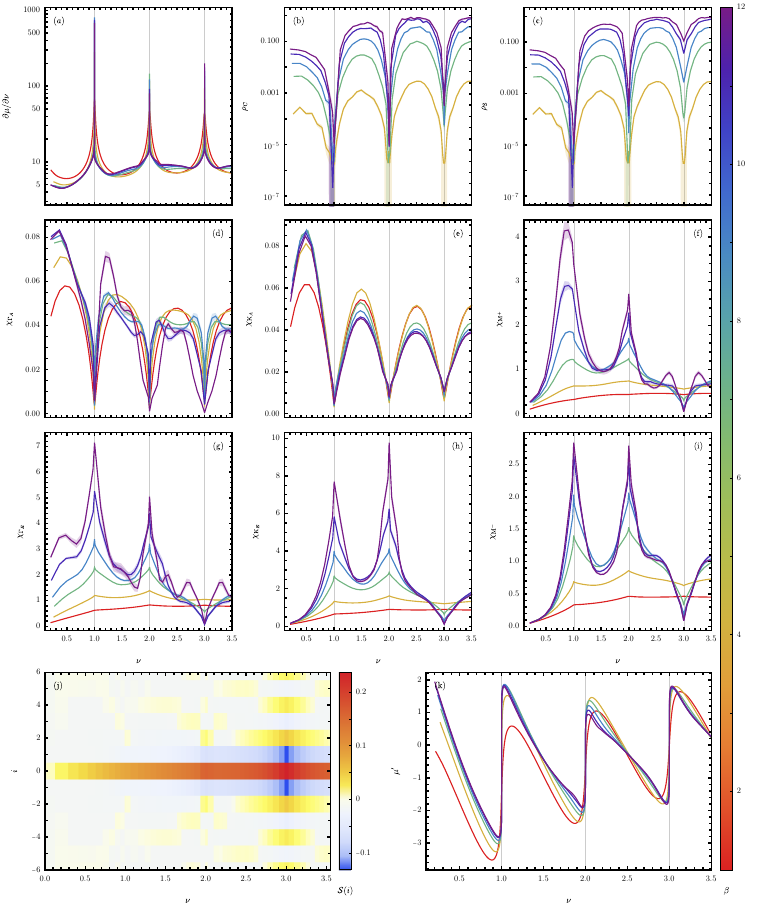}
\subfloat{\label{app:fig:allPlotSummary_20:a}}\subfloat{\label{app:fig:allPlotSummary_20:b}}\subfloat{\label{app:fig:allPlotSummary_20:c}}\subfloat{\label{app:fig:allPlotSummary_20:d}}\subfloat{\label{app:fig:allPlotSummary_20:e}}\subfloat{\label{app:fig:allPlotSummary_20:f}}\subfloat{\label{app:fig:allPlotSummary_20:g}}\subfloat{\label{app:fig:allPlotSummary_20:h}}\subfloat{\label{app:fig:allPlotSummary_20:i}}\subfloat{\label{app:fig:allPlotSummary_20:j}}\subfloat{\label{app:fig:allPlotSummary_20:k}}\caption{Numerical results for the AA t-\ch{SnSe2}{} Hamiltonian with only on-site interaction and $U = 12$ and $U' = 11.4$. (a) shows the inverse compressibility $\pdv{\mu}{\nu}$. The charge and spin stiffnesses, $\rho_{\text{C}}$ and $\rho_{\text{S}}$, are displayed in (b) and (c), respectively. The various components of the charge-charge susceptibility defined in \crefrange{app:eqn:def_susc_hsm_gamma_a}{app:eqn:def_susc_hsm_m_+} are shown in (d)-(i). In (j), we plot the spin-spin correlation function along a chain, $\mathcal{S}(i)$, at $\beta=12$ as a function of $\nu$, while (k) shows the electrochemical potential $\mu'$. We consider $\beta \in \left\{1, 2, 4, 6, 9, 12\right\}$, except for the charge- and spin-stiffness plots, where we restrict to $\beta \geq 2$.}
\label{app:fig:allPlotSummary_20}
\end{figure} 
\clearpage

\section{Benchmarking}\label{app:sec:benchmarking}

In this \siSection{}, we provide several technical benchmarks of our SSE QMC algorithm and code implementation. We begin by comparing our results with independent numerical calculations, specifically exact diagonalization and determinantal QMC results, as detailed in the companion work~\cite{VAS26}. As explained in the main text in \cref{sec:SSE:inter-chain} and in \cref{app:sec:updates:inter-chain}, the conventional SSE updates are supplemented by an inter-chain update, which is used for all simulations presented here, and by replica-exchange sampling, which is used in the low-temperature simulations of \cref{fig:spin_corr_ins}. In this \siSection{}, we also provide a numerical justification for the inter-chain updates by comparing simulations performed with and without them. We then show that, at large interactions and low temperatures, replica-exchange sampling improves the quality of the results.

\subsection{Comparison with independent numerical data}\label{app:sec:benchmarking:indep}

We begin by comparing our SSE simulations with independent numerical data. For this purpose, we use exact diagonalization (ED) on small clusters, varying both the chemical potential $\mu$ and the inverse temperature $\beta$. Since ED is restricted to very small system sizes, we also compare our results with determinantal QMC (DQMC) simulations at half-filling, $\mu = 0$, where the system does not suffer from a sign problem. This allows us to benchmark the SSE results at different inverse temperatures $\beta$ and system sizes $\mathcal{N}$.

Defining the total particle number and the valley-resolved particle numbers as
\begin{equation}
	\hat{N}_{\eta} = \sum_{\vec{R}} \hat{N}_{\vec{R},\eta}, \quad \hat{N} = \sum_{\eta} \hat{N}_{\eta}, 
\end{equation}
we consider several thermodynamic observables, including the total energy $\left\langle H \right\rangle$, the total particle number $\left\langle \hat{N} \right\rangle$, its fluctuations $\left\langle \hat{N}^2 \right\rangle_c \equiv \left\langle  \hat{N}^2  \right\rangle - \left\langle \hat{N} \right\rangle^2$, and the fluctuations of the valley-imbalance
\begin{equation}
	\label{app:eqn:valley_imbalance_fluct_def}
	\beta^{-1} \chi_{\Gamma_E} = \frac{1}{\mathcal{N}^2} \left\langle \left( \hat{N}_{\eta} - \hat{N}_{\eta'} \right)^2 \right\rangle = \frac{1}{\mathcal{N}^2} \left\langle \left( \hat{N}_{0} - \hat{N}_{1} \right)^2 \right\rangle, \qq{for} \eta \neq \eta',
\end{equation}
where the susceptibility $\chi_{\Gamma_E}$ was defined in \cref{app:eqn:def_susc_hsm_gamma_e}. 

Directly comparing scatter plots can be difficult, especially when both datasets have statistical uncertainties, as in the SSE-DQMC comparison. To improve readability, we therefore plot the SSE data as lines, with the corresponding errors indicated by shaded regions around them. These lines are only linear interpolations between the actual simulated points and should not be interpreted as additional data. When the shaded region is not visible, the errors are smaller than the plotted markers. In contrast, the independent benchmarking data are shown as discrete points: ED data are plotted without error bars, while DQMC data are plotted with error bars.

\subsubsection{Comparison with ED}\label{app:sec:benchmarking:indep:ED}

\begin{figure}[!t]
	\centering
	\includegraphics[width=\textwidth]{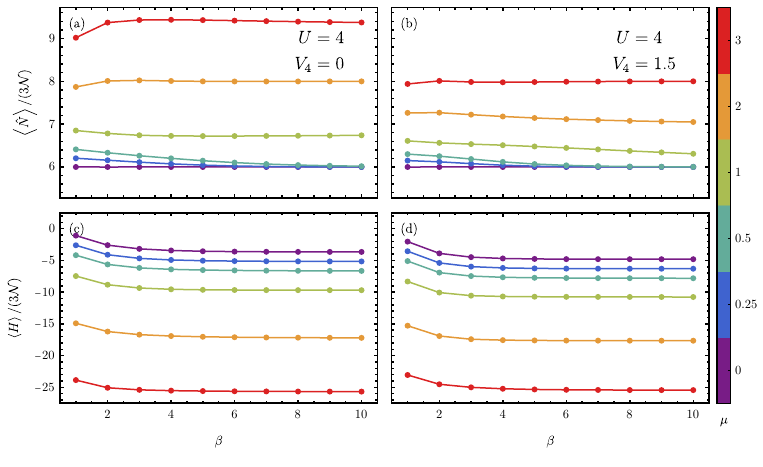}
	\caption{Benchmarking the intra-chain SSE updates. We compare SSE results for the Hamiltonian in \cref{app:eqn:final_qmc_hamiltonia} in the strict one-dimensional limit, obtained by setting $U' = 0$ and $V_i = 0$ for all $i \neq 4$. In this limit, the model reduces to a collection of $3 \mathcal{N}$ uncoupled extended one-dimensional Hubbard models of length $\mathcal{N}$. (a) and (b) show the particle number per chain, with lines denoting SSE results and dots denoting ED results, for $U=4$ with $V_4 = 0$ and $V_4=1.5$, respectively. Panels (c) and (d) show the corresponding energy per chain for the same parameters. The lines and dots are colored according to the chemical potential, as indicated by the color bar on the left. The errors are smaller than the line thickness. We take $\mathcal{N}=6$.}
	\label{app:fig:bench_ed_6_Hubbard}
\end{figure}

The first benchmarks validate the numerical implementation of the conventional intra-chain SSE updates, namely the diagonal and off-diagonal updates. To isolate this limit, we turn off all inter-chain interactions by setting $U' = 0$ and $V_i = 0$ for all $i \neq 4$ (see \cref{app:fig:model}). The resulting purely one-dimensional limit of our $\mathcal{N} \times \mathcal{N}$ system is equivalent to a collection of $3 \mathcal{N}$ uncoupled extended one-dimensional Hubbard models of length $\mathcal{N}$, with onsite repulsion $U$, nearest-neighbour repulsion $V_4$, and nearest-neighbour hopping $t$. These simulations can therefore be compared directly with one-dimensional ED calculations. \Cref{app:fig:bench_ed_6_Hubbard} compares the ED and SSE results for $\mathcal{N} = 6$ and shows excellent agreement for both the energy and particle number as functions of the chemical potential $\mu$ and inverse temperature $\beta$. In the ED calculations, we use the same parity-dependent boundary condition as in \cref{app:eqn:boundary_condition_ham} (the corresponding Hubbard model is equivalent to a Bose-Hubbard model with two species of hardcore bosons on each site, corresponding to the spin-$\uparrow$ and spin-$\downarrow$ fermions).

\begin{figure}[!t]
	\centering
	\includegraphics[width=\textwidth]{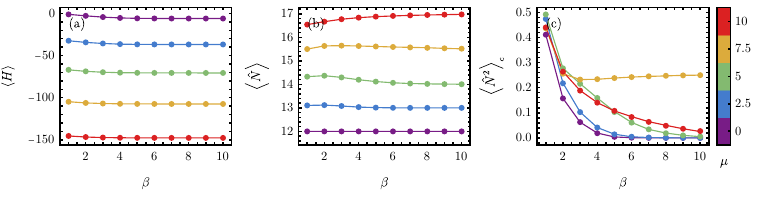}\subfloat{\label{app:fig:2D:a}}\subfloat{\label{app:fig:2D:b}}\subfloat{\label{app:fig:2D:c}}\caption{Benchmarking our SSE algorithm in the mixed-dimensional limit against ED results. We consider the total energy $\left\langle H \right\rangle$, the particle number $\left\langle \hat{N} \right\rangle$, and its fluctuations $\left\langle \hat{N}^2 \right\rangle_c$, shown respectively in (a) -- (c) as functions of the inverse temperature $\beta$. The lines and dots are colored according to the chemical potential, as indicated by the color bar on the right. The errors are smaller than the line thickness. We take $\mathcal{N}=2$, $t=1/2$, $U = U' = 5$, and $V_i = 1/2$.}
	\label{app:fig:2D}
\end{figure}

Having validated the one-dimensional limit of our model against ED simulations, we now turn to the full mixed-dimensional case. Because the Hilbert space is significantly larger in this case, we restrict the comparison to the small system size $\mathcal{N}=2$. As in all simulations presented in this work, unless stated otherwise, we include both intra- and inter-chain updates, but no replica-exchange sampling. We consider a system with typical parameters $t=1/2$, $U = U' = 5$, and $V_i=1/2$ for all $1 \leq i \leq 6$. Once again, our SSE simulations agree with ED within errors, as shown in \cref{app:fig:2D}. We note that for $\mathcal{N} = 2$, the hopping and nearest-neighbor interaction between any two atoms are effectively doubled by the periodic boundary conditions. Thus, a periodic $\mathcal{N} = 2$ system with $t=1/2$ and $V_i = 1/2$ is equivalent to an open-boundary system of the same size with $t=1$ and $V_i = 1$.

\subsubsection{Comparison with DQMC}\label{app:sec:benchmarking:indep:DQMC}

The advantage of benchmarking against ED is that ED can be performed at arbitrary chemical potential, rather than only at half-filling, as in the DQMC comparison. However, ED is limited to very small clusters because of the multi-orbital nature of our spinful model, whereas DQMC can access larger system sizes.

\begin{figure}[!t]
	\centering
	\includegraphics[width=0.333333\textwidth]{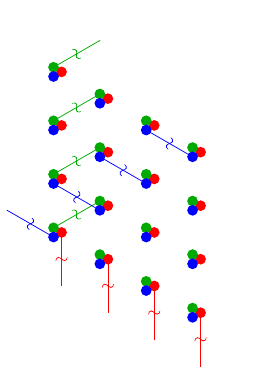}\caption{Open boundary conditions in the kinetic Hamiltonian of AA t-\ch{SnSe2}{}. When comparing the SSE and DQMC results at half-filling, $\mu=0$, we use open boundary conditions only for the kinetic term, implemented by removing the crossed hopping processes.}
	\label{app:fig:OBC_example}
\end{figure}

To further validate our simulations, we also compare against DQMC data obtained as described in our companion work~\cite{VAS26}. To ensure that the DQMC and SSE results agree to numerical precision on finite-size systems, it is essential to use the same boundary conditions. In this work, we employ parity-dependent periodic boundary conditions, whereas Ref.~\cite{VAS26} employs \textit{bona fide} periodic boundary conditions. To compare the two methods in a regime where both can be applied directly, we instead use \emph{open} boundary conditions for the kinetic term along each chain. Mathematically, this is equivalent to setting $\hat{\varphi}_{\alpha,s} = 0$ in \cref{app:eqn:boundary_condition_ham}, where the hopping phase operator $\hat{\varphi}_{\alpha,s}$ was defined in \cref{app:eqn:boundary_condition_twisted}. Graphically, this corresponds to removing the $3\mathcal{N}$ boundary hopping bonds in both the SSE and DQMC simulations, as shown in \cref{app:fig:OBC_example}. We emphasize that open boundary conditions are used only for the kinetic term; the nearest-neighbour repulsion retains periodic boundary conditions.

\begin{figure}[!t]
	\centering
	\includegraphics[width=0.696183\textwidth]{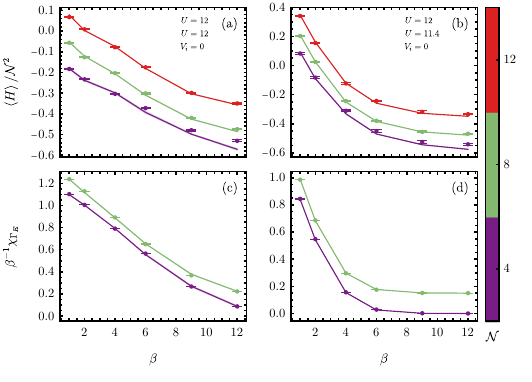}\subfloat{\label{app:fig:DQMC:a}}\subfloat{\label{app:fig:DQMC:b}}\subfloat{\label{app:fig:DQMC:c}}\subfloat{\label{app:fig:DQMC:d}}\caption{Benchmarking our SSE simulations against DQMC. We present results for the two idealized interaction parameter sets considered in \cref{app:tab:parameters_master} at half-filling, $\mu = 0$. (a) and (b) show the total energy per unit cell for the two parameter sets, as indicated in the inset of each panel. The corresponding valley-imbalance fluctuations $\beta^{-1} \chi_{\Gamma_E}$ are shown in (c) and (d). The lines denote SSE results, while the dots denote DQMC results; both are colored according to the system size $\mathcal{N}$, as indicated by the color map on the right. For ease of comparison, each curve is offset by $0.15 ( \mathcal{N} / 4 - 1)$.}
	\label{app:fig:DQMC}
\end{figure}

Our benchmark results are presented in \cref{app:fig:DQMC}. We compare both the total energy and the valley-imbalance fluctuations $\beta^{-1} \chi_{\Gamma_E}$. The energy agrees within one standard deviation between DQMC and SSE, except for the lowest three temperatures, $\beta \in \{6,9,12\}$, for the smallest system size considered, $\mathcal{N} = 4$. This discrepancy is due to an underestimation of the DQMC error bars for the kinetic contribution to the total energy. The valley-imbalance fluctuations agree within one standard deviation throughout. These simulations employ inter-chain updates, but no replica-exchange sampling. We focus only on the regime where neither DQMC nor SSE suffers from a sign problem, namely $\mu = 0$ and no inter-chain hopping.

\subsection{Effect of updates beyond conventional SSE} \label{app:sec:benchmarking:beyondSSE}

As explained in \cref{app:sec:updates}, in this work we employ two types of updates beyond conventional SSE implementations. The inter-chain update introduced in \cref{app:sec:updates:inter-chain} is used in all simulations presented in this work, unless explicitly stated otherwise, whereas replica-exchange sampling is used only at very low temperatures, $\beta > 12$, deep in the Mott regime. In this section, we briefly discuss the effects of these updates.

\subsubsection{Benchmarking the inter-chain updates}\label{app:sec:benchmarking:beyondSSE:interchain}

To show that inter-chain updates are essential for obtaining the correct physics in the mixed-dimensional limit of AA t-\ch{SnSe2}{}, we temporarily turn them off and compare the resulting simulations against DQMC. More specifically, we reproduce \cref{app:fig:DQMC}, but without any inter-chain updates, using only the intra-chain updates familiar from conventional SSE implementations.

\begin{figure}[!t]
	\centering
	\includegraphics[width=0.696183\textwidth]{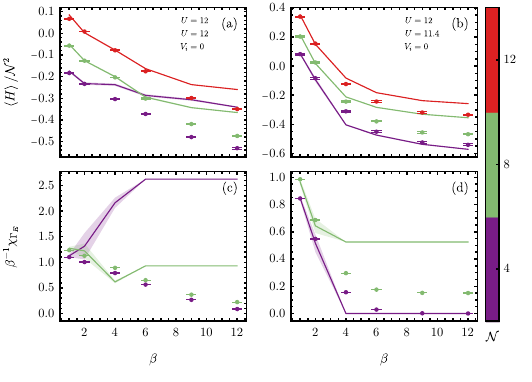}\subfloat{\label{app:fig:VF:a}}\subfloat{\label{app:fig:VF:b}}\subfloat{\label{app:fig:VF:c}}\subfloat{\label{app:fig:VF:d}}\caption{SSE simulations without inter-chain updates. The layout and notation are identical to those in \cref{app:fig:DQMC}, but here each QMC step in the SSE simulations uses only intra-chain updates, with no inter-chain updates.}
	\label{app:fig:VF}
\end{figure}

The results in \cref{app:fig:VF} show that, without inter-chain updates, the SSE simulations fail to reproduce the DQMC results even qualitatively, except at high temperatures, $\beta \leq 2$. This is consistent with expectations: at such high temperatures, charge fluctuations are not completely suppressed, the particle number along each chain can still fluctuate, and the system can equilibrate through the grand-canonical ensemble. In contrast, as the temperature is lowered, the overall charge fluctuations freeze out. Without inter-chain updates, no process remains that can transfer particles between chains, and the system therefore fails to equilibrate.

\subsubsection{Benchmarking replica-exchange sampling}\label{app:sec:benchmarking:beyondSSE:PT}

Finally, we consider the effects of replica-exchange sampling. To do so, we choose a regime with strongly suppressed charge fluctuations, taking $t=1/2$, $U = U' = 20$, and $V_i = 1/2$. This regime is much more strongly correlated than any of the physical AA t-\ch{SnSe2}{} parameter sets considered in this work, and is chosen specifically to illustrate the importance of replica-exchange sampling at very low temperatures, $\beta \bar{U} \gg 1$. 

We again focus on the small cluster $\mathcal{N} = 2$ studied in \cref{app:sec:benchmarking:indep:ED}, and compare our results with ED both with and without replica-exchange sampling. Inter-chain updates are always included. We find that the observables agree with ED within errors in both cases. However, the total charge fluctuations $\left\langle \hat{N}^2 \right\rangle_c$ obtained with replica-exchange sampling are significantly less noisy at low temperatures. This justifies our use of replica-exchange sampling at low temperatures deep in the Mott regime. We use replica-exchange sampling only for the simulations shown in \cref{fig:spin_corr_ins}.

\begin{figure}[!t]
	\centering
	\includegraphics[width=\textwidth]{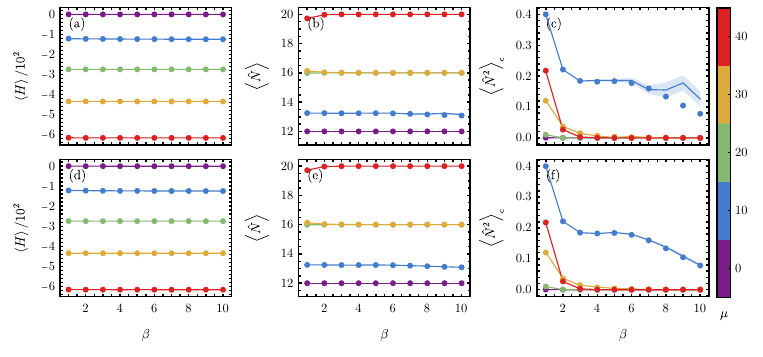}\subfloat{\label{app:fig:PT:a}}\subfloat{\label{app:fig:PT:b}}\subfloat{\label{app:fig:PT:c}}\subfloat{\label{app:fig:PT:d}}\subfloat{\label{app:fig:PT:e}}\subfloat{\label{app:fig:PT:f}}\caption{Effects of replica-exchange sampling in SSE simulations. We consider a system with $\mathcal{N} = 2$, $t = 1/2$, $U = U' = 20$, and $V_i = 1/2$. The meaning and layout of each row of panels are identical to those in \cref{app:fig:2D}. (a) -- (c) show results without replica-exchange sampling, (d) -- (f) show the corresponding results with replica-exchange sampling. In (d) -- (f), we simulate approximately $100$ replicas, introducing intermediate temperatures to ensure that adjacent replicas have non-vanishing exchange probabilities.}
	\label{app:fig:PT}
\end{figure}


\begin{thebibliography}{166}%
\makeatletter
\providecommand \@ifxundefined [1]{%
 \@ifx{#1\undefined}
}%
\providecommand \@ifnum [1]{%
 \ifnum #1\expandafter \@firstoftwo
 \else \expandafter \@secondoftwo
 \fi
}%
\providecommand \@ifx [1]{%
 \ifx #1\expandafter \@firstoftwo
 \else \expandafter \@secondoftwo
 \fi
}%
\providecommand \natexlab [1]{#1}%
\providecommand \enquote  [1]{``#1''}%
\providecommand \bibnamefont  [1]{#1}%
\providecommand \bibfnamefont [1]{#1}%
\providecommand \citenamefont [1]{#1}%
\providecommand \href@noop [0]{\@secondoftwo}%
\providecommand \href [0]{\begingroup \@sanitize@url \@href}%
\providecommand \@href[1]{\@@startlink{#1}\@@href}%
\providecommand \@@href[1]{\endgroup#1\@@endlink}%
\providecommand \@sanitize@url [0]{\catcode `\\12\catcode `\$12\catcode
  `\&12\catcode `\#12\catcode `\^12\catcode `\_12\catcode `\%12\relax}%
\providecommand \@@startlink[1]{}%
\providecommand \@@endlink[0]{}%
\providecommand \url  [0]{\begingroup\@sanitize@url \@url }%
\providecommand \@url [1]{\endgroup\@href {#1}{\urlprefix }}%
\providecommand \urlprefix  [0]{URL }%
\providecommand \Eprint [0]{\href }%
\providecommand \doibase [0]{https://doi.org/}%
\providecommand \selectlanguage [0]{\@gobble}%
\providecommand \bibinfo  [0]{\@secondoftwo}%
\providecommand \bibfield  [0]{\@secondoftwo}%
\providecommand \translation [1]{[#1]}%
\providecommand \BibitemOpen [0]{}%
\providecommand \bibitemStop [0]{}%
\providecommand \bibitemNoStop [0]{.\EOS\space}%
\providecommand \EOS [0]{\spacefactor3000\relax}%
\providecommand \BibitemShut  [1]{\csname bibitem#1\endcsname}%
\let\auto@bib@innerbib\@empty
\bibitem [{\citenamefont {Kondo}(1964)}]{KON64}%
  \BibitemOpen
  \bibfield  {author} {\bibinfo {author} {\bibfnamefont {J.}~\bibnamefont
  {Kondo}},\ }\href {https://doi.org/10.1143/PTP.32.37} {\bibfield  {journal}
  {\bibinfo  {journal} {Prog. Theor. Phys.}\ }\textbf {\bibinfo {volume}
  {32}},\ \bibinfo {pages} {37} (\bibinfo {year} {1964})}\BibitemShut {NoStop}%
\bibitem [{\citenamefont {Wilson}\ and\ \citenamefont {Fisher}(1972)}]{WIL72}%
  \BibitemOpen
  \bibfield  {author} {\bibinfo {author} {\bibfnamefont {K.~G.}\ \bibnamefont
  {Wilson}}\ and\ \bibinfo {author} {\bibfnamefont {M.~E.}\ \bibnamefont
  {Fisher}},\ }\href {https://doi.org/10.1103/PhysRevLett.28.240} {\bibfield
  {journal} {\bibinfo  {journal} {Phys. Rev. Lett.}\ }\textbf {\bibinfo
  {volume} {28}},\ \bibinfo {pages} {240} (\bibinfo {year} {1972})}\BibitemShut
  {NoStop}%
\bibitem [{\citenamefont {Wilson}\ and\ \citenamefont {Kogut}(1974)}]{WIL74}%
  \BibitemOpen
  \bibfield  {author} {\bibinfo {author} {\bibfnamefont {K.~G.}\ \bibnamefont
  {Wilson}}\ and\ \bibinfo {author} {\bibfnamefont {J.}~\bibnamefont {Kogut}},\
  }\href {https://doi.org/10.1016/0370-1573(74)90023-4} {\bibfield  {journal}
  {\bibinfo  {journal} {Phys. Rep.}\ }\textbf {\bibinfo {volume} {12}},\
  \bibinfo {pages} {75} (\bibinfo {year} {1974})}\BibitemShut {NoStop}%
\bibitem [{\citenamefont {Wilson}(1975)}]{WIL75a}%
  \BibitemOpen
  \bibfield  {author} {\bibinfo {author} {\bibfnamefont {K.~G.}\ \bibnamefont
  {Wilson}},\ }\href {https://doi.org/10.1103/RevModPhys.47.773} {\bibfield
  {journal} {\bibinfo  {journal} {Rev. Mod. Phys.}\ }\textbf {\bibinfo {volume}
  {47}},\ \bibinfo {pages} {773} (\bibinfo {year} {1975})}\BibitemShut
  {NoStop}%
\bibitem [{\citenamefont {Thouless}\ \emph {et~al.}(1982)\citenamefont
  {Thouless}, \citenamefont {Kohmoto}, \citenamefont {Nightingale},\ and\
  \citenamefont {{den Nijs}}}]{THO82}%
  \BibitemOpen
  \bibfield  {author} {\bibinfo {author} {\bibfnamefont {D.~J.}\ \bibnamefont
  {Thouless}}, \bibinfo {author} {\bibfnamefont {M.}~\bibnamefont {Kohmoto}},
  \bibinfo {author} {\bibfnamefont {M.~P.}\ \bibnamefont {Nightingale}},\ and\
  \bibinfo {author} {\bibfnamefont {M.}~\bibnamefont {{den Nijs}}},\ }\href
  {https://doi.org/10.1103/PhysRevLett.49.405} {\bibfield  {journal} {\bibinfo
  {journal} {Phys. Rev. Lett.}\ }\textbf {\bibinfo {volume} {49}},\ \bibinfo
  {pages} {405} (\bibinfo {year} {1982})}\BibitemShut {NoStop}%
\bibitem [{\citenamefont {Tsui}\ \emph {et~al.}(1982)\citenamefont {Tsui},
  \citenamefont {Stormer},\ and\ \citenamefont {Gossard}}]{TSU82a}%
  \BibitemOpen
  \bibfield  {author} {\bibinfo {author} {\bibfnamefont {D.~C.}\ \bibnamefont
  {Tsui}}, \bibinfo {author} {\bibfnamefont {H.~L.}\ \bibnamefont {Stormer}},\
  and\ \bibinfo {author} {\bibfnamefont {A.~C.}\ \bibnamefont {Gossard}},\
  }\href {https://doi.org/10.1103/PhysRevLett.48.1559} {\bibfield  {journal}
  {\bibinfo  {journal} {Phys. Rev. Lett.}\ }\textbf {\bibinfo {volume} {48}},\
  \bibinfo {pages} {1559} (\bibinfo {year} {1982})}\BibitemShut {NoStop}%
\bibitem [{\citenamefont {Laughlin}(1983)}]{LAU83a}%
  \BibitemOpen
  \bibfield  {author} {\bibinfo {author} {\bibfnamefont {R.~B.}\ \bibnamefont
  {Laughlin}},\ }\href {https://doi.org/10.1103/PhysRevLett.50.1395} {\bibfield
   {journal} {\bibinfo  {journal} {Phys. Rev. Lett.}\ }\textbf {\bibinfo
  {volume} {50}},\ \bibinfo {pages} {1395} (\bibinfo {year}
  {1983})}\BibitemShut {NoStop}%
\bibitem [{\citenamefont {Metzner}\ and\ \citenamefont
  {Vollhardt}(1989)}]{MET89a}%
  \BibitemOpen
  \bibfield  {author} {\bibinfo {author} {\bibfnamefont {W.}~\bibnamefont
  {Metzner}}\ and\ \bibinfo {author} {\bibfnamefont {D.}~\bibnamefont
  {Vollhardt}},\ }\href {https://doi.org/10.1103/PhysRevLett.62.324} {\bibfield
   {journal} {\bibinfo  {journal} {Phys. Rev. Lett.}\ }\textbf {\bibinfo
  {volume} {62}},\ \bibinfo {pages} {324} (\bibinfo {year} {1989})}\BibitemShut
  {NoStop}%
\bibitem [{\citenamefont {Georges}\ and\ \citenamefont
  {Krauth}(1992)}]{GEO92a}%
  \BibitemOpen
  \bibfield  {author} {\bibinfo {author} {\bibfnamefont {A.}~\bibnamefont
  {Georges}}\ and\ \bibinfo {author} {\bibfnamefont {W.}~\bibnamefont
  {Krauth}},\ }\href {https://doi.org/10.1103/PhysRevLett.69.1240} {\bibfield
  {journal} {\bibinfo  {journal} {Phys. Rev. Lett.}\ }\textbf {\bibinfo
  {volume} {69}},\ \bibinfo {pages} {1240} (\bibinfo {year}
  {1992})}\BibitemShut {NoStop}%
\bibitem [{\citenamefont {Georges}\ \emph {et~al.}(1996)\citenamefont
  {Georges}, \citenamefont {Kotliar}, \citenamefont {Krauth},\ and\
  \citenamefont {Rozenberg}}]{GEO96}%
  \BibitemOpen
  \bibfield  {author} {\bibinfo {author} {\bibfnamefont {A.}~\bibnamefont
  {Georges}}, \bibinfo {author} {\bibfnamefont {G.}~\bibnamefont {Kotliar}},
  \bibinfo {author} {\bibfnamefont {W.}~\bibnamefont {Krauth}},\ and\ \bibinfo
  {author} {\bibfnamefont {M.~J.}\ \bibnamefont {Rozenberg}},\ }\href
  {https://doi.org/10.1103/RevModPhys.68.13} {\bibfield  {journal} {\bibinfo
  {journal} {Rev. Mod. Phys.}\ }\textbf {\bibinfo {volume} {68}},\ \bibinfo
  {pages} {13} (\bibinfo {year} {1996})}\BibitemShut {NoStop}%
\bibitem [{\citenamefont {White}(1992)}]{WHI92}%
  \BibitemOpen
  \bibfield  {author} {\bibinfo {author} {\bibfnamefont {S.~R.}\ \bibnamefont
  {White}},\ }\href {https://doi.org/10.1103/PhysRevLett.69.2863} {\bibfield
  {journal} {\bibinfo  {journal} {Phys. Rev. Lett.}\ }\textbf {\bibinfo
  {volume} {69}},\ \bibinfo {pages} {2863} (\bibinfo {year}
  {1992})}\BibitemShut {NoStop}%
\bibitem [{\citenamefont {Schollw{\"o}ck}(2011)}]{SCH11a}%
  \BibitemOpen
  \bibfield  {author} {\bibinfo {author} {\bibfnamefont {U.}~\bibnamefont
  {Schollw{\"o}ck}},\ }\href {https://doi.org/10.1016/j.aop.2010.09.012}
  {\bibfield  {journal} {\bibinfo  {journal} {Ann. Phys.}\ }\bibinfo {series}
  {January 2011 {{Special Issue}}},\ \textbf {\bibinfo {volume} {326}},\
  \bibinfo {pages} {96} (\bibinfo {year} {2011})}\BibitemShut {NoStop}%
\bibitem [{\citenamefont {C{\u a}lug{\u a}ru}\ \emph
  {et~al.}(2025)\citenamefont {C{\u a}lug{\u a}ru}, \citenamefont {Jiang},
  \citenamefont {Hu}, \citenamefont {Pi}, \citenamefont {Yu}, \citenamefont
  {Vergniory}, \citenamefont {Shan}, \citenamefont {Felser}, \citenamefont
  {Schoop}, \citenamefont {Efetov}, \citenamefont {Mak},\ and\ \citenamefont
  {Bernevig}}]{CAL25b}%
  \BibitemOpen
  \bibfield  {author} {\bibinfo {author} {\bibfnamefont {D.}~\bibnamefont {C{\u
  a}lug{\u a}ru}}, \bibinfo {author} {\bibfnamefont {Y.}~\bibnamefont {Jiang}},
  \bibinfo {author} {\bibfnamefont {H.}~\bibnamefont {Hu}}, \bibinfo {author}
  {\bibfnamefont {H.}~\bibnamefont {Pi}}, \bibinfo {author} {\bibfnamefont
  {J.}~\bibnamefont {Yu}}, \bibinfo {author} {\bibfnamefont {M.~G.}\
  \bibnamefont {Vergniory}}, \bibinfo {author} {\bibfnamefont {J.}~\bibnamefont
  {Shan}}, \bibinfo {author} {\bibfnamefont {C.}~\bibnamefont {Felser}},
  \bibinfo {author} {\bibfnamefont {L.~M.}\ \bibnamefont {Schoop}}, \bibinfo
  {author} {\bibfnamefont {D.~K.}\ \bibnamefont {Efetov}}, \bibinfo {author}
  {\bibfnamefont {K.~F.}\ \bibnamefont {Mak}},\ and\ \bibinfo {author}
  {\bibfnamefont {B.~A.}\ \bibnamefont {Bernevig}},\ }\href
  {https://doi.org/10.1038/s41586-025-09187-5} {\bibfield  {journal} {\bibinfo
  {journal} {Nature}\ }\textbf {\bibinfo {volume} {643}},\ \bibinfo {pages}
  {376} (\bibinfo {year} {2025})}\BibitemShut {NoStop}%
\bibitem [{\citenamefont {Lei}\ \emph {et~al.}(2025)\citenamefont {Lei},
  \citenamefont {Mahon},\ and\ \citenamefont {MacDonald}}]{MAH24}%
  \BibitemOpen
  \bibfield  {author} {\bibinfo {author} {\bibfnamefont {C.}~\bibnamefont
  {Lei}}, \bibinfo {author} {\bibfnamefont {P.~T.}\ \bibnamefont {Mahon}},\
  and\ \bibinfo {author} {\bibfnamefont {A.~H.}\ \bibnamefont {MacDonald}},\
  }\href {https://doi.org/10.1103/5zt2-scbg} {\bibfield  {journal} {\bibinfo
  {journal} {Phys. Rev. Lett.}\ }\textbf {\bibinfo {volume} {135}},\ \bibinfo
  {pages} {196402} (\bibinfo {year} {2025})}\BibitemShut {NoStop}%
\bibitem [{\citenamefont {Jiang}\ \emph {et~al.}(2024)\citenamefont {Jiang},
  \citenamefont {Petralanda}, \citenamefont {Skorupskii}, \citenamefont {Xu},
  \citenamefont {Pi}, \citenamefont {C{\u a}lug{\u a}ru}, \citenamefont {Hu},
  \citenamefont {Xie}, \citenamefont {Mustaf}, \citenamefont {H{\"o}hn},
  \citenamefont {Haase}, \citenamefont {Vergniory}, \citenamefont {Claassen},
  \citenamefont {Elcoro}, \citenamefont {Regnault}, \citenamefont {Shan},
  \citenamefont {Mak}, \citenamefont {Efetov}, \citenamefont {Morosan},
  \citenamefont {Kennes}, \citenamefont {Rubio}, \citenamefont {Xian},
  \citenamefont {Felser}, \citenamefont {Schoop},\ and\ \citenamefont
  {Bernevig}}]{JIA24b}%
  \BibitemOpen
  \bibfield  {author} {\bibinfo {author} {\bibfnamefont {Y.}~\bibnamefont
  {Jiang}}, \bibinfo {author} {\bibfnamefont {U.}~\bibnamefont {Petralanda}},
  \bibinfo {author} {\bibfnamefont {G.}~\bibnamefont {Skorupskii}}, \bibinfo
  {author} {\bibfnamefont {Q.}~\bibnamefont {Xu}}, \bibinfo {author}
  {\bibfnamefont {H.}~\bibnamefont {Pi}}, \bibinfo {author} {\bibfnamefont
  {D.}~\bibnamefont {C{\u a}lug{\u a}ru}}, \bibinfo {author} {\bibfnamefont
  {H.}~\bibnamefont {Hu}}, \bibinfo {author} {\bibfnamefont {J.}~\bibnamefont
  {Xie}}, \bibinfo {author} {\bibfnamefont {R.~A.}\ \bibnamefont {Mustaf}},
  \bibinfo {author} {\bibfnamefont {P.}~\bibnamefont {H{\"o}hn}}, \bibinfo
  {author} {\bibfnamefont {V.}~\bibnamefont {Haase}}, \bibinfo {author}
  {\bibfnamefont {M.~G.}\ \bibnamefont {Vergniory}}, \bibinfo {author}
  {\bibfnamefont {M.}~\bibnamefont {Claassen}}, \bibinfo {author}
  {\bibfnamefont {L.}~\bibnamefont {Elcoro}}, \bibinfo {author} {\bibfnamefont
  {N.}~\bibnamefont {Regnault}}, \bibinfo {author} {\bibfnamefont
  {J.}~\bibnamefont {Shan}}, \bibinfo {author} {\bibfnamefont {K.~F.}\
  \bibnamefont {Mak}}, \bibinfo {author} {\bibfnamefont {D.~K.}\ \bibnamefont
  {Efetov}}, \bibinfo {author} {\bibfnamefont {E.}~\bibnamefont {Morosan}},
  \bibinfo {author} {\bibfnamefont {D.~M.}\ \bibnamefont {Kennes}}, \bibinfo
  {author} {\bibfnamefont {A.}~\bibnamefont {Rubio}}, \bibinfo {author}
  {\bibfnamefont {L.}~\bibnamefont {Xian}}, \bibinfo {author} {\bibfnamefont
  {C.}~\bibnamefont {Felser}}, \bibinfo {author} {\bibfnamefont {L.~M.}\
  \bibnamefont {Schoop}},\ and\ \bibinfo {author} {\bibfnamefont {B.~A.}\
  \bibnamefont {Bernevig}},\ }\href {https://doi.org/10.48550/arXiv.2411.09741}
  {\bibinfo {title} {{{2D Theoretically Twistable Material Database}}}}
  (\bibinfo {year} {2024}),\ \Eprint {https://arxiv.org/abs/2411.09741}
  {arXiv:2411.09741 [cond-mat.mtrl-sci]} \BibitemShut {NoStop}%
\bibitem [{\citenamefont {Sandvik}(1999)}]{SAN99}%
  \BibitemOpen
  \bibfield  {author} {\bibinfo {author} {\bibfnamefont {A.~W.}\ \bibnamefont
  {Sandvik}},\ }\href {https://doi.org/10.1103/PhysRevB.59.R14157} {\bibfield
  {journal} {\bibinfo  {journal} {Phys. Rev. B}\ }\textbf {\bibinfo {volume}
  {59}},\ \bibinfo {pages} {R14157} (\bibinfo {year} {1999})}\BibitemShut
  {NoStop}%
\bibitem [{\citenamefont {Sengupta}\ \emph {et~al.}(2002)\citenamefont
  {Sengupta}, \citenamefont {Sandvik},\ and\ \citenamefont {Campbell}}]{SEN02}%
  \BibitemOpen
  \bibfield  {author} {\bibinfo {author} {\bibfnamefont {P.}~\bibnamefont
  {Sengupta}}, \bibinfo {author} {\bibfnamefont {A.~W.}\ \bibnamefont
  {Sandvik}},\ and\ \bibinfo {author} {\bibfnamefont {D.~K.}\ \bibnamefont
  {Campbell}},\ }\href {https://doi.org/10.1103/PhysRevB.65.155113} {\bibfield
  {journal} {\bibinfo  {journal} {Phys. Rev. B}\ }\textbf {\bibinfo {volume}
  {65}},\ \bibinfo {pages} {155113} (\bibinfo {year} {2002})}\BibitemShut
  {NoStop}%
\bibitem [{\citenamefont {Sandvik}(2019)}]{SAN19}%
  \BibitemOpen
  \bibfield  {author} {\bibinfo {author} {\bibfnamefont {A.~W.}\ \bibnamefont
  {Sandvik}},\ }\href {https://doi.org/10.48550/arXiv.1909.10591} {\bibinfo
  {title} {Stochastic {{Series Expansion Methods}}}} (\bibinfo {year} {2019}),\
  \Eprint {https://arxiv.org/abs/1909.10591} {arXiv:1909.10591
  [cond-mat.str-el]} \BibitemShut {NoStop}%
\bibitem [{\citenamefont {Li}\ \emph {et~al.}(2025)\citenamefont {Li},
  \citenamefont {Calugaru}, \citenamefont {Jiang}, \citenamefont {Pi},
  \citenamefont {Fischer}, \citenamefont {Schl{\"o}mer}, \citenamefont {Klebl},
  \citenamefont {Vergniory}, \citenamefont {Kennes}, \citenamefont
  {Parameswaran}, \citenamefont {Yao}, \citenamefont {Bernevig},\ and\
  \citenamefont {Hu}}]{LI25}%
  \BibitemOpen
  \bibfield  {author} {\bibinfo {author} {\bibfnamefont {M.-R.}\ \bibnamefont
  {Li}}, \bibinfo {author} {\bibfnamefont {D.}~\bibnamefont {Calugaru}},
  \bibinfo {author} {\bibfnamefont {Y.}~\bibnamefont {Jiang}}, \bibinfo
  {author} {\bibfnamefont {H.}~\bibnamefont {Pi}}, \bibinfo {author}
  {\bibfnamefont {A.}~\bibnamefont {Fischer}}, \bibinfo {author} {\bibfnamefont
  {H.}~\bibnamefont {Schl{\"o}mer}}, \bibinfo {author} {\bibfnamefont
  {L.}~\bibnamefont {Klebl}}, \bibinfo {author} {\bibfnamefont {M.~G.}\
  \bibnamefont {Vergniory}}, \bibinfo {author} {\bibfnamefont {D.~M.}\
  \bibnamefont {Kennes}}, \bibinfo {author} {\bibfnamefont {S.~A.}\
  \bibnamefont {Parameswaran}}, \bibinfo {author} {\bibfnamefont
  {H.}~\bibnamefont {Yao}}, \bibinfo {author} {\bibfnamefont {B.~A.}\
  \bibnamefont {Bernevig}},\ and\ \bibinfo {author} {\bibfnamefont
  {H.}~\bibnamefont {Hu}},\ }\href {https://doi.org/10.48550/arXiv.2508.10098}
  {\bibinfo {title} {Emergent {{Interacting Phases}} in the {{Strong Coupling
  Limit}} of {{Twisted M-Valley Moir\'e Systems}}: {{Application}} to
  {{SnSe}}$_2$}} (\bibinfo {year} {2025}),\ \Eprint
  {https://arxiv.org/abs/2508.10098} {arXiv:2508.10098 [cond-mat.str-el]}
  \BibitemShut {NoStop}%
\bibitem [{\citenamefont {Andrei}\ \emph {et~al.}(2021)\citenamefont {Andrei},
  \citenamefont {Efetov}, \citenamefont {{Jarillo-Herrero}}, \citenamefont
  {MacDonald}, \citenamefont {Mak}, \citenamefont {Senthil}, \citenamefont
  {Tutuc}, \citenamefont {Yazdani},\ and\ \citenamefont {Young}}]{AND21}%
  \BibitemOpen
  \bibfield  {author} {\bibinfo {author} {\bibfnamefont {E.~Y.}\ \bibnamefont
  {Andrei}}, \bibinfo {author} {\bibfnamefont {D.~K.}\ \bibnamefont {Efetov}},
  \bibinfo {author} {\bibfnamefont {P.}~\bibnamefont {{Jarillo-Herrero}}},
  \bibinfo {author} {\bibfnamefont {A.~H.}\ \bibnamefont {MacDonald}}, \bibinfo
  {author} {\bibfnamefont {K.~F.}\ \bibnamefont {Mak}}, \bibinfo {author}
  {\bibfnamefont {T.}~\bibnamefont {Senthil}}, \bibinfo {author} {\bibfnamefont
  {E.}~\bibnamefont {Tutuc}}, \bibinfo {author} {\bibfnamefont
  {A.}~\bibnamefont {Yazdani}},\ and\ \bibinfo {author} {\bibfnamefont {A.~F.}\
  \bibnamefont {Young}},\ }\href {https://doi.org/10.1038/s41578-021-00284-1}
  {\bibfield  {journal} {\bibinfo  {journal} {Nat. Rev. Mater.}\ }\textbf
  {\bibinfo {volume} {6}},\ \bibinfo {pages} {201} (\bibinfo {year}
  {2021})}\BibitemShut {NoStop}%
\bibitem [{\citenamefont {Kennes}\ \emph {et~al.}(2021)\citenamefont {Kennes},
  \citenamefont {Claassen}, \citenamefont {Xian}, \citenamefont {Georges},
  \citenamefont {Millis}, \citenamefont {Hone}, \citenamefont {Dean},
  \citenamefont {Basov}, \citenamefont {Pasupathy},\ and\ \citenamefont
  {Rubio}}]{KEN21}%
  \BibitemOpen
  \bibfield  {author} {\bibinfo {author} {\bibfnamefont {D.~M.}\ \bibnamefont
  {Kennes}}, \bibinfo {author} {\bibfnamefont {M.}~\bibnamefont {Claassen}},
  \bibinfo {author} {\bibfnamefont {L.}~\bibnamefont {Xian}}, \bibinfo {author}
  {\bibfnamefont {A.}~\bibnamefont {Georges}}, \bibinfo {author} {\bibfnamefont
  {A.~J.}\ \bibnamefont {Millis}}, \bibinfo {author} {\bibfnamefont
  {J.}~\bibnamefont {Hone}}, \bibinfo {author} {\bibfnamefont {C.~R.}\
  \bibnamefont {Dean}}, \bibinfo {author} {\bibfnamefont {D.~N.}\ \bibnamefont
  {Basov}}, \bibinfo {author} {\bibfnamefont {A.~N.}\ \bibnamefont
  {Pasupathy}},\ and\ \bibinfo {author} {\bibfnamefont {A.}~\bibnamefont
  {Rubio}},\ }\href {https://doi.org/10.1038/s41567-020-01154-3} {\bibfield
  {journal} {\bibinfo  {journal} {Nat. Phys.}\ }\textbf {\bibinfo {volume}
  {17}},\ \bibinfo {pages} {155} (\bibinfo {year} {2021})}\BibitemShut
  {NoStop}%
\bibitem [{\citenamefont {Nuckolls}\ and\ \citenamefont
  {Yazdani}(2024)}]{NUC24}%
  \BibitemOpen
  \bibfield  {author} {\bibinfo {author} {\bibfnamefont {K.~P.}\ \bibnamefont
  {Nuckolls}}\ and\ \bibinfo {author} {\bibfnamefont {A.}~\bibnamefont
  {Yazdani}},\ }\href {https://doi.org/10.1038/s41578-024-00682-1} {\bibfield
  {journal} {\bibinfo  {journal} {Nat. Rev. Mater.}\ }\textbf {\bibinfo
  {volume} {9}},\ \bibinfo {pages} {460} (\bibinfo {year} {2024})}\BibitemShut
  {NoStop}%
\bibitem [{\citenamefont {Cao}\ \emph {et~al.}(2018{\natexlab{a}})\citenamefont
  {Cao}, \citenamefont {Fatemi}, \citenamefont {Demir}, \citenamefont {Fang},
  \citenamefont {Tomarken}, \citenamefont {Luo}, \citenamefont
  {{Sanchez-Yamagishi}}, \citenamefont {Watanabe}, \citenamefont {Taniguchi},
  \citenamefont {Kaxiras}, \citenamefont {Ashoori},\ and\ \citenamefont
  {{Jarillo-Herrero}}}]{CAO18}%
  \BibitemOpen
  \bibfield  {author} {\bibinfo {author} {\bibfnamefont {Y.}~\bibnamefont
  {Cao}}, \bibinfo {author} {\bibfnamefont {V.}~\bibnamefont {Fatemi}},
  \bibinfo {author} {\bibfnamefont {A.}~\bibnamefont {Demir}}, \bibinfo
  {author} {\bibfnamefont {S.}~\bibnamefont {Fang}}, \bibinfo {author}
  {\bibfnamefont {S.~L.}\ \bibnamefont {Tomarken}}, \bibinfo {author}
  {\bibfnamefont {J.~Y.}\ \bibnamefont {Luo}}, \bibinfo {author} {\bibfnamefont
  {J.~D.}\ \bibnamefont {{Sanchez-Yamagishi}}}, \bibinfo {author}
  {\bibfnamefont {K.}~\bibnamefont {Watanabe}}, \bibinfo {author}
  {\bibfnamefont {T.}~\bibnamefont {Taniguchi}}, \bibinfo {author}
  {\bibfnamefont {E.}~\bibnamefont {Kaxiras}}, \bibinfo {author} {\bibfnamefont
  {R.~C.}\ \bibnamefont {Ashoori}},\ and\ \bibinfo {author} {\bibfnamefont
  {P.}~\bibnamefont {{Jarillo-Herrero}}},\ }\href
  {https://doi.org/10.1038/nature26154} {\bibfield  {journal} {\bibinfo
  {journal} {Nature}\ }\textbf {\bibinfo {volume} {556}},\ \bibinfo {pages}
  {80} (\bibinfo {year} {2018}{\natexlab{a}})}\BibitemShut {NoStop}%
\bibitem [{\citenamefont {Cao}\ \emph {et~al.}(2018{\natexlab{b}})\citenamefont
  {Cao}, \citenamefont {Fatemi}, \citenamefont {Fang}, \citenamefont
  {Watanabe}, \citenamefont {Taniguchi}, \citenamefont {Kaxiras},\ and\
  \citenamefont {{Jarillo-Herrero}}}]{CAO18a}%
  \BibitemOpen
  \bibfield  {author} {\bibinfo {author} {\bibfnamefont {Y.}~\bibnamefont
  {Cao}}, \bibinfo {author} {\bibfnamefont {V.}~\bibnamefont {Fatemi}},
  \bibinfo {author} {\bibfnamefont {S.}~\bibnamefont {Fang}}, \bibinfo {author}
  {\bibfnamefont {K.}~\bibnamefont {Watanabe}}, \bibinfo {author}
  {\bibfnamefont {T.}~\bibnamefont {Taniguchi}}, \bibinfo {author}
  {\bibfnamefont {E.}~\bibnamefont {Kaxiras}},\ and\ \bibinfo {author}
  {\bibfnamefont {P.}~\bibnamefont {{Jarillo-Herrero}}},\ }\href
  {https://doi.org/10.1038/nature26160} {\bibfield  {journal} {\bibinfo
  {journal} {Nature}\ }\textbf {\bibinfo {volume} {556}},\ \bibinfo {pages}
  {43} (\bibinfo {year} {2018}{\natexlab{b}})}\BibitemShut {NoStop}%
\bibitem [{\citenamefont {Kerelsky}\ \emph {et~al.}(2019)\citenamefont
  {Kerelsky}, \citenamefont {McGilly}, \citenamefont {Kennes}, \citenamefont
  {Xian}, \citenamefont {Yankowitz}, \citenamefont {Chen}, \citenamefont
  {Watanabe}, \citenamefont {Taniguchi}, \citenamefont {Hone}, \citenamefont
  {Dean}, \citenamefont {Rubio},\ and\ \citenamefont {Pasupathy}}]{KER19}%
  \BibitemOpen
  \bibfield  {author} {\bibinfo {author} {\bibfnamefont {A.}~\bibnamefont
  {Kerelsky}}, \bibinfo {author} {\bibfnamefont {L.~J.}\ \bibnamefont
  {McGilly}}, \bibinfo {author} {\bibfnamefont {D.~M.}\ \bibnamefont {Kennes}},
  \bibinfo {author} {\bibfnamefont {L.}~\bibnamefont {Xian}}, \bibinfo {author}
  {\bibfnamefont {M.}~\bibnamefont {Yankowitz}}, \bibinfo {author}
  {\bibfnamefont {S.}~\bibnamefont {Chen}}, \bibinfo {author} {\bibfnamefont
  {K.}~\bibnamefont {Watanabe}}, \bibinfo {author} {\bibfnamefont
  {T.}~\bibnamefont {Taniguchi}}, \bibinfo {author} {\bibfnamefont
  {J.}~\bibnamefont {Hone}}, \bibinfo {author} {\bibfnamefont {C.}~\bibnamefont
  {Dean}}, \bibinfo {author} {\bibfnamefont {A.}~\bibnamefont {Rubio}},\ and\
  \bibinfo {author} {\bibfnamefont {A.~N.}\ \bibnamefont {Pasupathy}},\ }\href
  {https://doi.org/10.1038/s41586-019-1431-9} {\bibfield  {journal} {\bibinfo
  {journal} {Nature}\ }\textbf {\bibinfo {volume} {572}},\ \bibinfo {pages}
  {95} (\bibinfo {year} {2019})}\BibitemShut {NoStop}%
\bibitem [{\citenamefont {Xie}\ \emph {et~al.}(2019)\citenamefont {Xie},
  \citenamefont {Lian}, \citenamefont {J{\"a}ck}, \citenamefont {Liu},
  \citenamefont {Chiu}, \citenamefont {Watanabe}, \citenamefont {Taniguchi},
  \citenamefont {Bernevig},\ and\ \citenamefont {Yazdani}}]{XIE19}%
  \BibitemOpen
  \bibfield  {author} {\bibinfo {author} {\bibfnamefont {Y.}~\bibnamefont
  {Xie}}, \bibinfo {author} {\bibfnamefont {B.}~\bibnamefont {Lian}}, \bibinfo
  {author} {\bibfnamefont {B.}~\bibnamefont {J{\"a}ck}}, \bibinfo {author}
  {\bibfnamefont {X.}~\bibnamefont {Liu}}, \bibinfo {author} {\bibfnamefont
  {C.-L.}\ \bibnamefont {Chiu}}, \bibinfo {author} {\bibfnamefont
  {K.}~\bibnamefont {Watanabe}}, \bibinfo {author} {\bibfnamefont
  {T.}~\bibnamefont {Taniguchi}}, \bibinfo {author} {\bibfnamefont {B.~A.}\
  \bibnamefont {Bernevig}},\ and\ \bibinfo {author} {\bibfnamefont
  {A.}~\bibnamefont {Yazdani}},\ }\href
  {https://doi.org/10.1038/s41586-019-1422-x} {\bibfield  {journal} {\bibinfo
  {journal} {Nature}\ }\textbf {\bibinfo {volume} {572}},\ \bibinfo {pages}
  {101} (\bibinfo {year} {2019})}\BibitemShut {NoStop}%
\bibitem [{\citenamefont {Choi}\ \emph {et~al.}(2019)\citenamefont {Choi},
  \citenamefont {Kemmer}, \citenamefont {Peng}, \citenamefont {Thomson},
  \citenamefont {Arora}, \citenamefont {Polski}, \citenamefont {Zhang},
  \citenamefont {Ren}, \citenamefont {Alicea}, \citenamefont {Refael},
  \citenamefont {{von Oppen}}, \citenamefont {Watanabe}, \citenamefont
  {Taniguchi},\ and\ \citenamefont {{Nadj-Perge}}}]{CHO19}%
  \BibitemOpen
  \bibfield  {author} {\bibinfo {author} {\bibfnamefont {Y.}~\bibnamefont
  {Choi}}, \bibinfo {author} {\bibfnamefont {J.}~\bibnamefont {Kemmer}},
  \bibinfo {author} {\bibfnamefont {Y.}~\bibnamefont {Peng}}, \bibinfo {author}
  {\bibfnamefont {A.}~\bibnamefont {Thomson}}, \bibinfo {author} {\bibfnamefont
  {H.}~\bibnamefont {Arora}}, \bibinfo {author} {\bibfnamefont
  {R.}~\bibnamefont {Polski}}, \bibinfo {author} {\bibfnamefont
  {Y.}~\bibnamefont {Zhang}}, \bibinfo {author} {\bibfnamefont
  {H.}~\bibnamefont {Ren}}, \bibinfo {author} {\bibfnamefont {J.}~\bibnamefont
  {Alicea}}, \bibinfo {author} {\bibfnamefont {G.}~\bibnamefont {Refael}},
  \bibinfo {author} {\bibfnamefont {F.}~\bibnamefont {{von Oppen}}}, \bibinfo
  {author} {\bibfnamefont {K.}~\bibnamefont {Watanabe}}, \bibinfo {author}
  {\bibfnamefont {T.}~\bibnamefont {Taniguchi}},\ and\ \bibinfo {author}
  {\bibfnamefont {S.}~\bibnamefont {{Nadj-Perge}}},\ }\href
  {https://doi.org/10.1038/s41567-019-0606-5} {\bibfield  {journal} {\bibinfo
  {journal} {Nat. Phys.}\ }\textbf {\bibinfo {volume} {15}},\ \bibinfo {pages}
  {1174} (\bibinfo {year} {2019})}\BibitemShut {NoStop}%
\bibitem [{\citenamefont {Tomarken}\ \emph {et~al.}(2019)\citenamefont
  {Tomarken}, \citenamefont {Cao}, \citenamefont {Demir}, \citenamefont
  {Watanabe}, \citenamefont {Taniguchi}, \citenamefont {{Jarillo-Herrero}},\
  and\ \citenamefont {Ashoori}}]{TOM19}%
  \BibitemOpen
  \bibfield  {author} {\bibinfo {author} {\bibfnamefont {S.~L.}\ \bibnamefont
  {Tomarken}}, \bibinfo {author} {\bibfnamefont {Y.}~\bibnamefont {Cao}},
  \bibinfo {author} {\bibfnamefont {A.}~\bibnamefont {Demir}}, \bibinfo
  {author} {\bibfnamefont {K.}~\bibnamefont {Watanabe}}, \bibinfo {author}
  {\bibfnamefont {T.}~\bibnamefont {Taniguchi}}, \bibinfo {author}
  {\bibfnamefont {P.}~\bibnamefont {{Jarillo-Herrero}}},\ and\ \bibinfo
  {author} {\bibfnamefont {R.~C.}\ \bibnamefont {Ashoori}},\ }\href
  {https://doi.org/10.1103/PhysRevLett.123.046601} {\bibfield  {journal}
  {\bibinfo  {journal} {Phys. Rev. Lett.}\ }\textbf {\bibinfo {volume} {123}},\
  \bibinfo {pages} {046601} (\bibinfo {year} {2019})}\BibitemShut {NoStop}%
\bibitem [{\citenamefont {Zondiner}\ \emph {et~al.}(2020)\citenamefont
  {Zondiner}, \citenamefont {Rozen}, \citenamefont {{Rodan-Legrain}},
  \citenamefont {Cao}, \citenamefont {Queiroz}, \citenamefont {Taniguchi},
  \citenamefont {Watanabe}, \citenamefont {Oreg}, \citenamefont {{von Oppen}},
  \citenamefont {Stern}, \citenamefont {Berg}, \citenamefont
  {{Jarillo-Herrero}},\ and\ \citenamefont {Ilani}}]{ZON20}%
  \BibitemOpen
  \bibfield  {author} {\bibinfo {author} {\bibfnamefont {U.}~\bibnamefont
  {Zondiner}}, \bibinfo {author} {\bibfnamefont {A.}~\bibnamefont {Rozen}},
  \bibinfo {author} {\bibfnamefont {D.}~\bibnamefont {{Rodan-Legrain}}},
  \bibinfo {author} {\bibfnamefont {Y.}~\bibnamefont {Cao}}, \bibinfo {author}
  {\bibfnamefont {R.}~\bibnamefont {Queiroz}}, \bibinfo {author} {\bibfnamefont
  {T.}~\bibnamefont {Taniguchi}}, \bibinfo {author} {\bibfnamefont
  {K.}~\bibnamefont {Watanabe}}, \bibinfo {author} {\bibfnamefont
  {Y.}~\bibnamefont {Oreg}}, \bibinfo {author} {\bibfnamefont {F.}~\bibnamefont
  {{von Oppen}}}, \bibinfo {author} {\bibfnamefont {A.}~\bibnamefont {Stern}},
  \bibinfo {author} {\bibfnamefont {E.}~\bibnamefont {Berg}}, \bibinfo {author}
  {\bibfnamefont {P.}~\bibnamefont {{Jarillo-Herrero}}},\ and\ \bibinfo
  {author} {\bibfnamefont {S.}~\bibnamefont {Ilani}},\ }\href
  {https://doi.org/10.1038/s41586-020-2373-y} {\bibfield  {journal} {\bibinfo
  {journal} {Nature}\ }\textbf {\bibinfo {volume} {582}},\ \bibinfo {pages}
  {203} (\bibinfo {year} {2020})}\BibitemShut {NoStop}%
\bibitem [{\citenamefont {Lisi}\ \emph {et~al.}(2021)\citenamefont {Lisi},
  \citenamefont {Lu}, \citenamefont {Benschop}, \citenamefont {{de Jong}},
  \citenamefont {Stepanov}, \citenamefont {Duran}, \citenamefont {Margot},
  \citenamefont {Cucchi}, \citenamefont {Cappelli}, \citenamefont {Hunter},
  \citenamefont {Tamai}, \citenamefont {Kandyba}, \citenamefont {Giampietri},
  \citenamefont {Barinov}, \citenamefont {Jobst}, \citenamefont {Stalman},
  \citenamefont {Leeuwenhoek}, \citenamefont {Watanabe}, \citenamefont
  {Taniguchi}, \citenamefont {Rademaker}, \citenamefont {{van der Molen}},
  \citenamefont {Allan}, \citenamefont {Efetov},\ and\ \citenamefont
  {Baumberger}}]{LIS21}%
  \BibitemOpen
  \bibfield  {author} {\bibinfo {author} {\bibfnamefont {S.}~\bibnamefont
  {Lisi}}, \bibinfo {author} {\bibfnamefont {X.}~\bibnamefont {Lu}}, \bibinfo
  {author} {\bibfnamefont {T.}~\bibnamefont {Benschop}}, \bibinfo {author}
  {\bibfnamefont {T.~A.}\ \bibnamefont {{de Jong}}}, \bibinfo {author}
  {\bibfnamefont {P.}~\bibnamefont {Stepanov}}, \bibinfo {author}
  {\bibfnamefont {J.~R.}\ \bibnamefont {Duran}}, \bibinfo {author}
  {\bibfnamefont {F.}~\bibnamefont {Margot}}, \bibinfo {author} {\bibfnamefont
  {I.}~\bibnamefont {Cucchi}}, \bibinfo {author} {\bibfnamefont
  {E.}~\bibnamefont {Cappelli}}, \bibinfo {author} {\bibfnamefont
  {A.}~\bibnamefont {Hunter}}, \bibinfo {author} {\bibfnamefont
  {A.}~\bibnamefont {Tamai}}, \bibinfo {author} {\bibfnamefont
  {V.}~\bibnamefont {Kandyba}}, \bibinfo {author} {\bibfnamefont
  {A.}~\bibnamefont {Giampietri}}, \bibinfo {author} {\bibfnamefont
  {A.}~\bibnamefont {Barinov}}, \bibinfo {author} {\bibfnamefont
  {J.}~\bibnamefont {Jobst}}, \bibinfo {author} {\bibfnamefont
  {V.}~\bibnamefont {Stalman}}, \bibinfo {author} {\bibfnamefont
  {M.}~\bibnamefont {Leeuwenhoek}}, \bibinfo {author} {\bibfnamefont
  {K.}~\bibnamefont {Watanabe}}, \bibinfo {author} {\bibfnamefont
  {T.}~\bibnamefont {Taniguchi}}, \bibinfo {author} {\bibfnamefont
  {L.}~\bibnamefont {Rademaker}}, \bibinfo {author} {\bibfnamefont {S.~J.}\
  \bibnamefont {{van der Molen}}}, \bibinfo {author} {\bibfnamefont {M.~P.}\
  \bibnamefont {Allan}}, \bibinfo {author} {\bibfnamefont {D.~K.}\ \bibnamefont
  {Efetov}},\ and\ \bibinfo {author} {\bibfnamefont {F.}~\bibnamefont
  {Baumberger}},\ }\href {https://doi.org/10.1038/s41567-020-01041-x}
  {\bibfield  {journal} {\bibinfo  {journal} {Nat. Phys.}\ }\textbf {\bibinfo
  {volume} {17}},\ \bibinfo {pages} {189} (\bibinfo {year} {2021})}\BibitemShut
  {NoStop}%
\bibitem [{\citenamefont {Chen}\ \emph {et~al.}(2024)\citenamefont {Chen},
  \citenamefont {Nuckolls}, \citenamefont {Ding}, \citenamefont {Miao},
  \citenamefont {Wong}, \citenamefont {Oh}, \citenamefont {Lee}, \citenamefont
  {He}, \citenamefont {Peng}, \citenamefont {Pei}, \citenamefont {Li},
  \citenamefont {Hao}, \citenamefont {Yan}, \citenamefont {Xiao}, \citenamefont
  {Gao}, \citenamefont {Li}, \citenamefont {Zhang}, \citenamefont {Liu},
  \citenamefont {He}, \citenamefont {Watanabe}, \citenamefont {Taniguchi},
  \citenamefont {Jozwiak}, \citenamefont {Bostwick}, \citenamefont {Rotenberg},
  \citenamefont {Li}, \citenamefont {Han}, \citenamefont {Pan}, \citenamefont
  {Liu}, \citenamefont {Dai}, \citenamefont {Liu}, \citenamefont {Bernevig},
  \citenamefont {Wang}, \citenamefont {Yazdani},\ and\ \citenamefont
  {Chen}}]{CHE24b}%
  \BibitemOpen
  \bibfield  {author} {\bibinfo {author} {\bibfnamefont {C.}~\bibnamefont
  {Chen}}, \bibinfo {author} {\bibfnamefont {K.~P.}\ \bibnamefont {Nuckolls}},
  \bibinfo {author} {\bibfnamefont {S.}~\bibnamefont {Ding}}, \bibinfo {author}
  {\bibfnamefont {W.}~\bibnamefont {Miao}}, \bibinfo {author} {\bibfnamefont
  {D.}~\bibnamefont {Wong}}, \bibinfo {author} {\bibfnamefont {M.}~\bibnamefont
  {Oh}}, \bibinfo {author} {\bibfnamefont {R.~L.}\ \bibnamefont {Lee}},
  \bibinfo {author} {\bibfnamefont {S.}~\bibnamefont {He}}, \bibinfo {author}
  {\bibfnamefont {C.}~\bibnamefont {Peng}}, \bibinfo {author} {\bibfnamefont
  {D.}~\bibnamefont {Pei}}, \bibinfo {author} {\bibfnamefont {Y.}~\bibnamefont
  {Li}}, \bibinfo {author} {\bibfnamefont {C.}~\bibnamefont {Hao}}, \bibinfo
  {author} {\bibfnamefont {H.}~\bibnamefont {Yan}}, \bibinfo {author}
  {\bibfnamefont {H.}~\bibnamefont {Xiao}}, \bibinfo {author} {\bibfnamefont
  {H.}~\bibnamefont {Gao}}, \bibinfo {author} {\bibfnamefont {Q.}~\bibnamefont
  {Li}}, \bibinfo {author} {\bibfnamefont {S.}~\bibnamefont {Zhang}}, \bibinfo
  {author} {\bibfnamefont {J.}~\bibnamefont {Liu}}, \bibinfo {author}
  {\bibfnamefont {L.}~\bibnamefont {He}}, \bibinfo {author} {\bibfnamefont
  {K.}~\bibnamefont {Watanabe}}, \bibinfo {author} {\bibfnamefont
  {T.}~\bibnamefont {Taniguchi}}, \bibinfo {author} {\bibfnamefont
  {C.}~\bibnamefont {Jozwiak}}, \bibinfo {author} {\bibfnamefont
  {A.}~\bibnamefont {Bostwick}}, \bibinfo {author} {\bibfnamefont
  {E.}~\bibnamefont {Rotenberg}}, \bibinfo {author} {\bibfnamefont
  {C.}~\bibnamefont {Li}}, \bibinfo {author} {\bibfnamefont {X.}~\bibnamefont
  {Han}}, \bibinfo {author} {\bibfnamefont {D.}~\bibnamefont {Pan}}, \bibinfo
  {author} {\bibfnamefont {Z.}~\bibnamefont {Liu}}, \bibinfo {author}
  {\bibfnamefont {X.}~\bibnamefont {Dai}}, \bibinfo {author} {\bibfnamefont
  {C.}~\bibnamefont {Liu}}, \bibinfo {author} {\bibfnamefont {B.~A.}\
  \bibnamefont {Bernevig}}, \bibinfo {author} {\bibfnamefont {Y.}~\bibnamefont
  {Wang}}, \bibinfo {author} {\bibfnamefont {A.}~\bibnamefont {Yazdani}},\ and\
  \bibinfo {author} {\bibfnamefont {Y.}~\bibnamefont {Chen}},\ }\href
  {https://doi.org/10.1038/s41586-024-08227-w} {\bibfield  {journal} {\bibinfo
  {journal} {Nature}\ }\textbf {\bibinfo {volume} {636}},\ \bibinfo {pages}
  {342} (\bibinfo {year} {2024})}\BibitemShut {NoStop}%
\bibitem [{\citenamefont {Seyler}\ \emph {et~al.}(2019)\citenamefont {Seyler},
  \citenamefont {Rivera}, \citenamefont {Yu}, \citenamefont {Wilson},
  \citenamefont {Ray}, \citenamefont {Mandrus}, \citenamefont {Yan},
  \citenamefont {Yao},\ and\ \citenamefont {Xu}}]{SEY19}%
  \BibitemOpen
  \bibfield  {author} {\bibinfo {author} {\bibfnamefont {K.~L.}\ \bibnamefont
  {Seyler}}, \bibinfo {author} {\bibfnamefont {P.}~\bibnamefont {Rivera}},
  \bibinfo {author} {\bibfnamefont {H.}~\bibnamefont {Yu}}, \bibinfo {author}
  {\bibfnamefont {N.~P.}\ \bibnamefont {Wilson}}, \bibinfo {author}
  {\bibfnamefont {E.~L.}\ \bibnamefont {Ray}}, \bibinfo {author} {\bibfnamefont
  {D.~G.}\ \bibnamefont {Mandrus}}, \bibinfo {author} {\bibfnamefont
  {J.}~\bibnamefont {Yan}}, \bibinfo {author} {\bibfnamefont {W.}~\bibnamefont
  {Yao}},\ and\ \bibinfo {author} {\bibfnamefont {X.}~\bibnamefont {Xu}},\
  }\href {https://doi.org/10.1038/s41586-019-0957-1} {\bibfield  {journal}
  {\bibinfo  {journal} {Nature}\ }\textbf {\bibinfo {volume} {567}},\ \bibinfo
  {pages} {66} (\bibinfo {year} {2019})}\BibitemShut {NoStop}%
\bibitem [{\citenamefont {Tran}\ \emph {et~al.}(2019)\citenamefont {Tran},
  \citenamefont {Moody}, \citenamefont {Wu}, \citenamefont {Lu}, \citenamefont
  {Choi}, \citenamefont {Kim}, \citenamefont {Rai}, \citenamefont {Sanchez},
  \citenamefont {Quan}, \citenamefont {Singh}, \citenamefont {Embley},
  \citenamefont {Zepeda}, \citenamefont {Campbell}, \citenamefont {Autry},
  \citenamefont {Taniguchi}, \citenamefont {Watanabe}, \citenamefont {Lu},
  \citenamefont {Banerjee}, \citenamefont {Silverman}, \citenamefont {Kim},
  \citenamefont {Tutuc}, \citenamefont {Yang}, \citenamefont {MacDonald},\ and\
  \citenamefont {Li}}]{TRA19}%
  \BibitemOpen
  \bibfield  {author} {\bibinfo {author} {\bibfnamefont {K.}~\bibnamefont
  {Tran}}, \bibinfo {author} {\bibfnamefont {G.}~\bibnamefont {Moody}},
  \bibinfo {author} {\bibfnamefont {F.}~\bibnamefont {Wu}}, \bibinfo {author}
  {\bibfnamefont {X.}~\bibnamefont {Lu}}, \bibinfo {author} {\bibfnamefont
  {J.}~\bibnamefont {Choi}}, \bibinfo {author} {\bibfnamefont {K.}~\bibnamefont
  {Kim}}, \bibinfo {author} {\bibfnamefont {A.}~\bibnamefont {Rai}}, \bibinfo
  {author} {\bibfnamefont {D.~A.}\ \bibnamefont {Sanchez}}, \bibinfo {author}
  {\bibfnamefont {J.}~\bibnamefont {Quan}}, \bibinfo {author} {\bibfnamefont
  {A.}~\bibnamefont {Singh}}, \bibinfo {author} {\bibfnamefont
  {J.}~\bibnamefont {Embley}}, \bibinfo {author} {\bibfnamefont
  {A.}~\bibnamefont {Zepeda}}, \bibinfo {author} {\bibfnamefont
  {M.}~\bibnamefont {Campbell}}, \bibinfo {author} {\bibfnamefont
  {T.}~\bibnamefont {Autry}}, \bibinfo {author} {\bibfnamefont
  {T.}~\bibnamefont {Taniguchi}}, \bibinfo {author} {\bibfnamefont
  {K.}~\bibnamefont {Watanabe}}, \bibinfo {author} {\bibfnamefont
  {N.}~\bibnamefont {Lu}}, \bibinfo {author} {\bibfnamefont {S.~K.}\
  \bibnamefont {Banerjee}}, \bibinfo {author} {\bibfnamefont {K.~L.}\
  \bibnamefont {Silverman}}, \bibinfo {author} {\bibfnamefont {S.}~\bibnamefont
  {Kim}}, \bibinfo {author} {\bibfnamefont {E.}~\bibnamefont {Tutuc}}, \bibinfo
  {author} {\bibfnamefont {L.}~\bibnamefont {Yang}}, \bibinfo {author}
  {\bibfnamefont {A.~H.}\ \bibnamefont {MacDonald}},\ and\ \bibinfo {author}
  {\bibfnamefont {X.}~\bibnamefont {Li}},\ }\href
  {https://doi.org/10.1038/s41586-019-0975-z} {\bibfield  {journal} {\bibinfo
  {journal} {Nature}\ }\textbf {\bibinfo {volume} {567}},\ \bibinfo {pages}
  {71} (\bibinfo {year} {2019})}\BibitemShut {NoStop}%
\bibitem [{\citenamefont {Jin}\ \emph {et~al.}(2019)\citenamefont {Jin},
  \citenamefont {Regan}, \citenamefont {Yan}, \citenamefont {Iqbal
  Bakti~Utama}, \citenamefont {Wang}, \citenamefont {Zhao}, \citenamefont
  {Qin}, \citenamefont {Yang}, \citenamefont {Zheng}, \citenamefont {Shi},
  \citenamefont {Watanabe}, \citenamefont {Taniguchi}, \citenamefont {Tongay},
  \citenamefont {Zettl},\ and\ \citenamefont {Wang}}]{JIN19}%
  \BibitemOpen
  \bibfield  {author} {\bibinfo {author} {\bibfnamefont {C.}~\bibnamefont
  {Jin}}, \bibinfo {author} {\bibfnamefont {E.~C.}\ \bibnamefont {Regan}},
  \bibinfo {author} {\bibfnamefont {A.}~\bibnamefont {Yan}}, \bibinfo {author}
  {\bibfnamefont {M.}~\bibnamefont {Iqbal Bakti~Utama}}, \bibinfo {author}
  {\bibfnamefont {D.}~\bibnamefont {Wang}}, \bibinfo {author} {\bibfnamefont
  {S.}~\bibnamefont {Zhao}}, \bibinfo {author} {\bibfnamefont {Y.}~\bibnamefont
  {Qin}}, \bibinfo {author} {\bibfnamefont {S.}~\bibnamefont {Yang}}, \bibinfo
  {author} {\bibfnamefont {Z.}~\bibnamefont {Zheng}}, \bibinfo {author}
  {\bibfnamefont {S.}~\bibnamefont {Shi}}, \bibinfo {author} {\bibfnamefont
  {K.}~\bibnamefont {Watanabe}}, \bibinfo {author} {\bibfnamefont
  {T.}~\bibnamefont {Taniguchi}}, \bibinfo {author} {\bibfnamefont
  {S.}~\bibnamefont {Tongay}}, \bibinfo {author} {\bibfnamefont
  {A.}~\bibnamefont {Zettl}},\ and\ \bibinfo {author} {\bibfnamefont
  {F.}~\bibnamefont {Wang}},\ }\href
  {https://doi.org/10.1038/s41586-019-0976-y} {\bibfield  {journal} {\bibinfo
  {journal} {Nature}\ }\textbf {\bibinfo {volume} {567}},\ \bibinfo {pages}
  {76} (\bibinfo {year} {2019})}\BibitemShut {NoStop}%
\bibitem [{\citenamefont {Inbar}\ \emph {et~al.}(2023)\citenamefont {Inbar},
  \citenamefont {Birkbeck}, \citenamefont {Xiao}, \citenamefont {Taniguchi},
  \citenamefont {Watanabe}, \citenamefont {Yan}, \citenamefont {Oreg},
  \citenamefont {Stern}, \citenamefont {Berg},\ and\ \citenamefont
  {Ilani}}]{INB23}%
  \BibitemOpen
  \bibfield  {author} {\bibinfo {author} {\bibfnamefont {A.}~\bibnamefont
  {Inbar}}, \bibinfo {author} {\bibfnamefont {J.}~\bibnamefont {Birkbeck}},
  \bibinfo {author} {\bibfnamefont {J.}~\bibnamefont {Xiao}}, \bibinfo {author}
  {\bibfnamefont {T.}~\bibnamefont {Taniguchi}}, \bibinfo {author}
  {\bibfnamefont {K.}~\bibnamefont {Watanabe}}, \bibinfo {author}
  {\bibfnamefont {B.}~\bibnamefont {Yan}}, \bibinfo {author} {\bibfnamefont
  {Y.}~\bibnamefont {Oreg}}, \bibinfo {author} {\bibfnamefont {A.}~\bibnamefont
  {Stern}}, \bibinfo {author} {\bibfnamefont {E.}~\bibnamefont {Berg}},\ and\
  \bibinfo {author} {\bibfnamefont {S.}~\bibnamefont {Ilani}},\ }\href
  {https://doi.org/10.1038/s41586-022-05685-y} {\bibfield  {journal} {\bibinfo
  {journal} {Nature}\ }\textbf {\bibinfo {volume} {614}},\ \bibinfo {pages}
  {682} (\bibinfo {year} {2023})}\BibitemShut {NoStop}%
\bibitem [{\citenamefont {Xiao}\ \emph {et~al.}(2026)\citenamefont {Xiao},
  \citenamefont {Inbar}, \citenamefont {Birkbeck}, \citenamefont {Gershon},
  \citenamefont {Zamir}, \citenamefont {Vituri}, \citenamefont {Taniguchi},
  \citenamefont {Watanabe}, \citenamefont {Berg},\ and\ \citenamefont
  {Ilani}}]{XIA25}%
  \BibitemOpen
  \bibfield  {author} {\bibinfo {author} {\bibfnamefont {J.}~\bibnamefont
  {Xiao}}, \bibinfo {author} {\bibfnamefont {A.}~\bibnamefont {Inbar}},
  \bibinfo {author} {\bibfnamefont {J.}~\bibnamefont {Birkbeck}}, \bibinfo
  {author} {\bibfnamefont {N.}~\bibnamefont {Gershon}}, \bibinfo {author}
  {\bibfnamefont {Y.}~\bibnamefont {Zamir}}, \bibinfo {author} {\bibfnamefont
  {Y.}~\bibnamefont {Vituri}}, \bibinfo {author} {\bibfnamefont
  {T.}~\bibnamefont {Taniguchi}}, \bibinfo {author} {\bibfnamefont
  {K.}~\bibnamefont {Watanabe}}, \bibinfo {author} {\bibfnamefont
  {E.}~\bibnamefont {Berg}},\ and\ \bibinfo {author} {\bibfnamefont
  {S.}~\bibnamefont {Ilani}},\ }\href
  {https://doi.org/10.1038/s41586-026-10378-x} {\bibfield  {journal} {\bibinfo
  {journal} {Nature}\ }\textbf {\bibinfo {volume} {653}},\ \bibinfo {pages}
  {68} (\bibinfo {year} {2026})}\BibitemShut {NoStop}%
\bibitem [{\citenamefont {Angeli}\ and\ \citenamefont
  {MacDonald}(2021)}]{ANG21}%
  \BibitemOpen
  \bibfield  {author} {\bibinfo {author} {\bibfnamefont {M.}~\bibnamefont
  {Angeli}}\ and\ \bibinfo {author} {\bibfnamefont {A.~H.}\ \bibnamefont
  {MacDonald}},\ }\href {https://doi.org/10.1073/pnas.2021826118} {\bibfield
  {journal} {\bibinfo  {journal} {PNAS}\ }\textbf {\bibinfo {volume} {118}},\
  \bibinfo {pages} {e2021826118} (\bibinfo {year} {2021})}\BibitemShut
  {NoStop}%
\bibitem [{\citenamefont {Claassen}\ \emph {et~al.}(2022)\citenamefont
  {Claassen}, \citenamefont {Xian}, \citenamefont {Kennes},\ and\ \citenamefont
  {Rubio}}]{CLA22a}%
  \BibitemOpen
  \bibfield  {author} {\bibinfo {author} {\bibfnamefont {M.}~\bibnamefont
  {Claassen}}, \bibinfo {author} {\bibfnamefont {L.}~\bibnamefont {Xian}},
  \bibinfo {author} {\bibfnamefont {D.~M.}\ \bibnamefont {Kennes}},\ and\
  \bibinfo {author} {\bibfnamefont {A.}~\bibnamefont {Rubio}},\ }\href
  {https://doi.org/10.1038/s41467-022-31604-w} {\bibfield  {journal} {\bibinfo
  {journal} {Nat. Commun.}\ }\textbf {\bibinfo {volume} {13}},\ \bibinfo
  {pages} {4915} (\bibinfo {year} {2022})}\BibitemShut {NoStop}%
\bibitem [{\citenamefont {Pi}\ \emph {et~al.}(2026)\citenamefont {Pi},
  \citenamefont {Kwan}, \citenamefont {Hu}, \citenamefont {Jiang},
  \citenamefont {C{\u a}lug{\u a}ru}, \citenamefont {Shan}, \citenamefont
  {Mak}, \citenamefont {Ugeda}, \citenamefont {Efetov}, \citenamefont
  {Vergniory},\ and\ \citenamefont {Bernevig}}]{PI26}%
  \BibitemOpen
  \bibfield  {author} {\bibinfo {author} {\bibfnamefont {H.}~\bibnamefont
  {Pi}}, \bibinfo {author} {\bibfnamefont {Y.~H.}\ \bibnamefont {Kwan}},
  \bibinfo {author} {\bibfnamefont {H.}~\bibnamefont {Hu}}, \bibinfo {author}
  {\bibfnamefont {Y.}~\bibnamefont {Jiang}}, \bibinfo {author} {\bibfnamefont
  {D.}~\bibnamefont {C{\u a}lug{\u a}ru}}, \bibinfo {author} {\bibfnamefont
  {J.}~\bibnamefont {Shan}}, \bibinfo {author} {\bibfnamefont {K.~F.}\
  \bibnamefont {Mak}}, \bibinfo {author} {\bibfnamefont {M.~M.}\ \bibnamefont
  {Ugeda}}, \bibinfo {author} {\bibfnamefont {D.~K.}\ \bibnamefont {Efetov}},
  \bibinfo {author} {\bibfnamefont {M.~G.}\ \bibnamefont {Vergniory}},\ and\
  \bibinfo {author} {\bibfnamefont {B.~A.}\ \bibnamefont {Bernevig}},\ }\href
  {https://doi.org/10.48550/arXiv.2605.13984} {\bibinfo {title} {Engineering
  topological flat bands in ${{\Gamma}}$-valley moir\'e systems with
  {{Ising-type SOC}}: Twisted {{1T-ZrS}}$_2$ and {{1T-SnSe}}$_2$}} (\bibinfo
  {year} {2026}),\ \Eprint {https://arxiv.org/abs/2605.13984} {arXiv:2605.13984
  [cond-mat.mtrl-sci]} \BibitemShut {NoStop}%
\bibitem [{\citenamefont {Bistritzer}\ and\ \citenamefont
  {MacDonald}(2011)}]{BIS11}%
  \BibitemOpen
  \bibfield  {author} {\bibinfo {author} {\bibfnamefont {R.}~\bibnamefont
  {Bistritzer}}\ and\ \bibinfo {author} {\bibfnamefont {A.~H.}\ \bibnamefont
  {MacDonald}},\ }\href {https://doi.org/10.1073/pnas.1108174108} {\bibfield
  {journal} {\bibinfo  {journal} {PNAS}\ }\textbf {\bibinfo {volume} {108}},\
  \bibinfo {pages} {12233} (\bibinfo {year} {2011})}\BibitemShut {NoStop}%
\bibitem [{\citenamefont {Su{\'a}rez~Morell}\ \emph {et~al.}(2010)\citenamefont
  {Su{\'a}rez~Morell}, \citenamefont {Correa}, \citenamefont {Vargas},
  \citenamefont {Pacheco},\ and\ \citenamefont {Barticevic}}]{SUA10}%
  \BibitemOpen
  \bibfield  {author} {\bibinfo {author} {\bibfnamefont {E.}~\bibnamefont
  {Su{\'a}rez~Morell}}, \bibinfo {author} {\bibfnamefont {J.~D.}\ \bibnamefont
  {Correa}}, \bibinfo {author} {\bibfnamefont {P.}~\bibnamefont {Vargas}},
  \bibinfo {author} {\bibfnamefont {M.}~\bibnamefont {Pacheco}},\ and\ \bibinfo
  {author} {\bibfnamefont {Z.}~\bibnamefont {Barticevic}},\ }\href
  {https://doi.org/10.1103/PhysRevB.82.121407} {\bibfield  {journal} {\bibinfo
  {journal} {Phys. Rev. B}\ }\textbf {\bibinfo {volume} {82}},\ \bibinfo
  {pages} {121407} (\bibinfo {year} {2010})}\BibitemShut {NoStop}%
\bibitem [{\citenamefont {{Lopes~dos~Santos}}\ \emph
  {et~al.}(2007)\citenamefont {{Lopes~dos~Santos}}, \citenamefont {Peres},\
  and\ \citenamefont {Castro~Neto}}]{LOP07}%
  \BibitemOpen
  \bibfield  {author} {\bibinfo {author} {\bibfnamefont {J.~M.~B.}\
  \bibnamefont {{Lopes~dos~Santos}}}, \bibinfo {author} {\bibfnamefont
  {N.~M.~R.}\ \bibnamefont {Peres}},\ and\ \bibinfo {author} {\bibfnamefont
  {A.~H.}\ \bibnamefont {Castro~Neto}},\ }\href
  {https://doi.org/10.1103/PhysRevLett.99.256802} {\bibfield  {journal}
  {\bibinfo  {journal} {Phys. Rev. Lett.}\ }\textbf {\bibinfo {volume} {99}},\
  \bibinfo {pages} {256802} (\bibinfo {year} {2007})}\BibitemShut {NoStop}%
\bibitem [{\citenamefont {Song}\ and\ \citenamefont {Bernevig}(2022)}]{SON22}%
  \BibitemOpen
  \bibfield  {author} {\bibinfo {author} {\bibfnamefont {Z.-D.}\ \bibnamefont
  {Song}}\ and\ \bibinfo {author} {\bibfnamefont {B.~A.}\ \bibnamefont
  {Bernevig}},\ }\href {https://doi.org/10.1103/PhysRevLett.129.047601}
  {\bibfield  {journal} {\bibinfo  {journal} {Phys. Rev. Lett.}\ }\textbf
  {\bibinfo {volume} {129}},\ \bibinfo {pages} {047601} (\bibinfo {year}
  {2022})}\BibitemShut {NoStop}%
\bibitem [{\citenamefont {Lu}\ \emph {et~al.}(2024)\citenamefont {Lu},
  \citenamefont {Han}, \citenamefont {Yao}, \citenamefont {Reddy},
  \citenamefont {Yang}, \citenamefont {Seo}, \citenamefont {Watanabe},
  \citenamefont {Taniguchi}, \citenamefont {Fu},\ and\ \citenamefont
  {Ju}}]{LU24}%
  \BibitemOpen
  \bibfield  {author} {\bibinfo {author} {\bibfnamefont {Z.}~\bibnamefont
  {Lu}}, \bibinfo {author} {\bibfnamefont {T.}~\bibnamefont {Han}}, \bibinfo
  {author} {\bibfnamefont {Y.}~\bibnamefont {Yao}}, \bibinfo {author}
  {\bibfnamefont {A.~P.}\ \bibnamefont {Reddy}}, \bibinfo {author}
  {\bibfnamefont {J.}~\bibnamefont {Yang}}, \bibinfo {author} {\bibfnamefont
  {J.}~\bibnamefont {Seo}}, \bibinfo {author} {\bibfnamefont {K.}~\bibnamefont
  {Watanabe}}, \bibinfo {author} {\bibfnamefont {T.}~\bibnamefont {Taniguchi}},
  \bibinfo {author} {\bibfnamefont {L.}~\bibnamefont {Fu}},\ and\ \bibinfo
  {author} {\bibfnamefont {L.}~\bibnamefont {Ju}},\ }\href
  {https://doi.org/10.1038/s41586-023-07010-7} {\bibfield  {journal} {\bibinfo
  {journal} {Nature}\ }\textbf {\bibinfo {volume} {626}},\ \bibinfo {pages}
  {759} (\bibinfo {year} {2024})}\BibitemShut {NoStop}%
\bibitem [{\citenamefont {Dong}\ \emph
  {et~al.}(2024{\natexlab{a}})\citenamefont {Dong}, \citenamefont {Wang},
  \citenamefont {Wang}, \citenamefont {Soejima}, \citenamefont {Zaletel},
  \citenamefont {Vishwanath},\ and\ \citenamefont {Parker}}]{DON23}%
  \BibitemOpen
  \bibfield  {author} {\bibinfo {author} {\bibfnamefont {J.}~\bibnamefont
  {Dong}}, \bibinfo {author} {\bibfnamefont {T.}~\bibnamefont {Wang}}, \bibinfo
  {author} {\bibfnamefont {T.}~\bibnamefont {Wang}}, \bibinfo {author}
  {\bibfnamefont {T.}~\bibnamefont {Soejima}}, \bibinfo {author} {\bibfnamefont
  {M.~P.}\ \bibnamefont {Zaletel}}, \bibinfo {author} {\bibfnamefont
  {A.}~\bibnamefont {Vishwanath}},\ and\ \bibinfo {author} {\bibfnamefont
  {D.~E.}\ \bibnamefont {Parker}},\ }\href
  {https://doi.org/10.1103/PhysRevLett.133.206503} {\bibfield  {journal}
  {\bibinfo  {journal} {Phys. Rev. Lett.}\ }\textbf {\bibinfo {volume} {133}},\
  \bibinfo {pages} {206503} (\bibinfo {year} {2024}{\natexlab{a}})}\BibitemShut
  {NoStop}%
\bibitem [{\citenamefont {Dong}\ \emph
  {et~al.}(2024{\natexlab{b}})\citenamefont {Dong}, \citenamefont {Patri},\
  and\ \citenamefont {Senthil}}]{DON24}%
  \BibitemOpen
  \bibfield  {author} {\bibinfo {author} {\bibfnamefont {Z.}~\bibnamefont
  {Dong}}, \bibinfo {author} {\bibfnamefont {A.~S.}\ \bibnamefont {Patri}},\
  and\ \bibinfo {author} {\bibfnamefont {T.}~\bibnamefont {Senthil}},\ }\href
  {https://doi.org/10.1103/PhysRevLett.133.206502} {\bibfield  {journal}
  {\bibinfo  {journal} {Phys. Rev. Lett.}\ }\textbf {\bibinfo {volume} {133}},\
  \bibinfo {pages} {206502} (\bibinfo {year} {2024}{\natexlab{b}})}\BibitemShut
  {NoStop}%
\bibitem [{\citenamefont {{Herzog-Arbeitman}}\ \emph
  {et~al.}(2024)\citenamefont {{Herzog-Arbeitman}}, \citenamefont {Wang},
  \citenamefont {Liu}, \citenamefont {Tam}, \citenamefont {Qi}, \citenamefont
  {Jia}, \citenamefont {Efetov}, \citenamefont {Vafek}, \citenamefont
  {Regnault}, \citenamefont {Weng}, \citenamefont {Wu}, \citenamefont
  {Bernevig},\ and\ \citenamefont {Yu}}]{HER24b}%
  \BibitemOpen
  \bibfield  {author} {\bibinfo {author} {\bibfnamefont {J.}~\bibnamefont
  {{Herzog-Arbeitman}}}, \bibinfo {author} {\bibfnamefont {Y.}~\bibnamefont
  {Wang}}, \bibinfo {author} {\bibfnamefont {J.}~\bibnamefont {Liu}}, \bibinfo
  {author} {\bibfnamefont {P.~M.}\ \bibnamefont {Tam}}, \bibinfo {author}
  {\bibfnamefont {Z.}~\bibnamefont {Qi}}, \bibinfo {author} {\bibfnamefont
  {Y.}~\bibnamefont {Jia}}, \bibinfo {author} {\bibfnamefont {D.~K.}\
  \bibnamefont {Efetov}}, \bibinfo {author} {\bibfnamefont {O.}~\bibnamefont
  {Vafek}}, \bibinfo {author} {\bibfnamefont {N.}~\bibnamefont {Regnault}},
  \bibinfo {author} {\bibfnamefont {H.}~\bibnamefont {Weng}}, \bibinfo {author}
  {\bibfnamefont {Q.}~\bibnamefont {Wu}}, \bibinfo {author} {\bibfnamefont
  {B.~A.}\ \bibnamefont {Bernevig}},\ and\ \bibinfo {author} {\bibfnamefont
  {J.}~\bibnamefont {Yu}},\ }\href
  {https://doi.org/10.1103/PhysRevB.109.205122} {\bibfield  {journal} {\bibinfo
   {journal} {Phys. Rev. B}\ }\textbf {\bibinfo {volume} {109}},\ \bibinfo
  {pages} {205122} (\bibinfo {year} {2024})}\BibitemShut {NoStop}%
\bibitem [{\citenamefont {Han}\ \emph {et~al.}(2025)\citenamefont {Han},
  \citenamefont {Lu}, \citenamefont {Hadjri}, \citenamefont {Shi},
  \citenamefont {Wu}, \citenamefont {Xu}, \citenamefont {Yao}, \citenamefont
  {Cotten}, \citenamefont {Sharifi~Sedeh}, \citenamefont {Weldeyesus},
  \citenamefont {Yang}, \citenamefont {Seo}, \citenamefont {Ye}, \citenamefont
  {Zhou}, \citenamefont {Liu}, \citenamefont {Shi}, \citenamefont {Hua},
  \citenamefont {Watanabe}, \citenamefont {Taniguchi}, \citenamefont {Xiong},
  \citenamefont {Zumb{\"u}hl}, \citenamefont {Fu},\ and\ \citenamefont
  {Ju}}]{HAN25}%
  \BibitemOpen
  \bibfield  {author} {\bibinfo {author} {\bibfnamefont {T.}~\bibnamefont
  {Han}}, \bibinfo {author} {\bibfnamefont {Z.}~\bibnamefont {Lu}}, \bibinfo
  {author} {\bibfnamefont {Z.}~\bibnamefont {Hadjri}}, \bibinfo {author}
  {\bibfnamefont {L.}~\bibnamefont {Shi}}, \bibinfo {author} {\bibfnamefont
  {Z.}~\bibnamefont {Wu}}, \bibinfo {author} {\bibfnamefont {W.}~\bibnamefont
  {Xu}}, \bibinfo {author} {\bibfnamefont {Y.}~\bibnamefont {Yao}}, \bibinfo
  {author} {\bibfnamefont {A.~A.}\ \bibnamefont {Cotten}}, \bibinfo {author}
  {\bibfnamefont {O.}~\bibnamefont {Sharifi~Sedeh}}, \bibinfo {author}
  {\bibfnamefont {H.}~\bibnamefont {Weldeyesus}}, \bibinfo {author}
  {\bibfnamefont {J.}~\bibnamefont {Yang}}, \bibinfo {author} {\bibfnamefont
  {J.}~\bibnamefont {Seo}}, \bibinfo {author} {\bibfnamefont {S.}~\bibnamefont
  {Ye}}, \bibinfo {author} {\bibfnamefont {M.}~\bibnamefont {Zhou}}, \bibinfo
  {author} {\bibfnamefont {H.}~\bibnamefont {Liu}}, \bibinfo {author}
  {\bibfnamefont {G.}~\bibnamefont {Shi}}, \bibinfo {author} {\bibfnamefont
  {Z.}~\bibnamefont {Hua}}, \bibinfo {author} {\bibfnamefont {K.}~\bibnamefont
  {Watanabe}}, \bibinfo {author} {\bibfnamefont {T.}~\bibnamefont {Taniguchi}},
  \bibinfo {author} {\bibfnamefont {P.}~\bibnamefont {Xiong}}, \bibinfo
  {author} {\bibfnamefont {D.~M.}\ \bibnamefont {Zumb{\"u}hl}}, \bibinfo
  {author} {\bibfnamefont {L.}~\bibnamefont {Fu}},\ and\ \bibinfo {author}
  {\bibfnamefont {L.}~\bibnamefont {Ju}},\ }\href
  {https://doi.org/10.1038/s41586-025-09169-7} {\bibfield  {journal} {\bibinfo
  {journal} {Nature}\ }\textbf {\bibinfo {volume} {643}},\ \bibinfo {pages}
  {654} (\bibinfo {year} {2025})}\BibitemShut {NoStop}%
\bibitem [{\citenamefont {Wu}\ \emph {et~al.}(2018)\citenamefont {Wu},
  \citenamefont {Lovorn}, \citenamefont {Tutuc},\ and\ \citenamefont
  {MacDonald}}]{WU18c}%
  \BibitemOpen
  \bibfield  {author} {\bibinfo {author} {\bibfnamefont {F.}~\bibnamefont
  {Wu}}, \bibinfo {author} {\bibfnamefont {T.}~\bibnamefont {Lovorn}}, \bibinfo
  {author} {\bibfnamefont {E.}~\bibnamefont {Tutuc}},\ and\ \bibinfo {author}
  {\bibfnamefont {A.~H.}\ \bibnamefont {MacDonald}},\ }\href
  {https://doi.org/10.1103/PhysRevLett.121.026402} {\bibfield  {journal}
  {\bibinfo  {journal} {Phys. Rev. Lett.}\ }\textbf {\bibinfo {volume} {121}},\
  \bibinfo {pages} {026402} (\bibinfo {year} {2018})}\BibitemShut {NoStop}%
\bibitem [{\citenamefont {Tang}\ \emph {et~al.}(2020)\citenamefont {Tang},
  \citenamefont {Li}, \citenamefont {Li}, \citenamefont {Xu}, \citenamefont
  {Liu}, \citenamefont {Barmak}, \citenamefont {Watanabe}, \citenamefont
  {Taniguchi}, \citenamefont {MacDonald}, \citenamefont {Shan},\ and\
  \citenamefont {Mak}}]{TAN20}%
  \BibitemOpen
  \bibfield  {author} {\bibinfo {author} {\bibfnamefont {Y.}~\bibnamefont
  {Tang}}, \bibinfo {author} {\bibfnamefont {L.}~\bibnamefont {Li}}, \bibinfo
  {author} {\bibfnamefont {T.}~\bibnamefont {Li}}, \bibinfo {author}
  {\bibfnamefont {Y.}~\bibnamefont {Xu}}, \bibinfo {author} {\bibfnamefont
  {S.}~\bibnamefont {Liu}}, \bibinfo {author} {\bibfnamefont {K.}~\bibnamefont
  {Barmak}}, \bibinfo {author} {\bibfnamefont {K.}~\bibnamefont {Watanabe}},
  \bibinfo {author} {\bibfnamefont {T.}~\bibnamefont {Taniguchi}}, \bibinfo
  {author} {\bibfnamefont {A.~H.}\ \bibnamefont {MacDonald}}, \bibinfo {author}
  {\bibfnamefont {J.}~\bibnamefont {Shan}},\ and\ \bibinfo {author}
  {\bibfnamefont {K.~F.}\ \bibnamefont {Mak}},\ }\href
  {https://doi.org/10.1038/s41586-020-2085-3} {\bibfield  {journal} {\bibinfo
  {journal} {Nature}\ }\textbf {\bibinfo {volume} {579}},\ \bibinfo {pages}
  {353} (\bibinfo {year} {2020})}\BibitemShut {NoStop}%
\bibitem [{\citenamefont {Xu}\ \emph {et~al.}(2020)\citenamefont {Xu},
  \citenamefont {Liu}, \citenamefont {Rhodes}, \citenamefont {Watanabe},
  \citenamefont {Taniguchi}, \citenamefont {Hone}, \citenamefont {Elser},
  \citenamefont {Mak},\ and\ \citenamefont {Shan}}]{XU20a}%
  \BibitemOpen
  \bibfield  {author} {\bibinfo {author} {\bibfnamefont {Y.}~\bibnamefont
  {Xu}}, \bibinfo {author} {\bibfnamefont {S.}~\bibnamefont {Liu}}, \bibinfo
  {author} {\bibfnamefont {D.~A.}\ \bibnamefont {Rhodes}}, \bibinfo {author}
  {\bibfnamefont {K.}~\bibnamefont {Watanabe}}, \bibinfo {author}
  {\bibfnamefont {T.}~\bibnamefont {Taniguchi}}, \bibinfo {author}
  {\bibfnamefont {J.}~\bibnamefont {Hone}}, \bibinfo {author} {\bibfnamefont
  {V.}~\bibnamefont {Elser}}, \bibinfo {author} {\bibfnamefont {K.~F.}\
  \bibnamefont {Mak}},\ and\ \bibinfo {author} {\bibfnamefont {J.}~\bibnamefont
  {Shan}},\ }\href {https://doi.org/10.1038/s41586-020-2868-6} {\bibfield
  {journal} {\bibinfo  {journal} {Nature}\ }\textbf {\bibinfo {volume} {587}},\
  \bibinfo {pages} {214} (\bibinfo {year} {2020})}\BibitemShut {NoStop}%
\bibitem [{\citenamefont {Li}\ \emph {et~al.}(2021)\citenamefont {Li},
  \citenamefont {Jiang}, \citenamefont {Li}, \citenamefont {Zhang},
  \citenamefont {Kang}, \citenamefont {Zhu}, \citenamefont {Watanabe},
  \citenamefont {Taniguchi}, \citenamefont {Chowdhury}, \citenamefont {Fu},
  \citenamefont {Shan},\ and\ \citenamefont {Mak}}]{LI21e}%
  \BibitemOpen
  \bibfield  {author} {\bibinfo {author} {\bibfnamefont {T.}~\bibnamefont
  {Li}}, \bibinfo {author} {\bibfnamefont {S.}~\bibnamefont {Jiang}}, \bibinfo
  {author} {\bibfnamefont {L.}~\bibnamefont {Li}}, \bibinfo {author}
  {\bibfnamefont {Y.}~\bibnamefont {Zhang}}, \bibinfo {author} {\bibfnamefont
  {K.}~\bibnamefont {Kang}}, \bibinfo {author} {\bibfnamefont {J.}~\bibnamefont
  {Zhu}}, \bibinfo {author} {\bibfnamefont {K.}~\bibnamefont {Watanabe}},
  \bibinfo {author} {\bibfnamefont {T.}~\bibnamefont {Taniguchi}}, \bibinfo
  {author} {\bibfnamefont {D.}~\bibnamefont {Chowdhury}}, \bibinfo {author}
  {\bibfnamefont {L.}~\bibnamefont {Fu}}, \bibinfo {author} {\bibfnamefont
  {J.}~\bibnamefont {Shan}},\ and\ \bibinfo {author} {\bibfnamefont {K.~F.}\
  \bibnamefont {Mak}},\ }\href {https://doi.org/10.1038/s41586-021-03853-0}
  {\bibfield  {journal} {\bibinfo  {journal} {Nature}\ }\textbf {\bibinfo
  {volume} {597}},\ \bibinfo {pages} {350} (\bibinfo {year}
  {2021})}\BibitemShut {NoStop}%
\bibitem [{\citenamefont {Wang}\ \emph {et~al.}(2022)\citenamefont {Wang},
  \citenamefont {Yu}, \citenamefont {Kwan}, \citenamefont {Jia}, \citenamefont
  {Lei}, \citenamefont {Klemenz}, \citenamefont {Cevallos}, \citenamefont
  {Singha}, \citenamefont {Devakul}, \citenamefont {Watanabe}, \citenamefont
  {Taniguchi}, \citenamefont {Sondhi}, \citenamefont {Cava}, \citenamefont
  {Schoop}, \citenamefont {Parameswaran},\ and\ \citenamefont {Wu}}]{WAN22}%
  \BibitemOpen
  \bibfield  {author} {\bibinfo {author} {\bibfnamefont {P.}~\bibnamefont
  {Wang}}, \bibinfo {author} {\bibfnamefont {G.}~\bibnamefont {Yu}}, \bibinfo
  {author} {\bibfnamefont {Y.~H.}\ \bibnamefont {Kwan}}, \bibinfo {author}
  {\bibfnamefont {Y.}~\bibnamefont {Jia}}, \bibinfo {author} {\bibfnamefont
  {S.}~\bibnamefont {Lei}}, \bibinfo {author} {\bibfnamefont {S.}~\bibnamefont
  {Klemenz}}, \bibinfo {author} {\bibfnamefont {F.~A.}\ \bibnamefont
  {Cevallos}}, \bibinfo {author} {\bibfnamefont {R.}~\bibnamefont {Singha}},
  \bibinfo {author} {\bibfnamefont {T.}~\bibnamefont {Devakul}}, \bibinfo
  {author} {\bibfnamefont {K.}~\bibnamefont {Watanabe}}, \bibinfo {author}
  {\bibfnamefont {T.}~\bibnamefont {Taniguchi}}, \bibinfo {author}
  {\bibfnamefont {S.~L.}\ \bibnamefont {Sondhi}}, \bibinfo {author}
  {\bibfnamefont {R.~J.}\ \bibnamefont {Cava}}, \bibinfo {author}
  {\bibfnamefont {L.~M.}\ \bibnamefont {Schoop}}, \bibinfo {author}
  {\bibfnamefont {S.~A.}\ \bibnamefont {Parameswaran}},\ and\ \bibinfo {author}
  {\bibfnamefont {S.}~\bibnamefont {Wu}},\ }\href
  {https://doi.org/10.1038/s41586-022-04514-6} {\bibfield  {journal} {\bibinfo
  {journal} {Nature}\ }\textbf {\bibinfo {volume} {605}},\ \bibinfo {pages}
  {57} (\bibinfo {year} {2022})}\BibitemShut {NoStop}%
\bibitem [{\citenamefont {Zhao}\ \emph {et~al.}(2023)\citenamefont {Zhao},
  \citenamefont {Shen}, \citenamefont {Tao}, \citenamefont {Han}, \citenamefont
  {Kang}, \citenamefont {Watanabe}, \citenamefont {Taniguchi}, \citenamefont
  {Mak},\ and\ \citenamefont {Shan}}]{ZHA23c}%
  \BibitemOpen
  \bibfield  {author} {\bibinfo {author} {\bibfnamefont {W.}~\bibnamefont
  {Zhao}}, \bibinfo {author} {\bibfnamefont {B.}~\bibnamefont {Shen}}, \bibinfo
  {author} {\bibfnamefont {Z.}~\bibnamefont {Tao}}, \bibinfo {author}
  {\bibfnamefont {Z.}~\bibnamefont {Han}}, \bibinfo {author} {\bibfnamefont
  {K.}~\bibnamefont {Kang}}, \bibinfo {author} {\bibfnamefont {K.}~\bibnamefont
  {Watanabe}}, \bibinfo {author} {\bibfnamefont {T.}~\bibnamefont {Taniguchi}},
  \bibinfo {author} {\bibfnamefont {K.~F.}\ \bibnamefont {Mak}},\ and\ \bibinfo
  {author} {\bibfnamefont {J.}~\bibnamefont {Shan}},\ }\href
  {https://doi.org/10.1038/s41586-023-05800-7} {\bibfield  {journal} {\bibinfo
  {journal} {Nature}\ }\textbf {\bibinfo {volume} {616}},\ \bibinfo {pages}
  {61} (\bibinfo {year} {2023})}\BibitemShut {NoStop}%
\bibitem [{\citenamefont {Xia}\ \emph {et~al.}(2025)\citenamefont {Xia},
  \citenamefont {Han}, \citenamefont {Watanabe}, \citenamefont {Taniguchi},
  \citenamefont {Shan},\ and\ \citenamefont {Mak}}]{XIA24a}%
  \BibitemOpen
  \bibfield  {author} {\bibinfo {author} {\bibfnamefont {Y.}~\bibnamefont
  {Xia}}, \bibinfo {author} {\bibfnamefont {Z.}~\bibnamefont {Han}}, \bibinfo
  {author} {\bibfnamefont {K.}~\bibnamefont {Watanabe}}, \bibinfo {author}
  {\bibfnamefont {T.}~\bibnamefont {Taniguchi}}, \bibinfo {author}
  {\bibfnamefont {J.}~\bibnamefont {Shan}},\ and\ \bibinfo {author}
  {\bibfnamefont {K.~F.}\ \bibnamefont {Mak}},\ }\href
  {https://doi.org/10.1038/s41586-024-08116-2} {\bibfield  {journal} {\bibinfo
  {journal} {Nature}\ }\textbf {\bibinfo {volume} {637}},\ \bibinfo {pages}
  {833} (\bibinfo {year} {2025})}\BibitemShut {NoStop}%
\bibitem [{\citenamefont {Guo}\ \emph {et~al.}(2025)\citenamefont {Guo},
  \citenamefont {Pack}, \citenamefont {Swann}, \citenamefont {Holtzman},
  \citenamefont {Cothrine}, \citenamefont {Watanabe}, \citenamefont
  {Taniguchi}, \citenamefont {Mandrus}, \citenamefont {Barmak}, \citenamefont
  {Hone}, \citenamefont {Millis}, \citenamefont {Pasupathy},\ and\
  \citenamefont {Dean}}]{GUO24}%
  \BibitemOpen
  \bibfield  {author} {\bibinfo {author} {\bibfnamefont {Y.}~\bibnamefont
  {Guo}}, \bibinfo {author} {\bibfnamefont {J.}~\bibnamefont {Pack}}, \bibinfo
  {author} {\bibfnamefont {J.}~\bibnamefont {Swann}}, \bibinfo {author}
  {\bibfnamefont {L.}~\bibnamefont {Holtzman}}, \bibinfo {author}
  {\bibfnamefont {M.}~\bibnamefont {Cothrine}}, \bibinfo {author}
  {\bibfnamefont {K.}~\bibnamefont {Watanabe}}, \bibinfo {author}
  {\bibfnamefont {T.}~\bibnamefont {Taniguchi}}, \bibinfo {author}
  {\bibfnamefont {D.~G.}\ \bibnamefont {Mandrus}}, \bibinfo {author}
  {\bibfnamefont {K.}~\bibnamefont {Barmak}}, \bibinfo {author} {\bibfnamefont
  {J.}~\bibnamefont {Hone}}, \bibinfo {author} {\bibfnamefont {A.~J.}\
  \bibnamefont {Millis}}, \bibinfo {author} {\bibfnamefont {A.}~\bibnamefont
  {Pasupathy}},\ and\ \bibinfo {author} {\bibfnamefont {C.~R.}\ \bibnamefont
  {Dean}},\ }\href {https://doi.org/10.1038/s41586-024-08381-1} {\bibfield
  {journal} {\bibinfo  {journal} {Nature}\ }\textbf {\bibinfo {volume} {637}},\
  \bibinfo {pages} {839} (\bibinfo {year} {2025})}\BibitemShut {NoStop}%
\bibitem [{\citenamefont {Xia}\ \emph {et~al.}(2026)\citenamefont {Xia},
  \citenamefont {Han}, \citenamefont {Zhu}, \citenamefont {Zhang},
  \citenamefont {Kn{\"u}ppel}, \citenamefont {Watanabe}, \citenamefont
  {Taniguchi}, \citenamefont {Mak},\ and\ \citenamefont {Shan}}]{XIA26}%
  \BibitemOpen
  \bibfield  {author} {\bibinfo {author} {\bibfnamefont {Y.}~\bibnamefont
  {Xia}}, \bibinfo {author} {\bibfnamefont {Z.}~\bibnamefont {Han}}, \bibinfo
  {author} {\bibfnamefont {J.}~\bibnamefont {Zhu}}, \bibinfo {author}
  {\bibfnamefont {Y.}~\bibnamefont {Zhang}}, \bibinfo {author} {\bibfnamefont
  {P.}~\bibnamefont {Kn{\"u}ppel}}, \bibinfo {author} {\bibfnamefont
  {K.}~\bibnamefont {Watanabe}}, \bibinfo {author} {\bibfnamefont
  {T.}~\bibnamefont {Taniguchi}}, \bibinfo {author} {\bibfnamefont {K.~F.}\
  \bibnamefont {Mak}},\ and\ \bibinfo {author} {\bibfnamefont {J.}~\bibnamefont
  {Shan}},\ }\href {https://doi.org/10.1038/s41586-025-10049-3} {\bibfield
  {journal} {\bibinfo  {journal} {Nature}\ }\textbf {\bibinfo {volume} {650}},\
  \bibinfo {pages} {585} (\bibinfo {year} {2026})}\BibitemShut {NoStop}%
\bibitem [{\citenamefont {Wu}\ \emph {et~al.}(2019)\citenamefont {Wu},
  \citenamefont {Lovorn}, \citenamefont {Tutuc}, \citenamefont {Martin},\ and\
  \citenamefont {MacDonald}}]{WU19b}%
  \BibitemOpen
  \bibfield  {author} {\bibinfo {author} {\bibfnamefont {F.}~\bibnamefont
  {Wu}}, \bibinfo {author} {\bibfnamefont {T.}~\bibnamefont {Lovorn}}, \bibinfo
  {author} {\bibfnamefont {E.}~\bibnamefont {Tutuc}}, \bibinfo {author}
  {\bibfnamefont {I.}~\bibnamefont {Martin}},\ and\ \bibinfo {author}
  {\bibfnamefont {A.~H.}\ \bibnamefont {MacDonald}},\ }\href
  {https://doi.org/10.1103/PhysRevLett.122.086402} {\bibfield  {journal}
  {\bibinfo  {journal} {Phys. Rev. Lett.}\ }\textbf {\bibinfo {volume} {122}},\
  \bibinfo {pages} {086402} (\bibinfo {year} {2019})}\BibitemShut {NoStop}%
\bibitem [{\citenamefont {Devakul}\ \emph {et~al.}(2021)\citenamefont
  {Devakul}, \citenamefont {Cr{\'e}pel}, \citenamefont {Zhang},\ and\
  \citenamefont {Fu}}]{DEV21}%
  \BibitemOpen
  \bibfield  {author} {\bibinfo {author} {\bibfnamefont {T.}~\bibnamefont
  {Devakul}}, \bibinfo {author} {\bibfnamefont {V.}~\bibnamefont {Cr{\'e}pel}},
  \bibinfo {author} {\bibfnamefont {Y.}~\bibnamefont {Zhang}},\ and\ \bibinfo
  {author} {\bibfnamefont {L.}~\bibnamefont {Fu}},\ }\href
  {https://doi.org/10.1038/s41467-021-27042-9} {\bibfield  {journal} {\bibinfo
  {journal} {Nat. Commun.}\ }\textbf {\bibinfo {volume} {12}},\ \bibinfo
  {pages} {6730} (\bibinfo {year} {2021})}\BibitemShut {NoStop}%
\bibitem [{\citenamefont {Zeng}\ \emph {et~al.}(2023)\citenamefont {Zeng},
  \citenamefont {Xia}, \citenamefont {Kang}, \citenamefont {Zhu}, \citenamefont
  {Kn{\"u}ppel}, \citenamefont {Vaswani}, \citenamefont {Watanabe},
  \citenamefont {Taniguchi}, \citenamefont {Mak},\ and\ \citenamefont
  {Shan}}]{ZEN23}%
  \BibitemOpen
  \bibfield  {author} {\bibinfo {author} {\bibfnamefont {Y.}~\bibnamefont
  {Zeng}}, \bibinfo {author} {\bibfnamefont {Z.}~\bibnamefont {Xia}}, \bibinfo
  {author} {\bibfnamefont {K.}~\bibnamefont {Kang}}, \bibinfo {author}
  {\bibfnamefont {J.}~\bibnamefont {Zhu}}, \bibinfo {author} {\bibfnamefont
  {P.}~\bibnamefont {Kn{\"u}ppel}}, \bibinfo {author} {\bibfnamefont
  {C.}~\bibnamefont {Vaswani}}, \bibinfo {author} {\bibfnamefont
  {K.}~\bibnamefont {Watanabe}}, \bibinfo {author} {\bibfnamefont
  {T.}~\bibnamefont {Taniguchi}}, \bibinfo {author} {\bibfnamefont {K.~F.}\
  \bibnamefont {Mak}},\ and\ \bibinfo {author} {\bibfnamefont {J.}~\bibnamefont
  {Shan}},\ }\href {https://doi.org/10.1038/s41586-023-06452-3} {\bibfield
  {journal} {\bibinfo  {journal} {Nature}\ }\textbf {\bibinfo {volume} {622}},\
  \bibinfo {pages} {69} (\bibinfo {year} {2023})}\BibitemShut {NoStop}%
\bibitem [{\citenamefont {Wang}\ \emph
  {et~al.}(2024{\natexlab{a}})\citenamefont {Wang}, \citenamefont {Zhang},
  \citenamefont {Liu}, \citenamefont {He}, \citenamefont {Xu}, \citenamefont
  {Ran}, \citenamefont {Cao},\ and\ \citenamefont {Xiao}}]{WAN24a}%
  \BibitemOpen
  \bibfield  {author} {\bibinfo {author} {\bibfnamefont {C.}~\bibnamefont
  {Wang}}, \bibinfo {author} {\bibfnamefont {X.-W.}\ \bibnamefont {Zhang}},
  \bibinfo {author} {\bibfnamefont {X.}~\bibnamefont {Liu}}, \bibinfo {author}
  {\bibfnamefont {Y.}~\bibnamefont {He}}, \bibinfo {author} {\bibfnamefont
  {X.}~\bibnamefont {Xu}}, \bibinfo {author} {\bibfnamefont {Y.}~\bibnamefont
  {Ran}}, \bibinfo {author} {\bibfnamefont {T.}~\bibnamefont {Cao}},\ and\
  \bibinfo {author} {\bibfnamefont {D.}~\bibnamefont {Xiao}},\ }\href
  {https://doi.org/10.1103/PhysRevLett.132.036501} {\bibfield  {journal}
  {\bibinfo  {journal} {Phys. Rev. Lett.}\ }\textbf {\bibinfo {volume} {132}},\
  \bibinfo {pages} {036501} (\bibinfo {year} {2024}{\natexlab{a}})}\BibitemShut
  {NoStop}%
\bibitem [{\citenamefont {Yu}\ \emph {et~al.}(2024)\citenamefont {Yu},
  \citenamefont {{Herzog-Arbeitman}}, \citenamefont {Wang}, \citenamefont
  {Vafek}, \citenamefont {Bernevig},\ and\ \citenamefont {Regnault}}]{YU24a}%
  \BibitemOpen
  \bibfield  {author} {\bibinfo {author} {\bibfnamefont {J.}~\bibnamefont
  {Yu}}, \bibinfo {author} {\bibfnamefont {J.}~\bibnamefont
  {{Herzog-Arbeitman}}}, \bibinfo {author} {\bibfnamefont {M.}~\bibnamefont
  {Wang}}, \bibinfo {author} {\bibfnamefont {O.}~\bibnamefont {Vafek}},
  \bibinfo {author} {\bibfnamefont {B.~A.}\ \bibnamefont {Bernevig}},\ and\
  \bibinfo {author} {\bibfnamefont {N.}~\bibnamefont {Regnault}},\ }\href
  {https://doi.org/10.1103/PhysRevB.109.045147} {\bibfield  {journal} {\bibinfo
   {journal} {Phys. Rev. B}\ }\textbf {\bibinfo {volume} {109}},\ \bibinfo
  {pages} {045147} (\bibinfo {year} {2024})}\BibitemShut {NoStop}%
\bibitem [{\citenamefont {Jia}\ \emph {et~al.}(2024)\citenamefont {Jia},
  \citenamefont {Yu}, \citenamefont {Liu}, \citenamefont {{Herzog-Arbeitman}},
  \citenamefont {Qi}, \citenamefont {Pi}, \citenamefont {Regnault},
  \citenamefont {Weng}, \citenamefont {Bernevig},\ and\ \citenamefont
  {Wu}}]{JIA24}%
  \BibitemOpen
  \bibfield  {author} {\bibinfo {author} {\bibfnamefont {Y.}~\bibnamefont
  {Jia}}, \bibinfo {author} {\bibfnamefont {J.}~\bibnamefont {Yu}}, \bibinfo
  {author} {\bibfnamefont {J.}~\bibnamefont {Liu}}, \bibinfo {author}
  {\bibfnamefont {J.}~\bibnamefont {{Herzog-Arbeitman}}}, \bibinfo {author}
  {\bibfnamefont {Z.}~\bibnamefont {Qi}}, \bibinfo {author} {\bibfnamefont
  {H.}~\bibnamefont {Pi}}, \bibinfo {author} {\bibfnamefont {N.}~\bibnamefont
  {Regnault}}, \bibinfo {author} {\bibfnamefont {H.}~\bibnamefont {Weng}},
  \bibinfo {author} {\bibfnamefont {B.~A.}\ \bibnamefont {Bernevig}},\ and\
  \bibinfo {author} {\bibfnamefont {Q.}~\bibnamefont {Wu}},\ }\href
  {https://doi.org/10.1103/PhysRevB.109.205121} {\bibfield  {journal} {\bibinfo
   {journal} {Phys. Rev. B}\ }\textbf {\bibinfo {volume} {109}},\ \bibinfo
  {pages} {205121} (\bibinfo {year} {2024})}\BibitemShut {NoStop}%
\bibitem [{\citenamefont {Zhang}\ \emph {et~al.}(2024)\citenamefont {Zhang},
  \citenamefont {Wang}, \citenamefont {Liu}, \citenamefont {Fan}, \citenamefont
  {Cao},\ and\ \citenamefont {Xiao}}]{ZHA24a}%
  \BibitemOpen
  \bibfield  {author} {\bibinfo {author} {\bibfnamefont {X.-W.}\ \bibnamefont
  {Zhang}}, \bibinfo {author} {\bibfnamefont {C.}~\bibnamefont {Wang}},
  \bibinfo {author} {\bibfnamefont {X.}~\bibnamefont {Liu}}, \bibinfo {author}
  {\bibfnamefont {Y.}~\bibnamefont {Fan}}, \bibinfo {author} {\bibfnamefont
  {T.}~\bibnamefont {Cao}},\ and\ \bibinfo {author} {\bibfnamefont
  {D.}~\bibnamefont {Xiao}},\ }\href
  {https://doi.org/10.1038/s41467-024-48511-x} {\bibfield  {journal} {\bibinfo
  {journal} {Nat. Commun.}\ }\textbf {\bibinfo {volume} {15}},\ \bibinfo
  {pages} {4223} (\bibinfo {year} {2024})}\BibitemShut {NoStop}%
\bibitem [{\citenamefont {Sheng}\ \emph {et~al.}(2024)\citenamefont {Sheng},
  \citenamefont {Reddy}, \citenamefont {Abouelkomsan}, \citenamefont
  {Bergholtz},\ and\ \citenamefont {Fu}}]{SHE24}%
  \BibitemOpen
  \bibfield  {author} {\bibinfo {author} {\bibfnamefont {D.~N.}\ \bibnamefont
  {Sheng}}, \bibinfo {author} {\bibfnamefont {A.~P.}\ \bibnamefont {Reddy}},
  \bibinfo {author} {\bibfnamefont {A.}~\bibnamefont {Abouelkomsan}}, \bibinfo
  {author} {\bibfnamefont {E.~J.}\ \bibnamefont {Bergholtz}},\ and\ \bibinfo
  {author} {\bibfnamefont {L.}~\bibnamefont {Fu}},\ }\href
  {https://doi.org/10.1103/PhysRevLett.133.066601} {\bibfield  {journal}
  {\bibinfo  {journal} {Phys. Rev. Lett.}\ }\textbf {\bibinfo {volume} {133}},\
  \bibinfo {pages} {066601} (\bibinfo {year} {2024})}\BibitemShut {NoStop}%
\bibitem [{\citenamefont {Ingham}\ \emph {et~al.}(2025)\citenamefont {Ingham},
  \citenamefont {Scheurer},\ and\ \citenamefont {Scammell}}]{ING25}%
  \BibitemOpen
  \bibfield  {author} {\bibinfo {author} {\bibfnamefont {J.}~\bibnamefont
  {Ingham}}, \bibinfo {author} {\bibfnamefont {M.~S.}\ \bibnamefont
  {Scheurer}},\ and\ \bibinfo {author} {\bibfnamefont {H.~D.}\ \bibnamefont
  {Scammell}},\ }\href {https://doi.org/10.48550/arXiv.2503.11754} {\bibinfo
  {title} {Moir\'e {{M-valley}} bilayers: Quasi-one-dimensional physics,
  unconventional spin textures and twisted van {{Hove}} singularities}}
  (\bibinfo {year} {2025}),\ \Eprint {https://arxiv.org/abs/2503.11754}
  {arXiv:2503.11754 [cond-mat.str-el]} \BibitemShut {NoStop}%
\bibitem [{\citenamefont {Bao}\ \emph {et~al.}(2025)\citenamefont {Bao},
  \citenamefont {Wang}, \citenamefont {Liu},\ and\ \citenamefont
  {Wang}}]{BAO25}%
  \BibitemOpen
  \bibfield  {author} {\bibinfo {author} {\bibfnamefont {K.}~\bibnamefont
  {Bao}}, \bibinfo {author} {\bibfnamefont {H.}~\bibnamefont {Wang}}, \bibinfo
  {author} {\bibfnamefont {Z.}~\bibnamefont {Liu}},\ and\ \bibinfo {author}
  {\bibfnamefont {J.}~\bibnamefont {Wang}},\ }\href
  {https://doi.org/10.1103/l88b-67d2} {\bibfield  {journal} {\bibinfo
  {journal} {Phys. Rev. B}\ }\textbf {\bibinfo {volume} {112}},\ \bibinfo
  {pages} {L041406} (\bibinfo {year} {2025})}\BibitemShut {NoStop}%
\bibitem [{\citenamefont {Kariyado}\ and\ \citenamefont
  {Vishwanath}(2019)}]{KAR19}%
  \BibitemOpen
  \bibfield  {author} {\bibinfo {author} {\bibfnamefont {T.}~\bibnamefont
  {Kariyado}}\ and\ \bibinfo {author} {\bibfnamefont {A.}~\bibnamefont
  {Vishwanath}},\ }\href {https://doi.org/10.1103/PhysRevResearch.1.033076}
  {\bibfield  {journal} {\bibinfo  {journal} {Phys. Rev. Res.}\ }\textbf
  {\bibinfo {volume} {1}},\ \bibinfo {pages} {033076} (\bibinfo {year}
  {2019})}\BibitemShut {NoStop}%
\bibitem [{\citenamefont {Fujimoto}\ \emph {et~al.}(2022)\citenamefont
  {Fujimoto}, \citenamefont {Kawakami},\ and\ \citenamefont {Koshino}}]{FUJ22}%
  \BibitemOpen
  \bibfield  {author} {\bibinfo {author} {\bibfnamefont {M.}~\bibnamefont
  {Fujimoto}}, \bibinfo {author} {\bibfnamefont {T.}~\bibnamefont {Kawakami}},\
  and\ \bibinfo {author} {\bibfnamefont {M.}~\bibnamefont {Koshino}},\ }\href
  {https://doi.org/10.1103/PhysRevResearch.4.043209} {\bibfield  {journal}
  {\bibinfo  {journal} {Phys. Rev. Res.}\ }\textbf {\bibinfo {volume} {4}},\
  \bibinfo {pages} {043209} (\bibinfo {year} {2022})}\BibitemShut {NoStop}%
\bibitem [{\citenamefont {Kariyado}(2023)}]{KAR23}%
  \BibitemOpen
  \bibfield  {author} {\bibinfo {author} {\bibfnamefont {T.}~\bibnamefont
  {Kariyado}},\ }\href {https://doi.org/10.1103/PhysRevB.107.085127} {\bibfield
   {journal} {\bibinfo  {journal} {Phys. Rev. B}\ }\textbf {\bibinfo {volume}
  {107}},\ \bibinfo {pages} {085127} (\bibinfo {year} {2023})}\BibitemShut
  {NoStop}%
\bibitem [{\citenamefont {Emery}\ \emph {et~al.}(2000)\citenamefont {Emery},
  \citenamefont {Fradkin}, \citenamefont {Kivelson},\ and\ \citenamefont
  {Lubensky}}]{EME00}%
  \BibitemOpen
  \bibfield  {author} {\bibinfo {author} {\bibfnamefont {V.~J.}\ \bibnamefont
  {Emery}}, \bibinfo {author} {\bibfnamefont {E.}~\bibnamefont {Fradkin}},
  \bibinfo {author} {\bibfnamefont {S.~A.}\ \bibnamefont {Kivelson}},\ and\
  \bibinfo {author} {\bibfnamefont {T.~C.}\ \bibnamefont {Lubensky}},\ }\href
  {https://doi.org/10.1103/PhysRevLett.85.2160} {\bibfield  {journal} {\bibinfo
   {journal} {Phys. Rev. Lett.}\ }\textbf {\bibinfo {volume} {85}},\ \bibinfo
  {pages} {2160} (\bibinfo {year} {2000})}\BibitemShut {NoStop}%
\bibitem [{\citenamefont {Vishwanath}\ and\ \citenamefont
  {Carpentier}(2001)}]{VIS01}%
  \BibitemOpen
  \bibfield  {author} {\bibinfo {author} {\bibfnamefont {A.}~\bibnamefont
  {Vishwanath}}\ and\ \bibinfo {author} {\bibfnamefont {D.}~\bibnamefont
  {Carpentier}},\ }\href {https://doi.org/10.1103/PhysRevLett.86.676}
  {\bibfield  {journal} {\bibinfo  {journal} {Phys. Rev. Lett.}\ }\textbf
  {\bibinfo {volume} {86}},\ \bibinfo {pages} {676} (\bibinfo {year}
  {2001})}\BibitemShut {NoStop}%
\bibitem [{\citenamefont {Mukhopadhyay}\ \emph
  {et~al.}(2001{\natexlab{a}})\citenamefont {Mukhopadhyay}, \citenamefont
  {Kane},\ and\ \citenamefont {Lubensky}}]{MUK01a}%
  \BibitemOpen
  \bibfield  {author} {\bibinfo {author} {\bibfnamefont {R.}~\bibnamefont
  {Mukhopadhyay}}, \bibinfo {author} {\bibfnamefont {C.~L.}\ \bibnamefont
  {Kane}},\ and\ \bibinfo {author} {\bibfnamefont {T.~C.}\ \bibnamefont
  {Lubensky}},\ }\href {https://doi.org/10.1103/PhysRevB.64.045120} {\bibfield
  {journal} {\bibinfo  {journal} {Phys. Rev. B}\ }\textbf {\bibinfo {volume}
  {64}},\ \bibinfo {pages} {045120} (\bibinfo {year}
  {2001}{\natexlab{a}})}\BibitemShut {NoStop}%
\bibitem [{\citenamefont {Mukhopadhyay}\ \emph
  {et~al.}(2001{\natexlab{b}})\citenamefont {Mukhopadhyay}, \citenamefont
  {Kane},\ and\ \citenamefont {Lubensky}}]{MUK01}%
  \BibitemOpen
  \bibfield  {author} {\bibinfo {author} {\bibfnamefont {R.}~\bibnamefont
  {Mukhopadhyay}}, \bibinfo {author} {\bibfnamefont {C.~L.}\ \bibnamefont
  {Kane}},\ and\ \bibinfo {author} {\bibfnamefont {T.~C.}\ \bibnamefont
  {Lubensky}},\ }\href {https://doi.org/10.1103/PhysRevB.63.081103} {\bibfield
  {journal} {\bibinfo  {journal} {Phys. Rev. B}\ }\textbf {\bibinfo {volume}
  {63}},\ \bibinfo {pages} {081103} (\bibinfo {year}
  {2001}{\natexlab{b}})}\BibitemShut {NoStop}%
\bibitem [{\citenamefont {Beule}\ \emph {et~al.}(2025)\citenamefont {Beule},
  \citenamefont {Peng}, \citenamefont {Mele},\ and\ \citenamefont
  {Adam}}]{BEU25}%
  \BibitemOpen
  \bibfield  {author} {\bibinfo {author} {\bibfnamefont {C.~D.}\ \bibnamefont
  {Beule}}, \bibinfo {author} {\bibfnamefont {L.}~\bibnamefont {Peng}},
  \bibinfo {author} {\bibfnamefont {E.~J.}\ \bibnamefont {Mele}},\ and\
  \bibinfo {author} {\bibfnamefont {S.}~\bibnamefont {Adam}},\ }\href
  {https://doi.org/10.48550/arXiv.2508.14283} {\bibinfo {title} {Role of
  electron-electron interactions in {{M-valley}} twisted transition metal
  dichalcogenides}} (\bibinfo {year} {2025}),\ \Eprint
  {https://arxiv.org/abs/2508.14283} {arXiv:2508.14283 [cond-mat.mes-hall]}
  \BibitemShut {NoStop}%
\bibitem [{\citenamefont {Vasiliou}\ \emph {et~al.}(2026)\citenamefont
  {Vasiliou}, \citenamefont {C{\u a}lug{\u a}ru}, \citenamefont {Hofmann},\
  and\ \citenamefont {Parameswaran}}]{VAS26}%
  \BibitemOpen
  \bibfield  {author} {\bibinfo {author} {\bibfnamefont {K.}~\bibnamefont
  {Vasiliou}}, \bibinfo {author} {\bibfnamefont {D.}~\bibnamefont {C{\u
  a}lug{\u a}ru}}, \bibinfo {author} {\bibfnamefont {J.~S.}\ \bibnamefont
  {Hofmann}},\ and\ \bibinfo {author} {\bibfnamefont {S.~A.}\ \bibnamefont
  {Parameswaran}},\ }\href@noop {} {\bibfield  {journal} {\bibinfo  {journal}
  {arXiv:2606.XXXXX [cond-mat]}\ } (\bibinfo {year} {2026})}\BibitemShut
  {NoStop}%
\bibitem [{\citenamefont {Blankenbecler}\ \emph {et~al.}(1981)\citenamefont
  {Blankenbecler}, \citenamefont {Scalapino},\ and\ \citenamefont
  {Sugar}}]{BLA81}%
  \BibitemOpen
  \bibfield  {author} {\bibinfo {author} {\bibfnamefont {R.}~\bibnamefont
  {Blankenbecler}}, \bibinfo {author} {\bibfnamefont {D.~J.}\ \bibnamefont
  {Scalapino}},\ and\ \bibinfo {author} {\bibfnamefont {R.~L.}\ \bibnamefont
  {Sugar}},\ }\href {https://doi.org/10.1103/PhysRevD.24.2278} {\bibfield
  {journal} {\bibinfo  {journal} {Phys. Rev. D}\ }\textbf {\bibinfo {volume}
  {24}},\ \bibinfo {pages} {2278} (\bibinfo {year} {1981})}\BibitemShut
  {NoStop}%
\bibitem [{\citenamefont {Scalapino}\ and\ \citenamefont
  {Sugar}(1981)}]{SCA81}%
  \BibitemOpen
  \bibfield  {author} {\bibinfo {author} {\bibfnamefont {D.~J.}\ \bibnamefont
  {Scalapino}}\ and\ \bibinfo {author} {\bibfnamefont {R.~L.}\ \bibnamefont
  {Sugar}},\ }\href {https://doi.org/10.1103/PhysRevB.24.4295} {\bibfield
  {journal} {\bibinfo  {journal} {Phys. Rev. B}\ }\textbf {\bibinfo {volume}
  {24}},\ \bibinfo {pages} {4295} (\bibinfo {year} {1981})}\BibitemShut
  {NoStop}%
\bibitem [{\citenamefont {White}\ \emph {et~al.}(1989)\citenamefont {White},
  \citenamefont {Scalapino}, \citenamefont {Sugar}, \citenamefont {Loh},
  \citenamefont {Gubernatis},\ and\ \citenamefont {Scalettar}}]{WHI89}%
  \BibitemOpen
  \bibfield  {author} {\bibinfo {author} {\bibfnamefont {S.~R.}\ \bibnamefont
  {White}}, \bibinfo {author} {\bibfnamefont {D.~J.}\ \bibnamefont
  {Scalapino}}, \bibinfo {author} {\bibfnamefont {R.~L.}\ \bibnamefont
  {Sugar}}, \bibinfo {author} {\bibfnamefont {E.~Y.}\ \bibnamefont {Loh}},
  \bibinfo {author} {\bibfnamefont {J.~E.}\ \bibnamefont {Gubernatis}},\ and\
  \bibinfo {author} {\bibfnamefont {R.~T.}\ \bibnamefont {Scalettar}},\ }\href
  {https://doi.org/10.1103/PhysRevB.40.506} {\bibfield  {journal} {\bibinfo
  {journal} {Phys. Rev. B}\ }\textbf {\bibinfo {volume} {40}},\ \bibinfo
  {pages} {506} (\bibinfo {year} {1989})}\BibitemShut {NoStop}%
\bibitem [{\citenamefont {Assaad}\ and\ \citenamefont {Evertz}(2008)}]{ASS08}%
  \BibitemOpen
  \bibfield  {author} {\bibinfo {author} {\bibfnamefont {F.}~\bibnamefont
  {Assaad}}\ and\ \bibinfo {author} {\bibfnamefont {H.}~\bibnamefont
  {Evertz}},\ }in\ \href {https://doi.org/10.1007/978-3-540-74686-7_10} {\emph
  {\bibinfo {booktitle} {Computational {{Many-Particle Physics}}}}},\ \bibinfo
  {editor} {edited by\ \bibinfo {editor} {\bibfnamefont {H.}~\bibnamefont
  {Fehske}}, \bibinfo {editor} {\bibfnamefont {R.}~\bibnamefont {Schneider}},\
  and\ \bibinfo {editor} {\bibfnamefont {A.}~\bibnamefont {Wei{\ss}e}}}\
  (\bibinfo  {publisher} {Springer},\ \bibinfo {address} {Berlin, Heidelberg},\
  \bibinfo {year} {2008})\ pp.\ \bibinfo {pages} {277--356}\BibitemShut
  {NoStop}%
\bibitem [{\citenamefont {Sandvik}(1992)}]{SAN92}%
  \BibitemOpen
  \bibfield  {author} {\bibinfo {author} {\bibfnamefont {A.~W.}\ \bibnamefont
  {Sandvik}},\ }\href {https://doi.org/10.1088/0305-4470/25/13/017} {\bibfield
  {journal} {\bibinfo  {journal} {J. Phys. A: Math. Gen.}\ }\textbf {\bibinfo
  {volume} {25}},\ \bibinfo {pages} {3667} (\bibinfo {year}
  {1992})}\BibitemShut {NoStop}%
\bibitem [{\citenamefont {Troyer}\ and\ \citenamefont {Wiese}(2005)}]{TRO05}%
  \BibitemOpen
  \bibfield  {author} {\bibinfo {author} {\bibfnamefont {M.}~\bibnamefont
  {Troyer}}\ and\ \bibinfo {author} {\bibfnamefont {U.-J.}\ \bibnamefont
  {Wiese}},\ }\href {https://doi.org/10.1103/PhysRevLett.94.170201} {\bibfield
  {journal} {\bibinfo  {journal} {Phys. Rev. Lett.}\ }\textbf {\bibinfo
  {volume} {94}},\ \bibinfo {pages} {170201} (\bibinfo {year}
  {2005})}\BibitemShut {NoStop}%
\bibitem [{\citenamefont {Li}\ and\ \citenamefont {Yao}(2019)}]{LI19b}%
  \BibitemOpen
  \bibfield  {author} {\bibinfo {author} {\bibfnamefont {Z.-X.}\ \bibnamefont
  {Li}}\ and\ \bibinfo {author} {\bibfnamefont {H.}~\bibnamefont {Yao}},\
  }\href {https://doi.org/10.1146/annurev-conmatphys-033117-054307} {\bibfield
  {journal} {\bibinfo  {journal} {Annu. Rev. Condens. Matter Phys.}\ }\textbf
  {\bibinfo {volume} {10}},\ \bibinfo {pages} {337} (\bibinfo {year}
  {2019})}\BibitemShut {NoStop}%
\bibitem [{\citenamefont {Wu}\ and\ \citenamefont {Zhang}(2005)}]{WU05}%
  \BibitemOpen
  \bibfield  {author} {\bibinfo {author} {\bibfnamefont {C.}~\bibnamefont
  {Wu}}\ and\ \bibinfo {author} {\bibfnamefont {S.-C.}\ \bibnamefont {Zhang}},\
  }\href {https://doi.org/10.1103/PhysRevB.71.155115} {\bibfield  {journal}
  {\bibinfo  {journal} {Phys. Rev. B}\ }\textbf {\bibinfo {volume} {71}},\
  \bibinfo {pages} {155115} (\bibinfo {year} {2005})}\BibitemShut {NoStop}%
\bibitem [{\citenamefont {Berg}\ \emph {et~al.}(2012)\citenamefont {Berg},
  \citenamefont {Metlitski},\ and\ \citenamefont {Sachdev}}]{BER12}%
  \BibitemOpen
  \bibfield  {author} {\bibinfo {author} {\bibfnamefont {E.}~\bibnamefont
  {Berg}}, \bibinfo {author} {\bibfnamefont {M.~A.}\ \bibnamefont
  {Metlitski}},\ and\ \bibinfo {author} {\bibfnamefont {S.}~\bibnamefont
  {Sachdev}},\ }\href {https://doi.org/10.1126/science.1227769} {\bibfield
  {journal} {\bibinfo  {journal} {Science}\ }\textbf {\bibinfo {volume}
  {338}},\ \bibinfo {pages} {1606} (\bibinfo {year} {2012})}\BibitemShut
  {NoStop}%
\bibitem [{\citenamefont {Christensen}\ \emph {et~al.}(2020)\citenamefont
  {Christensen}, \citenamefont {Wang}, \citenamefont {Schattner}, \citenamefont
  {Berg},\ and\ \citenamefont {Fernandes}}]{CHR20a}%
  \BibitemOpen
  \bibfield  {author} {\bibinfo {author} {\bibfnamefont {M.~H.}\ \bibnamefont
  {Christensen}}, \bibinfo {author} {\bibfnamefont {X.}~\bibnamefont {Wang}},
  \bibinfo {author} {\bibfnamefont {Y.}~\bibnamefont {Schattner}}, \bibinfo
  {author} {\bibfnamefont {E.}~\bibnamefont {Berg}},\ and\ \bibinfo {author}
  {\bibfnamefont {R.~M.}\ \bibnamefont {Fernandes}},\ }\href
  {https://doi.org/10.1103/PhysRevLett.125.247001} {\bibfield  {journal}
  {\bibinfo  {journal} {Phys. Rev. Lett.}\ }\textbf {\bibinfo {volume} {125}},\
  \bibinfo {pages} {247001} (\bibinfo {year} {2020})}\BibitemShut {NoStop}%
\bibitem [{\citenamefont {Wang}\ \emph {et~al.}(2021)\citenamefont {Wang},
  \citenamefont {Christensen}, \citenamefont {Berg},\ and\ \citenamefont
  {Fernandes}}]{WAN21h}%
  \BibitemOpen
  \bibfield  {author} {\bibinfo {author} {\bibfnamefont {X.}~\bibnamefont
  {Wang}}, \bibinfo {author} {\bibfnamefont {M.~H.}\ \bibnamefont
  {Christensen}}, \bibinfo {author} {\bibfnamefont {E.}~\bibnamefont {Berg}},\
  and\ \bibinfo {author} {\bibfnamefont {R.~M.}\ \bibnamefont {Fernandes}},\
  }\href {https://doi.org/10.1016/j.aop.2021.168522} {\bibfield  {journal}
  {\bibinfo  {journal} {Ann. Phys.}\ }\bibinfo {series} {Special Issue on
  {{Philip W}}. {{Anderson}}},\ \textbf {\bibinfo {volume} {435}},\ \bibinfo
  {pages} {168522} (\bibinfo {year} {2021})}\BibitemShut {NoStop}%
\bibitem [{\citenamefont {Xu}\ \emph {et~al.}(2015)\citenamefont {Xu},
  \citenamefont {Li},\ and\ \citenamefont {Wu}}]{XU15}%
  \BibitemOpen
  \bibfield  {author} {\bibinfo {author} {\bibfnamefont {S.}~\bibnamefont
  {Xu}}, \bibinfo {author} {\bibfnamefont {Y.}~\bibnamefont {Li}},\ and\
  \bibinfo {author} {\bibfnamefont {C.}~\bibnamefont {Wu}},\ }\href
  {https://doi.org/10.1103/PhysRevX.5.021032} {\bibfield  {journal} {\bibinfo
  {journal} {Phys. Rev. X}\ }\textbf {\bibinfo {volume} {5}},\ \bibinfo {pages}
  {021032} (\bibinfo {year} {2015})}\BibitemShut {NoStop}%
\bibitem [{\citenamefont {Li}\ \emph {et~al.}(2015)\citenamefont {Li},
  \citenamefont {Jiang},\ and\ \citenamefont {Yao}}]{LI15}%
  \BibitemOpen
  \bibfield  {author} {\bibinfo {author} {\bibfnamefont {Z.-X.}\ \bibnamefont
  {Li}}, \bibinfo {author} {\bibfnamefont {Y.-F.}\ \bibnamefont {Jiang}},\ and\
  \bibinfo {author} {\bibfnamefont {H.}~\bibnamefont {Yao}},\ }\href
  {https://doi.org/10.1103/PhysRevB.91.241117} {\bibfield  {journal} {\bibinfo
  {journal} {Phys. Rev. B}\ }\textbf {\bibinfo {volume} {91}},\ \bibinfo
  {pages} {241117} (\bibinfo {year} {2015})}\BibitemShut {NoStop}%
\bibitem [{\citenamefont {Li}\ \emph {et~al.}(2016)\citenamefont {Li},
  \citenamefont {Jiang},\ and\ \citenamefont {Yao}}]{LI16}%
  \BibitemOpen
  \bibfield  {author} {\bibinfo {author} {\bibfnamefont {Z.-X.}\ \bibnamefont
  {Li}}, \bibinfo {author} {\bibfnamefont {Y.-F.}\ \bibnamefont {Jiang}},\ and\
  \bibinfo {author} {\bibfnamefont {H.}~\bibnamefont {Yao}},\ }\href
  {https://doi.org/10.1103/PhysRevLett.117.267002} {\bibfield  {journal}
  {\bibinfo  {journal} {Phys. Rev. Lett.}\ }\textbf {\bibinfo {volume} {117}},\
  \bibinfo {pages} {267002} (\bibinfo {year} {2016})}\BibitemShut {NoStop}%
\bibitem [{\citenamefont {Han}\ \emph {et~al.}(2024)\citenamefont {Han},
  \citenamefont {{Herzog-Arbeitman}}, \citenamefont {Bernevig},\ and\
  \citenamefont {Kivelson}}]{HAN24}%
  \BibitemOpen
  \bibfield  {author} {\bibinfo {author} {\bibfnamefont {Z.}~\bibnamefont
  {Han}}, \bibinfo {author} {\bibfnamefont {J.}~\bibnamefont
  {{Herzog-Arbeitman}}}, \bibinfo {author} {\bibfnamefont {B.~A.}\ \bibnamefont
  {Bernevig}},\ and\ \bibinfo {author} {\bibfnamefont {S.~A.}\ \bibnamefont
  {Kivelson}},\ }\href {https://doi.org/10.1103/PhysRevX.14.041004} {\bibfield
  {journal} {\bibinfo  {journal} {Phys. Rev. X}\ }\textbf {\bibinfo {volume}
  {14}},\ \bibinfo {pages} {041004} (\bibinfo {year} {2024})}\BibitemShut
  {NoStop}%
\bibitem [{\citenamefont {Kugel'}\ and\ \citenamefont
  {Khomskii}(1982)}]{KUG82}%
  \BibitemOpen
  \bibfield  {author} {\bibinfo {author} {\bibfnamefont {K.~I.}\ \bibnamefont
  {Kugel'}}\ and\ \bibinfo {author} {\bibfnamefont {D.~I.}\ \bibnamefont
  {Khomskii}},\ }\href@noop {} {\bibfield  {journal} {\bibinfo  {journal}
  {Physics-Uspekhi}\ }\textbf {\bibinfo {volume} {25}},\ \bibinfo {pages} {231}
  (\bibinfo {year} {1982})}\BibitemShut {NoStop}%
\bibitem [{\citenamefont {Khaliullin}(2005)}]{KHA05}%
  \BibitemOpen
  \bibfield  {author} {\bibinfo {author} {\bibfnamefont {G.}~\bibnamefont
  {Khaliullin}},\ }\href {https://doi.org/10.1143/PTPS.160.155} {\bibfield
  {journal} {\bibinfo  {journal} {Prog. Theor. Phys.}\ }\textbf {\bibinfo
  {volume} {160}},\ \bibinfo {pages} {155} (\bibinfo {year}
  {2005})}\BibitemShut {NoStop}%
\bibitem [{\citenamefont {Streltsov}\ and\ \citenamefont
  {Khomskii}(2017)}]{STR17}%
  \BibitemOpen
  \bibfield  {author} {\bibinfo {author} {\bibfnamefont {S.~V.}\ \bibnamefont
  {Streltsov}}\ and\ \bibinfo {author} {\bibfnamefont {D.~I.}\ \bibnamefont
  {Khomskii}},\ }\href@noop {} {\bibfield  {journal} {\bibinfo  {journal}
  {Physics-Uspekhi}\ }\textbf {\bibinfo {volume} {60}},\ \bibinfo {pages}
  {1121} (\bibinfo {year} {2017})}\BibitemShut {NoStop}%
\bibitem [{\citenamefont {Nussinov}\ and\ \citenamefont {{van den
  Brink}}(2015)}]{NUS15}%
  \BibitemOpen
  \bibfield  {author} {\bibinfo {author} {\bibfnamefont {Z.}~\bibnamefont
  {Nussinov}}\ and\ \bibinfo {author} {\bibfnamefont {J.}~\bibnamefont {{van
  den Brink}}},\ }\href {https://doi.org/10.1103/RevModPhys.87.1} {\bibfield
  {journal} {\bibinfo  {journal} {Rev. Mod. Phys.}\ }\textbf {\bibinfo {volume}
  {87}},\ \bibinfo {pages} {1} (\bibinfo {year} {2015})}\BibitemShut {NoStop}%
\bibitem [{\citenamefont {Grusdt}\ \emph {et~al.}(2018)\citenamefont {Grusdt},
  \citenamefont {Zhu}, \citenamefont {Shi},\ and\ \citenamefont
  {Demler}}]{GRU18}%
  \BibitemOpen
  \bibfield  {author} {\bibinfo {author} {\bibfnamefont {F.}~\bibnamefont
  {Grusdt}}, \bibinfo {author} {\bibfnamefont {Z.}~\bibnamefont {Zhu}},
  \bibinfo {author} {\bibfnamefont {T.}~\bibnamefont {Shi}},\ and\ \bibinfo
  {author} {\bibfnamefont {E.}~\bibnamefont {Demler}},\ }\href
  {https://doi.org/10.21468/SciPostPhys.5.6.057} {\bibfield  {journal}
  {\bibinfo  {journal} {SciPost Phys.}\ }\textbf {\bibinfo {volume} {5}},\
  \bibinfo {pages} {057} (\bibinfo {year} {2018})}\BibitemShut {NoStop}%
\bibitem [{\citenamefont {Grusdt}\ and\ \citenamefont {Pollet}(2020)}]{GRU20}%
  \BibitemOpen
  \bibfield  {author} {\bibinfo {author} {\bibfnamefont {F.}~\bibnamefont
  {Grusdt}}\ and\ \bibinfo {author} {\bibfnamefont {L.}~\bibnamefont
  {Pollet}},\ }\href {https://doi.org/10.1103/PhysRevLett.125.256401}
  {\bibfield  {journal} {\bibinfo  {journal} {Phys. Rev. Lett.}\ }\textbf
  {\bibinfo {volume} {125}},\ \bibinfo {pages} {256401} (\bibinfo {year}
  {2020})}\BibitemShut {NoStop}%
\bibitem [{\citenamefont {Schl{\"o}mer}\ \emph {et~al.}(2023)\citenamefont
  {Schl{\"o}mer}, \citenamefont {Bohrdt}, \citenamefont {Pollet}, \citenamefont
  {Schollw{\"o}ck},\ and\ \citenamefont {Grusdt}}]{SCH23}%
  \BibitemOpen
  \bibfield  {author} {\bibinfo {author} {\bibfnamefont {H.}~\bibnamefont
  {Schl{\"o}mer}}, \bibinfo {author} {\bibfnamefont {A.}~\bibnamefont
  {Bohrdt}}, \bibinfo {author} {\bibfnamefont {L.}~\bibnamefont {Pollet}},
  \bibinfo {author} {\bibfnamefont {U.}~\bibnamefont {Schollw{\"o}ck}},\ and\
  \bibinfo {author} {\bibfnamefont {F.}~\bibnamefont {Grusdt}},\ }\href
  {https://doi.org/10.1103/PhysRevResearch.5.L022027} {\bibfield  {journal}
  {\bibinfo  {journal} {Phys. Rev. Res.}\ }\textbf {\bibinfo {volume} {5}},\
  \bibinfo {pages} {L022027} (\bibinfo {year} {2023})}\BibitemShut {NoStop}%
\bibitem [{\citenamefont {Schl{\"o}mer}\ \emph {et~al.}(2024)\citenamefont
  {Schl{\"o}mer}, \citenamefont {Lange}, \citenamefont {Franz}, \citenamefont
  {Chalopin}, \citenamefont {Bojovi{\'c}}, \citenamefont {Wang}, \citenamefont
  {Bloch}, \citenamefont {Hilker}, \citenamefont {Grusdt},\ and\ \citenamefont
  {Bohrdt}}]{SCH24}%
  \BibitemOpen
  \bibfield  {author} {\bibinfo {author} {\bibfnamefont {H.}~\bibnamefont
  {Schl{\"o}mer}}, \bibinfo {author} {\bibfnamefont {H.}~\bibnamefont {Lange}},
  \bibinfo {author} {\bibfnamefont {T.}~\bibnamefont {Franz}}, \bibinfo
  {author} {\bibfnamefont {T.}~\bibnamefont {Chalopin}}, \bibinfo {author}
  {\bibfnamefont {P.}~\bibnamefont {Bojovi{\'c}}}, \bibinfo {author}
  {\bibfnamefont {S.}~\bibnamefont {Wang}}, \bibinfo {author} {\bibfnamefont
  {I.}~\bibnamefont {Bloch}}, \bibinfo {author} {\bibfnamefont {T.~A.}\
  \bibnamefont {Hilker}}, \bibinfo {author} {\bibfnamefont {F.}~\bibnamefont
  {Grusdt}},\ and\ \bibinfo {author} {\bibfnamefont {A.}~\bibnamefont
  {Bohrdt}},\ }\href {https://doi.org/10.1103/PRXQuantum.5.040341} {\bibfield
  {journal} {\bibinfo  {journal} {PRX Quantum}\ }\textbf {\bibinfo {volume}
  {5}},\ \bibinfo {pages} {040341} (\bibinfo {year} {2024})}\BibitemShut
  {NoStop}%
\bibitem [{\citenamefont {Kudrynskyi}\ \emph {et~al.}(2020)\citenamefont
  {Kudrynskyi}, \citenamefont {Wang}, \citenamefont {Sutcliffe}, \citenamefont
  {Bhuiyan}, \citenamefont {Fu}, \citenamefont {Yang}, \citenamefont
  {Makarovsky}, \citenamefont {Eaves}, \citenamefont {Solomon}, \citenamefont
  {Maslyuk}, \citenamefont {Kovalyuk}, \citenamefont {Zhang},\ and\
  \citenamefont {Patan{\`e}}}]{KUD20}%
  \BibitemOpen
  \bibfield  {author} {\bibinfo {author} {\bibfnamefont {Z.~R.}\ \bibnamefont
  {Kudrynskyi}}, \bibinfo {author} {\bibfnamefont {X.}~\bibnamefont {Wang}},
  \bibinfo {author} {\bibfnamefont {J.}~\bibnamefont {Sutcliffe}}, \bibinfo
  {author} {\bibfnamefont {M.~A.}\ \bibnamefont {Bhuiyan}}, \bibinfo {author}
  {\bibfnamefont {Y.}~\bibnamefont {Fu}}, \bibinfo {author} {\bibfnamefont
  {Z.}~\bibnamefont {Yang}}, \bibinfo {author} {\bibfnamefont {O.}~\bibnamefont
  {Makarovsky}}, \bibinfo {author} {\bibfnamefont {L.}~\bibnamefont {Eaves}},
  \bibinfo {author} {\bibfnamefont {A.}~\bibnamefont {Solomon}}, \bibinfo
  {author} {\bibfnamefont {V.~T.}\ \bibnamefont {Maslyuk}}, \bibinfo {author}
  {\bibfnamefont {Z.~D.}\ \bibnamefont {Kovalyuk}}, \bibinfo {author}
  {\bibfnamefont {L.}~\bibnamefont {Zhang}},\ and\ \bibinfo {author}
  {\bibfnamefont {A.}~\bibnamefont {Patan{\`e}}},\ }\href
  {https://doi.org/10.1002/adfm.201908092} {\bibfield  {journal} {\bibinfo
  {journal} {Adv. Funct. Mater.}\ }\textbf {\bibinfo {volume} {30}},\ \bibinfo
  {pages} {1908092} (\bibinfo {year} {2020})}\BibitemShut {NoStop}%
\bibitem [{\citenamefont {Kafle}\ \emph {et~al.}(2020)\citenamefont {Kafle},
  \citenamefont {Heil}, \citenamefont {Paudyal},\ and\ \citenamefont
  {Margine}}]{KAF20}%
  \BibitemOpen
  \bibfield  {author} {\bibinfo {author} {\bibfnamefont {G.~P.}\ \bibnamefont
  {Kafle}}, \bibinfo {author} {\bibfnamefont {C.}~\bibnamefont {Heil}},
  \bibinfo {author} {\bibfnamefont {H.}~\bibnamefont {Paudyal}},\ and\ \bibinfo
  {author} {\bibfnamefont {E.~R.}\ \bibnamefont {Margine}},\ }\href
  {https://doi.org/10.1039/D0TC04356G} {\bibfield  {journal} {\bibinfo
  {journal} {J. Mater. Chem. C}\ }\textbf {\bibinfo {volume} {8}},\ \bibinfo
  {pages} {16404} (\bibinfo {year} {2020})}\BibitemShut {NoStop}%
\bibitem [{\citenamefont {Fan}\ \emph {et~al.}(2026)\citenamefont {Fan},
  \citenamefont {Qiu}, \citenamefont {Gong}, \citenamefont {Liu},\ and\
  \citenamefont {Lu}}]{FAN26}%
  \BibitemOpen
  \bibfield  {author} {\bibinfo {author} {\bibfnamefont {J.}~\bibnamefont
  {Fan}}, \bibinfo {author} {\bibfnamefont {X.-L.}\ \bibnamefont {Qiu}},
  \bibinfo {author} {\bibfnamefont {B.-C.}\ \bibnamefont {Gong}}, \bibinfo
  {author} {\bibfnamefont {K.}~\bibnamefont {Liu}},\ and\ \bibinfo {author}
  {\bibfnamefont {Z.-Y.}\ \bibnamefont {Lu}},\ }\href
  {https://doi.org/10.1088/0256-307X/43/1/010711} {\bibfield  {journal}
  {\bibinfo  {journal} {Chinese Phys. Lett.}\ }\textbf {\bibinfo {volume}
  {43}},\ \bibinfo {pages} {010711} (\bibinfo {year} {2026})}\BibitemShut
  {NoStop}%
\bibitem [{\citenamefont {Zeng}\ \emph {et~al.}(2018)\citenamefont {Zeng},
  \citenamefont {Liu}, \citenamefont {Fu}, \citenamefont {Chen}, \citenamefont
  {Pan}, \citenamefont {Wang}, \citenamefont {Wang}, \citenamefont {Wang},
  \citenamefont {Xu}, \citenamefont {Cai}, \citenamefont {Yan}, \citenamefont
  {Wang}, \citenamefont {Liu}, \citenamefont {Wang}, \citenamefont {Liang},
  \citenamefont {Cui}, \citenamefont {Hwang}, \citenamefont {Yuan},\ and\
  \citenamefont {Miao}}]{ZEN18}%
  \BibitemOpen
  \bibfield  {author} {\bibinfo {author} {\bibfnamefont {J.}~\bibnamefont
  {Zeng}}, \bibinfo {author} {\bibfnamefont {E.}~\bibnamefont {Liu}}, \bibinfo
  {author} {\bibfnamefont {Y.}~\bibnamefont {Fu}}, \bibinfo {author}
  {\bibfnamefont {Z.}~\bibnamefont {Chen}}, \bibinfo {author} {\bibfnamefont
  {C.}~\bibnamefont {Pan}}, \bibinfo {author} {\bibfnamefont {C.}~\bibnamefont
  {Wang}}, \bibinfo {author} {\bibfnamefont {M.}~\bibnamefont {Wang}}, \bibinfo
  {author} {\bibfnamefont {Y.}~\bibnamefont {Wang}}, \bibinfo {author}
  {\bibfnamefont {K.}~\bibnamefont {Xu}}, \bibinfo {author} {\bibfnamefont
  {S.}~\bibnamefont {Cai}}, \bibinfo {author} {\bibfnamefont {X.}~\bibnamefont
  {Yan}}, \bibinfo {author} {\bibfnamefont {Y.}~\bibnamefont {Wang}}, \bibinfo
  {author} {\bibfnamefont {X.}~\bibnamefont {Liu}}, \bibinfo {author}
  {\bibfnamefont {P.}~\bibnamefont {Wang}}, \bibinfo {author} {\bibfnamefont
  {S.-J.}\ \bibnamefont {Liang}}, \bibinfo {author} {\bibfnamefont
  {Y.}~\bibnamefont {Cui}}, \bibinfo {author} {\bibfnamefont {H.~Y.}\
  \bibnamefont {Hwang}}, \bibinfo {author} {\bibfnamefont {H.}~\bibnamefont
  {Yuan}},\ and\ \bibinfo {author} {\bibfnamefont {F.}~\bibnamefont {Miao}},\
  }\href {https://doi.org/10.1021/acs.nanolett.7b05157} {\bibfield  {journal}
  {\bibinfo  {journal} {Nano Lett.}\ }\textbf {\bibinfo {volume} {18}},\
  \bibinfo {pages} {1410} (\bibinfo {year} {2018})}\BibitemShut {NoStop}%
\bibitem [{\citenamefont {Ying}\ \emph {et~al.}(2018)\citenamefont {Ying},
  \citenamefont {Paudyal}, \citenamefont {Heil}, \citenamefont {Chen},
  \citenamefont {Struzhkin},\ and\ \citenamefont {Margine}}]{YIN18}%
  \BibitemOpen
  \bibfield  {author} {\bibinfo {author} {\bibfnamefont {J.}~\bibnamefont
  {Ying}}, \bibinfo {author} {\bibfnamefont {H.}~\bibnamefont {Paudyal}},
  \bibinfo {author} {\bibfnamefont {C.}~\bibnamefont {Heil}}, \bibinfo {author}
  {\bibfnamefont {X.-J.}\ \bibnamefont {Chen}}, \bibinfo {author}
  {\bibfnamefont {V.~V.}\ \bibnamefont {Struzhkin}},\ and\ \bibinfo {author}
  {\bibfnamefont {E.~R.}\ \bibnamefont {Margine}},\ }\href
  {https://doi.org/10.1103/PhysRevLett.121.027003} {\bibfield  {journal}
  {\bibinfo  {journal} {Phys. Rev. Lett.}\ }\textbf {\bibinfo {volume} {121}},\
  \bibinfo {pages} {027003} (\bibinfo {year} {2018})}\BibitemShut {NoStop}%
\bibitem [{\citenamefont {Kwan}\ \emph {et~al.}(2024)\citenamefont {Kwan},
  \citenamefont {Wagner}, \citenamefont {Bultinck}, \citenamefont {Simon},
  \citenamefont {Berg},\ and\ \citenamefont {Parameswaran}}]{KWA23}%
  \BibitemOpen
  \bibfield  {author} {\bibinfo {author} {\bibfnamefont {Y.~H.}\ \bibnamefont
  {Kwan}}, \bibinfo {author} {\bibfnamefont {G.}~\bibnamefont {Wagner}},
  \bibinfo {author} {\bibfnamefont {N.}~\bibnamefont {Bultinck}}, \bibinfo
  {author} {\bibfnamefont {S.~H.}\ \bibnamefont {Simon}}, \bibinfo {author}
  {\bibfnamefont {E.}~\bibnamefont {Berg}},\ and\ \bibinfo {author}
  {\bibfnamefont {S.~A.}\ \bibnamefont {Parameswaran}},\ }\href
  {https://doi.org/10.1103/PhysRevB.110.085160} {\bibfield  {journal} {\bibinfo
   {journal} {Phys. Rev. B}\ }\textbf {\bibinfo {volume} {110}},\ \bibinfo
  {pages} {085160} (\bibinfo {year} {2024})}\BibitemShut {NoStop}%
\bibitem [{\citenamefont {Wang}\ \emph
  {et~al.}(2024{\natexlab{b}})\citenamefont {Wang}, \citenamefont {Zhou},
  \citenamefont {Peng}, \citenamefont {Lian},\ and\ \citenamefont
  {Song}}]{WAN24}%
  \BibitemOpen
  \bibfield  {author} {\bibinfo {author} {\bibfnamefont {Y.-J.}\ \bibnamefont
  {Wang}}, \bibinfo {author} {\bibfnamefont {G.-D.}\ \bibnamefont {Zhou}},
  \bibinfo {author} {\bibfnamefont {S.-Y.}\ \bibnamefont {Peng}}, \bibinfo
  {author} {\bibfnamefont {B.}~\bibnamefont {Lian}},\ and\ \bibinfo {author}
  {\bibfnamefont {Z.-D.}\ \bibnamefont {Song}},\ }\href
  {https://doi.org/10.1103/PhysRevLett.133.146001} {\bibfield  {journal}
  {\bibinfo  {journal} {Phys. Rev. Lett.}\ }\textbf {\bibinfo {volume} {133}},\
  \bibinfo {pages} {146001} (\bibinfo {year} {2024}{\natexlab{b}})}\BibitemShut
  {NoStop}%
\bibitem [{\citenamefont {Liu}\ \emph {et~al.}(2024)\citenamefont {Liu},
  \citenamefont {Chen}, \citenamefont {Yazdani},\ and\ \citenamefont
  {Bernevig}}]{LIU24e}%
  \BibitemOpen
  \bibfield  {author} {\bibinfo {author} {\bibfnamefont {C.-X.}\ \bibnamefont
  {Liu}}, \bibinfo {author} {\bibfnamefont {Y.}~\bibnamefont {Chen}}, \bibinfo
  {author} {\bibfnamefont {A.}~\bibnamefont {Yazdani}},\ and\ \bibinfo {author}
  {\bibfnamefont {B.~A.}\ \bibnamefont {Bernevig}},\ }\href
  {https://doi.org/10.1103/PhysRevB.110.045133} {\bibfield  {journal} {\bibinfo
   {journal} {Phys. Rev. B}\ }\textbf {\bibinfo {volume} {110}},\ \bibinfo
  {pages} {045133} (\bibinfo {year} {2024})}\BibitemShut {NoStop}%
\bibitem [{\citenamefont {Youn}\ \emph {et~al.}(2024)\citenamefont {Youn},
  \citenamefont {Goh}, \citenamefont {Zhou}, \citenamefont {Song},\ and\
  \citenamefont {Lee}}]{YOU24}%
  \BibitemOpen
  \bibfield  {author} {\bibinfo {author} {\bibfnamefont {S.}~\bibnamefont
  {Youn}}, \bibinfo {author} {\bibfnamefont {B.}~\bibnamefont {Goh}}, \bibinfo
  {author} {\bibfnamefont {G.-D.}\ \bibnamefont {Zhou}}, \bibinfo {author}
  {\bibfnamefont {Z.-D.}\ \bibnamefont {Song}},\ and\ \bibinfo {author}
  {\bibfnamefont {S.-S.~B.}\ \bibnamefont {Lee}},\ }\href
  {https://doi.org/10.48550/arXiv.2412.03108} {\bibinfo {title} {Hundness in
  twisted bilayer graphene: Correlated gaps and pairing}} (\bibinfo {year}
  {2024}),\ \Eprint {https://arxiv.org/abs/2412.03108} {arXiv:2412.03108
  [cond-mat.str-el]} \BibitemShut {NoStop}%
\bibitem [{\citenamefont {Wang}\ \emph
  {et~al.}(2025{\natexlab{a}})\citenamefont {Wang}, \citenamefont {Wagner},
  \citenamefont {Kwan}, \citenamefont {Bultinck}, \citenamefont {Simon},\ and\
  \citenamefont {Parameswaran}}]{WAN25}%
  \BibitemOpen
  \bibfield  {author} {\bibinfo {author} {\bibfnamefont {Z.}~\bibnamefont
  {Wang}}, \bibinfo {author} {\bibfnamefont {G.}~\bibnamefont {Wagner}},
  \bibinfo {author} {\bibfnamefont {Y.~H.}\ \bibnamefont {Kwan}}, \bibinfo
  {author} {\bibfnamefont {N.}~\bibnamefont {Bultinck}}, \bibinfo {author}
  {\bibfnamefont {S.~H.}\ \bibnamefont {Simon}},\ and\ \bibinfo {author}
  {\bibfnamefont {S.~A.}\ \bibnamefont {Parameswaran}},\ }\href
  {https://doi.org/10.48550/arXiv.2509.12320} {\bibinfo {title} {Putting a new
  spin on the incommensurate {{Kekul\'e}} spiral: From spin-valley locking and
  collective modes to fermiology and implications for superconductivity}}
  (\bibinfo {year} {2025}{\natexlab{a}}),\ \Eprint
  {https://arxiv.org/abs/2509.12320} {arXiv:2509.12320 [cond-mat.str-el]}
  \BibitemShut {NoStop}%
\bibitem [{\citenamefont {Wang}\ \emph
  {et~al.}(2025{\natexlab{b}})\citenamefont {Wang}, \citenamefont {Zhou},
  \citenamefont {Lian},\ and\ \citenamefont {Song}}]{WAN25a}%
  \BibitemOpen
  \bibfield  {author} {\bibinfo {author} {\bibfnamefont {Y.-J.}\ \bibnamefont
  {Wang}}, \bibinfo {author} {\bibfnamefont {G.-D.}\ \bibnamefont {Zhou}},
  \bibinfo {author} {\bibfnamefont {B.}~\bibnamefont {Lian}},\ and\ \bibinfo
  {author} {\bibfnamefont {Z.-D.}\ \bibnamefont {Song}},\ }\href
  {https://doi.org/10.1103/PhysRevB.111.035110} {\bibfield  {journal} {\bibinfo
   {journal} {Phys. Rev. B}\ }\textbf {\bibinfo {volume} {111}},\ \bibinfo
  {pages} {035110} (\bibinfo {year} {2025}{\natexlab{b}})}\BibitemShut
  {NoStop}%
\bibitem [{\citenamefont {Sch{\"u}ler}\ \emph {et~al.}(2013)\citenamefont
  {Sch{\"u}ler}, \citenamefont {R{\"o}sner}, \citenamefont {Wehling},
  \citenamefont {Lichtenstein},\ and\ \citenamefont {Katsnelson}}]{SCH13}%
  \BibitemOpen
  \bibfield  {author} {\bibinfo {author} {\bibfnamefont {M.}~\bibnamefont
  {Sch{\"u}ler}}, \bibinfo {author} {\bibfnamefont {M.}~\bibnamefont
  {R{\"o}sner}}, \bibinfo {author} {\bibfnamefont {T.~O.}\ \bibnamefont
  {Wehling}}, \bibinfo {author} {\bibfnamefont {A.~I.}\ \bibnamefont
  {Lichtenstein}},\ and\ \bibinfo {author} {\bibfnamefont {M.~I.}\ \bibnamefont
  {Katsnelson}},\ }\href {https://doi.org/10.1103/PhysRevLett.111.036601}
  {\bibfield  {journal} {\bibinfo  {journal} {Phys. Rev. Lett.}\ }\textbf
  {\bibinfo {volume} {111}},\ \bibinfo {pages} {036601} (\bibinfo {year}
  {2013})}\BibitemShut {NoStop}%
\bibitem [{\citenamefont {Huang}\ \emph {et~al.}(2014)\citenamefont {Huang},
  \citenamefont {Ayral}, \citenamefont {Biermann},\ and\ \citenamefont
  {Werner}}]{HUA14}%
  \BibitemOpen
  \bibfield  {author} {\bibinfo {author} {\bibfnamefont {L.}~\bibnamefont
  {Huang}}, \bibinfo {author} {\bibfnamefont {T.}~\bibnamefont {Ayral}},
  \bibinfo {author} {\bibfnamefont {S.}~\bibnamefont {Biermann}},\ and\
  \bibinfo {author} {\bibfnamefont {P.}~\bibnamefont {Werner}},\ }\href
  {https://doi.org/10.1103/PhysRevB.90.195114} {\bibfield  {journal} {\bibinfo
  {journal} {Phys. Rev. B}\ }\textbf {\bibinfo {volume} {90}},\ \bibinfo
  {pages} {195114} (\bibinfo {year} {2014})}\BibitemShut {NoStop}%
\bibitem [{\citenamefont {Morath}\ \emph {et~al.}(2016)\citenamefont {Morath},
  \citenamefont {Sedlmayr}, \citenamefont {Sirker},\ and\ \citenamefont
  {Eggert}}]{MOR16a}%
  \BibitemOpen
  \bibfield  {author} {\bibinfo {author} {\bibfnamefont {D.}~\bibnamefont
  {Morath}}, \bibinfo {author} {\bibfnamefont {N.}~\bibnamefont {Sedlmayr}},
  \bibinfo {author} {\bibfnamefont {J.}~\bibnamefont {Sirker}},\ and\ \bibinfo
  {author} {\bibfnamefont {S.}~\bibnamefont {Eggert}},\ }\href
  {https://doi.org/10.1103/PhysRevB.94.115162} {\bibfield  {journal} {\bibinfo
  {journal} {Phys. Rev. B}\ }\textbf {\bibinfo {volume} {94}},\ \bibinfo
  {pages} {115162} (\bibinfo {year} {2016})}\BibitemShut {NoStop}%
\bibitem [{\citenamefont {Dorneich}\ and\ \citenamefont
  {Troyer}(2001)}]{DOR01}%
  \BibitemOpen
  \bibfield  {author} {\bibinfo {author} {\bibfnamefont {A.}~\bibnamefont
  {Dorneich}}\ and\ \bibinfo {author} {\bibfnamefont {M.}~\bibnamefont
  {Troyer}},\ }\href {https://doi.org/10.1103/PhysRevE.64.066701} {\bibfield
  {journal} {\bibinfo  {journal} {Phys. Rev. E}\ }\textbf {\bibinfo {volume}
  {64}},\ \bibinfo {pages} {066701} (\bibinfo {year} {2001})}\BibitemShut
  {NoStop}%
\bibitem [{\citenamefont {Dorneich}\ \emph {et~al.}(2002)\citenamefont
  {Dorneich}, \citenamefont {Hanke}, \citenamefont {Arrigoni}, \citenamefont
  {Troyer},\ and\ \citenamefont {Zhang}}]{DOR02}%
  \BibitemOpen
  \bibfield  {author} {\bibinfo {author} {\bibfnamefont {A.}~\bibnamefont
  {Dorneich}}, \bibinfo {author} {\bibfnamefont {W.}~\bibnamefont {Hanke}},
  \bibinfo {author} {\bibfnamefont {E.}~\bibnamefont {Arrigoni}}, \bibinfo
  {author} {\bibfnamefont {M.}~\bibnamefont {Troyer}},\ and\ \bibinfo {author}
  {\bibfnamefont {S.~C.}\ \bibnamefont {Zhang}},\ }\href
  {https://doi.org/10.1103/PhysRevLett.88.057003} {\bibfield  {journal}
  {\bibinfo  {journal} {Phys. Rev. Lett.}\ }\textbf {\bibinfo {volume} {88}},\
  \bibinfo {pages} {057003} (\bibinfo {year} {2002})}\BibitemShut {NoStop}%
\bibitem [{\citenamefont {Majumder}\ and\ \citenamefont {Garg}(2016)}]{MAJ16}%
  \BibitemOpen
  \bibfield  {author} {\bibinfo {author} {\bibfnamefont {G.}~\bibnamefont
  {Majumder}}\ and\ \bibinfo {author} {\bibfnamefont {A.}~\bibnamefont
  {Garg}},\ }\href {https://doi.org/10.1103/PhysRevB.94.134508} {\bibfield
  {journal} {\bibinfo  {journal} {Phys. Rev. B}\ }\textbf {\bibinfo {volume}
  {94}},\ \bibinfo {pages} {134508} (\bibinfo {year} {2016})}\BibitemShut
  {NoStop}%
\bibitem [{\citenamefont {Sandvik}\ and\ \citenamefont
  {Kurkij{\"a}rvi}(1991)}]{SAN91}%
  \BibitemOpen
  \bibfield  {author} {\bibinfo {author} {\bibfnamefont {A.~W.}\ \bibnamefont
  {Sandvik}}\ and\ \bibinfo {author} {\bibfnamefont {J.}~\bibnamefont
  {Kurkij{\"a}rvi}},\ }\href {https://doi.org/10.1103/PhysRevB.43.5950}
  {\bibfield  {journal} {\bibinfo  {journal} {Phys. Rev. B}\ }\textbf {\bibinfo
  {volume} {43}},\ \bibinfo {pages} {5950} (\bibinfo {year}
  {1991})}\BibitemShut {NoStop}%
\bibitem [{\citenamefont {Sandvik}(1997)}]{SAN97a}%
  \BibitemOpen
  \bibfield  {author} {\bibinfo {author} {\bibfnamefont {A.~W.}\ \bibnamefont
  {Sandvik}},\ }\href {https://doi.org/10.1103/PhysRevB.56.11678} {\bibfield
  {journal} {\bibinfo  {journal} {Phys. Rev. B}\ }\textbf {\bibinfo {volume}
  {56}},\ \bibinfo {pages} {11678} (\bibinfo {year} {1997})}\BibitemShut
  {NoStop}%
\bibitem [{\citenamefont {Sandvik}\ \emph {et~al.}(1997)\citenamefont
  {Sandvik}, \citenamefont {Singh},\ and\ \citenamefont {Campbell}}]{SAN97}%
  \BibitemOpen
  \bibfield  {author} {\bibinfo {author} {\bibfnamefont {A.~W.}\ \bibnamefont
  {Sandvik}}, \bibinfo {author} {\bibfnamefont {R.~R.~P.}\ \bibnamefont
  {Singh}},\ and\ \bibinfo {author} {\bibfnamefont {D.~K.}\ \bibnamefont
  {Campbell}},\ }\href {https://doi.org/10.1103/PhysRevB.56.14510} {\bibfield
  {journal} {\bibinfo  {journal} {Phys. Rev. B}\ }\textbf {\bibinfo {volume}
  {56}},\ \bibinfo {pages} {14510} (\bibinfo {year} {1997})}\BibitemShut
  {NoStop}%
\bibitem [{\citenamefont {Sylju{\aa}sen}\ and\ \citenamefont
  {Sandvik}(2002)}]{SYL02}%
  \BibitemOpen
  \bibfield  {author} {\bibinfo {author} {\bibfnamefont {O.~F.}\ \bibnamefont
  {Sylju{\aa}sen}}\ and\ \bibinfo {author} {\bibfnamefont {A.~W.}\ \bibnamefont
  {Sandvik}},\ }\href {https://doi.org/10.1103/PhysRevE.66.046701} {\bibfield
  {journal} {\bibinfo  {journal} {Phys. Rev. E}\ }\textbf {\bibinfo {volume}
  {66}},\ \bibinfo {pages} {046701} (\bibinfo {year} {2002})}\BibitemShut
  {NoStop}%
\bibitem [{\citenamefont {Sandvik}(2003)}]{SAN03}%
  \BibitemOpen
  \bibfield  {author} {\bibinfo {author} {\bibfnamefont {A.~W.}\ \bibnamefont
  {Sandvik}},\ }\href {https://doi.org/10.1103/PhysRevE.68.056701} {\bibfield
  {journal} {\bibinfo  {journal} {Phys. Rev. E}\ }\textbf {\bibinfo {volume}
  {68}},\ \bibinfo {pages} {056701} (\bibinfo {year} {2003})}\BibitemShut
  {NoStop}%
\bibitem [{\citenamefont {Melko}(2007)}]{MEL07}%
  \BibitemOpen
  \bibfield  {author} {\bibinfo {author} {\bibfnamefont {R.~G.}\ \bibnamefont
  {Melko}},\ }\href {https://doi.org/10.1088/0953-8984/19/14/145203} {\bibfield
   {journal} {\bibinfo  {journal} {J. Phys.: Condens. Matter}\ }\textbf
  {\bibinfo {volume} {19}},\ \bibinfo {pages} {145203} (\bibinfo {year}
  {2007})}\BibitemShut {NoStop}%
\bibitem [{\citenamefont {Grossjohann}\ and\ \citenamefont
  {Brenig}(2009)}]{GRO09}%
  \BibitemOpen
  \bibfield  {author} {\bibinfo {author} {\bibfnamefont {S.}~\bibnamefont
  {Grossjohann}}\ and\ \bibinfo {author} {\bibfnamefont {W.}~\bibnamefont
  {Brenig}},\ }\href {https://doi.org/10.1103/PhysRevB.79.094409} {\bibfield
  {journal} {\bibinfo  {journal} {Phys. Rev. B}\ }\textbf {\bibinfo {volume}
  {79}},\ \bibinfo {pages} {094409} (\bibinfo {year} {2009})}\BibitemShut
  {NoStop}%
\bibitem [{\citenamefont {Swendsen}\ and\ \citenamefont {Wang}(1986)}]{SWE86}%
  \BibitemOpen
  \bibfield  {author} {\bibinfo {author} {\bibfnamefont {R.~H.}\ \bibnamefont
  {Swendsen}}\ and\ \bibinfo {author} {\bibfnamefont {J.-S.}\ \bibnamefont
  {Wang}},\ }\href {https://doi.org/10.1103/PhysRevLett.57.2607} {\bibfield
  {journal} {\bibinfo  {journal} {Phys. Rev. Lett.}\ }\textbf {\bibinfo
  {volume} {57}},\ \bibinfo {pages} {2607} (\bibinfo {year}
  {1986})}\BibitemShut {NoStop}%
\bibitem [{\citenamefont {Hukushima}\ and\ \citenamefont
  {Nemoto}(1996)}]{HUK96}%
  \BibitemOpen
  \bibfield  {author} {\bibinfo {author} {\bibfnamefont {K.}~\bibnamefont
  {Hukushima}}\ and\ \bibinfo {author} {\bibfnamefont {K.}~\bibnamefont
  {Nemoto}},\ }\href {https://doi.org/10.1143/JPSJ.65.1604} {\bibfield
  {journal} {\bibinfo  {journal} {J. Phys. Soc. Jpn.}\ }\textbf {\bibinfo
  {volume} {65}},\ \bibinfo {pages} {1604} (\bibinfo {year}
  {1996})}\BibitemShut {NoStop}%
\bibitem [{\citenamefont {Prokof'ev}\ and\ \citenamefont
  {Svistunov}(2001)}]{PRO01}%
  \BibitemOpen
  \bibfield  {author} {\bibinfo {author} {\bibfnamefont {N.}~\bibnamefont
  {Prokof'ev}}\ and\ \bibinfo {author} {\bibfnamefont {B.}~\bibnamefont
  {Svistunov}},\ }\href {https://doi.org/10.1103/PhysRevLett.87.160601}
  {\bibfield  {journal} {\bibinfo  {journal} {Phys. Rev. Lett.}\ }\textbf
  {\bibinfo {volume} {87}},\ \bibinfo {pages} {160601} (\bibinfo {year}
  {2001})}\BibitemShut {NoStop}%
\bibitem [{\citenamefont {Scalapino}\ \emph {et~al.}(1992)\citenamefont
  {Scalapino}, \citenamefont {White},\ and\ \citenamefont {Zhang}}]{SCA92}%
  \BibitemOpen
  \bibfield  {author} {\bibinfo {author} {\bibfnamefont {D.~J.}\ \bibnamefont
  {Scalapino}}, \bibinfo {author} {\bibfnamefont {S.~R.}\ \bibnamefont
  {White}},\ and\ \bibinfo {author} {\bibfnamefont {S.~C.}\ \bibnamefont
  {Zhang}},\ }\href {https://doi.org/10.1103/PhysRevLett.68.2830} {\bibfield
  {journal} {\bibinfo  {journal} {Phys. Rev. Lett.}\ }\textbf {\bibinfo
  {volume} {68}},\ \bibinfo {pages} {2830} (\bibinfo {year}
  {1992})}\BibitemShut {NoStop}%
\bibitem [{\citenamefont {Scalapino}\ \emph {et~al.}(1993)\citenamefont
  {Scalapino}, \citenamefont {White},\ and\ \citenamefont {Zhang}}]{SCA93}%
  \BibitemOpen
  \bibfield  {author} {\bibinfo {author} {\bibfnamefont {D.~J.}\ \bibnamefont
  {Scalapino}}, \bibinfo {author} {\bibfnamefont {S.~R.}\ \bibnamefont
  {White}},\ and\ \bibinfo {author} {\bibfnamefont {S.}~\bibnamefont {Zhang}},\
  }\href {https://doi.org/10.1103/PhysRevB.47.7995} {\bibfield  {journal}
  {\bibinfo  {journal} {Phys. Rev. B}\ }\textbf {\bibinfo {volume} {47}},\
  \bibinfo {pages} {7995} (\bibinfo {year} {1993})}\BibitemShut {NoStop}%
\bibitem [{\citenamefont {Rozenberg}\ \emph {et~al.}(1992)\citenamefont
  {Rozenberg}, \citenamefont {Zhang},\ and\ \citenamefont {Kotliar}}]{ROZ92}%
  \BibitemOpen
  \bibfield  {author} {\bibinfo {author} {\bibfnamefont {M.~J.}\ \bibnamefont
  {Rozenberg}}, \bibinfo {author} {\bibfnamefont {X.~Y.}\ \bibnamefont
  {Zhang}},\ and\ \bibinfo {author} {\bibfnamefont {G.}~\bibnamefont
  {Kotliar}},\ }\href {https://doi.org/10.1103/PhysRevLett.69.1236} {\bibfield
  {journal} {\bibinfo  {journal} {Phys. Rev. Lett.}\ }\textbf {\bibinfo
  {volume} {69}},\ \bibinfo {pages} {1236} (\bibinfo {year}
  {1992})}\BibitemShut {NoStop}%
\bibitem [{\citenamefont {Florens}\ and\ \citenamefont
  {Georges}(2002)}]{FLO02}%
  \BibitemOpen
  \bibfield  {author} {\bibinfo {author} {\bibfnamefont {S.}~\bibnamefont
  {Florens}}\ and\ \bibinfo {author} {\bibfnamefont {A.}~\bibnamefont
  {Georges}},\ }\href {https://doi.org/10.1103/PhysRevB.66.165111} {\bibfield
  {journal} {\bibinfo  {journal} {Phys. Rev. B}\ }\textbf {\bibinfo {volume}
  {66}},\ \bibinfo {pages} {165111} (\bibinfo {year} {2002})}\BibitemShut
  {NoStop}%
\bibitem [{\citenamefont {Florens}\ and\ \citenamefont
  {Georges}(2004)}]{FLO04}%
  \BibitemOpen
  \bibfield  {author} {\bibinfo {author} {\bibfnamefont {S.}~\bibnamefont
  {Florens}}\ and\ \bibinfo {author} {\bibfnamefont {A.}~\bibnamefont
  {Georges}},\ }\href {https://doi.org/10.1103/PhysRevB.70.035114} {\bibfield
  {journal} {\bibinfo  {journal} {Phys. Rev. B}\ }\textbf {\bibinfo {volume}
  {70}},\ \bibinfo {pages} {035114} (\bibinfo {year} {2004})}\BibitemShut
  {NoStop}%
\bibitem [{\citenamefont {Lu}(1994)}]{LU94}%
  \BibitemOpen
  \bibfield  {author} {\bibinfo {author} {\bibfnamefont {J.~P.}\ \bibnamefont
  {Lu}},\ }\href {https://doi.org/10.1103/PhysRevB.49.5687} {\bibfield
  {journal} {\bibinfo  {journal} {Phys. Rev. B}\ }\textbf {\bibinfo {volume}
  {49}},\ \bibinfo {pages} {5687} (\bibinfo {year} {1994})}\BibitemShut
  {NoStop}%
\bibitem [{\citenamefont {Lee}(1997)}]{LEE97}%
  \BibitemOpen
  \bibfield  {author} {\bibinfo {author} {\bibfnamefont {D.~M.}\ \bibnamefont
  {Lee}},\ }\href {https://doi.org/10.1103/RevModPhys.69.645} {\bibfield
  {journal} {\bibinfo  {journal} {Rev. Mod. Phys.}\ }\textbf {\bibinfo {volume}
  {69}},\ \bibinfo {pages} {645} (\bibinfo {year} {1997})}\BibitemShut
  {NoStop}%
\bibitem [{\citenamefont {Rozen}\ \emph {et~al.}(2021)\citenamefont {Rozen},
  \citenamefont {Park}, \citenamefont {Zondiner}, \citenamefont {Cao},
  \citenamefont {{Rodan-Legrain}}, \citenamefont {Taniguchi}, \citenamefont
  {Watanabe}, \citenamefont {Oreg}, \citenamefont {Stern}, \citenamefont
  {Berg}, \citenamefont {{Jarillo-Herrero}},\ and\ \citenamefont
  {Ilani}}]{ROZ21}%
  \BibitemOpen
  \bibfield  {author} {\bibinfo {author} {\bibfnamefont {A.}~\bibnamefont
  {Rozen}}, \bibinfo {author} {\bibfnamefont {J.~M.}\ \bibnamefont {Park}},
  \bibinfo {author} {\bibfnamefont {U.}~\bibnamefont {Zondiner}}, \bibinfo
  {author} {\bibfnamefont {Y.}~\bibnamefont {Cao}}, \bibinfo {author}
  {\bibfnamefont {D.}~\bibnamefont {{Rodan-Legrain}}}, \bibinfo {author}
  {\bibfnamefont {T.}~\bibnamefont {Taniguchi}}, \bibinfo {author}
  {\bibfnamefont {K.}~\bibnamefont {Watanabe}}, \bibinfo {author}
  {\bibfnamefont {Y.}~\bibnamefont {Oreg}}, \bibinfo {author} {\bibfnamefont
  {A.}~\bibnamefont {Stern}}, \bibinfo {author} {\bibfnamefont
  {E.}~\bibnamefont {Berg}}, \bibinfo {author} {\bibfnamefont {P.}~\bibnamefont
  {{Jarillo-Herrero}}},\ and\ \bibinfo {author} {\bibfnamefont
  {S.}~\bibnamefont {Ilani}},\ }\href
  {https://doi.org/10.1038/s41586-021-03319-3} {\bibfield  {journal} {\bibinfo
  {journal} {Nature}\ }\textbf {\bibinfo {volume} {592}},\ \bibinfo {pages}
  {214} (\bibinfo {year} {2021})}\BibitemShut {NoStop}%
\bibitem [{\citenamefont {Saito}\ \emph {et~al.}(2021)\citenamefont {Saito},
  \citenamefont {Yang}, \citenamefont {Ge}, \citenamefont {Liu}, \citenamefont
  {Taniguchi}, \citenamefont {Watanabe}, \citenamefont {Li}, \citenamefont
  {Berg},\ and\ \citenamefont {Young}}]{SAI21a}%
  \BibitemOpen
  \bibfield  {author} {\bibinfo {author} {\bibfnamefont {Y.}~\bibnamefont
  {Saito}}, \bibinfo {author} {\bibfnamefont {F.}~\bibnamefont {Yang}},
  \bibinfo {author} {\bibfnamefont {J.}~\bibnamefont {Ge}}, \bibinfo {author}
  {\bibfnamefont {X.}~\bibnamefont {Liu}}, \bibinfo {author} {\bibfnamefont
  {T.}~\bibnamefont {Taniguchi}}, \bibinfo {author} {\bibfnamefont
  {K.}~\bibnamefont {Watanabe}}, \bibinfo {author} {\bibfnamefont {J.~I.~A.}\
  \bibnamefont {Li}}, \bibinfo {author} {\bibfnamefont {E.}~\bibnamefont
  {Berg}},\ and\ \bibinfo {author} {\bibfnamefont {A.~F.}\ \bibnamefont
  {Young}},\ }\href {https://doi.org/10.1038/s41586-021-03409-2} {\bibfield
  {journal} {\bibinfo  {journal} {Nature}\ }\textbf {\bibinfo {volume} {592}},\
  \bibinfo {pages} {220} (\bibinfo {year} {2021})}\BibitemShut {NoStop}%
\bibitem [{\citenamefont {Khaliullin}\ and\ \citenamefont
  {Oudovenko}(1997)}]{KHA97}%
  \BibitemOpen
  \bibfield  {author} {\bibinfo {author} {\bibfnamefont {G.}~\bibnamefont
  {Khaliullin}}\ and\ \bibinfo {author} {\bibfnamefont {V.}~\bibnamefont
  {Oudovenko}},\ }\href {https://doi.org/10.1103/PhysRevB.56.R14243} {\bibfield
   {journal} {\bibinfo  {journal} {Phys. Rev. B}\ }\textbf {\bibinfo {volume}
  {56}},\ \bibinfo {pages} {R14243} (\bibinfo {year} {1997})}\BibitemShut
  {NoStop}%
\bibitem [{\citenamefont {Frenkel}\ and\ \citenamefont {Smit}(2001)}]{FRE01}%
  \BibitemOpen
  \bibfield  {author} {\bibinfo {author} {\bibfnamefont {D.}~\bibnamefont
  {Frenkel}}\ and\ \bibinfo {author} {\bibfnamefont {B.}~\bibnamefont {Smit}},\
  }\href@noop {} {\emph {\bibinfo {title} {Understanding {{Molecular
  Simulation}}: {{From Algorithms}} to {{Applications}}}}}\ (\bibinfo
  {publisher} {Elsevier},\ \bibinfo {year} {2001})\BibitemShut {NoStop}%
\bibitem [{\citenamefont {Li}\ \emph {et~al.}(2014)\citenamefont {Li},
  \citenamefont {Lieb},\ and\ \citenamefont {Wu}}]{LI14}%
  \BibitemOpen
  \bibfield  {author} {\bibinfo {author} {\bibfnamefont {Y.}~\bibnamefont
  {Li}}, \bibinfo {author} {\bibfnamefont {E.~H.}\ \bibnamefont {Lieb}},\ and\
  \bibinfo {author} {\bibfnamefont {C.}~\bibnamefont {Wu}},\ }\href
  {https://doi.org/10.1103/PhysRevLett.112.217201} {\bibfield  {journal}
  {\bibinfo  {journal} {Phys. Rev. Lett.}\ }\textbf {\bibinfo {volume} {112}},\
  \bibinfo {pages} {217201} (\bibinfo {year} {2014})}\BibitemShut {NoStop}%
\bibitem [{\citenamefont {Lu}\ \emph {et~al.}(2019)\citenamefont {Lu},
  \citenamefont {Stepanov}, \citenamefont {Yang}, \citenamefont {Xie},
  \citenamefont {Aamir}, \citenamefont {Das}, \citenamefont {Urgell},
  \citenamefont {Watanabe}, \citenamefont {Taniguchi}, \citenamefont {Zhang},
  \citenamefont {Bachtold}, \citenamefont {MacDonald},\ and\ \citenamefont
  {Efetov}}]{LU19}%
  \BibitemOpen
  \bibfield  {author} {\bibinfo {author} {\bibfnamefont {X.}~\bibnamefont
  {Lu}}, \bibinfo {author} {\bibfnamefont {P.}~\bibnamefont {Stepanov}},
  \bibinfo {author} {\bibfnamefont {W.}~\bibnamefont {Yang}}, \bibinfo {author}
  {\bibfnamefont {M.}~\bibnamefont {Xie}}, \bibinfo {author} {\bibfnamefont
  {M.~A.}\ \bibnamefont {Aamir}}, \bibinfo {author} {\bibfnamefont
  {I.}~\bibnamefont {Das}}, \bibinfo {author} {\bibfnamefont {C.}~\bibnamefont
  {Urgell}}, \bibinfo {author} {\bibfnamefont {K.}~\bibnamefont {Watanabe}},
  \bibinfo {author} {\bibfnamefont {T.}~\bibnamefont {Taniguchi}}, \bibinfo
  {author} {\bibfnamefont {G.}~\bibnamefont {Zhang}}, \bibinfo {author}
  {\bibfnamefont {A.}~\bibnamefont {Bachtold}}, \bibinfo {author}
  {\bibfnamefont {A.~H.}\ \bibnamefont {MacDonald}},\ and\ \bibinfo {author}
  {\bibfnamefont {D.~K.}\ \bibnamefont {Efetov}},\ }\href
  {https://doi.org/10.1038/s41586-019-1695-0} {\bibfield  {journal} {\bibinfo
  {journal} {Nature}\ }\textbf {\bibinfo {volume} {574}},\ \bibinfo {pages}
  {653} (\bibinfo {year} {2019})}\BibitemShut {NoStop}%
\bibitem [{\citenamefont {Pierce}\ \emph {et~al.}(2021)\citenamefont {Pierce},
  \citenamefont {Xie}, \citenamefont {Park}, \citenamefont {Khalaf},
  \citenamefont {Lee}, \citenamefont {Cao}, \citenamefont {Parker},
  \citenamefont {Forrester}, \citenamefont {Chen}, \citenamefont {Watanabe},
  \citenamefont {Taniguchi}, \citenamefont {Vishwanath}, \citenamefont
  {{Jarillo-Herrero}},\ and\ \citenamefont {Yacoby}}]{PIE21}%
  \BibitemOpen
  \bibfield  {author} {\bibinfo {author} {\bibfnamefont {A.~T.}\ \bibnamefont
  {Pierce}}, \bibinfo {author} {\bibfnamefont {Y.}~\bibnamefont {Xie}},
  \bibinfo {author} {\bibfnamefont {J.~M.}\ \bibnamefont {Park}}, \bibinfo
  {author} {\bibfnamefont {E.}~\bibnamefont {Khalaf}}, \bibinfo {author}
  {\bibfnamefont {S.~H.}\ \bibnamefont {Lee}}, \bibinfo {author} {\bibfnamefont
  {Y.}~\bibnamefont {Cao}}, \bibinfo {author} {\bibfnamefont {D.~E.}\
  \bibnamefont {Parker}}, \bibinfo {author} {\bibfnamefont {P.~R.}\
  \bibnamefont {Forrester}}, \bibinfo {author} {\bibfnamefont {S.}~\bibnamefont
  {Chen}}, \bibinfo {author} {\bibfnamefont {K.}~\bibnamefont {Watanabe}},
  \bibinfo {author} {\bibfnamefont {T.}~\bibnamefont {Taniguchi}}, \bibinfo
  {author} {\bibfnamefont {A.}~\bibnamefont {Vishwanath}}, \bibinfo {author}
  {\bibfnamefont {P.}~\bibnamefont {{Jarillo-Herrero}}},\ and\ \bibinfo
  {author} {\bibfnamefont {A.}~\bibnamefont {Yacoby}},\ }\href
  {https://doi.org/10.1038/s41567-021-01347-4} {\bibfield  {journal} {\bibinfo
  {journal} {Nat. Phys.}\ }\textbf {\bibinfo {volume} {17}},\ \bibinfo {pages}
  {1210} (\bibinfo {year} {2021})}\BibitemShut {NoStop}%
\bibitem [{\citenamefont {Jaoui}\ \emph {et~al.}(2022)\citenamefont {Jaoui},
  \citenamefont {Das}, \citenamefont {Di~Battista}, \citenamefont
  {{D{\'i}ez-M{\'e}rida}}, \citenamefont {Lu}, \citenamefont {Watanabe},
  \citenamefont {Taniguchi}, \citenamefont {Ishizuka}, \citenamefont
  {Levitov},\ and\ \citenamefont {Efetov}}]{JAO22}%
  \BibitemOpen
  \bibfield  {author} {\bibinfo {author} {\bibfnamefont {A.}~\bibnamefont
  {Jaoui}}, \bibinfo {author} {\bibfnamefont {I.}~\bibnamefont {Das}}, \bibinfo
  {author} {\bibfnamefont {G.}~\bibnamefont {Di~Battista}}, \bibinfo {author}
  {\bibfnamefont {J.}~\bibnamefont {{D{\'i}ez-M{\'e}rida}}}, \bibinfo {author}
  {\bibfnamefont {X.}~\bibnamefont {Lu}}, \bibinfo {author} {\bibfnamefont
  {K.}~\bibnamefont {Watanabe}}, \bibinfo {author} {\bibfnamefont
  {T.}~\bibnamefont {Taniguchi}}, \bibinfo {author} {\bibfnamefont
  {H.}~\bibnamefont {Ishizuka}}, \bibinfo {author} {\bibfnamefont
  {L.}~\bibnamefont {Levitov}},\ and\ \bibinfo {author} {\bibfnamefont {D.~K.}\
  \bibnamefont {Efetov}},\ }\href {https://doi.org/10.1038/s41567-022-01556-5}
  {\bibfield  {journal} {\bibinfo  {journal} {Nat. Phys.}\ }\textbf {\bibinfo
  {volume} {18}},\ \bibinfo {pages} {633} (\bibinfo {year} {2022})}\BibitemShut
  {NoStop}%
\bibitem [{\citenamefont {Liu}\ \emph {et~al.}(2022)\citenamefont {Liu},
  \citenamefont {Zhang}, \citenamefont {Watanabe}, \citenamefont {Taniguchi},\
  and\ \citenamefont {Li}}]{LIU22b}%
  \BibitemOpen
  \bibfield  {author} {\bibinfo {author} {\bibfnamefont {X.}~\bibnamefont
  {Liu}}, \bibinfo {author} {\bibfnamefont {N.~J.}\ \bibnamefont {Zhang}},
  \bibinfo {author} {\bibfnamefont {K.}~\bibnamefont {Watanabe}}, \bibinfo
  {author} {\bibfnamefont {T.}~\bibnamefont {Taniguchi}},\ and\ \bibinfo
  {author} {\bibfnamefont {J.~I.~A.}\ \bibnamefont {Li}},\ }\href
  {https://doi.org/10.1038/s41567-022-01515-0} {\bibfield  {journal} {\bibinfo
  {journal} {Nat. Phys.}\ }\textbf {\bibinfo {volume} {18}},\ \bibinfo {pages}
  {522} (\bibinfo {year} {2022})}\BibitemShut {NoStop}%
\bibitem [{\citenamefont {Zhang}\ \emph {et~al.}(2025)\citenamefont {Zhang},
  \citenamefont {Wu}, \citenamefont {C{\u a}lug{\u a}ru}, \citenamefont {Hu},
  \citenamefont {Taniguchi}, \citenamefont {Wanatabe}, \citenamefont
  {Bernevig},\ and\ \citenamefont {Andrei}}]{ZHA25a}%
  \BibitemOpen
  \bibfield  {author} {\bibinfo {author} {\bibfnamefont {Z.}~\bibnamefont
  {Zhang}}, \bibinfo {author} {\bibfnamefont {S.}~\bibnamefont {Wu}}, \bibinfo
  {author} {\bibfnamefont {D.}~\bibnamefont {C{\u a}lug{\u a}ru}}, \bibinfo
  {author} {\bibfnamefont {H.}~\bibnamefont {Hu}}, \bibinfo {author}
  {\bibfnamefont {T.}~\bibnamefont {Taniguchi}}, \bibinfo {author}
  {\bibfnamefont {K.}~\bibnamefont {Wanatabe}}, \bibinfo {author}
  {\bibfnamefont {A.~B.}\ \bibnamefont {Bernevig}},\ and\ \bibinfo {author}
  {\bibfnamefont {E.~Y.}\ \bibnamefont {Andrei}},\ }\href
  {https://doi.org/10.48550/arXiv.2503.17875} {\bibinfo {title} {Heavy
  fermions, mass renormalization and local moments in magic-angle twisted
  bilayer graphene via planar tunneling spectroscopy}} (\bibinfo {year}
  {2025}),\ \Eprint {https://arxiv.org/abs/2503.17875} {arXiv:2503.17875
  [cond-mat.mes-hall]} \BibitemShut {NoStop}%
\bibitem [{\citenamefont {Klein}\ \emph {et~al.}(2026)\citenamefont {Klein},
  \citenamefont {Zondiner}, \citenamefont {Keren}, \citenamefont {Birkbeck},
  \citenamefont {Inbar}, \citenamefont {Xiao}, \citenamefont {Zamir},
  \citenamefont {Sidorova}, \citenamefont {Al~Ezzi}, \citenamefont {Peng},
  \citenamefont {Watanabe}, \citenamefont {Taniguchi}, \citenamefont {Adam},\
  and\ \citenamefont {Ilani}}]{KLE26}%
  \BibitemOpen
  \bibfield  {author} {\bibinfo {author} {\bibfnamefont {D.~R.}\ \bibnamefont
  {Klein}}, \bibinfo {author} {\bibfnamefont {U.}~\bibnamefont {Zondiner}},
  \bibinfo {author} {\bibfnamefont {A.}~\bibnamefont {Keren}}, \bibinfo
  {author} {\bibfnamefont {J.}~\bibnamefont {Birkbeck}}, \bibinfo {author}
  {\bibfnamefont {A.}~\bibnamefont {Inbar}}, \bibinfo {author} {\bibfnamefont
  {J.}~\bibnamefont {Xiao}}, \bibinfo {author} {\bibfnamefont {Y.}~\bibnamefont
  {Zamir}}, \bibinfo {author} {\bibfnamefont {M.}~\bibnamefont {Sidorova}},
  \bibinfo {author} {\bibfnamefont {M.~M.}\ \bibnamefont {Al~Ezzi}}, \bibinfo
  {author} {\bibfnamefont {L.}~\bibnamefont {Peng}}, \bibinfo {author}
  {\bibfnamefont {K.}~\bibnamefont {Watanabe}}, \bibinfo {author}
  {\bibfnamefont {T.}~\bibnamefont {Taniguchi}}, \bibinfo {author}
  {\bibfnamefont {S.}~\bibnamefont {Adam}},\ and\ \bibinfo {author}
  {\bibfnamefont {S.}~\bibnamefont {Ilani}},\ }\href
  {https://doi.org/10.1038/s41586-025-10085-z} {\bibfield  {journal} {\bibinfo
  {journal} {Nature}\ }\textbf {\bibinfo {volume} {650}},\ \bibinfo {pages}
  {875} (\bibinfo {year} {2026})}\BibitemShut {NoStop}%
\bibitem [{\citenamefont {Lee}\ \emph {et~al.}(2026)\citenamefont {Lee},
  \citenamefont {Das}, \citenamefont {{Herzog-Arbeitman}}, \citenamefont
  {Papp}, \citenamefont {Li}, \citenamefont {Daschner}, \citenamefont {Zhou},
  \citenamefont {Bhatt}, \citenamefont {Currle}, \citenamefont {Yu},
  \citenamefont {Jiang}, \citenamefont {Becherer}, \citenamefont {Mittermeier},
  \citenamefont {Altpeter}, \citenamefont {Obermayer}, \citenamefont {Lorenz},
  \citenamefont {Chavez}, \citenamefont {Le}, \citenamefont {Williams},
  \citenamefont {Watanabe}, \citenamefont {Taniguchi}, \citenamefont
  {Bernevig},\ and\ \citenamefont {Efetov}}]{LEE26}%
  \BibitemOpen
  \bibfield  {author} {\bibinfo {author} {\bibfnamefont {M.}~\bibnamefont
  {Lee}}, \bibinfo {author} {\bibfnamefont {I.}~\bibnamefont {Das}}, \bibinfo
  {author} {\bibfnamefont {J.}~\bibnamefont {{Herzog-Arbeitman}}}, \bibinfo
  {author} {\bibfnamefont {J.}~\bibnamefont {Papp}}, \bibinfo {author}
  {\bibfnamefont {J.}~\bibnamefont {Li}}, \bibinfo {author} {\bibfnamefont
  {M.}~\bibnamefont {Daschner}}, \bibinfo {author} {\bibfnamefont
  {Z.}~\bibnamefont {Zhou}}, \bibinfo {author} {\bibfnamefont {M.}~\bibnamefont
  {Bhatt}}, \bibinfo {author} {\bibfnamefont {M.}~\bibnamefont {Currle}},
  \bibinfo {author} {\bibfnamefont {J.}~\bibnamefont {Yu}}, \bibinfo {author}
  {\bibfnamefont {Y.}~\bibnamefont {Jiang}}, \bibinfo {author} {\bibfnamefont
  {M.}~\bibnamefont {Becherer}}, \bibinfo {author} {\bibfnamefont
  {R.}~\bibnamefont {Mittermeier}}, \bibinfo {author} {\bibfnamefont
  {P.}~\bibnamefont {Altpeter}}, \bibinfo {author} {\bibfnamefont
  {C.}~\bibnamefont {Obermayer}}, \bibinfo {author} {\bibfnamefont
  {H.}~\bibnamefont {Lorenz}}, \bibinfo {author} {\bibfnamefont
  {G.}~\bibnamefont {Chavez}}, \bibinfo {author} {\bibfnamefont {B.~T.}\
  \bibnamefont {Le}}, \bibinfo {author} {\bibfnamefont {J.}~\bibnamefont
  {Williams}}, \bibinfo {author} {\bibfnamefont {K.}~\bibnamefont {Watanabe}},
  \bibinfo {author} {\bibfnamefont {T.}~\bibnamefont {Taniguchi}}, \bibinfo
  {author} {\bibfnamefont {B.~A.}\ \bibnamefont {Bernevig}},\ and\ \bibinfo
  {author} {\bibfnamefont {D.~K.}\ \bibnamefont {Efetov}},\ }\href
  {https://doi.org/10.1021/acs.nanolett.5c05015} {\bibfield  {journal}
  {\bibinfo  {journal} {Nano Lett.}\ }\textbf {\bibinfo {volume} {26}},\
  \bibinfo {pages} {4046} (\bibinfo {year} {2026})}\BibitemShut {NoStop}%
\bibitem [{\citenamefont {{Mendez-Valderrama}}\ and\ \citenamefont
  {Chowdhury}(2021)}]{MEN21}%
  \BibitemOpen
  \bibfield  {author} {\bibinfo {author} {\bibfnamefont {J.~F.}\ \bibnamefont
  {{Mendez-Valderrama}}}\ and\ \bibinfo {author} {\bibfnamefont
  {D.}~\bibnamefont {Chowdhury}},\ }\href
  {https://doi.org/10.1103/PhysRevB.103.195111} {\bibfield  {journal} {\bibinfo
   {journal} {Phys. Rev. B}\ }\textbf {\bibinfo {volume} {103}},\ \bibinfo
  {pages} {195111} (\bibinfo {year} {2021})}\BibitemShut {NoStop}%
\bibitem [{\citenamefont {Giamarchi}(2014)}]{GIA14}%
  \BibitemOpen
  \bibfield  {author} {\bibinfo {author} {\bibfnamefont {T.}~\bibnamefont
  {Giamarchi}},\ }\href@noop {} {\emph {\bibinfo {title} {Quantum {{Physics}}
  in {{One Dimension}}: 121}}}\ (\bibinfo  {publisher} {OUP Oxford},\ \bibinfo
  {address} {Oxford},\ \bibinfo {year} {2014})\BibitemShut {NoStop}%
\bibitem [{\citenamefont {Capone}\ \emph {et~al.}(2002)\citenamefont {Capone},
  \citenamefont {Fabrizio}, \citenamefont {Castellani},\ and\ \citenamefont
  {Tosatti}}]{CAP02}%
  \BibitemOpen
  \bibfield  {author} {\bibinfo {author} {\bibfnamefont {M.}~\bibnamefont
  {Capone}}, \bibinfo {author} {\bibfnamefont {M.}~\bibnamefont {Fabrizio}},
  \bibinfo {author} {\bibfnamefont {C.}~\bibnamefont {Castellani}},\ and\
  \bibinfo {author} {\bibfnamefont {E.}~\bibnamefont {Tosatti}},\ }\href
  {https://doi.org/10.1126/science.1071122} {\bibfield  {journal} {\bibinfo
  {journal} {Science}\ }\textbf {\bibinfo {volume} {296}},\ \bibinfo {pages}
  {2364} (\bibinfo {year} {2002})}\BibitemShut {NoStop}%
\bibitem [{\citenamefont {Han}\ \emph {et~al.}(2003)\citenamefont {Han},
  \citenamefont {Gunnarsson},\ and\ \citenamefont {Crespi}}]{HAN03}%
  \BibitemOpen
  \bibfield  {author} {\bibinfo {author} {\bibfnamefont {J.~E.}\ \bibnamefont
  {Han}}, \bibinfo {author} {\bibfnamefont {O.}~\bibnamefont {Gunnarsson}},\
  and\ \bibinfo {author} {\bibfnamefont {V.~H.}\ \bibnamefont {Crespi}},\
  }\href {https://doi.org/10.1103/PhysRevLett.90.167006} {\bibfield  {journal}
  {\bibinfo  {journal} {Phys. Rev. Lett.}\ }\textbf {\bibinfo {volume} {90}},\
  \bibinfo {pages} {167006} (\bibinfo {year} {2003})}\BibitemShut {NoStop}%
\bibitem [{\citenamefont {Nomura}\ \emph {et~al.}(2016)\citenamefont {Nomura},
  \citenamefont {Sakai}, \citenamefont {Capone},\ and\ \citenamefont
  {Arita}}]{NOM16}%
  \BibitemOpen
  \bibfield  {author} {\bibinfo {author} {\bibfnamefont {Y.}~\bibnamefont
  {Nomura}}, \bibinfo {author} {\bibfnamefont {S.}~\bibnamefont {Sakai}},
  \bibinfo {author} {\bibfnamefont {M.}~\bibnamefont {Capone}},\ and\ \bibinfo
  {author} {\bibfnamefont {R.}~\bibnamefont {Arita}},\ }\href
  {https://doi.org/10.1088/0953-8984/28/15/153001} {\bibfield  {journal}
  {\bibinfo  {journal} {J. Phys.: Condens. Matter}\ }\textbf {\bibinfo {volume}
  {28}},\ \bibinfo {pages} {153001} (\bibinfo {year} {2016})}\BibitemShut
  {NoStop}%
\bibitem [{\citenamefont {Nomura}\ \emph {et~al.}(2015)\citenamefont {Nomura},
  \citenamefont {Sakai}, \citenamefont {Capone},\ and\ \citenamefont
  {Arita}}]{NOM15}%
  \BibitemOpen
  \bibfield  {author} {\bibinfo {author} {\bibfnamefont {Y.}~\bibnamefont
  {Nomura}}, \bibinfo {author} {\bibfnamefont {S.}~\bibnamefont {Sakai}},
  \bibinfo {author} {\bibfnamefont {M.}~\bibnamefont {Capone}},\ and\ \bibinfo
  {author} {\bibfnamefont {R.}~\bibnamefont {Arita}},\ }\href
  {https://doi.org/10.1126/sciadv.1500568} {\bibfield  {journal} {\bibinfo
  {journal} {Sci. Adv.}\ }\textbf {\bibinfo {volume} {1}},\ \bibinfo {pages}
  {e1500568} (\bibinfo {year} {2015})}\BibitemShut {NoStop}%
\bibitem [{\citenamefont {Hoshino}\ and\ \citenamefont {Werner}(2017)}]{HOS17}%
  \BibitemOpen
  \bibfield  {author} {\bibinfo {author} {\bibfnamefont {S.}~\bibnamefont
  {Hoshino}}\ and\ \bibinfo {author} {\bibfnamefont {P.}~\bibnamefont
  {Werner}},\ }\href {https://doi.org/10.1103/PhysRevLett.118.177002}
  {\bibfield  {journal} {\bibinfo  {journal} {Phys. Rev. Lett.}\ }\textbf
  {\bibinfo {volume} {118}},\ \bibinfo {pages} {177002} (\bibinfo {year}
  {2017})}\BibitemShut {NoStop}%
\bibitem [{\citenamefont {Georges}\ \emph {et~al.}(2000)\citenamefont
  {Georges}, \citenamefont {Giamarchi},\ and\ \citenamefont {Sandler}}]{GEO00}%
  \BibitemOpen
  \bibfield  {author} {\bibinfo {author} {\bibfnamefont {A.}~\bibnamefont
  {Georges}}, \bibinfo {author} {\bibfnamefont {T.}~\bibnamefont {Giamarchi}},\
  and\ \bibinfo {author} {\bibfnamefont {N.}~\bibnamefont {Sandler}},\ }\href
  {https://doi.org/10.1103/PhysRevB.61.16393} {\bibfield  {journal} {\bibinfo
  {journal} {Phys. Rev. B}\ }\textbf {\bibinfo {volume} {61}},\ \bibinfo
  {pages} {16393} (\bibinfo {year} {2000})}\BibitemShut {NoStop}%
\bibitem [{\citenamefont {Biermann}\ \emph {et~al.}(2002)\citenamefont
  {Biermann}, \citenamefont {Georges}, \citenamefont {Giamarchi},\ and\
  \citenamefont {Lichtenstein}}]{BIE02}%
  \BibitemOpen
  \bibfield  {author} {\bibinfo {author} {\bibfnamefont {S.}~\bibnamefont
  {Biermann}}, \bibinfo {author} {\bibfnamefont {A.}~\bibnamefont {Georges}},
  \bibinfo {author} {\bibfnamefont {T.}~\bibnamefont {Giamarchi}},\ and\
  \bibinfo {author} {\bibfnamefont {A.}~\bibnamefont {Lichtenstein}},\ }\href
  {https://doi.org/10.48550/arXiv.cond-mat/0201542} {\bibinfo {title}
  {Quasi-one-dimensional organic conductors: Dimensional crossover and some
  puzzles}} (\bibinfo {year} {2002}),\ \Eprint
  {https://arxiv.org/abs/cond-mat/0201542} {arXiv:cond-mat/0201542}
  \BibitemShut {NoStop}%
\bibitem [{\citenamefont {Giamarchi}\ \emph {et~al.}(2004)\citenamefont
  {Giamarchi}, \citenamefont {Biermann}, \citenamefont {Georges},\ and\
  \citenamefont {Lichtenstein}}]{GIA04}%
  \BibitemOpen
  \bibfield  {author} {\bibinfo {author} {\bibfnamefont {T.}~\bibnamefont
  {Giamarchi}}, \bibinfo {author} {\bibfnamefont {S.}~\bibnamefont {Biermann}},
  \bibinfo {author} {\bibfnamefont {A.}~\bibnamefont {Georges}},\ and\ \bibinfo
  {author} {\bibfnamefont {A.}~\bibnamefont {Lichtenstein}},\ }\href
  {https://doi.org/10.1051/jp4:2004114004} {\bibfield  {journal} {\bibinfo
  {journal} {J. Phys. IV France}\ }\textbf {\bibinfo {volume} {114}},\ \bibinfo
  {pages} {23} (\bibinfo {year} {2004})}\BibitemShut {NoStop}%
\bibitem [{\citenamefont {Bradlyn}\ \emph {et~al.}(2017)\citenamefont
  {Bradlyn}, \citenamefont {Elcoro}, \citenamefont {Cano}, \citenamefont
  {Vergniory}, \citenamefont {Wang}, \citenamefont {Felser}, \citenamefont
  {Aroyo},\ and\ \citenamefont {Bernevig}}]{BRA17}%
  \BibitemOpen
  \bibfield  {author} {\bibinfo {author} {\bibfnamefont {B.}~\bibnamefont
  {Bradlyn}}, \bibinfo {author} {\bibfnamefont {L.}~\bibnamefont {Elcoro}},
  \bibinfo {author} {\bibfnamefont {J.}~\bibnamefont {Cano}}, \bibinfo {author}
  {\bibfnamefont {M.~G.}\ \bibnamefont {Vergniory}}, \bibinfo {author}
  {\bibfnamefont {Z.}~\bibnamefont {Wang}}, \bibinfo {author} {\bibfnamefont
  {C.}~\bibnamefont {Felser}}, \bibinfo {author} {\bibfnamefont {M.~I.}\
  \bibnamefont {Aroyo}},\ and\ \bibinfo {author} {\bibfnamefont {B.~A.}\
  \bibnamefont {Bernevig}},\ }\href {https://doi.org/10.1038/nature23268}
  {\bibfield  {journal} {\bibinfo  {journal} {Nature}\ }\textbf {\bibinfo
  {volume} {547}},\ \bibinfo {pages} {298} (\bibinfo {year}
  {2017})}\BibitemShut {NoStop}%
\bibitem [{\citenamefont {Xu}\ \emph {et~al.}(2024)\citenamefont {Xu},
  \citenamefont {Vergniory}, \citenamefont {Ma}, \citenamefont {Ma{\~n}es},
  \citenamefont {Song}, \citenamefont {Bernevig}, \citenamefont {Regnault},\
  and\ \citenamefont {Elcoro}}]{XU24a}%
  \BibitemOpen
  \bibfield  {author} {\bibinfo {author} {\bibfnamefont {Y.}~\bibnamefont
  {Xu}}, \bibinfo {author} {\bibfnamefont {M.~G.}\ \bibnamefont {Vergniory}},
  \bibinfo {author} {\bibfnamefont {D.-S.}\ \bibnamefont {Ma}}, \bibinfo
  {author} {\bibfnamefont {J.~L.}\ \bibnamefont {Ma{\~n}es}}, \bibinfo {author}
  {\bibfnamefont {Z.-D.}\ \bibnamefont {Song}}, \bibinfo {author}
  {\bibfnamefont {B.~A.}\ \bibnamefont {Bernevig}}, \bibinfo {author}
  {\bibfnamefont {N.}~\bibnamefont {Regnault}},\ and\ \bibinfo {author}
  {\bibfnamefont {L.}~\bibnamefont {Elcoro}},\ }\href
  {https://doi.org/10.1126/science.adf8458} {\bibfield  {journal} {\bibinfo
  {journal} {Science}\ }\textbf {\bibinfo {volume} {384}},\ \bibinfo {pages}
  {eadf8458} (\bibinfo {year} {2024})}\BibitemShut {NoStop}%
\bibitem [{\citenamefont {Petralanda}\ \emph {et~al.}(2024)\citenamefont
  {Petralanda}, \citenamefont {Jiang}, \citenamefont {Bernevig}, \citenamefont
  {Regnault},\ and\ \citenamefont {Elcoro}}]{PET24}%
  \BibitemOpen
  \bibfield  {author} {\bibinfo {author} {\bibfnamefont {U.}~\bibnamefont
  {Petralanda}}, \bibinfo {author} {\bibfnamefont {Y.}~\bibnamefont {Jiang}},
  \bibinfo {author} {\bibfnamefont {B.~A.}\ \bibnamefont {Bernevig}}, \bibinfo
  {author} {\bibfnamefont {N.}~\bibnamefont {Regnault}},\ and\ \bibinfo
  {author} {\bibfnamefont {L.}~\bibnamefont {Elcoro}},\ }\href
  {https://doi.org/10.48550/arXiv.2411.08950} {\bibinfo {title}
  {Two-dimensional {{Topological Quantum Chemistry}} and {{Catalog}} of
  {{Topological Materials}}}} (\bibinfo {year} {2024}),\ \Eprint
  {https://arxiv.org/abs/2411.08950} {arXiv:2411.08950 [cond-mat.mes-hall]}
  \BibitemShut {NoStop}%
\bibitem [{\citenamefont {Schrieffer}\ and\ \citenamefont
  {Wolff}(1966)}]{SCH66}%
  \BibitemOpen
  \bibfield  {author} {\bibinfo {author} {\bibfnamefont {J.~R.}\ \bibnamefont
  {Schrieffer}}\ and\ \bibinfo {author} {\bibfnamefont {P.~A.}\ \bibnamefont
  {Wolff}},\ }\href {https://doi.org/10.1103/PhysRev.149.491} {\bibfield
  {journal} {\bibinfo  {journal} {Phys. Rev.}\ }\textbf {\bibinfo {volume}
  {149}},\ \bibinfo {pages} {491} (\bibinfo {year} {1966})}\BibitemShut
  {NoStop}%
\bibitem [{\citenamefont {Tarnopolsky}\ \emph {et~al.}(2019)\citenamefont
  {Tarnopolsky}, \citenamefont {Kruchkov},\ and\ \citenamefont
  {Vishwanath}}]{TAR19}%
  \BibitemOpen
  \bibfield  {author} {\bibinfo {author} {\bibfnamefont {G.}~\bibnamefont
  {Tarnopolsky}}, \bibinfo {author} {\bibfnamefont {A.~J.}\ \bibnamefont
  {Kruchkov}},\ and\ \bibinfo {author} {\bibfnamefont {A.}~\bibnamefont
  {Vishwanath}},\ }\href {https://doi.org/10.1103/PhysRevLett.122.106405}
  {\bibfield  {journal} {\bibinfo  {journal} {Phys. Rev. Lett.}\ }\textbf
  {\bibinfo {volume} {122}},\ \bibinfo {pages} {106405} (\bibinfo {year}
  {2019})}\BibitemShut {NoStop}%
\bibitem [{\citenamefont {{de'Medici}}\ \emph {et~al.}(2005)\citenamefont
  {{de'Medici}}, \citenamefont {Georges},\ and\ \citenamefont
  {Biermann}}]{DE05}%
  \BibitemOpen
  \bibfield  {author} {\bibinfo {author} {\bibfnamefont {L.}~\bibnamefont
  {{de'Medici}}}, \bibinfo {author} {\bibfnamefont {A.}~\bibnamefont
  {Georges}},\ and\ \bibinfo {author} {\bibfnamefont {S.}~\bibnamefont
  {Biermann}},\ }\href {https://doi.org/10.1103/PhysRevB.72.205124} {\bibfield
  {journal} {\bibinfo  {journal} {Phys. Rev. B}\ }\textbf {\bibinfo {volume}
  {72}},\ \bibinfo {pages} {205124} (\bibinfo {year} {2005})}\BibitemShut
  {NoStop}%
\bibitem [{\citenamefont {Hassan}\ and\ \citenamefont {{de'
  Medici}}(2010)}]{HAS10}%
  \BibitemOpen
  \bibfield  {author} {\bibinfo {author} {\bibfnamefont {S.~R.}\ \bibnamefont
  {Hassan}}\ and\ \bibinfo {author} {\bibfnamefont {L.}~\bibnamefont {{de'
  Medici}}},\ }\href {https://doi.org/10.1103/PhysRevB.81.035106} {\bibfield
  {journal} {\bibinfo  {journal} {Phys. Rev. B}\ }\textbf {\bibinfo {volume}
  {81}},\ \bibinfo {pages} {035106} (\bibinfo {year} {2010})}\BibitemShut
  {NoStop}%
\bibitem [{\citenamefont {Yu}\ and\ \citenamefont {Si}(2012)}]{YU12}%
  \BibitemOpen
  \bibfield  {author} {\bibinfo {author} {\bibfnamefont {R.}~\bibnamefont
  {Yu}}\ and\ \bibinfo {author} {\bibfnamefont {Q.}~\bibnamefont {Si}},\ }\href
  {https://doi.org/10.1103/PhysRevB.86.085104} {\bibfield  {journal} {\bibinfo
  {journal} {Phys. Rev. B}\ }\textbf {\bibinfo {volume} {86}},\ \bibinfo
  {pages} {085104} (\bibinfo {year} {2012})}\BibitemShut {NoStop}%
\bibitem [{\citenamefont {Crispino}\ \emph {et~al.}(2023)\citenamefont
  {Crispino}, \citenamefont {Chatzieleftheriou}, \citenamefont {Gorni},\ and\
  \citenamefont {{de' Medici}}}]{CRI23}%
  \BibitemOpen
  \bibfield  {author} {\bibinfo {author} {\bibfnamefont {M.}~\bibnamefont
  {Crispino}}, \bibinfo {author} {\bibfnamefont {M.}~\bibnamefont
  {Chatzieleftheriou}}, \bibinfo {author} {\bibfnamefont {T.}~\bibnamefont
  {Gorni}},\ and\ \bibinfo {author} {\bibfnamefont {L.}~\bibnamefont {{de'
  Medici}}},\ }\href {https://doi.org/10.1103/PhysRevB.107.155149} {\bibfield
  {journal} {\bibinfo  {journal} {Phys. Rev. B}\ }\textbf {\bibinfo {volume}
  {107}},\ \bibinfo {pages} {155149} (\bibinfo {year} {2023})}\BibitemShut
  {NoStop}%
\bibitem [{\citenamefont {Rai}\ \emph {et~al.}(2024)\citenamefont {Rai},
  \citenamefont {Crippa}, \citenamefont {C{\u a}lug{\u a}ru}, \citenamefont
  {Hu}, \citenamefont {Paoletti}, \citenamefont {{de' Medici}}, \citenamefont
  {Georges}, \citenamefont {Bernevig}, \citenamefont {Valent{\'i}},
  \citenamefont {Sangiovanni},\ and\ \citenamefont {Wehling}}]{RAI23a}%
  \BibitemOpen
  \bibfield  {author} {\bibinfo {author} {\bibfnamefont {G.}~\bibnamefont
  {Rai}}, \bibinfo {author} {\bibfnamefont {L.}~\bibnamefont {Crippa}},
  \bibinfo {author} {\bibfnamefont {D.}~\bibnamefont {C{\u a}lug{\u a}ru}},
  \bibinfo {author} {\bibfnamefont {H.}~\bibnamefont {Hu}}, \bibinfo {author}
  {\bibfnamefont {F.}~\bibnamefont {Paoletti}}, \bibinfo {author}
  {\bibfnamefont {L.}~\bibnamefont {{de' Medici}}}, \bibinfo {author}
  {\bibfnamefont {A.}~\bibnamefont {Georges}}, \bibinfo {author} {\bibfnamefont
  {B.~A.}\ \bibnamefont {Bernevig}}, \bibinfo {author} {\bibfnamefont
  {R.}~\bibnamefont {Valent{\'i}}}, \bibinfo {author} {\bibfnamefont
  {G.}~\bibnamefont {Sangiovanni}},\ and\ \bibinfo {author} {\bibfnamefont
  {T.}~\bibnamefont {Wehling}},\ }\href
  {https://doi.org/10.1103/PhysRevX.14.031045} {\bibfield  {journal} {\bibinfo
  {journal} {Phys. Rev. X}\ }\textbf {\bibinfo {volume} {14}},\ \bibinfo
  {pages} {031045} (\bibinfo {year} {2024})}\BibitemShut {NoStop}%
\bibitem [{\citenamefont {Flyvbjerg}\ and\ \citenamefont
  {Petersen}(1989)}]{FLY89}%
  \BibitemOpen
  \bibfield  {author} {\bibinfo {author} {\bibfnamefont {H.}~\bibnamefont
  {Flyvbjerg}}\ and\ \bibinfo {author} {\bibfnamefont {H.~G.}\ \bibnamefont
  {Petersen}},\ }\href {https://doi.org/10.1063/1.457480} {\bibfield  {journal}
  {\bibinfo  {journal} {J. Chem. Phys.}\ }\textbf {\bibinfo {volume} {91}},\
  \bibinfo {pages} {461} (\bibinfo {year} {1989})}\BibitemShut {NoStop}%
\end{thebibliography}
\end{document}